\documentclass[preprint2]{aastex}

\usepackage{lineno}
\usepackage{geometry}               
\usepackage{graphicx}
\usepackage{amssymb}
\usepackage{amsmath}
\usepackage{epstopdf}
 \usepackage{hyperref}
\usepackage{lscape}
\usepackage{widetext}
\usepackage{color}
\usepackage{xspace}
\usepackage{placeins}
\usepackage{natbib}
\usepackage{enumerate}
\usepackage[normalem]{ulem}
\usepackage{textcomp}
\usepackage{footmisc}

\bibpunct{(}{)}{;}{a}{}{,} 

\xspace
\newcommand{\Fermi}{\emph{Fermi}\xspace}
\newcommand{\Swift}{\emph{Swift}\xspace}
\newcommand{\xspec}{\texttt{XSPEC}\xspace}

\newcommand{\be}{\begin{equation}}
\newcommand{\ee}{\end{equation}}
\newcommand{\ba}{\begin{eqnarray}}
\newcommand{\ea}{\end{eqnarray}}

\newcommand{\ltsima} {$\; \buildrel < \over \sim \;$}
\newcommand{\gtsima} {$\; \buildrel > \over \sim \;$}
\newcommand{\lta} {\lower.5ex\hbox{\ltsima}}
\newcommand{\gta} {\lower.5ex\hbox{\gtsima}}

\newcommand{\de}{$^\circ$\xspace}
\newcommand{\ded}{$\overset{\circ}{.}$}

\shorttitle{The First LAT GRB Catalog}
\shortauthors{\Fermi-LAT Collaboration}

\begin{document}

\title{The First \Fermi LAT Gamma-Ray Burst Catalog}

\author{
M.~Ackermann\altaffilmark{2}, 
M.~Ajello\altaffilmark{3}, 
K.~Asano\altaffilmark{4}, 
M.~Axelsson\altaffilmark{5,6,7}, 
L.~Baldini\altaffilmark{8}, 
J.~Ballet\altaffilmark{9}, 
G.~Barbiellini\altaffilmark{10,11}, 
D.~Bastieri\altaffilmark{12,13}, 
K.~Bechtol\altaffilmark{14}, 
R.~Bellazzini\altaffilmark{15}
% ...
}
% ...
\altaffiltext{2}{Deutsches Elektronen Synchrotron DESY, D-15738 Zeuthen, Germany}
\altaffiltext{3}{Space Sciences Laboratory, 7 Gauss Way, University of California, Berkeley, CA 94720-7450, USA, , USA}
\altaffiltext{4}{Interactive Research Center of Science, Tokyo Institute of Technology, Meguro City, Tokyo 152-8551, Japan}
\altaffiltext{5}{Department of Astronomy, Stockholm University, SE-106 91 Stockholm, Sweden}
\altaffiltext{6}{The Oskar Klein Centre for Cosmoparticle Physics, AlbaNova, SE-106 91 Stockholm, Sweden}
\altaffiltext{7}{Department of Physics, Royal Institute of Technology (KTH), AlbaNova, SE-106 91 Stockholm, Sweden}
\altaffiltext{8}{Universit\`a  di Pisa and Istituto Nazionale di Fisica Nucleare, Sezione di Pisa I-56127 Pisa, Italy}
\altaffiltext{9}{Laboratoire AIM, CEA-IRFU/CNRS/Universit\'e Paris Diderot, Service d'Astrophysique, CEA Saclay, 91191 Gif sur Yvette, France}
\altaffiltext{10}{Istituto Nazionale di Fisica Nucleare, Sezione di Trieste, I-34127 Trieste, Italy}
\altaffiltext{11}{Dipartimento di Fisica, Universit\`a di Trieste, I-34127 Trieste, Italy}
\altaffiltext{12}{Istituto Nazionale di Fisica Nucleare, Sezione di Padova, I-35131 Padova, Italy}
\altaffiltext{13}{Dipartimento di Fisica e Astronomia "G. Galilei", Universit\`a di Padova, I-35131 Padova, Italy}
\altaffiltext{14}{W. W. Hansen Experimental Physics Laboratory, Kavli Institute for Particle Astrophysics and Cosmology, Department of Physics and SLAC National Accelerator Laboratory, Stanford University, Stanford, CA 94305, USA}
\altaffiltext{15}{Istituto Nazionale di Fisica Nucleare, Sezione di Pisa, I-56127 Pisa, Italy}
% ...
%
%

\begin{abstract}

\date{\today}                                           
In three years of observations since the beginning of nominal science operations in August 2008, the Large Area Telescope (LAT) on board the \Fermi Gamma Ray Space Telescope has observed high-energy ($\gtrsim 20$~MeV) $\gamma$-ray emission from 35 gamma-ray bursts (GRBs).  Among these, 28 GRBs have been detected above 100~MeV and 7 GRBs above $\sim 20$~MeV. The first \Fermi-LAT catalog of GRBs is a compilation of these detections and provides a systematic study of high-energy emission from GRBs for the first time. To generate the catalog, we examined 733 GRBs detected by the Gamma-Ray Burst Monitor (GBM) on \Fermi and processed each of them using the same analysis sequence. Details of the methodology followed by the LAT collaboration for GRB analysis are provided. We summarize the temporal and spectral properties of the LAT-detected GRBs. We also discuss characteristics of LAT-detected emission such as its delayed
onset and longer duration compared to emission detected by the GBM, its power-law temporal decay at late times, and the fact that it is dominated by a power-law spectral component that appears in addition to the usual Band model.

\end{abstract}

\clearpage
\newpage
\thispagestyle{plain}

\onecolumn
\tableofcontents
\twocolumn


\section{Introduction}
\label{sec_introduction}

Prior to the \Fermi Gamma-ray Space Telescope mission, high-energy emission from gamma-ray bursts (GRBs) was observed with
the Energetic Gamma-Ray Experiment Telescope (EGRET) covering the energy range from 30~MeV to~30~GeV \citep{Hughes:80,Kanbach:88,Thompson:93,Esposito:99} on board the Compton Gamma-Ray Observatory ({\it CGRO}; 1991--2000) and, more recently, by the Gamma-Ray Imaging Detector (GRID) onboard the Astro-rivelatore Gamma a Immagini LEggero spacecraft \citep[AGILE;][]{giuliani08,tavani08,Tavani:09}.
Despite the effective area and dead-time limitations of EGRET, substantial emission above 100~MeV was detected for a few GRBs \citep{Sommer:94,Hurley:94,Gonzalez:03}, suggesting a diversity of temporal and spectral properties at high energies.
Of particular interest was GRB\,940217, for which delayed high-energy emission was detected by EGRET up to $\sim$90 minutes after the trigger provided by {\it CGRO}'s Burst And Transient Source Experiment (BATSE).\\

The \Fermi observatory was placed into orbit on 2008 June 11. It provides unprecedented breadth of energy coverage and sensitivity for advancing knowledge of GRB properties at high energies. It has two instruments: the Gamma-ray Burst Monitor \citep[GBM;][]{GBMinstrument} and the Large Area Telescope \citep[LAT;][]{LATinstrument}, which together cover more than 7 decades in energy.
The GBM comprises twelve sodium iodide (NaI) and two bismuth germanate (BGO) detectors sensitive in the 8~keV--1~MeV and 150~keV--40~MeV energy ranges, respectively. 
The NaI detectors are arranged in groups of three at each of the four edges of the spacecraft, and the two BGO detectors are placed 
symmetrically on opposite sides of the spacecraft, resulting in a field of view (FoV) of $\sim$9.5~sr. Triggering and localization are determined from the NaI detectors, while spectroscopy is performed using both the NaI and BGO detectors. Localization is performed using the relative event rates of detectors with different orientations with respect to the source and is typically accurate to a few degrees. The GBM covers roughly four decades in energy and provides a bridge from the low energies (below $\sim$1~MeV), where most of the GRB emission takes place, to the less explored energy range that is accessible to the LAT.

The LAT is a pair production telescope sensitive to $\gamma$ rays in the energy range from $\sim$20~MeV to more than 300~GeV.
The instrument and its on-orbit calibrations are described in detail in~\cite{LATinstrument} and~\cite{LATcalib}.
The telescope consists of a 4$\times$4 array of identical towers, each including a tracker of silicon strip planes with foils of tungsten converter interleaved, followed by a cesium iodide calorimeter with a hodoscopic layout.
This array is covered by a segmented anti-coincidence detector of plastic scintillators which is designed to efficiently identify and reject charged particle background events.
The wide FoV ($\sim$2.4~sr at 1~GeV) of the LAT, its high observing efficiency (obtained by keeping the FoV on the sky with scanning observations), its broad energy range, its large effective area ($>$1 GeV is $\sim$6500 cm$^2$ on axis), its low dead time per event ($\sim$27~$\mu$s), its efficient background rejection, and its good angular resolution ($\sim$0\de.8 at 1~GeV) are vastly improved in comparison with those of EGRET. As a result, the LAT provides more GRB detections, higher statistics per detection, and more accurate localizations ($\lesssim$1\de).


\Fermi has been routinely monitoring the $\gamma$-ray sky since 2008 August.
From this time until 2011 August, when a new event analysis \citep[``Pass 7'',][]{FermiPass7} was introduced, the GBM detected about 730 GRBs, approximately half of which occurred inside the LAT FoV.  In ground processing we search for LAT counterparts to known GRBs, following each trigger provided by the GBM and other instruments.
In addition, we also undertake blind searches for bursts not detected by other instruments on the whole sample of LAT data, with however no independent (i.e. not detected by other instruments) detections so far.


Owing to the detection of temporally extended emission by EGRET from GRB\,940217 and the interest in studying GRB afterglow emission at high energies, \Fermi was designed with the additional capability to repoint in the direction of a bright GRB and keep its position near the center of the FoV of the LAT (where the effective area to $\gamma$ rays is maximal) for several hours (5~hrs initially, 2.5~hrs since 2010 November 23), subject to Earth-limb constraints.
This repointing occurs autonomously in response to requests to the \Fermi spacecraft from either the GBM or the LAT (Autonomous Repoint Request, or ARR hereafter), with adjustable brightness thresholds, and has resulted in more
than 60 extended GRB observations between 2008 October 8, when the capability was enabled, and 2011 August 1.

This article presents the first catalog of LAT-detected GRBs. It covers a three-year period starting at the beginning of routine science operations in 2008 August. In \S~\ref{sec_data_preparation} we describe the data used in this study and the list of GRB triggers that we searched for LAT detections. In \S~\ref{sec_analysis_methods} we give a detailed description of the analysis methods that
we applied to detect and localize GRBs with the LAT, as well as the methodology which we followed
to characterize their temporal and spectral properties.
In \S~\ref{sec_results} and \S~\ref{sec_discussion}, we present and discuss our results, with a special emphasis both on the
most interesting bursts and on the common properties revealed by the LAT.
The physical implications of our observations are addressed in \S~\ref{sec_interpretation}, where we also discuss several open questions and topics of interest for future analysis.
In Appendix~\ref{sec_systematics}, we investigate the possible sources of systematic uncertainties via testing different instrument response functions and configurations for the analysis. Finally, in Appendix~\ref{sec_fermi_lat_grb}, we discuss each individual GRB in the catalog, reporting the details of its observation and considering it in the context of multiwavelength observations.

\section{Data Preparation}\label{sec_data_preparation}
In this section we describe the data analyzed in this study and the list of GRB triggers that we searched for LAT detections.

The results of this paper were produced using two sets of LAT events corresponding to different quality levels and corresponding instrument response functions (IRFs) in the event reconstruction: the Transient event class~\citep{LATinstrument}, which requires the presence of a signal in both the tracker and the calorimeter of the LAT, and the ``LAT Low Energy'' (LLE) event class~\citep{PelassaLLE}, which requires a signal in only the tracker and essentially consists of all the events that pass the onboard $\gamma$ filter having a reconstructed direction~\citep{2012ApJS..203....4A}.

The LAT event classes underwent many stages of refinement and were released as different versions (or ``passes'') of the data. This catalog uses the whole ``Pass 6'' event data set, in particular, the Pass 6 version 3 Transient event class (``P6\_V3\_TRANSIENT''). The LAT team has switched from using ``Pass 6'', which had been used since the beginning of science operations, to ``Pass 7'' data on the 1st of August 2011, the end of the time period covered by this catalog.

As cross checks, we repeated some of the Transient class analyses using instead the ``P6\_V3\_DIFFUSE'' event class to search for possible systematics that might arise from the choice of event selection. 
Both the Transient and Diffuse classes offer good energy and angular resolutions, along with large effective areas above 100~MeV and reasonable residual background rates\footnote{For more information on these event classes see \url{http://www.slac.stanford.edu/exp/glast/groups/canda/archive/pass6v3/lat_Performance.htm} .}. The Diffuse class uses a very selective set of cuts to keep the highest quality $\gamma$-ray candidates. As a result, it has a relatively narrow point-spread function (PSF; 68\% containment radius of several degrees at 100~MeV and $\sim$0\de.25 at 10~GeV) and a smaller background contamination with respect to the Transient class. On the other hand, the Transient class, which is defined with a less selective set of cuts, offers a significantly larger effective area, especially below
1~GeV. The LLE class corresponds to a much-loosened selection, compared to the other two classes, and is designed to provide a far larger effective area at lower energies (especially below 100~MeV) and at larger off-axis angles (especially above $\sim$60$^\circ$). The LLE PSF is wide (with a 68\% containment radius of $\sim$20$^\circ$, $\sim$13$^\circ$ and $\sim$7$^\circ$ at 20~MeV, 50~MeV and 100~MeV respectively) and has a much higher background contamination ($\sim$300 Hz over the whole FoV) than the other two event classes. Since the flux of a GRB is typically a decreasing function of the energy, the LLE class provides very good statistics, which are useful for detailed studies of the temporal structure of GRB emissions. It also allows us to examine GRBs with soft spectra or occurring at a high off-axis angle, which are not detectable with the other two event classes.

Our baseline LAT-only analysis (namely localization, detection, spectral fitting, and duration estimation) uses the Transient class
data. We use the LLE data only for source detection and duration measurement. As mentioned above, the LAT Diffuse data are used only as a
cross-check of some of the analysis results for Transient class.

We perform joint GBM-LAT spectral fitting using the LAT Transient class data, the GBM Time-Tagged Event (TTE) data and the GBM RSP/RSP2 response files\footnote{All available from \Fermi Science Support Center (FSSC)\url{http://fermi.gsfc.nasa.gov/ssc/data/access/gbm/}}. We also use GBM CSPEC data to produce our background model (see \S~\ref{subsubsec_GBM_BKG}).

All our analyses also use the LAT FT2 data, which contain information on the pointing history and the location of the \Fermi spacecraft around the Earth. We use FT2 files with 1~s binning.

\subsection{Data Cuts}
\subsubsection{LAT Data}
\label{subsection_cuts}
We select Transient class with reconstructed energies in the 100~MeV--100~GeV range. The lower limit is chosen to reject events with poorly reconstructed directions and energies. Moreover, for Pass 6, the LAT response is not adequately verified at E$<$100~MeV energies and the contamination from cosmic rays misclassified as gamma rays is also significantly increased. The upper limit was chosen at 100~GeV since we do not expect to detect GRB photons at such high energies due to the opacity of the Universe and the limited effective area of the LAT.
We select events in a circular region of interest (ROI) that is centered on the best available GRB localization.
The LAT PSF depends on the event energy and off-axis angle and has been studied using Monte Carlo simulations.
We use the resulting description of the PSF to increase the sensitivity of our analyses.
For the event-counting and joint spectral-fitting analyses, we select a variable ROI radius that depends on the event energy and the off-axis angle of the GRB in such a way as to select almost all the events compatible with the position of the GRB given our PSF while rejecting much of the residual cosmic-ray background, increasing the signal-to-noise ratio of the selected data. To accomplish this, we split the events in logarithmically-spaced bins in energy and for each bin we select only the events contained in a ROI around the source having a radius corresponding to the 95\% containment radius of the PSF evaluated at an energy equal to the geometric mean of the bin's energy range. For the duration estimation using Transient data we deal with longer time periods; thus we dynamically adjust the radii of the energy-dependent ROIs to follow the variation of the off-axis angle with time. On the other hand, for the LLE duration estimations and the joint GBM-LAT spectral analyses we use a single set of radii
calculated using the PSF corresponding to the GRB off-axis angle at trigger time. The exact dependence of the LLE PSF on the off-axis angle is not available yet. Instead, only two possible LLE PSFs are available for setting the ROI radii: one for observations with off-axis angles greater and the other for observations closer to the center of the FoV. Finally, for cases for which the GRB localization error is not negligible (i.e., for GBM or LAT localizations) we increase the radius of each ROI by setting it equal to the sum in quadrature of the localization error and the 95\% containment radius of the PSF. For GRBs localized by the \Fermi GBM we also added in quadrature a 3\de systematic error.
The maximum-likelihood analysis utilizes the PSF information internally while calculating the probability of each event being associated with the GRB; thus no optimization of the ROI radius, as above, is necessary. For the maximum-likelihood analyses, we use a fixed-radius ROI set at 12\de, a value larger than the 99\% containment radius of the Transient LAT PSF evaluated for a 100~MeV event on axis.

We apply a cut to limit the contamination from $\gamma$ rays produced by interactions of cosmic rays with the Earth's upper atmosphere. For our maximum-likelihood analysis we use the {\em gtmktime} \Fermi Science Tool\footnote{\url{http://www.slac.stanford.edu/exp/glast/wb/prod/pages/sciTools_gtmktime/gtmktime.htm}} to select only the time intervals (the ``Good Time Intervals'' or GTIs) in which no portion of the ROI is too close to the Earth's limb. Because the Earth's limb lies at a zenith angle of 113\de and to take into account the finite angular resolution of the detector, we exclude any events taken when the ROI is closer than 8\de to the Earth's limb or equivalently when it intersects the fiducial line at 105\de from the local Zenith. For special cases, when the position of the GRB is very close to the Earth's limb, we compensate the loss of exposure due to this cut by reducing the size of the ROI and simultaneously increasing the maximum zenith angle to 110\de. This increases the duration of the GTI
significantly, allowing deeper exposures for searches of late $\gamma$-ray activity. For all the other analyses (namely event-counting analyses and joint spectral fitting), we do not apply a cut to select GTIs as above, but rather we process the whole observation and instead reject individual events reconstructed farther than 105\de from the local Zenith.

\subsubsection{GBM Data}
The response of a GBM detector depends on the continuously-varying position of the GRB in its FoV, with its effective area decreasing as the angular distance between the detector boresight and the source ($\theta_{\rm GBM}$) increases. Because of this, when $\theta_{\rm GBM}$ is large, any systematic effects due to imperfect modeling of the spacecraft or the individual detectors become relatively important \citep{Goldstein+12}. For this reason we use the data from the GBM NaI detectors that have angles $\theta_{\rm GBM}<50$\de at the time of the trigger and the BGO detector facing the GRB at the time of the trigger.

We also exclude any detector occulted by other detectors or the spacecraft during any part of the analyzed time interval, as advised in \citet{Goldstein+12}.

Since $\theta_{\rm GBM}$ usually changes with time, the GBM Collaboration released RSP2 files which contain several response matrices corresponding to short consecutive time intervals (every 2\de of slew of the detector about the source).
With a suitable weighting scheme, as described in \S~\ref{subsub_dataPreparation_sp3}, these files provide an adequate description of the GRB detector responses.

Finally, in some cases, bright GRBs trigger an ARR, causing rapid variations of $\theta_{\rm GBM}$ with time for some of the GBM detectors. 
These variations create further variations in those detector responses and background rates.
In fact, due to its orbital and angular dependence the background of those detectors can be very hard to predict. Also, the RSP2 files might not be binned finely-enough in time to cover these rapid variations, we excluded data from detectors that have such rapid variations.

\subsection{Input GRB List}\label{subsec_input_GRB_list}
To search for GRBs in the LAT data we use as input a list comprising 733 bursts that triggered the GBM from 2008 August 4 to the 2011 August 1 (GBM triggers bn080804456 to bn110731465). We use the localizations provided by the GBM, unless a localization from the Swift observatory \citep{2004ApJ...611.1005G}, obtained either from the Burst Alert Telescope \citep[Swift-BAT,][]{2005SSRv..120..143B}, the X-Ray Telescope \citep[Swift-XRT,][]{2005SSRv..120..165B}, or the UV-Optical Telescope \citep[Swift-UVOT,][]{2004ApJ...611.1005G},  is available via the Gamma-Ray Burst Coordinates Network (GCN)\footnote{\url{http://gcn.gsfc.nasa.gov/}}.

We analyzed all GRBs in the input list whether or not they occurred in the LAT FoV at the time of the trigger, since a GRB that is initially outside the LAT FoV can be observable at later times due to an ARR or simply due to the standard scanning mode.  As a reference, 368 GBM bursts were in the LAT FoV at the time of the GBM trigger, with the FoV considered to have a 70\de angular radius. 
In 64 of these cases, an ARR was performed. It should be noted that the sensitivity of the LLE event class extends to larger off-axis angles $\theta\approx$90\de.

In order to characterize our detection algorithm, we also created a list of ``fake'' GBM triggers, by considering trigger times earlier than the true GBM trigger time by 11466\,s (approximately two orbits).
Since the most common observing mode for the \Fermi spacecraft is to rock between the northern and southern orbital hemispheres on alternate orbits, with the exception of ARRs, the burst triggers of the ``fake'' sample has the desirable property of having very similar background conditions as those of the true sample.

\section{Analysis Methods and Procedure}
\label{sec_analysis_methods}

We implemented a standard sequence of analysis steps for uniformity.
 The sequence consists of event-counting analyses performed on the Transient class and LLE data for source detection and duration estimation (\S~\ref{subsec_counting_analysis}), unbinned maximum likelihood analysis performed on the Transient class data for source detection, spectral fitting, localization (\S~\ref{subsec_likelihood_analysis}), and a spectral-fitting analysis performed jointly on the LAT Transient class and the GBM data (\S~\ref{subsec_lat_gbm_spectral_analysis}). Details of the implementation of the analysis sequence are given in \S~\ref{subsec_analysis_sequence}.  Estimation of the backgrounds is a central part of all the analyses and is described below.

\subsection{Background Estimation}\label{subsec_backgrounds}
\subsubsection{LAT}\label{subsubsec_LAT_bkg}
The background in the LAT data is composed of charged cosmic rays (CRs) misclassified as $\gamma$ rays, astrophysical-source $\gamma$ rays coming from Galactic and extragalactic diffuse and point sources, and $\gamma$ rays from the Earth's limb produced by interactions of CRs in the upper atmosphere. The backgrounds for the Transient class and LLE data are dominated by the CR component, while for the cleaner Diffuse class the backgrounds are dominated by astrophysical $\gamma$ rays. The CR component of the background depends primarily on the geomagnetic coordinates of the spacecraft and on the direction of the GRB in instrument coordinates (since the LAT's effective area varies strongly with the inclination angle). The component from the Earth's atmosphere depends on the angle between the GRB and the limb (i.e., on the zenith angle of the GRB) and is strongest toward the limb. Finally, the astrophysical
background $\gamma$-ray component depends on the GRB direction and is typically stronger at low Galactic latitudes.

For the Transient event class analyses, we use the Background Estimation tool (``BKGE'' hereafter), which was developed by the LAT collaboration and which takes into account all these dependencies. It can estimate the total expected backgrounds for any given ROI and period of time with an accuracy of $\sim$10-15\%~\citep{GRB080825C_LATpaper}. It also provides separate estimates for the Galactic diffuse emission and for everything else, namely the sum of CRs and extragalactic diffuse emission (``isotropic component''). Note that the BKGE cannot estimate the backgrounds from the Earth's limb. However, the zenith-angle cut described in \S\ref{sec_data_preparation} is very effective at reducing this component to negligible levels; thus this limitation does not generally constitute an obstacle.

Our maximum likelihood analysis of Transient class data uses a background model calculated by a combination of the isotropic component provided by the BKGE tool and the Galactic diffuse emission template provided by the LAT Collaboration\footnote{\url{http://fermi.gsfc.nasa.gov/ssc/data/access/lat/BackgroundModels.html}}.

The maximum likelihood analysis using the cleaner Diffuse class data, which were performed for validation studies (see Appendix \ref{sec_systematics}), uses the Galactic diffuse emission template plus the public template describing the ``isotropic background'' (extragalactic diffuse emission and CR background) as a single spectrum of the intensity averaged over the whole sky. The BKGE does not produce estimates for Diffuse class events.
For the time scales analyzed in this study, the contribution from point sources is typically negligible, so we do not take them into account in the background models.

\begin{figure}[!ht]
\begin{center}
\includegraphics[width=1.0\columnwidth]{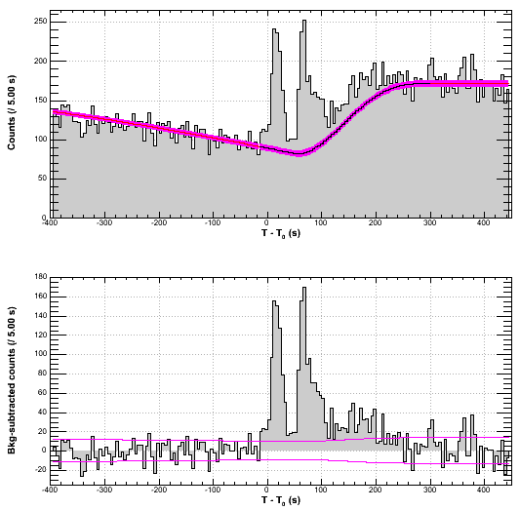}
\caption{LLE background estimation for GRB\,090323. The top panel shows the time history of the LLE count rate (histogram) and the background level estimated
  from a fit to the two off-pulse regions [-400 s, -15 s] and [300 s, 450 s] (curve). The bottom panel shows the background-subtracted LLE light
  curve. Magenta curves indicate the statistical error of the fitted background (top panel), and the statistical fluctuation of the
  background-subtracted signal in the null hypothesis (bottom panel).
}
\label{fig_LLE_BKG_EXAMPLE}
\end{center}
\end{figure}

For the joint GBM-LAT spectral analysis, we used as background for the LAT the estimates provided by BKGE of the total background in the energy-dependent ROI.
For technical reasons related to the broad PSF of the LLE class, we cannot use the BKGE to estimate the LLE background.
Instead, we evaluate it directly from the LLE data associated with each individual observation.
First, in order to ensure enough events in every time bin, we bin the LLE data in time with a coarse binning of 5~s, from well before the trigger time to well after the end of the burst as measured by the GBM.
We then fit the background rate as a function of time $b(t)$ by taking into account the variation of the exposure due to the changing orientation of the LAT.
Phenomenologically, we adopt the function $b(t)=p_0 + p_1~C(t)+ p_2~C(t)^2$, where $C(t)=\cos[\theta(t)]$ and $\theta$ is the off-axis angle.
The parameters $p_0$, $p_1$, and $p_2$ are obtained by fitting the ``pre-burst'' and ``post-burst'' time windows simultaneously.
We use a conservative definition of these time windows based on the burst duration as measured by the GBM. In particular, the ``post-burst'' data
start well after the end of the low-energy emission as seen by the GBM.
Finally, the fit parameters allow us to compute the background rate at any time during the burst, and we use the covariance matrix from the fit to evaluate the uncertainty of this prediction.
We compared this simple model to an alternative prescription $b(t)=pol(t)*C(t)$, where the degree of the polynomial function $pol(t)$ is increased until a good fit to the data is obtained. Typically a polynomial of degree 1 or 2 was sufficient, although in few cases a higher degree (3 or 4) was necessary.
The expression above is motivated by the fact that, as a first approximation, the effective area of the LAT to CRs scales as $\cos\left(\theta\right)$ and that we can model the CR contribution on-axis ($\theta = 0$) with a polynomial. 
The two prescriptions gave very similar results in all cases. An example of the standard prescription is shown in Fig.~\ref{fig_LLE_BKG_EXAMPLE}.
%

\subsubsection{GBM}\label{subsubsec_GBM_BKG}
\begin{figure*}[t!]
\begin{center}
\includegraphics[width=1.0\textwidth]{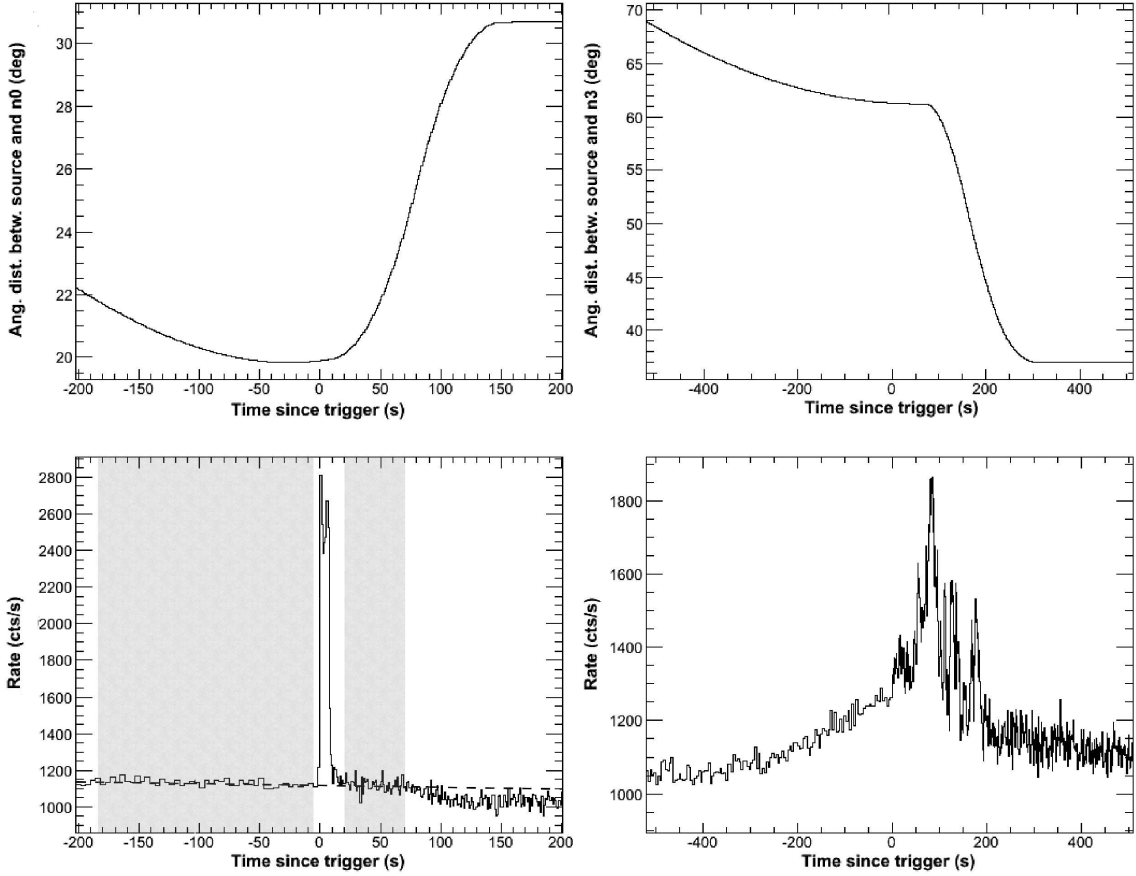}
\caption{{\it(left)} An example of a selected GBM detector (NaI$_{0}$) and its background fit (lower panel), and the angular distance between the axis of the detector and the GRB position (upper panel). The shaded regions mark the selected off-pulse intervals, while the dashed line is the best-fitting polynomial model (see text). {\it(right)} An example of an excluded detector (NaI$_{3}$): the change in angular distance between the detector axis and the source is too fast (upper panel), producing a change in the light curve which cannot be modeled satisfactorily with a polynomial model.}
\label{fig_backgroundARR}
\end{center}
\end{figure*}

We use the GBM CSPEC event data from before and after the GRB prompt phase to obtain a model for the background, similar to the procedure followed for the LLE data above. For each selected detector, we integrate the CSPEC spectra over all the energy channels
to obtain a light curve, and then select two off-pulse time intervals: one before and one after the GRB prompt emission (see left panel in Fig.~\ref{fig_backgroundARR}). We fit polynomial functions $f(t)$ of increasing degree $D$ to the data from these two time intervals, minimizing the $\chi^{2}$ statistic, until we reach a good fit (i.e., with a reduced $\chi^{2} \simeq 1$). Then, we consider the light curves corresponding to each of the 128 channels separately, again with data from the off-pulse intervals, and we fit them with a polynomial of degree $D$ by minimizing the Poisson log-likelihood function\footnote{Using \url{http://root.cern.ch/root/html/TH1.html\#TH1:Fit}}. After each fit, we check by eye that the residuals are
compatible with statistical fluctuations. If this is not the case, we repeat the procedure from the beginning, changing our choice for the off-pulse intervals, until a good fit has been achieved. The set of 128 polynomial functions constitutes our \textit{background model}, and the predicted number of background events $b_{i}$ in the i-th channel of the background spectrum is the integral of the corresponding polynomial function $f_{i}$ (describing the rate) between $t_{1}$ and $t_{2}$:
\[
b_{i} = \frac{\int_{t_{1}}^{t_{2}}f_{i}(t)dt}{t_{2}-t_{1}}.
\]
The statistical error of the integral is computed using the covariance matrix from the fit\footnote{Using \url{http://root.cern.ch/root/html/TF1.html\#TF1:IntegralError}}.
Since the background for GBM detectors is much less predictable than for LLE data, we determine the off-pulse regions manually. In order to minimize the statistical and systematic errors (hence ensure a reliable background estimate), the off-pulse time intervals must be close to the GRB's signal, have a long-enough duration, and also possibly have a smooth part of the light curve without bumps or other structures. Moreover, the number of counts in each channel is much smaller than the total number of counts used to determine $D$. Thus, the larger value of $D$ is, the more $f_{i}$ can pick up statistical fluctuations in some channels, giving a slightly wrong interpolation for those channels in the pulse region. Thus, we try to find off-pulse intervals well described by low-order polynomials (ideally $D=1$). Unfortunately, this is not always possible. For example, for GRBs triggering ARRs, the background can vary quickly in response to the change of pointing, requiring higher-order polynomials to describe it. This
effect
introduces some additional noise in the spectrum, but it is unlikely to introduce any bias in the fit results, given its random nature. Note that it is not possible to fix the shape of the polynomial, since the background shows spectral evolution and thus every channel needs to be considered independently. In some cases, even with high-order polynomials, fitting the model to the background can be difficult and even impossible without being completely arbitrary (see right panel in Fig.~\ref{fig_backgroundARR} for an example). In those cases we opt for excluding the problematic detector from the analysis. These issues are not solvable at present given our current understanding of the detectors and their backgrounds. More advanced techniques to deal with the backgrounds are currently under investigation by the \Fermi-GBM Collaboration \citep{2012SPIE.8443E..3BF}.

\subsection{Maximum Likelihood Analysis}
\label{subsec_likelihood_analysis}

We perform an unbinned maximum likelihood analysis using the tools in
the \Fermi ScienceTools software package, version
09-26-02\footnote{\url{http://fermi.gsfc.nasa.gov/ssc/data/analysis/scitools/ref\_likelihood.html}}.
An overview of the method and its application for this study is given
below. For more information see \cite{2009ApJ...701.1673B} and
references therein.

The unbinned analysis computes the log-likelihood of the data using
the reconstructed direction and energy of each individual gamma-ray
and the assumed sky model folded through the instrument response
functions of the LAT.  The sky model includes the GRB under
investigation modeled as a point source, typically with a power-law
spectrum, as well as other components that describe the other sources
that are expected to be present in the data.  For the short time
scales ($\la 10$--100s) considered these are predominantly diffuse
emission from the Galaxy and residual charged particle backgrounds,
though in principle, a bright, nearby point source, such as Vela may
be included.  To estimate the spectral properties of the GRB, the
model parameters are varied in order to maximize the log-likelihood
given the data.  Usually, the GRB coordinates are held fixed, but if
a localization using the LAT data is desired, those parameters can
also be varied.

The fitting in the \textit{Likelihood} tools is performed using an
underlying engine such as
MINUIT\footnote{\url{http://lcgapp.cern.ch/project/cls/work-packages/mathlibs/minuit/doc/doc.html}} 
to perform the maximization.
Currently, the unbinned analysis does not take into account energy
dispersion.  However, given the good energy resolution of the LAT
($\lesssim$15\% above 100~MeV), the moderate energy dependence of the
LAT effective area at the energies considered, and the simple
power-law spectral form that we consider, approximating the true
energy by the reconstructed one is justified.  The uncertainties of
the best-fit values of the parameters or any upper/lower limits are
estimated from the shape of the log-likelihood surface around the best-fit.

We apply the likelihood analysis to Transient class events, and as
cross check, we also analyze Diffuse class events, with the data cuts
described in \S~\ref{sec_data_preparation}.  We cannot apply a similar
unbinned maximum likelihood analysis to the LLE data, since the PSF,
energy dispersion, effective area for the LLE events and the expected
backgrounds are not adequately known and/or verified yet. The analysis
of LLE data is similar to that of the GBM data and is described below.

The background model is constructed as described in
\S~\ref{subsec_backgrounds}. The normalization of the ``isotropic
background'' provided by the BKGE, used for the analysis of Transient
class events, is one of the free parameters of the fit and has a
Gaussian prior of mean 1 and a width set to encompass any associated
statistical and systematic errors (typically around 15\%). The
normalization of the ``isotropic background'' template, used for the
analysis of Diffuse class data, is free to vary with no prior and no
constraints. To avoid increasing the number of free parameters, we
keep the normalization of the template for Galactic diffuse emission
fixed to 1 for the analyses based on both event classes.

\subsubsection{Source Detection}
\label{subsubsect_detection_like}
To determine the significance of the detections of sources using the maximum likelihood analysis, we consider the ``Test Statistic'' (TS) equal to twice the logarithm of the ratio of the maximum likelihood value produced with a model including the GRB over the maximum likelihood value of the null hypothesis, i.e., a model that does not include the GRB.  The probability distribution function (PDF) of the TS under the null model is given by the probability that a measured signal is compatible with statistical fluctuations.  The PDF in such a source-over-background model cannot, in general, be described by the usual asymptotic distributions expected from Wilks' theorem~\citep{Wilks:38,Protassov2002}.  However, it has been verified by dedicated Monte Carlo simulations~\citep{Mattox:96} that the cumulative PDF of the TS in the null hypothesis (i.e., integral of the TS PDF from some TS value to infinity) is approximately equal to a $\chi^2_{n_{dof}}/2$ distribution, where $n_{dof}$ is the number of degrees of freedom associated with the GRB.  The factor of \textonehalf in front of the TS PDF formula results from allowing only positive source fluxes.


Since we model the GRB spectrum as a power law with two degrees of freedom and we fix the localization, the TS distribution should nominally follow $(1/2)\chi^2_{2}$. This is formally correct if the localization of the GRB is provided by an independent data set (i.e., from another instrument). However, when the input localization is not sufficiently precise, we optimize it using the same data set used for detecting the source, thereby introducing two additional free parameters (R.A. and Dec.). In this case, the TS distribution should follow $(1/2)\chi^2_{4}$. In practice, the steps of detection and localization are iterated many times, and a detection step is performed using an ROI centered on the position found by a prior localization step. Therefore, the data sets used in each step are not exactly overlapping. For this reason, we expect some deviation from $(1/2)\chi^2_4$ distribution. For simplicity, we set a unique threshold of TS$_{min}$=20 for our analysis independent of the
origin of the localization. This formally corresponds to two slightly different one-sided Gaussian equivalent thresholds, 4.1$\sigma$ for $\chi^2_{2}$ and 3.5$\sigma$ for $\chi^2_{4}$. Additionally, we check the calibration of the detection algorithm on a sample of ``fake GBM triggers'' generated as described in \S~\ref{subsec_input_GRB_list}.  With the aforementioned value of TS$_{min}$ we obtain zero false detections on the ``fake GBM triggers'' sample (see \S\ref{subsec_sec_detections} for more details).

\subsubsection{Localization}
\label{subsubsec_likelihood_localization}

We compute the localizations with the LAT in two steps.
The first step provides a coarse estimation of the GRB position and is performed
using the {\em gtfindsrc} \Fermi ScienceTool. At this stage, we look
for an excess consistent with the LAT PSF, and we do not assume a particular
background model.
Although this method is quick and robust, it assumes that the likelihood function is parabolic and
symmetric in azimuth around the found position, and so the provided localization error can be
slightly underestimated. Therefore, this step is only used to obtain an initial seed for the follow-up analysis.

For a more accurate localization we use the \textit{gttsmap} \Fermi ScienceTool,
which starts from the best-fit background model obtained by the likelihood fit
and builds a map of the TS in a grid around the best available localization of the source. The GRB spectral parameters are fitted at each position in the grid, along with all free
parameters of the background model.
The grid size and spacing are set based on the localization error obtained in the first step.
The final LAT localization corresponds to the position of the maximum of the TS map.
Its statistical uncertainty is derived by examining the distribution of TS values
around it. Following~\citet{Mattox:96}, we interpret changes in the TS values in terms of a $\chi^2$ distribution with two degrees of freedom to account for the flux and spectral index of the GRB.
Specifically, a confidence-level (CL) uncertainty is
given by the TS map contour that corresponds to a decrease from the maximum
value by a value equal to the CL quantile of the
$\chi^2_2$ distribution. For example, the 90\% (68\%) confidence level corresponds to a
decrease of the TS from its maximum value by 4.61 (2.32).

\subsubsection{Event probability}
\label{susubbsec_event_probability}

We estimate the probability of each $\gamma$-ray being associated with the GRB by using
the {\em gtsrcprob} \Fermi ScienceTool. The probability computation takes into
account the spectral, spatial (extent), and temporal (flux) information of all
the components in the source model, and the response of the LAT (PSF and effective area) to the particular event.
The probabilities are assigned via the likelihood analysis and are computed
starting from the best-fit model.
In particular, the probability that a photon is produced by a component $i$ is proportional to $M_{i}$ given by
\begin{equation}
\label{gtsrcprob}
M_{i}(\epsilon',p', t)=\int d\epsilon dp\,S_{i}(\epsilon,p,t)\,R(\epsilon,p;\epsilon',p',t),
\end{equation}
where $S_{i}(\epsilon,p,t)$ is the predicted counts density from the component at energy $\epsilon$ and position $p$, and (observed) time $t$, and
and the integral is the convolution over the instrument response $R(\epsilon,p;\epsilon',p',t)$.
In general, the predicted count density is the sum of the different contributions  $S_{i}(\epsilon,p,t)$, including the extended backgrounds (such as the isotropic component and the Galactic diffuse emission), background point sources (nearby bright sources) and the GRB under study. Each contribution is described by a model, the parameters of which are optimized during the maximum likelihood fit. We simplify the calculation by not including nearby bright sources, as, in these short time scales, they do not contribute significantly to the total number of counts.
Once we compute the maximum likelihood model for the observed number of counts,
we assign to each event the probability of being associated to a particular component $i$.

Because the flux varies with time, we perform the calculation in several time bins so that the flux is never averaged over long time intervals. We tested schemes for defining the time intervals
including linear, logarithmic, and Bayesian-blocks \citep{2012arXiv1207.5578S} binnings, and the results were stable among
the different choices.
For consistency with the other parts of the analysis we chose the same logarithmically-spaced time bins used in the time-resolved spectral analysis described in \S~\ref{subsec_analysis_sequence} below.

\subsection{Event Counting Analyses}
\label{subsec_counting_analysis}

As mentioned in the previous section, the effective area of the Transient class decreases strongly for
off-axis angles greater than $\sim$70\de or for energies less than $\sim$100~MeV.  For this reason,
in addition to the maximum likelihood analysis applied to Transient class data described above, we search for sources using the LLE class. This class provides a significantly larger effective area below 100~MeV and a wider acceptance, although with a higher background level. We use it to obtain another duration measurement as well, which is dominated by events below 100~MeV and is thus complementary to the duration measurement obtained with Transient class data.

\subsubsection{Source Detection using LLE data}
\label{subsubsect_detection_LLE}

Consider a GRB as an impulse $f(t)$ superimposed on a
background signal $b(t)$. Depending on the
unknown shape of $f(t)$, there will be a particular time scale $\delta t$ and a
particular start time $t_0$ maximizing the quantity
\begin{equation}
S = \frac{\int^{t_{0}+\delta t}_{t_{0}}f(t)dt}{\sqrt{\int^{t_{0}+\delta
t}_{t_{0}}b(t)dt}},
\label{eq:lle}
\end{equation}
which is the significance of the signal in the Gaussian regime.
The pair $\delta t,t_{0}$ corresponds to the highest sensitivity to
the signal of this particular GRB. Our source-detection method searches for the closest pair to $\delta t,t_{0}$
by resizing and shifting the time bins and selecting the light curve that contains the single
bin with the highest significance. Since the typical event rate inside the LLE ROI
is not particularly large ($\sim$10-20~Hz for the background), the Gaussian approximation
implicit in Eq. \ref{eq:lle}
is not always justified. The significance $S$ in each bin is thus derived from the Poisson probability
of obtaining the observed number of counts given the expectation from the background, by
converting this probability to an equivalent sigma level for a one-sided Standard Normal distribution.
Our algorithm starts by defining a conservative window around the trigger
time, with a total duration depending on the GBM burst duration T$_{90}$. Then, a set
of 10 bin sizes $\delta t$ is defined depending on the T$_{90}$.
For each of these bin sizes, the algorithm computes 11 light curves with shifted bins, i.e.,
with the bins centered on $t_{0}+(i/20)\;\delta t$ ($i=0...10$).
For each of these 10$\times$11 light curves, the background function $b(t)$ is fitted to the
data outside the GRB window (as described in \S~\ref{subsec_backgrounds}), and the algorithm seeks
the bin with the largest significance $S$ inside the GRB window.
This value is then corrected for the number of trials, i.e., by the number of bins $N$ in the current light curve.
If $p$ is the probability corresponding to $S$, then the corrected-for-trials probability is
$p^{\prime} = 1-(1-p)^{N}$.
This new probability is converted to a Gaussian-equivalent significance $S^{\prime}$, and
the pre-trials significance for the detection of the GRB is defined as
$S_{pre}=\max(S^{\prime})$, where the maximum is computed over the 110 light curves.
Since the data have been rebinned multiple times, a post-trial probability is finally computed to account for these not-independent trials.
For this purpose, we performed $3\times10^6$ Monte Carlo simulations of the background, running
our algorithm and recording $S_{pre}$ for each realization. The resulting distribution of
$S_{pre}$ is well described by a Lorentzian function $1 + [ (x-x_{0}) /
r_{c} ]^2 ]^{-\beta}$, with $x_{0}=1.36$, $r_{c}=7.38$ and $\beta = 41.8$
($\chi^{2}=43.2$ with 38 d.o.f). We use this function to convert the
pre-trials significance $S_{pre}$ into a post-trials significance $S_{post}$.

We consider as LLE-detected the GRBs that have post-trial significances $S_{post}>4\sigma$, which correspond to chance probabilities $P < 3\times10^{-5}$.
We ran our algorithm on the 733 GRBs of the GBM sample (see \S\ref{subsec_input_GRB_list}), and so we expect no false-positive detections
using this arbitrary threshold.

\subsubsection{Duration measurement}
\label{subsubsect_duration}
We describe the duration of a GRB detected by the LAT using the
parameter T$_{90}$ \citep{Kouv93}. A simple measurement of T$_{90}$ starts with the construction of
the integral distribution of the number of background-subtracted events
accumulated since the trigger time. As the
GRB becomes progressively fainter, the distribution flattens and eventually
reaches a plateau.

The calculation of the duration of the emission consists of
finding the times where the integral distribution reaches the 5\% and
95\% levels of its total height (called T$_{05}$ and T$_{95}$ respectively), and
calculating their difference T$_{90}\equiv \rm{T}_{95}-\rm{T}_{05}$. Our duration estimation
method is based on the above simple prescription, but is also extended to
estimate the statistical uncertainty of the results, and accounts for the effects of
effective area variations over time (for its application to the Transient class events).

Because of the unavoidable statistical fluctuations involved in the process of detecting
an incoming GRB flux, a GRB observed under identical conditions by a number of
identical detectors will in general produce different detected light curves, and hence
different duration estimates.
Our method quantifies the uncertainties on the duration estimates associated to these
statistical fluctuations. In short, it accomplishes this by treating the
detected light curve as the true one (i.e., that of the incoming $\gamma$-ray flux),
producing a set of simulated light curves by applying Poissonian fluctuations on
the detected one, estimating the durations of the simulated light curves, and
calculating a single duration estimate and its uncertainty from the distribution
of simulated duration estimates.

Our method starts by constructing the integral distribution of the accumulated
background-subtracted events curve in small steps in time.
For each step, the number of expected background events is estimated and the number of detected events is counted.
At the end of each step, an algorithm checks for the presence of a plateau by searching for statistically significant increases in the average value of the points added last to the curve. If a certain number of steps does not increase the integral distribution, a plateau is reached and the construction stops. A set of simulated light curves are then produced, by adding Poisson noise to the observed light curve, and the corresponding integral distributions are produced. A duration estimation is made for each of the simulated light curves and the results (T$_{05}$, T$_{95}$, T$_{90}$) are recorded. After the durations of all the simulated light curves have been measured, the median and a (minimum-width) 68\% containment interval are calculated for each distribution, and used as our measurements and $\pm$1$\sigma$ errors. In case the light curve contains multiple peaks separated by quiescent periods, the algorithm, depending on the intensity of each peak and the duration of the intermittent quiescent periods, might set the beginning of the plateau at the end of the last peak or during a quiescent period. In the latter case, some of the late emission might not be fully accounted for by the produced duration. However, the returned statistical errors would be appropriately increased in both cases, indicating the uncertainty of identifying the end of the emission.

Any changes in the off-axis angle of the GRB during an observation will change the effective area of the LAT, affecting the  light curve. For example, a GRB
observation that involves an ARR will in general start with a moderate to large off-axis angle which will then rapidly decrease and stay small for most of the
rest of the observation. Because the effective area of this observation will be small before the ARR starts, the count rate will be artificially decreased and this would cause a bias in the measurement of the T$_{05}$ if it were simply based on counts.
To account for this effect, we weight the simulated light curves by the inverse of the exposure.

\begin{figure}[!ht]
\begin{center}
\includegraphics[width=0.85\columnwidth]{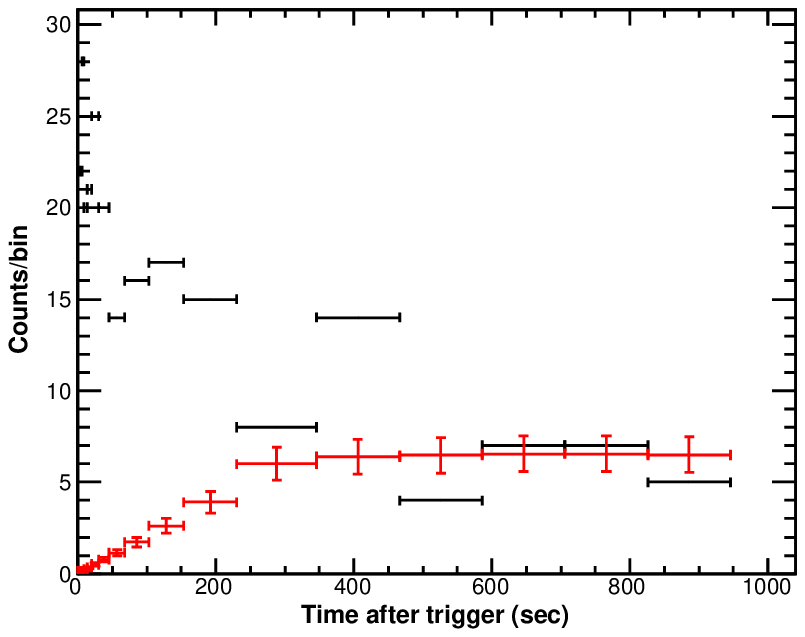}
\includegraphics[width=0.85\columnwidth]{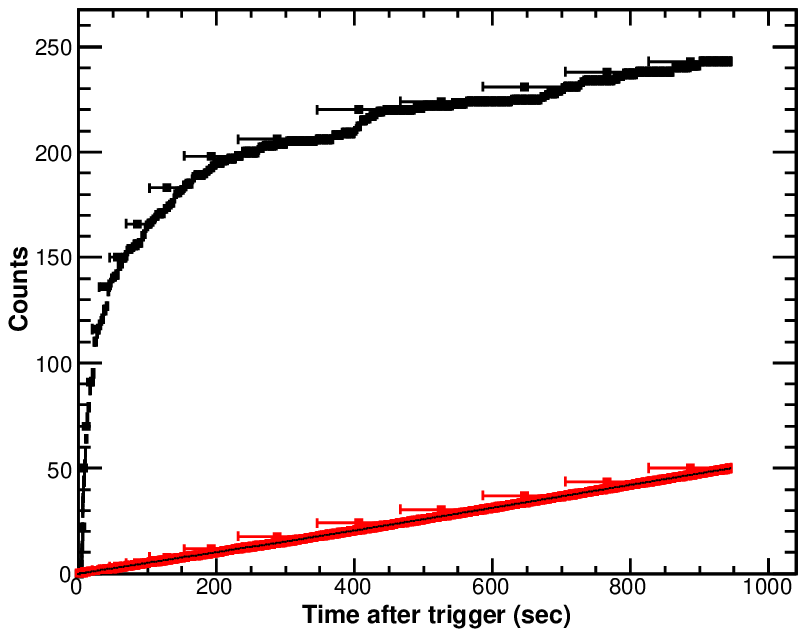}
\includegraphics[width=0.85\columnwidth]{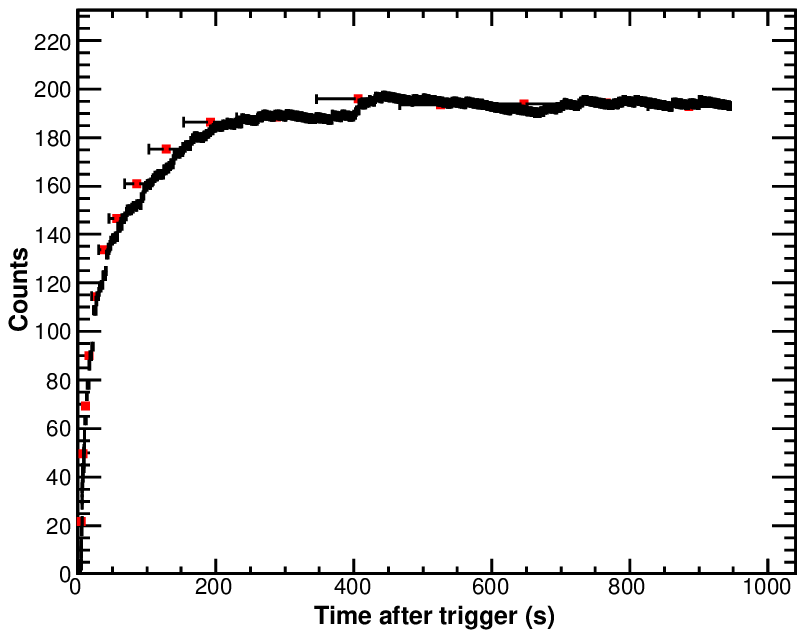}
\caption{Duration estimation of GRB\,080916C using Transient-class data. Top: number of detected counts (black) and estimated background (red) per time interval. Middle: accumulated number of detected counts (black) and expected background (red) since the trigger time. Bottom: accumulated number of events, background subtracted.}
\label{fig_080916c_dure_example}
\end{center}
\end{figure}

\begin{figure}[!ht]
\begin{center}
\includegraphics[width=0.85\columnwidth]{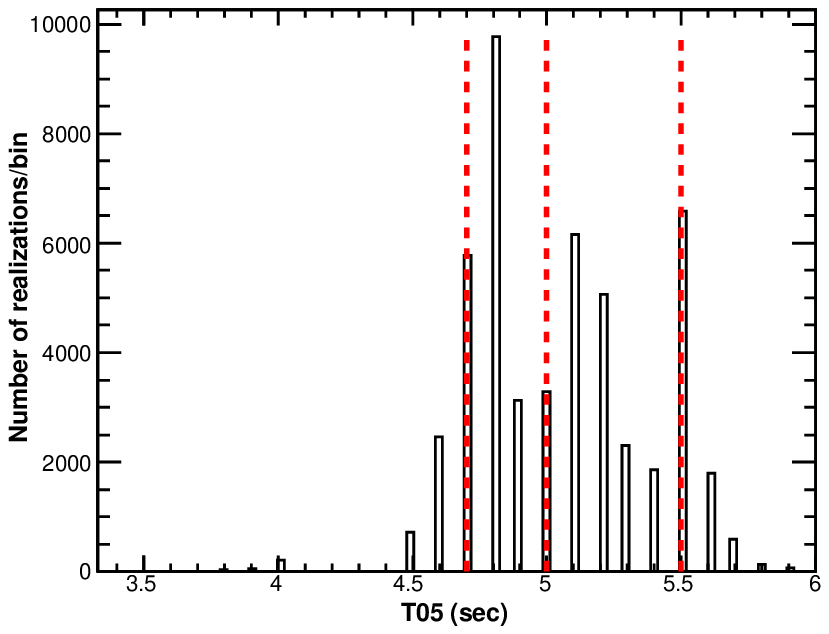}
\includegraphics[width=0.85\columnwidth]{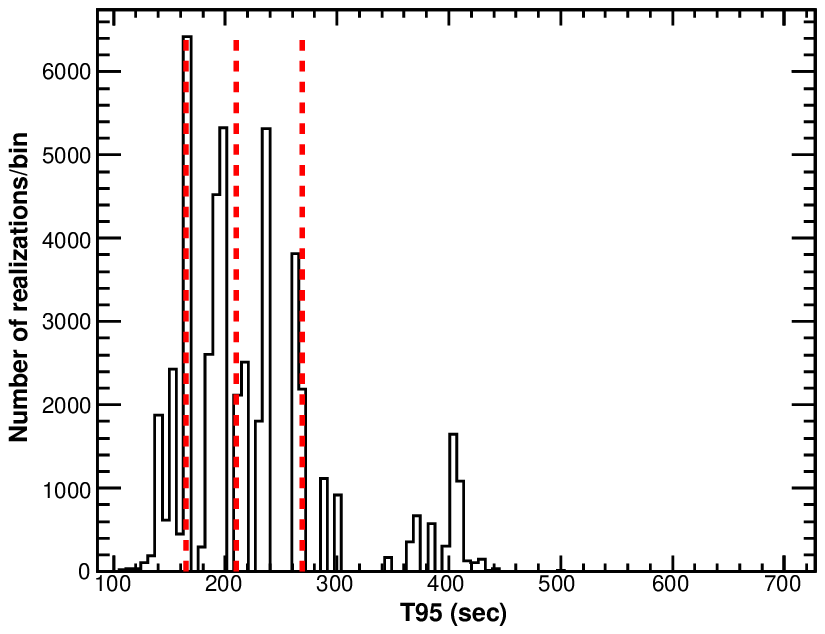}
\includegraphics[width=0.85\columnwidth]{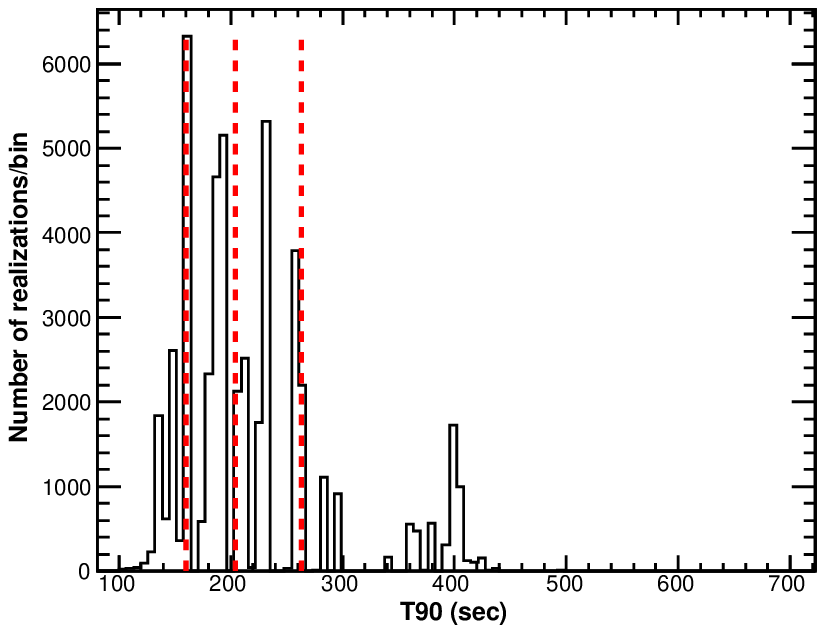}
\caption{Duration estimation of GRB\,080916C using Transient-class data. Curves: distributions of T$_{05}$ (top), T$_{95}$ (middle), and T$_{90}$ (bottom) as measured from the simulations. Middle vertical dashed lines: median of the distributions, constituting our best estimate of the duration. Left- and right-hand vertical dashed lines: 68\% containment intervals, constituting our estimated error for the duration.}
\label{fig_080916c_dure_example2}
\end{center}
\end{figure}

To illustrate this method we present in Fig.~\ref{fig_080916c_dure_example} the case of GRB\,080916C, and the duration measurement using the Transient class data.
 These curves are used as the basis for the simulations.
 Fig.~\ref{fig_080916c_dure_example2} shows the distribution of T$_{05}$, T$_{95}$, and T$_{90}$, as measured from the simulations.
These distributions are used to define the duration and  associated error. In this particular case  some excess events were observed at late times (about $\sim$400~s), as can be seen in Fig.~\ref{fig_080916c_dure_example}. Consequently, a small fraction of the simulated light curves gave T$_{90}$ and T$_{95}$ that were very
close to $\sim$400~s, which caused a small increase of the duration estimates and of the errors for positive fluctuations.

In some cases, a GRB observation can be interrupted before the GRB emission becomes too weak to be detectable (i.e., before reaching a plateau in the integral distribution).

Such interruptions can happen if the GRB exits the FoV of the LAT, it becomes occulted by the Earth, or the LAT enters the South Atlantic Anomaly (SAA), suspending observations. In these cases, only a lower limit on the duration can be obtained (with no errors), equal to the time interval between $\rm{T}_{05}$ and the interruption of the observation.

We apply this method to both Transient class data and LLE data. In the former case we use the BKGE to estimate the background, while for the latter case we use the polynomial fit, as described in \S~\ref{subsubsec_LAT_bkg}. Note however that in the calculation of the duration the exposure weighting is performed only for Transient class data, since the effective area for the LLE class has not been characterized yet.

As a cross check, we also apply a different algorithm on LLE data. We consider the light curve with the binning that gives the highest significance, as obtained by the algorithm explained in \S~\ref{subsubsect_detection_LLE}, and we measure $\rm{T}_{05}$, $\rm{T}_{95}$ and $\rm{T}_{90}$ on the integral distribution obtained from that light curve. We verified that the numbers obtained with this simple method are always within the errors obtained with the other method. Thus, we will only provide the set of results related to the first algorithm.

%

\subsection{Joint LAT-GBM Spectral Analysis}
\label{subsec_lat_gbm_spectral_analysis}
We performed joint GBM-LAT spectral fits for every GRB detected by the LAT.

\subsubsection{Data preparation}
\label{subsub_dataPreparation_sp3}
We start by selecting the GBM detectors as described in \S~\ref{sec_data_preparation} and estimate the expected backgrounds as described in \S~\ref{subsubsec_GBM_BKG}. We then use the \Fermi Science Tool \textit{gtbin} to extract the observed spectrum (source + background) from the GBM TTE event data. We obtain the response of a GBM detector in the interval to be analyzed ($t_{1}$--$t_{2}$) using the RSP2 file for the detector for the time interval. Because the RSP2 file contains several response matrices corresponding to consecutive time intervals that in general are shorter than $t_{1}$--$t_{2}$, we sum the matrices of all the sub-intervals included in $t_{1}$--$t_{2}$ using an appropriate weighting scheme. Specifically, if $c_{i}$ is the counts detected in the sub-interval covered by the i-th matrix, and $C=\sum_{j}c_{j}$ is the number of counts detected between $t_{1}$ and $t_{2}$, then the weight for the i-th response matrix is:
$$
w_{i}=\frac{c_{i}}{\sum_{j}c_{j}}.
$$
To sum the matrices we use the tool \textit{addrmf}\footnote{\url{http://heasarc.nasa.gov/ftools/caldb/addrmf.html}} part of NASA HEASARC's FTOOLS\footnote{\url{http://heasarc.nasa.gov/ftools}}.

For the analysis of LAT observations of all GRBs detected inside the LAT FoV, we use the Transient class events as described in \S~\ref{sec_data_preparation}. 
We bin the LAT data in 10 logarithmically-spaced energy bins between 100~MeV and 250~GeV, and use an energy-dependent ROI as described in $\S$~\ref{subsection_cuts}. We derive the observed spectrum and the response matrix using the \Fermi Science Tools \textit{gtbin} and \textit{gtrspgen}. We also use the BKGE to obtain a background spectrum containing the contributions from all the sources of background, as described in \S~\ref{subsubsec_LAT_bkg}.

Note that for GRBs detected by the LLE photon counting analysis outside the LAT FoV we used only GBM data for the spectral analysis.

\subsubsection{Spectral fit}
\label{subsub_joint_spectral_fit}
We load the spectra and response matrices in \xspec v.12.7\footnote{\url{http://heasarc.nasa.gov/docs/xanadu/xspec/}}. For GBM data, we exclude from the fit all of the NaI channels between 33~keV and 36~keV \citep[corresponding to the Iodine K-edge, see][]{2009arXiv0908.0450M}, and ignore the channels at the extremes of the spectra (channels below 8~keV and channels 127 and 128 for NaI ; channels 1, 2, 127, and 128 for BGO). We do not exclude any energy bin in the LAT spectrum, since we already selected the data before binning them. We jointly fit the GBM and LAT data with several models (described below), minimizing the negative log-likelihood. This likelihood function is derived from a joint probability distribution, obtained by modeling the spectral counts as a Poisson process and the background counts as Gaussian process. For the latter, the Gaussian standard deviation for the $i$-th channel is given by $\sigma_{i}= \sqrt{\sigma_{stat,i}^2+\sigma_{sys,i}^2}$, where $\sigma_{stat,i}^2$ and $\sigma_{sys,i}^2$ are the statistical
and the systematic variances respectively. The maximum likelihood principle assures that the derivatives of the likelihood function with respect to the parameters are null for the best-fitting set of parameters. Exploiting this, one can treat the means of the Gaussian functions describing the background counts as \textit{nuisance} parameters, and remove them from the fitting procedure by expressing them as functions of the other parameters. This is a rather standard statistical procedure, and leads to the formulation of a so-called \textit{profile} likelihood function. PG-stat is defined as the natural logarithm of such function (see the \xspec website\footnote{\url{http://heasarc.nasa.gov/xanadu/xspec/manual/XSappendixStatistics.html}} for more details). 
The fitting algorithms implemented in \xspec find local minima for the statistic, but they can fail to converge to the global minimum.
This is a known issue with gradient-descent algorithms \citep{handbook2011}. To mitigate this problem, we perform multiple fits (from 10 to 40) for each model, each time starting from a different set of values for the parameters, and we keep as the putative best fit the set giving the lowest overall value for the statistic. If while computing error contours for this set of parameters, the fitting algorithm finds an even better minimum for the statistic, we adopt that as the new putative best fit, and restart the error computation, iterating the procedure until no new minimum is found.

\onecolumn
\subsubsection{Spectral Models}
\label{sub_sub_spectral_models}

Traditionally, GRB spectra have been described using the phenomenological ``Band function'' \citep{Band:93} or a model consisting of a power law with an exponential cutoff (also called ``Comptonized model''). Another common choice is the smoothly broken power law (SBPL) \citep{1999ApL&C..39..281R}. Recently, the logarithmic parabola has been shown to be a good description the spectra of some GRBs, especially in time resolved analyses \citep{Massaro2010, Massaro2011}. We call these 4 spectral models \textit{main components}. One of the first results by \Fermi was the need for multi-component spectral models for some GRBs, showing an high-energy excess over the main component which has been modeled with an additional power law \citep{GRB090510:ApJ,090902B_PAPER}. In one case, \Fermi observed a high energy cutoff which required the addition of an exponential cutoff to the power law component in the spectral model \citep{GRB090926A:Fermi}, for a total of three components (Band, power law and exponential cutoff). In the
following we will call the power law and the exponential cutoff functions \textit{additional components}, to emphasize the fact that we add them to the main components when needed. Some authors have claimed the presence of a thermal component, modeled by a black body emission spectrum (see e.g., \cite{Guiriec2011,ZhangBing2011} and references therein). However, a careful time-resolved analysis is needed in order to investigate and characterize such a component, which is outside the scope of the present analysis. Thus we did not include the black body component in our spectral fits. Hereafter, $N(E)$ is the differential photon flux (in units of cm$^{-2}\,$s$^{-1}\,$keV$^{-1}\,$) expected from a model at a given energy $E$ (in~keV), and $k$ is a normalization constant whose units depend on the model. We have 4 main model components:

\begin{enumerate}[(I)]

 \item \textbf{Comptonized model} (a power law with an exponential cutoff):

$N(E)\equiv kE^{-\alpha}e^{-\frac{E}{E_{0}}}$,
where $\alpha$ is the photon index and $E_{0}$ is the cutoff energy.
 \item \textbf{Logarithmic parabola}, defined following equation 9 in~\cite{Massaro2010}:

$N(E) \equiv \frac{S_{p}}{E^{2}}10^{-b \log{E/E_{p}}^{2}}$,

where $S_{p}$ is the height of the SED at the peak frequency, $E_{p}$ is the peak energy and $b$ represents the curvature of the spectrum.
 \item\textbf{Band model}~\citep{Band:93}: two power laws joined by an exponential cutoff:
\begin{equation}
B(E) = N(E) \equiv k \left\{
\begin{array}{l}
  E^{\alpha}e^{-E/E_{0}}
  \hspace*{20pt}\mbox{when } E < (\alpha-\beta)E_{0}\\
  \left[(\alpha-\beta)\,E_{0}\right]^{\alpha-\beta}E^{\beta}e^{-(\alpha-\beta)}
 \hspace*{20pt}\mbox{when }  E > (\alpha-\beta)E_{0}\\
\end{array}
\right.
\end{equation}

Note that this is the representation that uses the $e$-folding energy $E_{0}$ (keV)
instead of the peak energy $E_{p}$, where $E_{p}=(2+\alpha)E_{0}$. $\alpha$ and
$\beta$ are respectively the (asymptotic) photon index at low energy and the photon index at high
energy.

\item \textbf{Smoothly broken power-law }~ \citep{1999ApL&C..39..281R}: two power laws joined by a hyperbolic tangent function with adjustable transition length:

\begin{equation}
N(E)\equiv k \left(\frac{E}{E_{piv}}\right)^{\frac{\alpha+\beta}{2}} \left[ \frac{\cosh{(\frac{\log{(E/E_{0})}}{\delta})}}{\cosh{(\frac{\log{(E_{piv}/E_{0})}}{\delta})}} \right]^{\frac{\alpha+\beta}{2}\delta\log_{e}{(10)}},
\end{equation}

where $E_{piv}$ is a fixed pivot energy, $\alpha$ and $\beta$ are respectively the photon index of the low-energy and of the high-energy power laws, $E_{0}$ is the $e$-folding energy and $\delta$ is the energy range over which the spectrum changes from one power law to the other.

\end{enumerate}

Here are the definitions of our additional model components:
\begin{enumerate}[(I)]
 \item \textbf{Power law}: $N(E)\equiv kE^{-\alpha}$, where $\alpha$ is the photon index.
 \item \textbf{Exponential cut off}: $e^{-\frac{E}{E_{0}}}$
\end{enumerate}
Because of the variety of spectral models, we have considered a number of functions, composed of one main component and one or more additional components: Band, Band + power law, Band + power law with exponential cutoff ($\equiv B(E)+kE^{-\alpha}e^{-E/E_{0}}$), Band with exponential cutoff ($\equiv B(E)e^{-E/E_{0}}$), Comptonized, Comptonized + power law, Comptonized + power law with exponential cutoff, Logarithmic Parabola, SBPL, SBPL + power law.

To take into account the relative unknown uncertainties in the inter-calibration between the different detectors, for bright bursts we also apply an effective area correction \citep{2011ApJ...733...97B}: we scale the model under examination by a multiplicative constant, with the constant being fixed to 1 for the LAT (taken as reference detector), but free to assume different values for all the other detectors.
For GRB for which we do not use LAT data, we choose one of the NaI detectors as the reference.
While for bright bursts adding such a correction changes the best fit parameters and the value of the statistic, for the other bursts it is essentially inconsequential, since in the latter cases statistical errors dominate over the inter-calibration uncertainties. For such spectra the multiplicative factors are unconstrained during the fit, and therefore we removed them.
After the best fit is found, we fix all the factors to their best fit values and we proceed with the error computation. The correction factors typically have values between 0.95 and 1.05 for NaI detectors, and between 0.75 and 1.25 for the BGO detectors.

\twocolumn
\subsubsection{Definition of a good fit and model selection}
\label{subsub_modelselection}
\begin{figure}[t!]
\begin{center}
\includegraphics[width=1.0\columnwidth,trim= 0 0 15 0, clip=true]{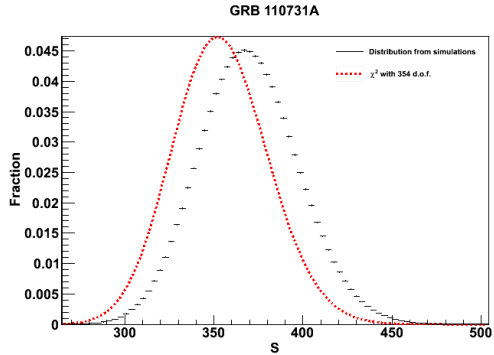}
\caption{Distribution for the PG-statistic as obtained from Monte Carlo simulations for GRB\,110731A using the Band model as null hypothesis (black points). We report the $\chi^{2}$ distribution for the same number of degrees of freedom for reference (red dashed line).}
\label{fig_GoodnessOfFit}
\end{center}
\end{figure}

The main focus of the spectral analysis performed here is to characterize the GRB spectrum, which requires selecting the most appropriate spectral model.  We define the best model for a given GRB as the simplest one giving a ``good'' value for the test statistic (PG-stat, $S$ in the following) and no evident structures in the residuals. Since $S$ is based on a Poisson likelihood, we do not have a simple goodness-of-fit test comparable to the $\chi^{2}$ test when minimizing the $\chi^{2}$ statistic. The actual expected value $S^{*}$ for the statistic $S$ is a function of the number of counts N in the spectrum and of the background model and its uncertainties, and can be estimated using Monte Carlo simulations. We assume a model $m_{0}(\vec p)$ (for example, the Band model) with the best fitting set of parameters $\vec p_{0}$ as the null hypothesis, and we generate 1 million realizations of $m_{0}(\vec p_{0})$ and the corresponding background spectrum using the \textit{fakeit} command of \xspec. Each realization
$r^{i}_{p_{0}}$ is obtained by adding Poisson noise to the count spectrum obtained by summing the observed background spectra and $m_{0}(\vec p_{0})$. Correspondingly, each realization of the background spectrum is obtained by adding Gaussian noise to the observed background spectrum, using a total variance composed of the statistical and the systematic variance of the observed background. Then we fit $m_{0}(\vec p)$ to $r^{i}_{p_{0}}$ and we record the value for the statistic $S_{i}$. In Fig.~\ref{fig_GoodnessOfFit} we show an example of a distribution for $S$ obtained using the Band model, and a $\chi^{2}$ distribution for the same number of degrees of freedom as reference. Note that depending on the case, the two distributions can be very different. We can now use the distribution for $S$ resulting from the simulations to compute the probability of obtaining the observed value for $S$ under the null hypothesis $m_{0}$. This approach requires a large number of simulations, so we applied it just
for the subsample of GRBs for which we claim the detection of an extra component (see below, and Section~\ref{subsub_extracomponents}).

\begin{figure*}[t!]
\begin{center}
\includegraphics[width=1.0\textwidth]{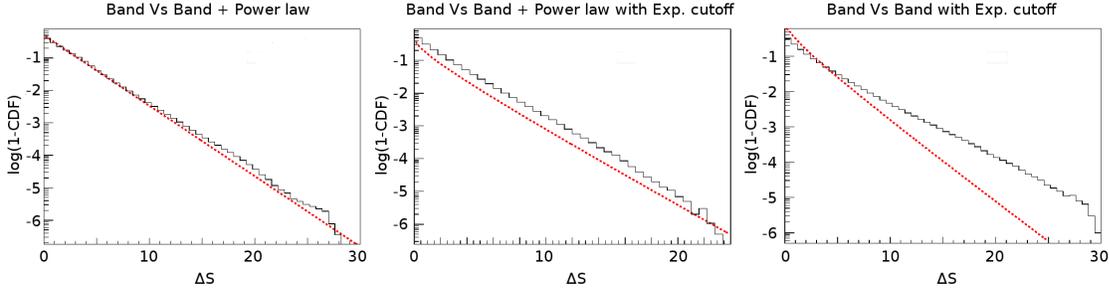}
\caption{Complementary Cumulative Distribution (1 $-$ CDF) for $\Delta S$, for three different pairs of models: Band versus Band + power law (left panel), Band versus Band + power law with exponential cutoff (center panel), and Band versus Band with exponential cutoff (right panel). The dashed line corresponds to the complementary CDF of a $\chi^2_{n}/2$ with $n=n_{0,dof}-n_{1,dof}$ (see text).} 
\label{fig_deltaS}
\end{center}
\end{figure*}

In order to compare different models, we considered them in pairs. Let us consider the model $m_{0}$ with $n_{0,dof}$ and $m_{1}$.
If $S_{0}<S_{1}$ and $n_{0,dof} \leq n_{1,dof}$ then $m_{0}$ describes the data better using fewer or the same number of parameters and we consider it a better fit following the definition given at the beginning of this section. If $S_{0} \simeq S_{1}$ and $n_{0,dof}=n_{1,dof}$ the two models are equivalent, and we should report the results for both the models. Anyway, this never happened in our analysis. 
On the other hand, if $S_{0}>S_{1}$ then $m_{1}$ better fits the data, and we have to decide if the improvement is significant enough to justify the added complexity. In the literature there are different ways to quantify this improvement, sometimes incorrectly (see for example discussion in \citealt{Protassov2002}). One of the standard methods is the likelihood-ratio test, which uses as test statistic the difference in $S$ between the two models $\Delta S$. In the case of nested
models $m_0$ and $m_1$,
Wilks' theorem~\citep{Wilks:38} assures under certain hypotheses that the quantity $\Delta S$ asymptotically follows a $\chi^2$ distribution with $n=n_{0,dof}-n_{1,dof}$ degrees of freedom. Unfortunately, in all the cases of interest here the theorem's hypotheses are not satisfied and the reference distribution for $\Delta S$ is not known. 
In general one should perform dedicated Monte Carlo simulations to obtain the reference distribution.
Performing such simulations for each pair of models is not practical. Thus, we select three cases of interest (i.e., Band versus Band + power law, Band versus Band with exponential cutoff, and Band+power law versus Band + power law with exponential cutoff) and we perform several million simulations to evaluate the reference distributions. We use the same procedure as above, using the simplest model as the null
hypothesis, but we fit both $m_0$ and $m_1$ to each
simulated data set,
recording $\Delta S$. At the end of the simulation, the distribution for $\Delta S$ is used to compute the probability $P$ of obtaining a $\Delta S$ greater than the observed value, which corresponds to the complement of the cumulative distribution function. In Fig.~\ref{fig_deltaS}, we plot this function for the three cases. We fix an arbitrary threshold at $P_{th}(>\Delta S)=1 \times 10^{-5}$, where the statistical error on the simulated distribution, visible toward the tail, is still low. $P_{th}$ corresponds to a significance level of $\sim 4.2 \sigma$, and defines a threshold for $\Delta S$ above which we claim a significant detection of an extra component. Specifically, $P_{th}$ corresponds to $\Delta S=25$ for Band versus Band + power law, $\Delta S=28$ for Band versus Band with exponential cutoff, and $\Delta S = 20$ for Band + power law versus Band + power law with exponential cutoff.

\subsection{Analysis Sequence}
\label{subsec_analysis_sequence}

\begin{figure}[!ht]
\begin{center}
\includegraphics[width=1.0\columnwidth]{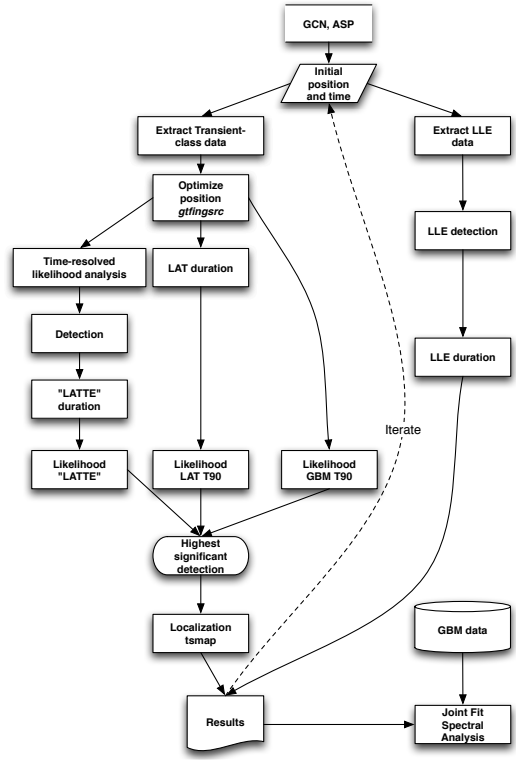}
\caption{Schematic representation of the analysis sequence adopted in this work.}
\label{fig_GRBCatalogAnalSequence}
\end{center}
\end{figure}
The sequence of analyses performed in this work is graphically represented in Fig.~\ref{fig_GRBCatalogAnalSequence}.
We start our analysis using the best available localization provided via GCN typically by Swift or the GBM and in some cases by other observatories.
Detections occurring in Automated Science Processing (ASP) of LAT data \citep{2009ApJ...701.1673B} are also used as inputs.
We then extract both Transient class and LLE data. We use the Transient class data to optimize the location of the GRB. However, if the reported position error is significantly smaller than the angular resolution of the LAT, there is no room for improvement and we  adopt the ``GCN'' position. This is the case for localizations provided by Swift or by optical observatories. On the other hand, if the reported position has an error larger than the characteristic size of the LAT Transient class PSF ($\sim$0.5~deg at 1~GeV) -- most notably those typically provided by the GBM -- we repeat most of the steps in our analysis sequence multiple times, starting each iteration with the best position obtained during the localization step of the previous one, until we cannot improve the localization further.
Typically we repeat the analysis 2-3 times until the localization obtained in the last step is within the error on the localization of the previous iteration. This introduce a small number of trials, which are also strongly correlated since they only involve small changes in the analysis configuration/data. High confidence localization errors (90\%--95\%) are not affected, and we therefore decided to ignore this trial factor.
The analysis of Transient class data consists of the following steps.
\begin{enumerate}[(I)]
\item{\bf{Duration Measurement}}

We apply the techniques described in \S~\ref{subsubsect_duration} to compute the duration (T$_{90}$) of the burst, using Transient class data.
We define the ``LAT interval'' as the time interval from T$_{05}$ to T$_{95}$ (of duration T$_{90}$= T$_{95}$-T$_{95}$) measured in this step.
In case of a non-detection, the value of the LAT\,T$_{90}$ is not available in the following steps.

\item{\bf{Time-resolved likelihood analysis}}

The next step consists of a time-resolved spectral analysis, which allows us to study the temporally extended emission systematically, one of the common characteristics of LAT GRBs. We analyze all data contained in Good Time Intervals (GTIs, see \S~\ref{subsection_cuts}) within 10\,ks from the GRB trigger, binning them in time. We tested several binning schemes, including linear, logarithmic, and Bayesian-blocks binning, and the resulting likelihood fit parameters were consistent among the different choices. The logarithmically-spaced binning provides constant-fluence bins when applied on a signal that decreases approximately as 1/time, such as the extended GRB emissions observed by the LAT. We adopt that scheme as the starting point, we start from a bin size containing at least N events, where N corresponds to the number of parameters in the model, plus 2, and then we merge consecutive time bins until obtaining a minimum TS value.

Specifically, we divide the data into logarithmically-spaced bins, truncating bins at the edges of excluded time intervals when necessary. Then we merge bins until each of them has a number of counts at least equal to the number of parameters of the likelihood model plus 2.  We then fit each bin using the maximum likelihood analysis described in \S~\ref{subsec_likelihood_analysis} obtaining the likelihood and the TS value corresponding to the best-fit source model.
If the resulting TS value is lower than an arbitrary threshold ($TS<16$ , corresponding to a pre-trials significance $ >\sim$3.2--3.8$\sigma$ depending on $n_{dof}$) we merge the corresponding time bin with the next one, and we repeat the likelihood analysis. This step is iterated until one of two conditions is satisfied: 1) we reach the end of a GTI before reaching $TS = 16$, in which case we compute the value of the 95\% CL upper limit (UL) for the flux evaluated using a photon index of 2\footnote{Conventionally the photon index for a GRB spectrum is defined as positive (i.e. dN/dE $\approx$ E$^{-\gamma}$)}; 2) we reach $TS > 16$, in which case we evaluate the best-fit values of the flux and the spectral index along with their 1$\sigma$ errors.

The time interval between the beginning of the first and the end of the last time bin for which TS$>16$, named ``LAT temporally extended time interval'' (hereafter ``LATTE''), constitutes a rough estimate of the time window where the GRB emission is detectable with at least a $\sim 3\sigma$ significance.

%

\item{\bf{Characterization of the extended emission}}\\
After having characterized the GRB in each time bins separately, we study the light curve as a whole. Specifically, we select the events contained in an energy-dependent ROI (see \S\ref{subsection_cuts}) in each time bin, building a light curve of the detected counts, and we estimate the background in each time bin using the BKGE. We also compute the exposure (in cm$^{2}$s) associated with each time bin, using the tool \textit{gtexposure}\footnote{\url{http://fermi.gsfc.nasa.gov/ssc/data/analysis/scitools/help/gtexposure.txt}} calculated in each energy-dependent ROI separately. 
This last step requires knowledge of the spectrum. For each time bin we use the corresponding best fit power-law model as found in the bin-by-bin analysis described before. We note here that in principle the uncertainty in the best fit parameters for the power law would translate into an uncertainty in the value of the exposure, because of the energy dependence of the effective collecting area of the LAT. In our case, such an error is typically of the order of 5\%, which is smaller than
the systematic uncertainty in the response of the LAT, and will be neglected.

Summarizing, for each time bin $i$ we have the observed number of counts $N_{i}$ (in the energy-dependent ROI), the corresponding background estimate $B_{i}$, and the corresponding exposure $A_{i}$. Assuming a given model for the light curve $M(t)$ (for example a power law), we compute the expected number of observed counts in the i-th bin between $t_{i,1}$ and $t_{i,2}$ as:
\[
N_{i,pred} = \left(\int_{t_{i,1}}^{t_{i,2}} M(t)dt\right) \times A_{i}+ B_{i}
\]

We compare $N_{i,pred}$ with $N_{i}$ and look for the best fit parameters for the model $M(t)$, minimizing a Poisson log-likelihood function. We actually used the PG-stat log-likelihood function implemented in Xspec v.12, which takes into account the uncertainty on $B_{i}$ (see \S\ref{subsub_joint_spectral_fit} for details). This technique, which might seem unnecessarily complex, provides a natural way of including in the fit the time intervals during which the source is barely detected, or not detected at all. Indeed, they can be treated exactly like all the others, by comparing $N_{i,pred}$ with $N_{i}$, even if $N_{i} \simeq B_{i}$. As a consistency check, we also have used the more conventional technique of fitting $M(t)$ to the count-flux light curve as obtained from the likelihood analysis, minimizing $\chi^{2}$.
To incorporate information from upper limits on the flux computed from the unbinned analysis\footnote{To calculate UL we use the \textit{python} interface to the \textit{Likelihood} package, as described here: \url{http://fermi.gsfc.nasa.gov/ssc/data/analysis/scitools/python\_usage\_notes.html\#UpperLimit}}, we first rescaled the one-sided 95\% CL UL to two-sided 68\% CL confidence intervals under the assumption that the errors are normally
distributed. Then, we replaced the value of the UL with the value of a point that would have the 68\% CL correspond to the value of the UL. To obtain reliable values from the fit, we required at least one positive detection after the peak flux (in addition to ULs). The two methods gave virtually identical results, and so we provide only the values from the second method, the fit of the count light curve.

We consider two models for the light curve: a simple power-law model:
\[
F(t)=F_{0}\times(t/t_{p})^{-\alpha},
\]
where F$_{0}$ and $\alpha$ are the free parameters, and a broken power-law model:
\begin{multline*}
 F(t)=F_{0}\times( H(t>t_{b})\times(t/t_{b})^{-\alpha_1}+\\ H(t<t_{b})\times(t/t_{b})^{-\alpha_2} ),
\end{multline*}
  where both indices ($\alpha_1$ and $\alpha_2$) are left free, the normalization is $F_{0}$ and the break time $t_{b}$. We measure the time $t_{p}$ at which the detected flux reaches its maximum value $F_{p}$ (the ``peak flux'') as the center of the time bin with the maximum count flux. We then consider two time intervals starting respectively at the peak t$\ge{t_{p}}$ and after the end of the prompt emission t$>$ GBM T$_{\rm{95}}$. For each time interval, we fit the power law and the broken power-law models and we compare them by performing Monte Carlo simulations similarly to the procedures described in \S~\ref{subsub_modelselection}. We consider a break significantly detected when its chance probability is smaller than $10^{-3}$. In the above, all times are with respect to the GBM trigger time.


\item{\bf{Time-integrated likelihood analysis}}

We now perform the likelihood analysis on different time intervals, defined in Table~\ref{tab_intervals_fit}. These intervals are defined using combinations of the GBM durations reported in \cite{Paciesas+12}, the Transient class durations, and the ``LATTE'' time window. If we obtain a $TS > 20$ in any of these time intervals, we consider the GRB detected.

\item{\bf{Localization}}

We select the interval where the GRB is detected with the largest significance among those considered in the previous step, along with the corresponding likelihood model, and we generate an improved localization using the second method described in \S~\ref{subsubsec_likelihood_localization}. If the new localization has a greater significance and a smaller error than the current one, we repeat the analysis chain from the beginning, adopting the new improved value. Otherwise, we select the old localization and all the results of the last iteration of the analysis chain as the final ones and proceed to the next step. Note that we typically perform a few iterations of the whole chain.

\item{\bf{LLE analysis}}

In parallel, we execute the LLE analysis, which consists of three steps. We first extract LLE data, then run the detection algorithm on LLE class data
(see \S~\ref{subsubsect_detection_LLE}). Finally, if the GRB is detected ($S_{post}>4\sigma$), we evaluate its duration (see
\S~\ref{subsubsect_duration}). Note that this part of the analysis is performed again when an improved localization is obtained using LAT Transient
class data.

\item{\bf{LAT-GBM joint spectral fits of the prompt emission}}

We use the best available position to extract the spectrum of the GRB across the whole energy range covered by \Fermi. We fit the spectrum, following the procedure described in \S~\ref{subsec_lat_gbm_spectral_analysis}. We perform a spectral analysis in two time intervals: the ``GBM'' time interval defined in Table~\ref{tab_intervals_fit} and the time interval starting when the first LAT photon is detected in the GRB ROI and extending up to the GBM T$_{95}$ instant.

\end{enumerate}

\section{Results}\label{sec_results}

In this section we describe the results from our analysis;
all tables are collected in \S~\ref{sec_tables} and detailed discussions for each detected GRB are in Appendix~\ref{sec_fermi_lat_grb}.
According to the standard definition, GRBs with GBM T$_{90}>$2\,s are defined long, while short-duration GRBs have GBM T$_{90}<$2\,s.
Any upper limits from the maximum likelihood analysis are for a 95\% confidence level and are calculated using a photon index of 2.
We quote fluences in two Earth reference frame energy ranges: 10~keV--1~MeV and 100~MeV--100~GeV, appropriate to characterize the GRB emission as measured by the GBM and LAT respectively.
For all of the quantities a subscript (``LAT, GBM, EXT'') is added, to indicate the time interval used to perform the spectral analysis.
Low-energy (10~keV--1~MeV) fluences of non-LAT-detected GRBs are from the GBM spectral catalog \citep{Goldstein+12} and of LAT-detected GRBs from our joint GBM-LAT spectral analysis.
A discussion on how the LAT-detected bursts fluences compare with the distribution of fluences for all the GBM-detected bursts are left for the next section.

\subsection{LAT Detections}
\label{subsec_sec_detections}

We searched for high-energy emission with the LAT for the 733 GRBs described in
\S~\ref{subsec_input_GRB_list} and detected 35, using the detection criteria described in \S~\ref{subsubsect_detection_LLE} and \S~\ref{subsubsect_detection_like}. Among them, 28
were detected by our maximum likelihood analysis at energies above 100~MeV
 and 21 were detected using event-counting methods applied to
the LLE data.
Among the GCN circulars issued by the LAT team, three GRBs (listed below) were not included in this catalog as they were below the significance threshold,
while we also discovered five not previously claimed bursts (GRBs 081006, 090227B, 090531B, 100620A, and 101123A).
Thirty of our detected GRBs are of the long-duration class and five are of the short-duration class (GRBs 081024B, 090227B, 090510, 090531B, and
110529A).

We list the LAT-detected GRBs in Table~\ref{tab_GRBs} and report their trigger times,
off-axis angles at trigger
time, best available localizations with errors, redshifts, and
references to GCN circulars. In the table we also report whether these GRBs were detected by
the LLE and the maximum likelihood analyses. The LLE detection
significances and the likelihood TS values can be found in
Table~\ref{tab_durations}.

\begin{figure}[ht!]
\begin{center}
\includegraphics[width=1.0\columnwidth,trim=10 10 30 10, clip=true]{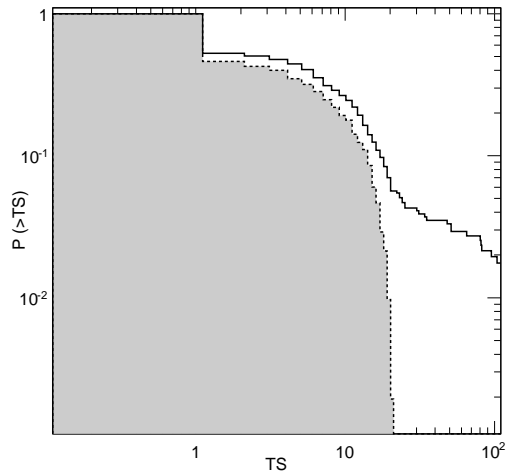}
\caption{Top: Normalized cumulative distribution of the maximum value of the Test Statistic (TS) obtained by performing likelihood analysis in different time windows. The dotted line with shaded grey area is the distribution of TS for a sample of fake GBM triggers, and the solid black line is the distribution for the sample of real GBM triggers.}
\label{fig_CumulativeTSDistribution}
\end{center}
\end{figure}

As a cross-check of our adopted detection thresholds and to estimate the
rate of false detections in our sample, we repeated the analysis
on a sample of  ``fake GBM triggers''. We generated the list of fake GBM triggers by changing the real trigger times T$_0$ of the input list to T$_0$ $-$ 11466~s, corresponding to approximately two orbits before the true trigger.  The standard operating mode for the \Fermi spacecraft is to change the rocking angle every orbit, viewing alternately the northern and southern orbital hemispheres. Thus, with the exception of ARRs, the ``fake'' sample has very similar background conditions with respect to the true sample.
Excluding the ARR, for each fake GRB trigger, we computed the TS value in a series of time intervals (of 1, 3, 10, 30, 100, 300, and 1000~s duration), kept the highest TS value we obtained for each fake GRB, and compiled them into a cumulative distribution. Figure~\ref{fig_CumulativeTSDistribution} compares the cumulative distribution
with the same distribution for the true GBM trigger sample.
Both distributions have been normalized to unity for TS=0 [i.e. P(TS$>$0)=1].
For the fake triggers, we did not obtain any value for the TS greater than TS$_{min}$=20 (our nominal detection threshold). The excess of the TS distribution of the true GRB sample with respect to the null distribution for TS$>$20 is evident.
It is important to note that the full analysis chain performed on the actual data and described in the previous section also optimizes the time window to compute the likelihood analysis, a task which is not included here.

As mentioned above, in addition to the GRBs reported here, the LAT team has reported detections of 3 other GRBs
  via GCN, but for the reasons explained below we have not included these events in the
  final table as they were formally below the detection threshold set for this catalog. These are:
\begin{itemize}
\item{GRB\,081224} for which a tentative onboard localization with the LAT was delivered via GCN \citep{2008GCN..8723....1W}.
Further on-ground analysis did not confirm the signal excess found in the LAT data, and a retraction GCN notice was issued
\citep{2008GCN..8726....1M}.
Whereas the GBM light curve is a broad single pulse event lasting $\sim$17~s, the LLE light curve shows a narrow spike at T$_{0}$ which is not
associated with the main pulse in the GBM, with a low significance of 3.1$\sigma$ only.

\item{GRB\,100707A} which had a significance of 3.7 $\sigma$ using the LLE data.
This result confirms the early detection \citep{2010GCN..10945...1P} obtained with a dedicated event selection which was required by the burst
inclination of $\sim$90\de at trigger time.

\item{GRB\,081215} which was similarly observed at a large
off-axis angle and the LAT team detection for the GCN circular was
by means of a dedicated event selection~\citep{2008GCN..8684....1M}.
However, this burst was not detected by either of our methods here, having a very low significance in both the LLE and standard likelihood analyses.
\end{itemize}

Using matched-filter techniques~\citet{Akerlof+10}, \citet{Akerlof+11} and \citet{2012arXiv1203.5113Z} reported that GRBs\,080905A, 090228A, 091208B, 100718A and 110709A are possibly detected by the LAT.
By means of a counting method based on the LAT Diffuse class events, \citet{Rubtsov:12} also claimed the detection of 4 new candidates: GRBs 081009, 090720B, 100911 and 100728A.
We concur on some of these GRBs:

\begin{itemize}

\item GRB\,080905A (localized by Swift \citealt{GCN080905A}) corresponded to only a marginal significance (TS=16.8), lower than our detection threshold.
Additionally, no signal was detected in LLE data.

\item GRB\,081009 is a GBM-detected burst, which was not detected by Swift. In our analysis, the final value of the TS is 14, which is below our detection threshold.
Also the GRB is not detected in the LLE data above our detection threshold of 4\,$\sigma$, likely due to the high inclination of 94\ded5 at the trigger time.

\item GRB\,090228A has TS=20 after optimization of its position. However,
  the TS map is entirely driven by two 5-GeV events in spatial coincidence,
  instead of being due to several events.
  Moreover our LLE analysis returned a null
  detection. In order to accommodate two high energy events
  and essentially no events at low energy the photon index of this GRB
  should have been greater (harder spectrum) than $1$, which
  is not very realistic as the energy (and the number of events at high energy) would tend to infinity.

\item GRB\,090720B is also found by our likelihood analysis, is not seen in LLE data, and will be discussed in more detail in subsequent sections.

\item GRB\,091208B is localized by Swift and our analysis
  finds the maximum TS=20. It is a marginal detection with only
  three events associated to the burst location. However, in this case
  the TS value reaches the threshold and the spectral shape is convincing, so we
  consider this a detection for the catalog.

\item GRB\,100718A is a GBM-detected burst, which was not detected by Swift. We note that the location of this GRB is only 0\de.5 (with an uncertainly of approximately 6\de) from the position of the Vela pulsar \citep{LATVela, LATVela2}, which is the brightest steady $\gamma$-ray point source in the sky. The reported GBM localization error is approximately 6\de, compatible with the location of Vela.
Including a point source at the position of Vela, with the flux fixed to the value reported in \citet{2012ApJS..199...31N,2012yCat..21990031N}, the final value of the TS is well below our threshold. Also the LLE lightcurve doesn't show any structure above threshold.

\item GRB\,100728A is found by our pipeline during the ``LATTE'' time interval with a TS=32 selecting the time interval between 5.6 and 749.9 seconds after the GBM trigger.
In addition, a dedicated article has already been published \citep{GRB100728A} by the LAT and GBM collaborations.

\item GRB\,100911A was detected by the GBM when the direction of the burst was very close to the Earth, with an angle from the local zenith of approximately 105\de.
In order to minimize contamination from the bright limb of the Earth, we rejected any data taken during intervals for which the ROI intersected the Earth's limb, a cut that is more conservative than requiring that the GRB is not occulted by the Earth. As a consequence GRB\,100911A was not detected.

\item GRB\,110709A is also found by our likelihood analysis, is not seen in LLE data, and will be discussed in more detail in subsequent sections.
\end{itemize}

\subsection{Emission Onset Time and Duration in the LAT}
\label{subsec_res_durations}

We applied our duration measurement algorithms to all of the
significantly-detected GRBs as described in
\S~\ref{subsubsect_duration}.  Our results are shown in
Table~\ref{tab_durations}.

Referring to the durations reported in the GBM
GRB catalog \citep{Paciesas+12}, we report in the second column whether the
GRB was categorized as long (L) or short (S) as determined from the measured
T$_{90}$ in the 50~keV--300~keV energy bands. Our results consist of
two sets of T$_{05}$ and T$_{95}$, determined using the
Transient event class (denoted as ``LAT'') and the LLE event class
(denoted as ``LLE'').
We report a duration measured with an event
class only if the GRB was also detected using that same event
class. In two cases (GRBs~090926A and 100116A), the burst emission
persisted long enough that our algorithm failed to detect a plateau
before the end of the first continuous segment of observation. For these
cases, we report lower limits for the LAT T$_{95}$ and T$_{05}$. This
work produced the first-ever set of GRB durations measured at high
(MeV/GeV) energies.

\begin{figure}[ht!]
\begin{center}
\includegraphics[width=1.0\columnwidth,trim=10 10 30 50,clip=true]{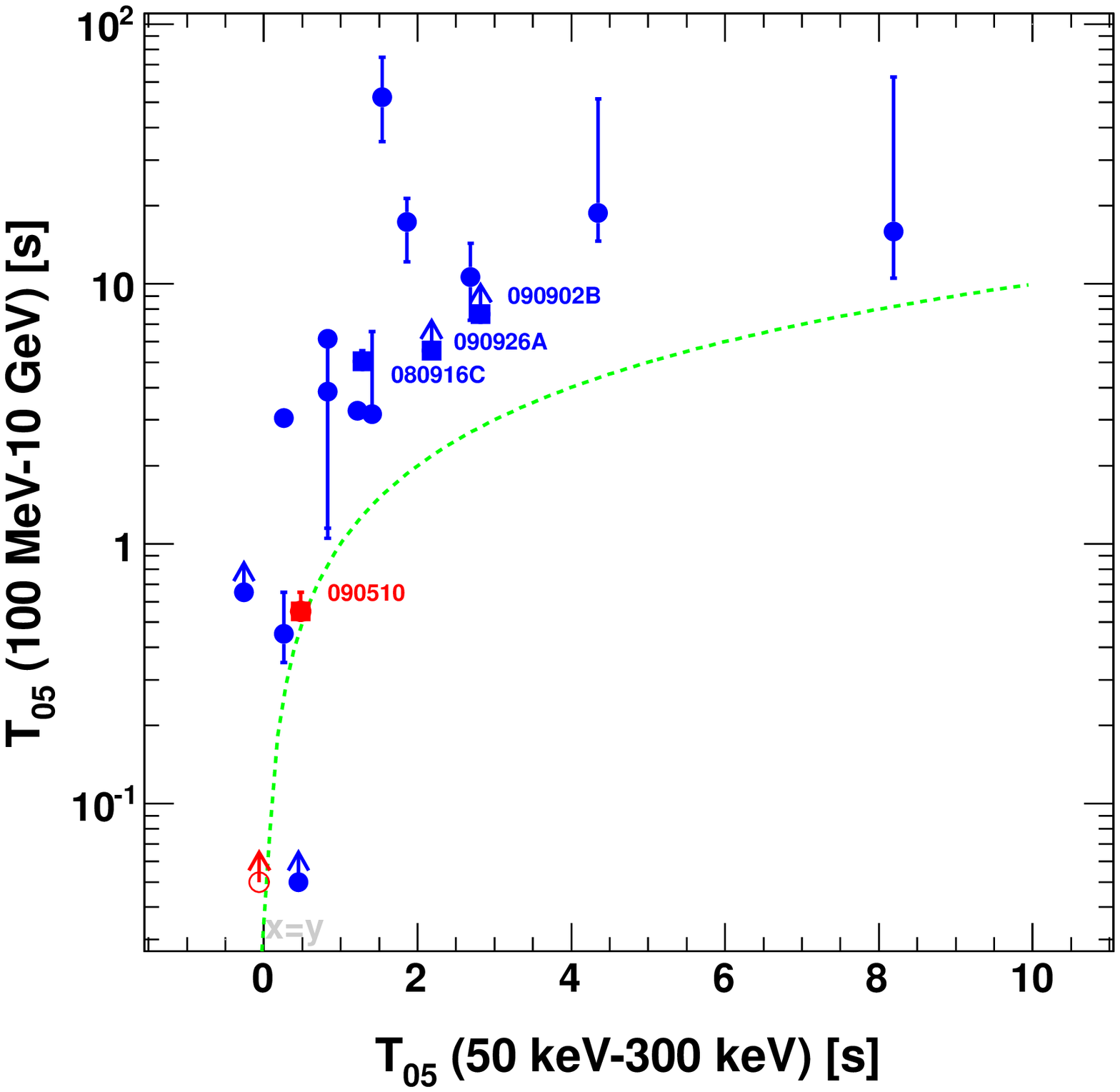}
\includegraphics[width=1.0\columnwidth,trim=10 10 30 50,clip=true]{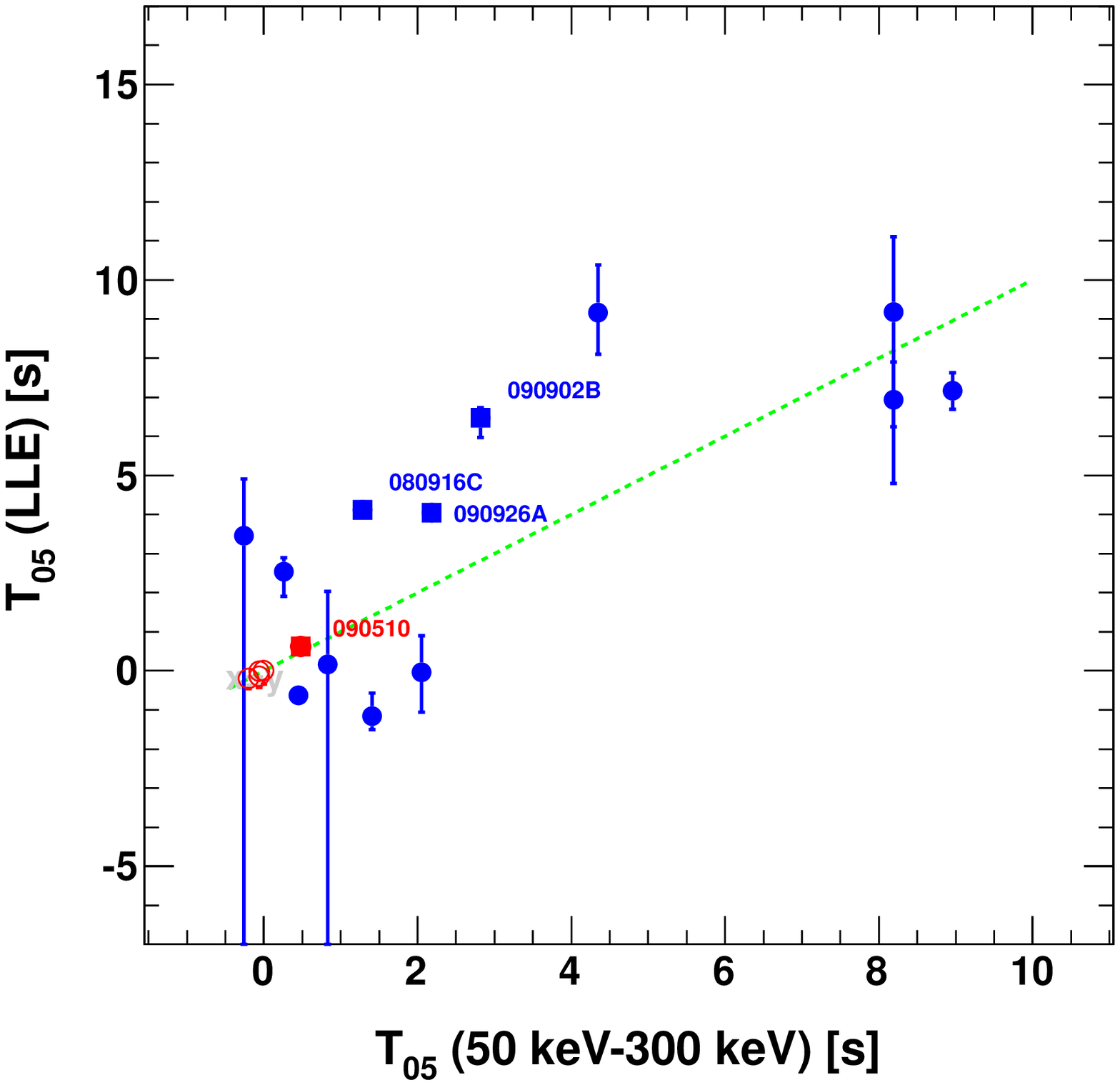}
\caption{Top: Comparison between the $>$100~MeV T$_{05}$ as measured using the LAT Transient class events and the  50~keV--300~keV T$_{05}$  as measured by the GBM. Bottom: Comparison between the LLE  T$_{05}$ and GBM T$_{05}$. The dashed line indicates equality.
Long duration GRBs are plotted with blue symbols, and short GRBs are plotted in red.  The 4 brightest LAT-detected bursts are plotted with square symbols and labeled.}
\label{fig_Onset}
\end{center}
\end{figure}
\begin{figure}[ht!]
\begin{center}
\includegraphics[width=1.0\columnwidth,trim=10 10 30 50,clip=true]{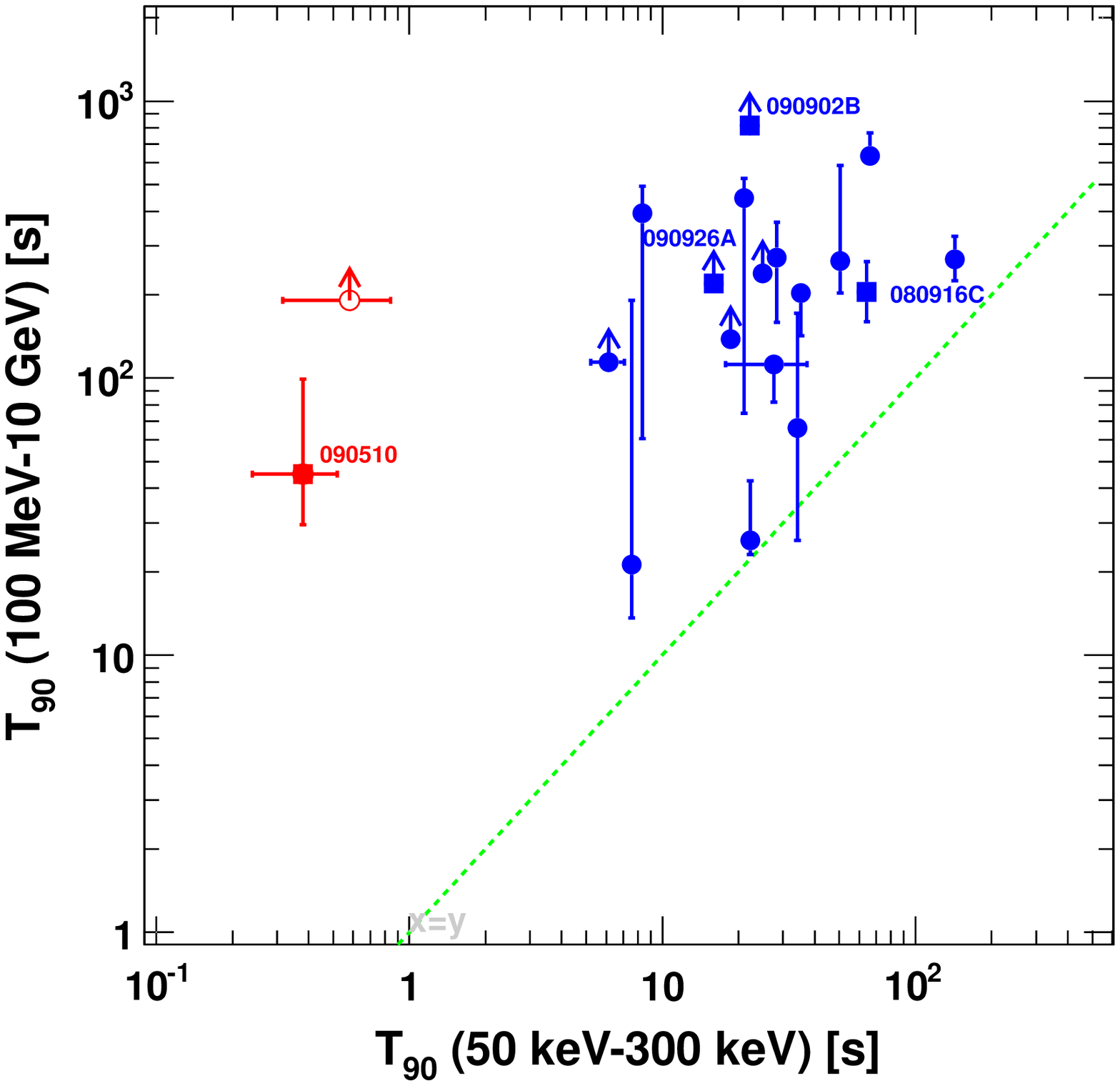}
\includegraphics[width=1.0\columnwidth,trim=10 10 30 50,clip=true]{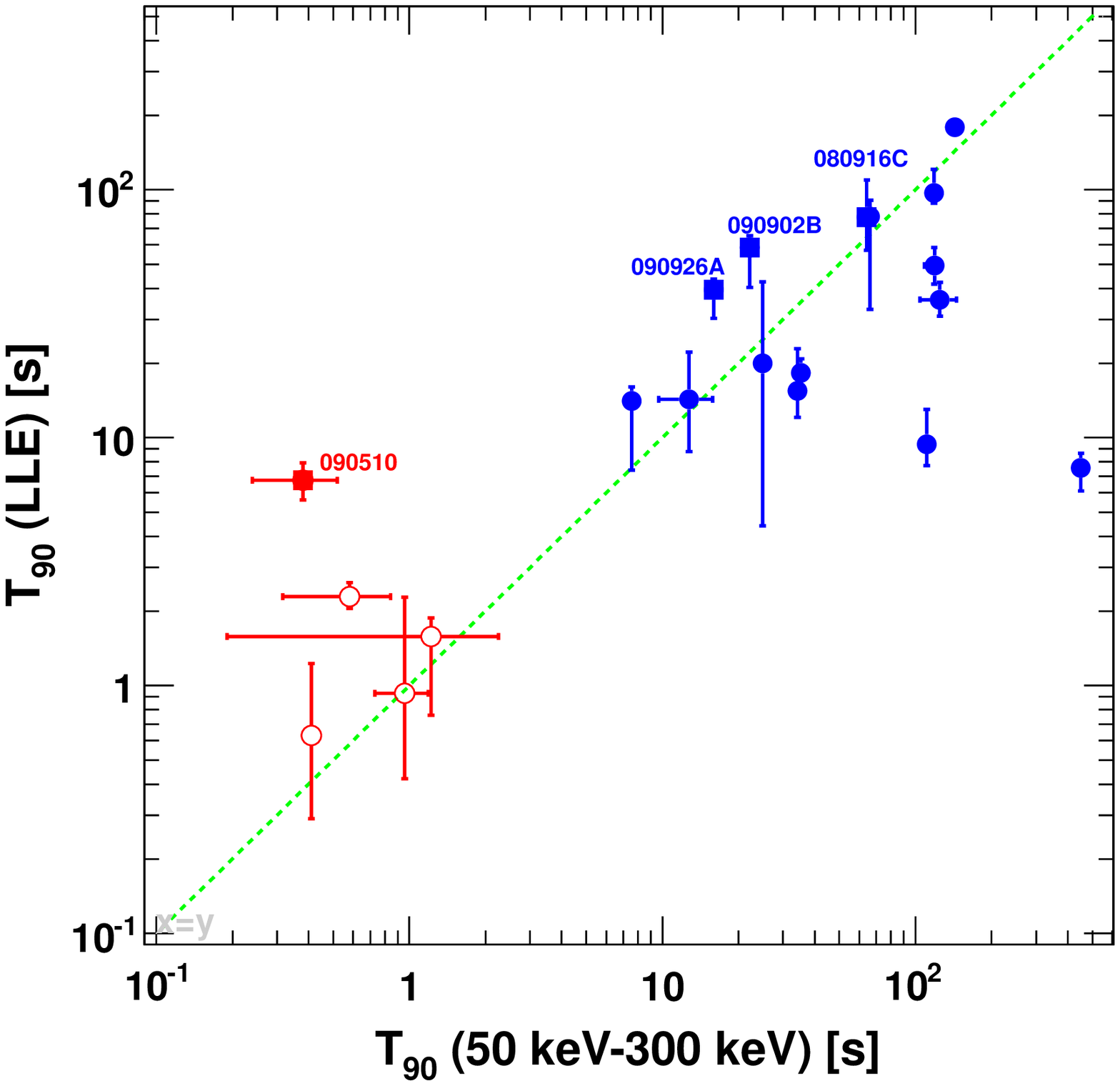}
\caption{Top: Comparison between the $>$100~MeV T$_{90}$ as measured using the LAT Transient class events versus the 50~keV--300~keV T$_{90}$ as measured by the GBM \citep{Paciesas+12}. Bottom: Comparison between the LLE  T$_{90}$ and GBM T$_{90}$. The dashed lines correspond to LAT T$_{90}$=GBM T$_{90}$ and LLE T$_{90}$=GBM T$_{90}$, respectively, in the top and bottom panels. The symbol convention is the same as in Fig.~\ref{fig_Onset}.}
\label{fig_Duration}
\end{center}
\end{figure}

The quantities compared in Table~\ref{tab_durations} are the onset time (reported here as T$_{05}$ and shown in Fig.~\ref{fig_Onset}) and the duration of the GRB emission (reported here as T$_{90}$ and shown in Fig.~\ref{fig_Duration}). In the top panels of Figs.~\ref{fig_Onset} and~\ref{fig_Duration} we compare the $>$100~MeV LAT Transient class duration measurements to the GBM results (in the 50~keV--300~keV energy band), while in the bottom panels we compare the tens-of-MeV LLE duration measurements to the GBM results. As shown in the top panels of both figures, the LAT-detected $>$100~MeV emission systematically starts later and has longer duration with respect to the GBM-detected emission. On the other hand, the bottom panels of both figures show that the durations measured using the LLE data are in better agreement with those measured by the GBM. Any deviation from the equal-duration line of the LLE versus GBM plots can be at least partially explained as due to spectral variations during the time of the
GRB emission, something that can be easily observed in the light curves reported in Appendix~\ref{sec_fermi_lat_grb}.

As was mentioned in \S~\ref{subsubsect_duration}, the duration
estimates are sensitive to the level of the background. Thus different
detectors, such as the GBM and LAT, or different event selections,
such as the LAT Transient and LLE class events, could produce
different duration estimates as a consequence of their very different
signal-to-noise ratios. This can partially explain the
systematically-longer durations (T$_{90}$) estimated using the LAT
Transient class events, but would not explain the systematically
later onset times (T$_{05}$).
We also note that a possible selection effect could arise owing to the typical GRB
off-axis angles at the trigger time. Bursts that are initially at the edge of
or outside the LAT FoV (i.e., having high $>60^{\circ}$ off-axis angles $\theta$)
enter the LAT FoV after some time (of the order of a few seconds),
thus introducing a delay between the onset of the GBM and
LAT observed signals. Even though we weight the LAT detected signal by
the inverse of the exposure to ameliorate this effect, we cannot
eliminate it since the weighting is not effective for the cases in
which no GRB Transient class events are detected at all by the LAT. This
effect might partially explain the delays of GRBs 090323 and
090328A. For most of the other cases, however, the GRB has a small enough
off-axis angle at onset to permit sufficiently sensitive prompt observations
(as shown by the $\theta$ column in Table~\ref{tab_GRBs}).\\

\subsection{Maximum Likelihood Analysis}

We split GRB observations into 6 time intervals listed in
Table~\ref{tab_intervals_fit} and performed a LAT-only spectral analysis using
the maximum likelihood technique described in
\S~\ref{subsec_likelihood_analysis}.
Since in the ``PRE'' interval the GRB is not detectable (by construction), we omit reporting results from this interval, and we focus on the five remaining time windows.
The results of this analysis, namely the TS, the best-fit photon index, the flux and
fluence for the 100~MeV--10~GeV energy range are presented in
Table~\ref{tab_likelihoods}. When possible, we also compute the
isotropic equivalent energy $\rm E_{iso}$ in the 100~MeV--1~GeV rest-frame energy band. In the same table, we also report the number of detected
events (originating from both the GRB and any possible background
components), and the number of events from the GRB as predicted by the
likelihood fit. These numbers are for the 100~MeV--10~GeV energy
range in the observer frame.

\subsubsection{Fluxes and Fluences}

\begin{figure*}[ht!]
\begin{center}
\includegraphics[width=0.5\textwidth,trim=10 0 10 10]{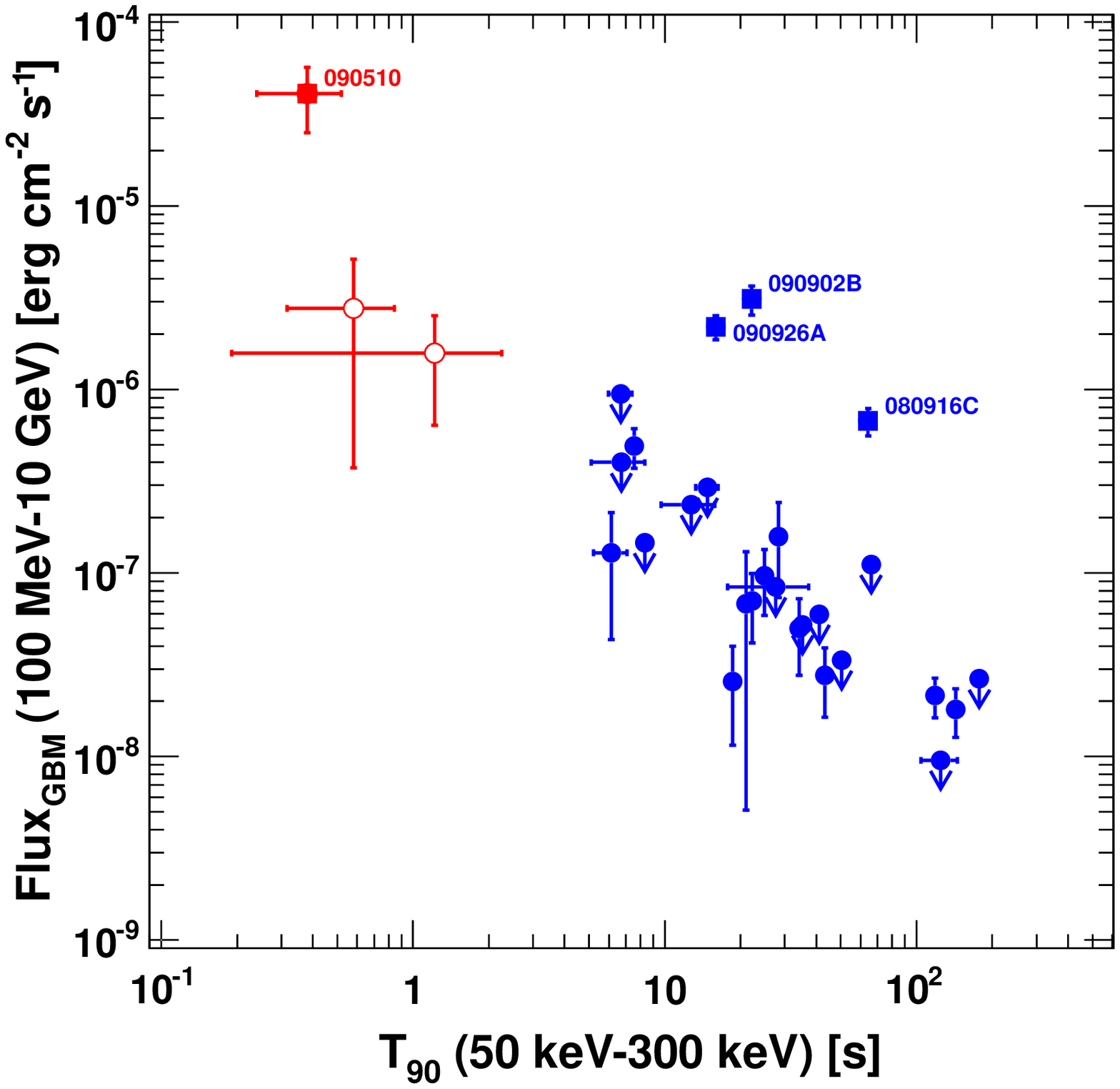}\includegraphics[width=0.5\textwidth,trim=10 0 10 10]{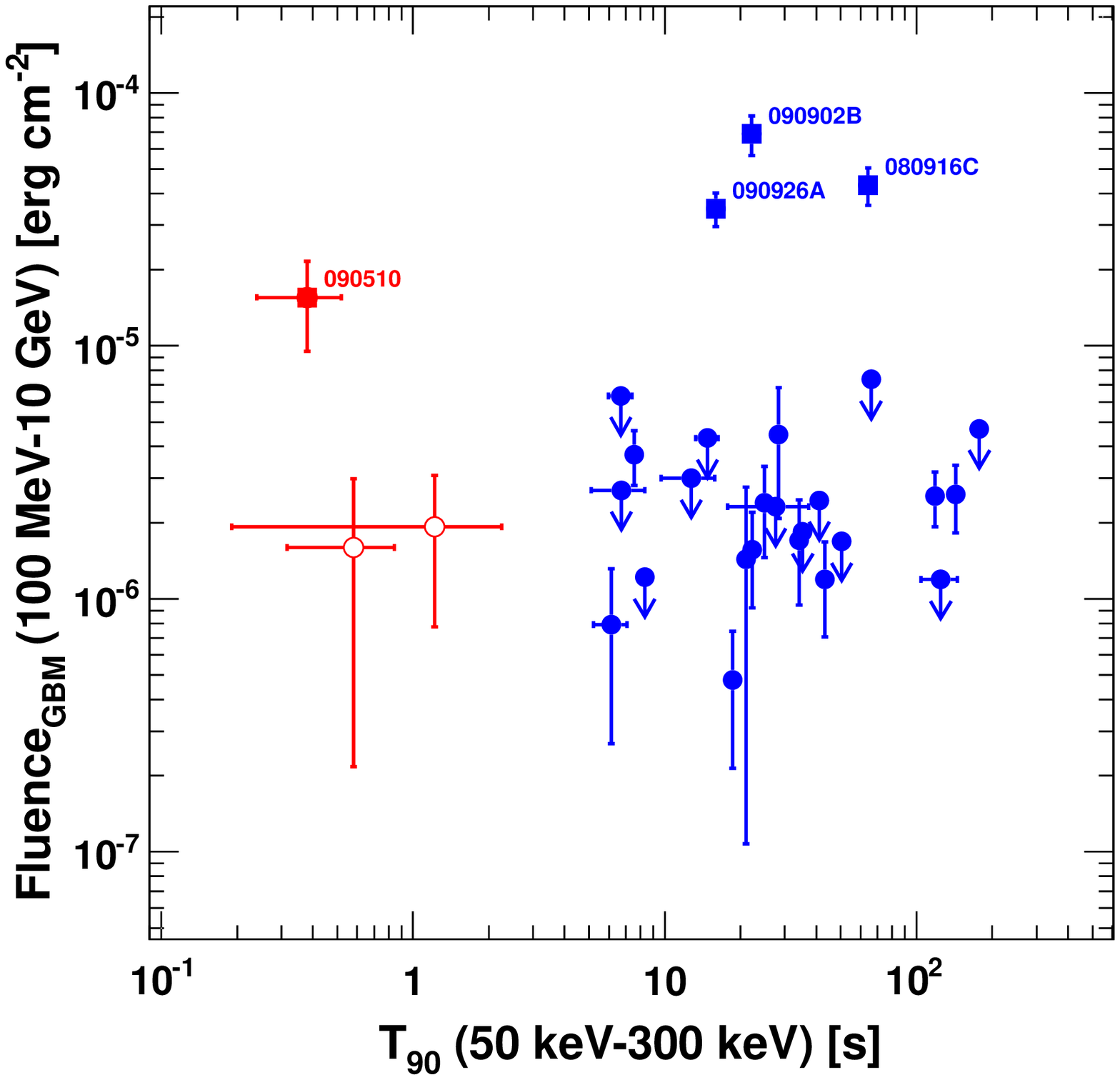}
\includegraphics[width=0.5\textwidth,trim=10 10 10 10]{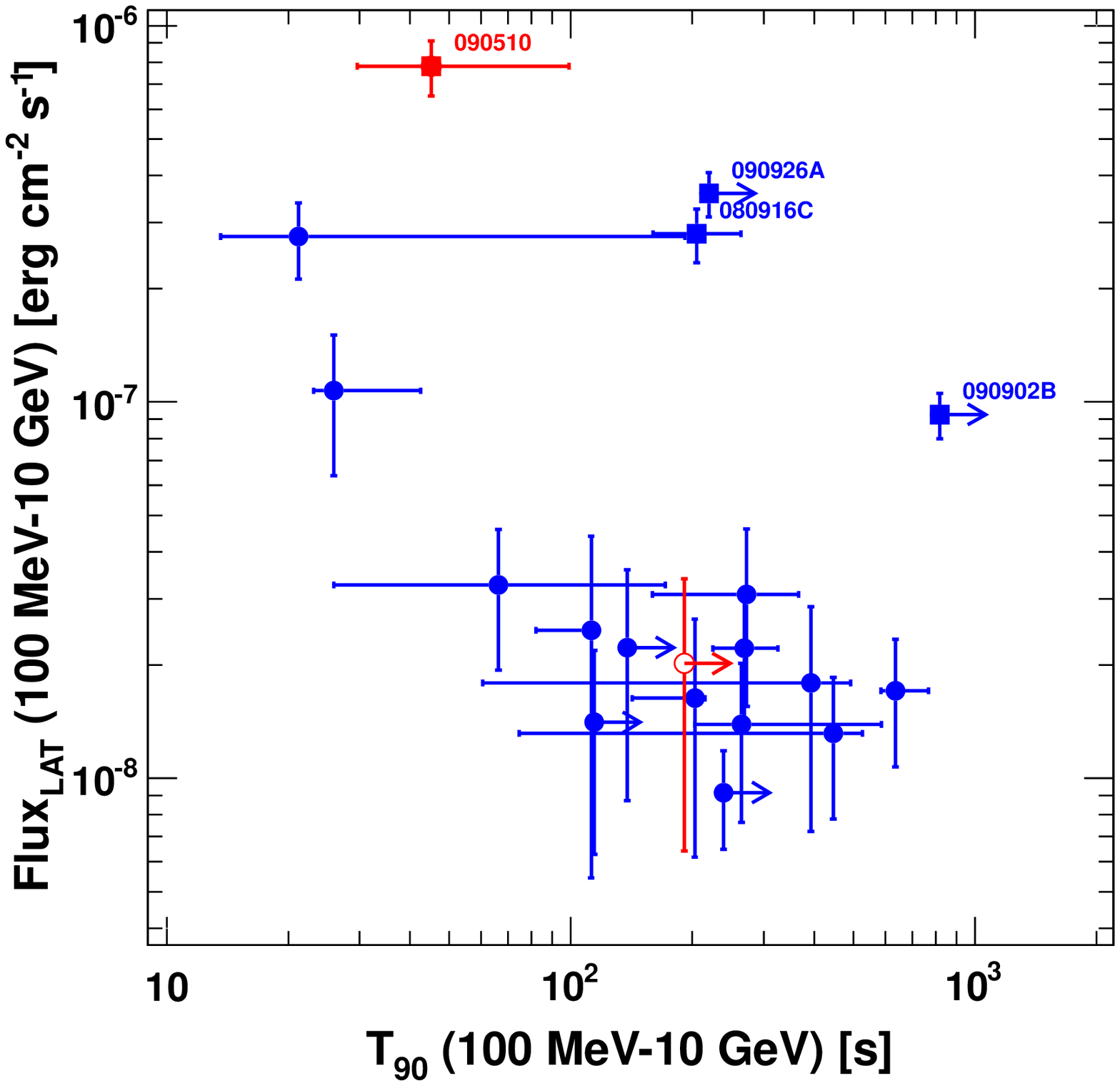}\includegraphics[width=0.5\textwidth,trim=10 10 10 10]{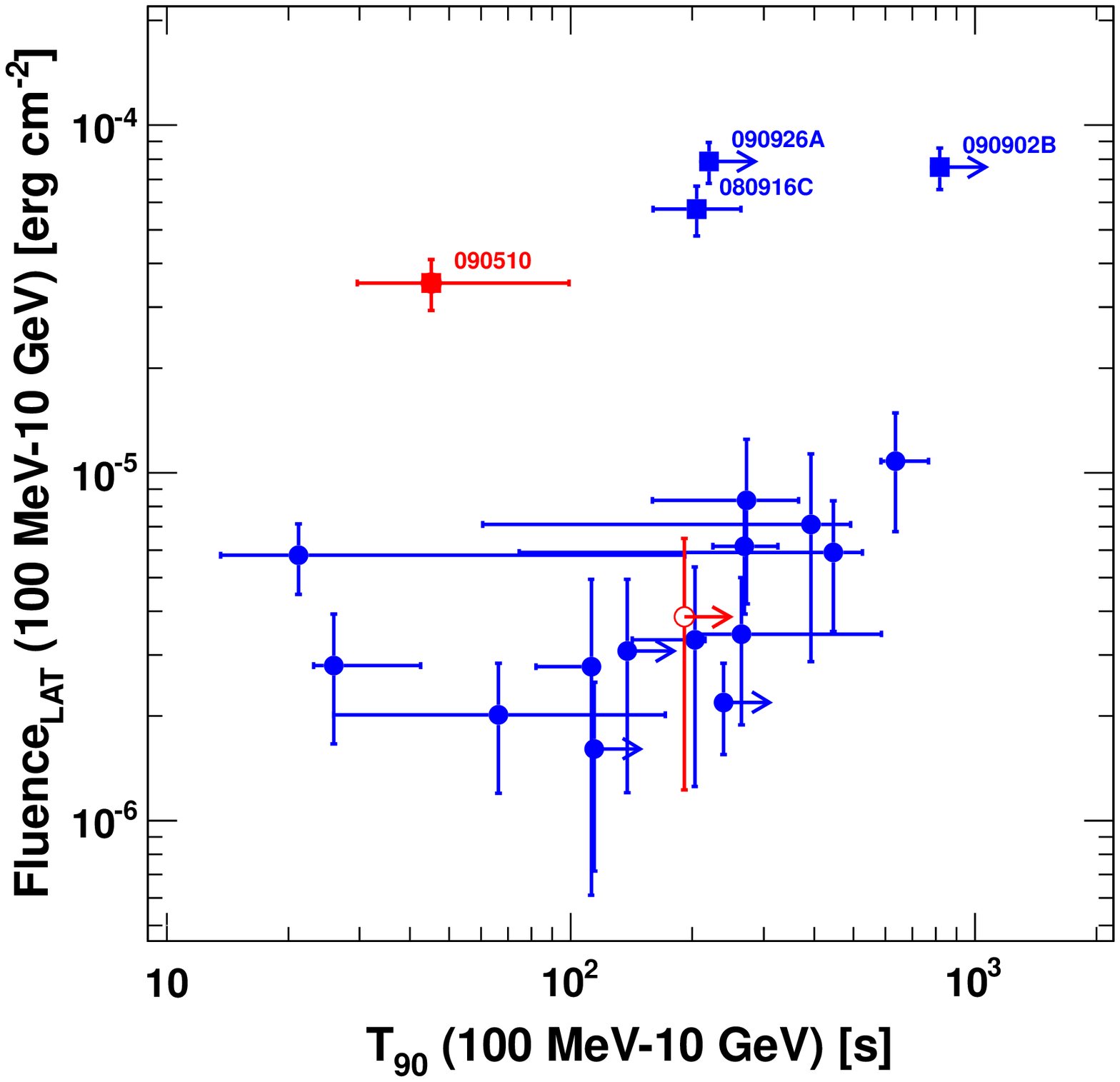}
\caption{Flux (left-hand column) and fluence (right-hand column) in the 100~MeV--10~GeV energy range for the ``GBM'' (top row) and ``LAT'' (bottom row) time intervals as functions of the durations of these intervals. The symbol convention is the same as in Fig.~\ref{fig_Onset}.}
\label{fig_FluxFluence_Duration}
\end{center}
\end{figure*}

Figure~\ref{fig_FluxFluence_Duration} shows the flux and fluence
measured by the LAT in the ``GBM'' (top two panels) and ``LAT'' (bottom two panels)
time intervals as a function of the durations of these time intervals
(i.e., GBM and LAT T$_{90}$ respectively).
The fluxes and fluences presented in these figures are for the 100~MeV--10~GeV energy range.
Interestingly and as can be seen in the bottom right panel of Fig.~\ref{fig_FluxFluence_Duration}, within the first 3 years of operations the LAT has detected 4 very-high fluence bursts GRBs 080916C, 090510, 090902B, and 090926A that are outliers with respect to the main distribution of the LAT-detected GRBs.
We will revisit these hyper-fluent bursts in \S~\ref{subsec_energetics}, where we discuss the energetics of \Fermi-LAT detected GRBs.


%

\subsubsection{LAT Localizations}
We evaluate localizations from the LAT for all GRBs detected by the
maximum likelihood analysis by searching for the maximum of the TS map
according to the procedure described in
\S~\ref{subsubsec_likelihood_localization}. We present our results in
Table~\ref{tab_localizations}, in which we report the position of the
maximum of the TS map (i.e., the LAT localization) along with its 68\%,
90\%, and 95\% statistical errors.

\subsubsection{High-Energy Photon Events}
\label{subsubsec_res_HEevents}
We report the energies and arrival times of a set of interesting high-energy photons
that, according to our likelihood analysis (as described in \S~\ref{susubbsec_event_probability}), have a high probability (P$>$0.9) of being associated with the GRBs.
Specifically, we give information for the following events:

\begin{itemize}

\item The highest-energy Transient class LAT $\gamma$-ray in the ``GBM'' time window (Table~\ref{tab_energymax_gbm});

\item The highest-energy Transient class LAT $\gamma$-ray in the interval starting from GBM
T$_{95}$ and ending at the end of the ``EXT'' window
(i.e. from the end of the measured duration in the GBM data up to the end of the LAT measured duration, Table~\ref{tab_energymax_ext});

\item The highest-energy Transient class LAT $\gamma$-ray detected in the time-resolved likelihood analysis (Table~\ref{tab_energymax_all}).

\end{itemize}

\begin{figure}[ht!]
\begin{center}
\includegraphics[width=1.0\columnwidth,trim=10 10 50 20, clip=true]{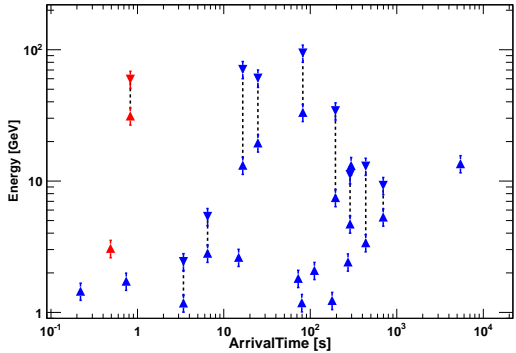}
\caption{Observed (upward triangles) and rest frame (downward triangles) energy and arrival time for highest-energy events associated with long (blue) and short (red) LAT detected GRBs. Vertical dashed lines connect the observed and the rest frame energy for the same burst. Data points are from Table~\ref{tab_energymax_all}.}
\label{fig_EnergyMax}
\end{center}
\end{figure}

The results are shown in Tables \ref{tab_energymax_gbm},
\ref{tab_energymax_ext}, and \ref{tab_energymax_all}.
These results show that the detection
     of high energy events with GRB point source probabilities P$>$0.9 is
     not strongly correlated with features in the GBM light curve.  In a few
     cases, such as GRB 090510, such events are coincident with bright pulses
     in the GBM light curve, but more often the most energetic event is detected after the intense low-energy emission, as with the 33.39~GeV event detected at
  T$_{0}$+81.75~s from GRB\,090902B, which is the highest energy ever observed from a burst.
GRB\,100728A is particularly interesting since a 13.54~GeV event was detected $\sim$90 minutes after the trigger time.
This is the only case in which we observe such a late event, and it can potentially confirm that high-energy $\gamma$-rays can arise very late in time, as
observed from GRB\,940217 by EGRET \citep{1994Natur.372..652H}.
On the other hand, GRB\,100728A is not significantly detected at the time the highest-energy event is observed (similarly to GRBs\,090217 and 100116A
  reported in Table~\ref{tab_energymax_all}), thus the probability P=0.987 that the 13.54~GeV event is associated with the burst must be taken with
  caution.
Considering the trials factors, this probability would be further reduced, weakening the case for hours-scale high-energy emission.
A detailed analysis of the probability corrected by the trials factors would be non-trivial as the background strongly varies as a function of the location in orbit, and it is beyond the scope of this paper.
Figure \ref{fig_EnergyMax} shows the energies and arrival times for the highest-energy $\gamma$-rays associated with LAT GRBs. The estimated errors are computed from the energy dispersion in the Instrument Response Functions and it is of the order of 10\% for energies $>$1GeV. When possible, we also indicate the source frame energy.

\subsubsection{Temporally Extended Emission}
\label{subsubsec_temporallyextended_ch4}

\begin{figure}[ht!]
\begin{center}
\includegraphics[width=1.0\columnwidth,trim=10 10 50 20, clip=true]{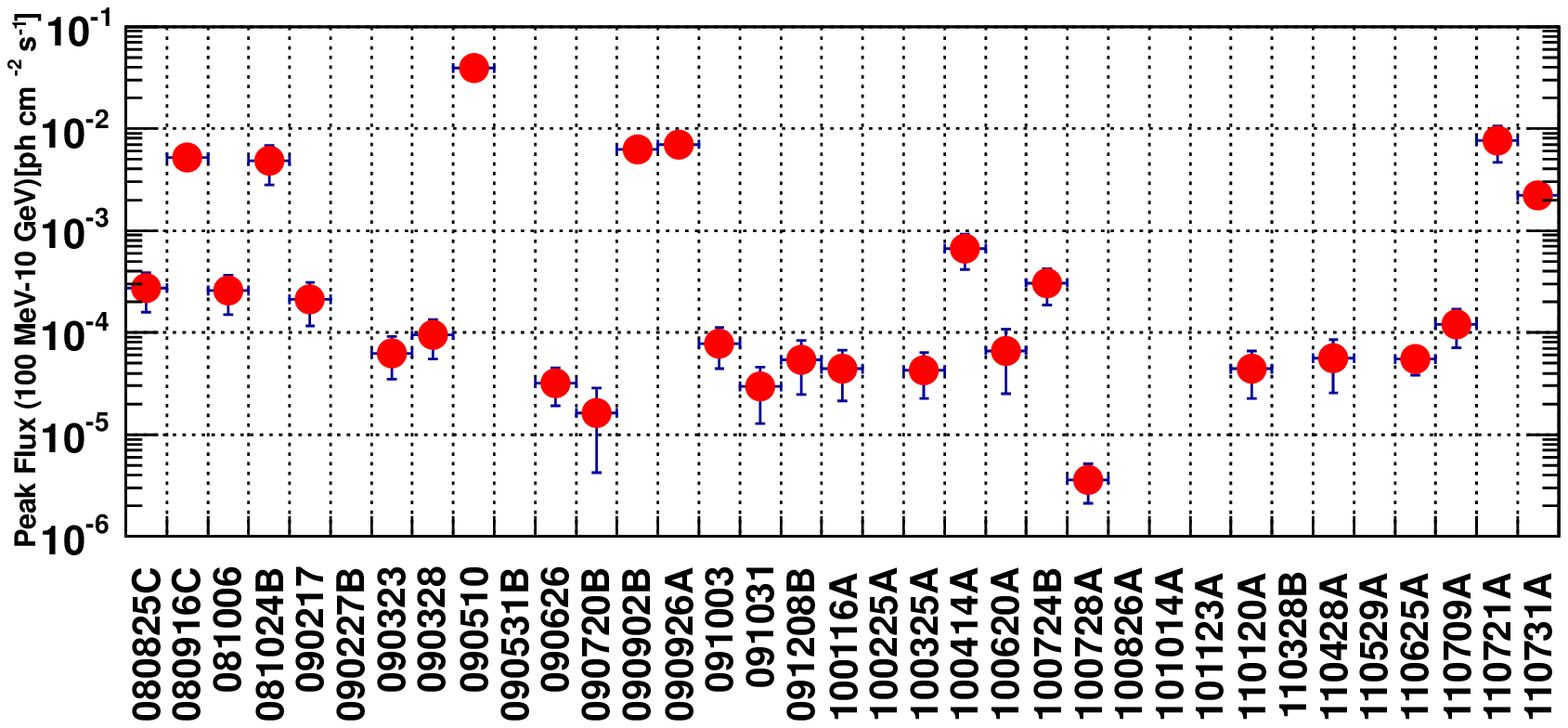}
\includegraphics[width=1.0\columnwidth,trim=10 10 50 20, clip=true]{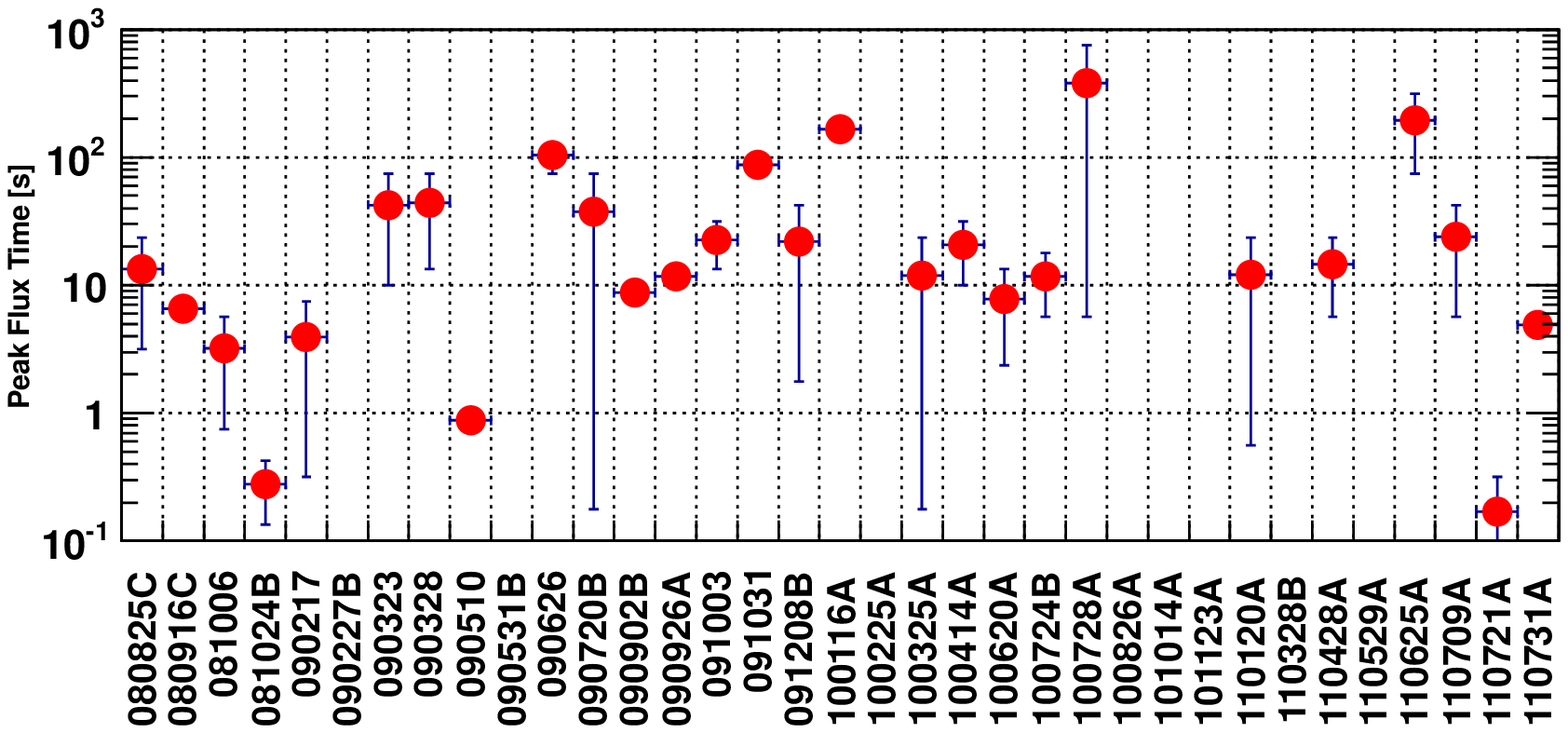}
\includegraphics[width=1.0\columnwidth,trim=10 10 50 20, clip=true]{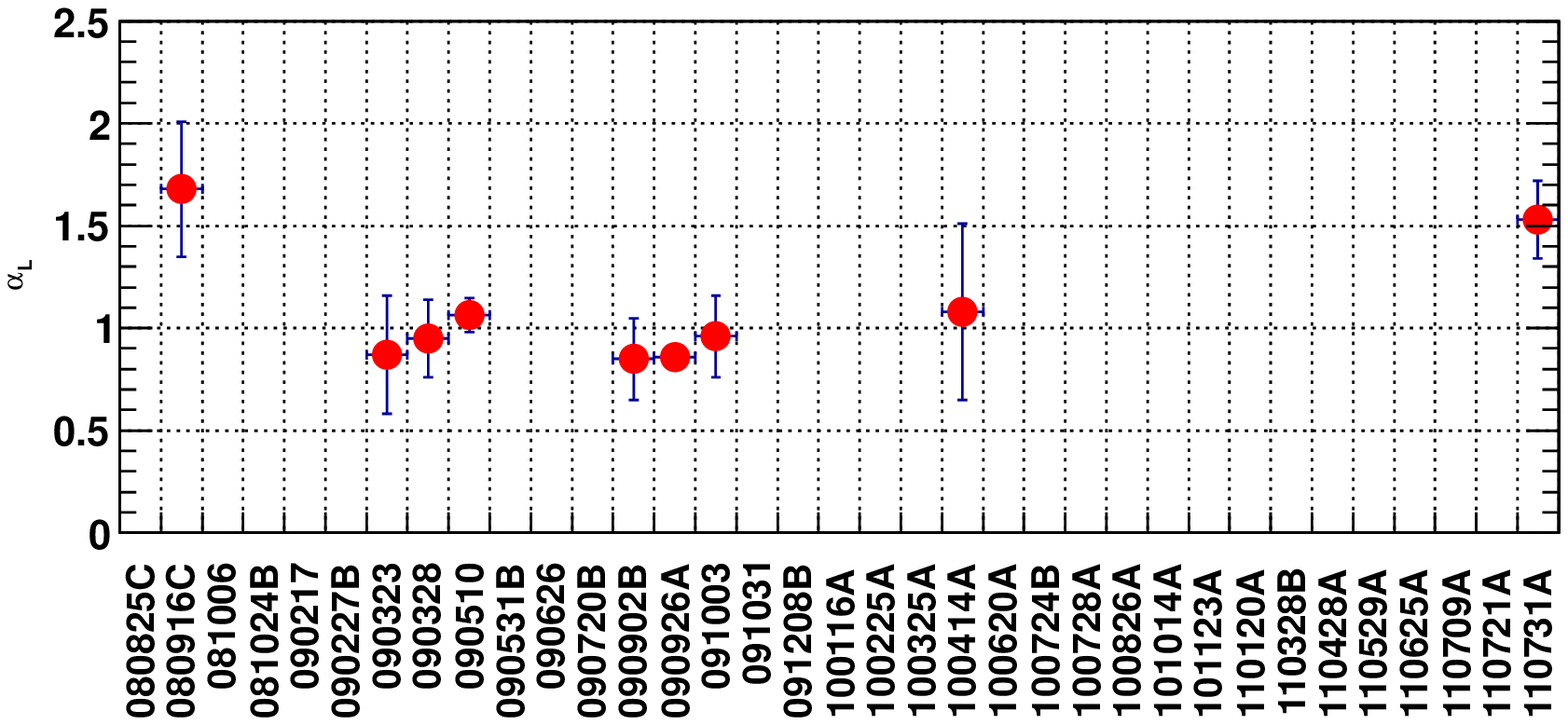}
\caption{Quantities characterizing extended high-energy emissions detected by the LAT. Top: peak flux, middle: time of the peak flux and bottom: temporal-decay index $\alpha_{L}$.}
\label{fig_ExtraComponent}
\end{center}
\end{figure}

To study the temporal decay of the extended emission detected by the LAT, we utilized the time-resolved analysis described in \S~\ref{subsec_analysis_sequence}. We first visualized
any detected extended emission using flux light curves (shown in
Appendix~\ref{sec_fermi_lat_grb}), and then calculated the peak-flux value
$F_{p}$ and the time of the peak flux $t_{p}$, quantities shown in the two top panels of Fig.~\ref{fig_ExtraComponent}. In the time-resolved analysis we adaptively changed the size of the time bin width in order to significantly detect the source, so $F_{p}$ corresponds to the average flux in the time bin of the maximum, and as a result it is more precise, (i.e., with a smaller uncertainty) for bright GRBs than for faint GRBs.

The 4 most luminous bursts detected by the LAT have some of the highest peak fluxes in the ensemble, all exceeding 10$^{-3}$\,cm$^{-2}$\,s$^{-1}$. Among the rest of the bursts, GRBs 081024B and 110721 also have notably high peak fluxes.
GRB\,100728A was at the edge of the FoV at the time of the GBM trigger and was detected only at later times. It has by far the lowest peak flux of all GRBs, at least an order of magnitude lower than the rest of the population; however, its value is possibly affected by large systematic errors.

\begin{figure}[ht!]
\begin{center}
\includegraphics[width=1.0\columnwidth]{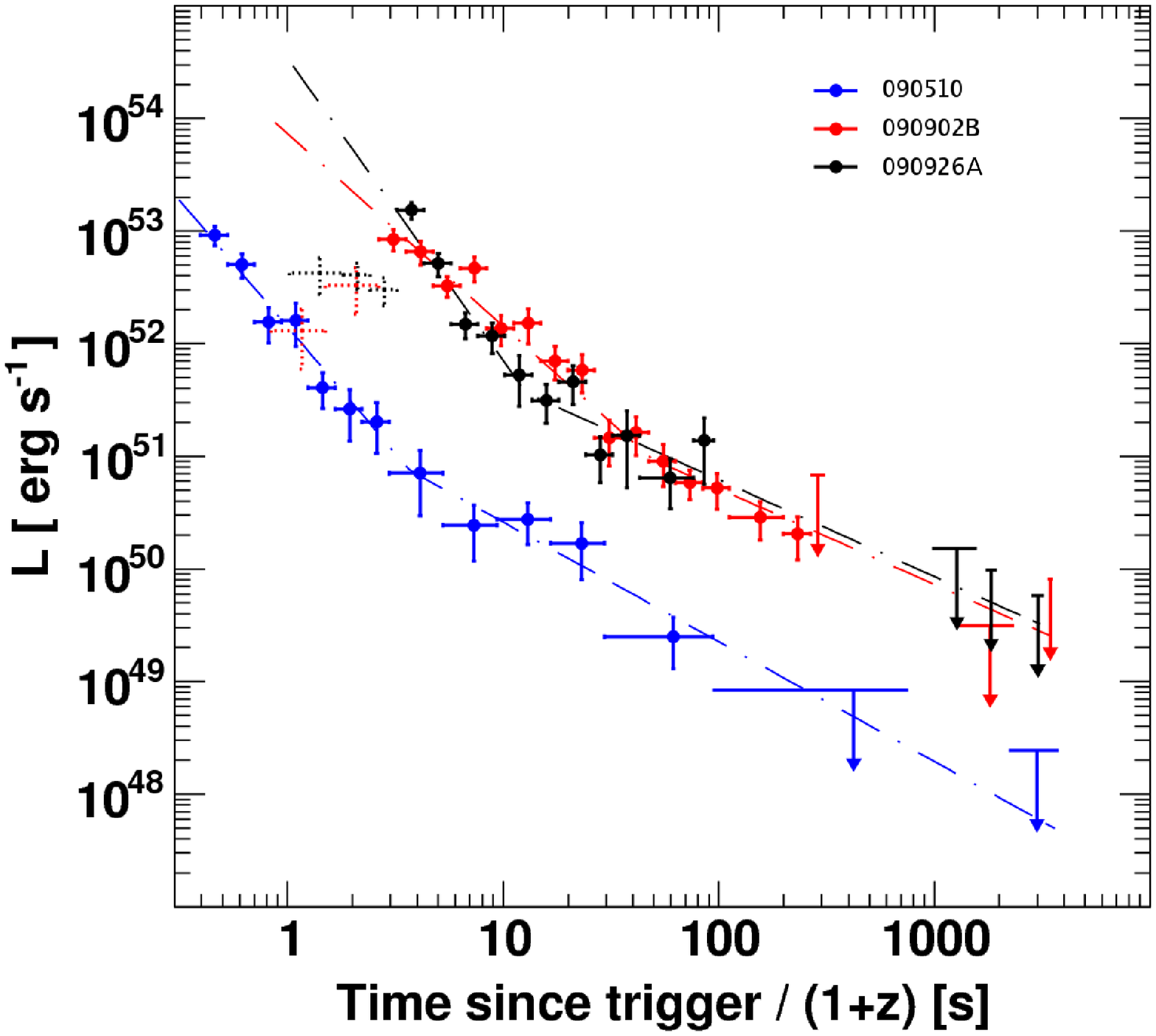}
\includegraphics[width=1.0\columnwidth]{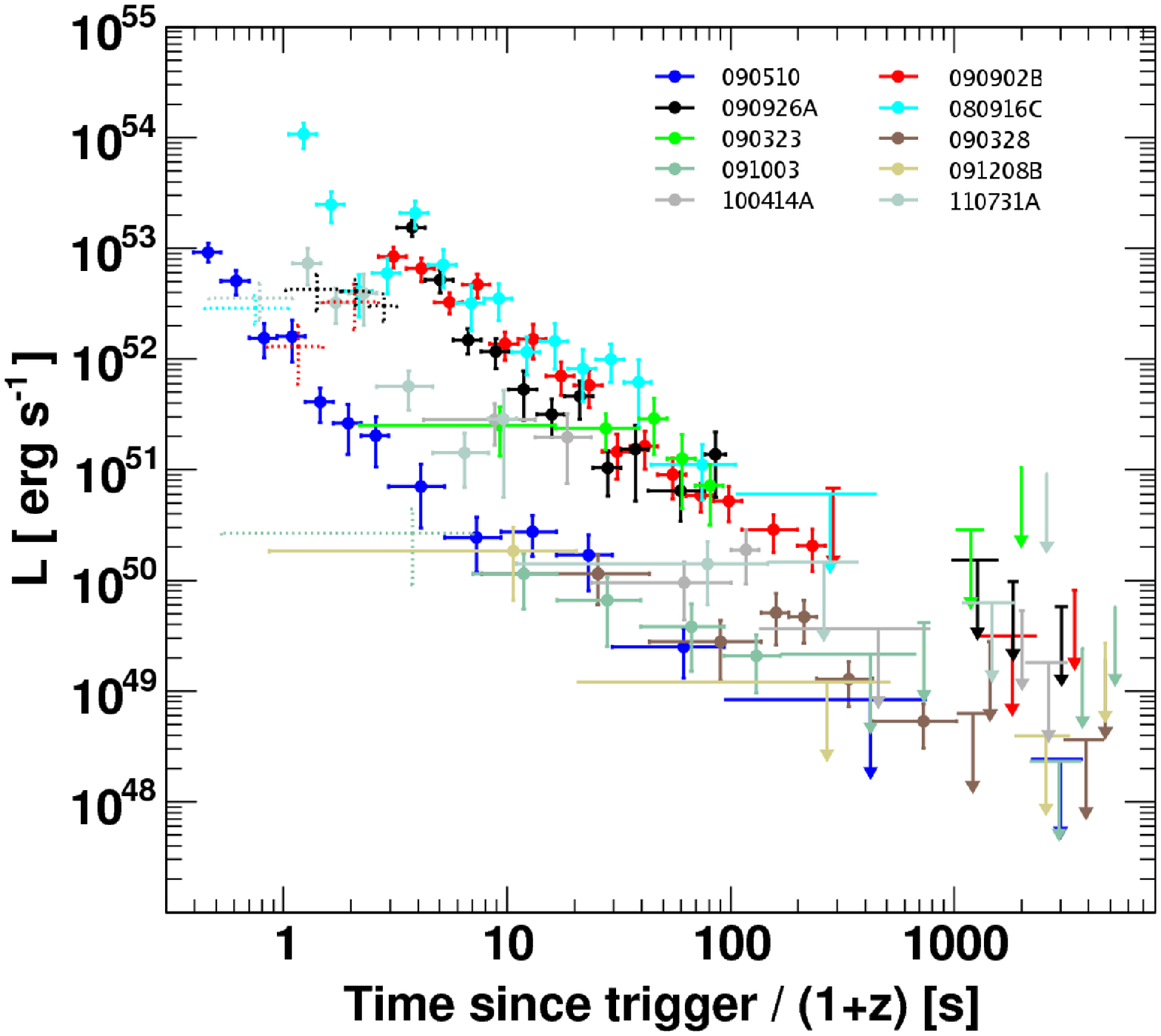}
\caption{Top: The decay of the luminosity L with time measured in the rest
frame for the 3 GRBs in which we detect a significant time break. Dashed-dotted lines
are the best fits of the broken power law model to each GRB, while dashed
crosses are the luminosities before the peak times, which have
not been used in the fits (see text). Bottom: the same quantities for all the
GRBs with detected extended emission.}
\label{fig_temporalBreaksLuminosity}
\end{center}
\end{figure}

We also applied the methods described in \S~\ref{subsec_analysis_sequence} to
the subsample of GRBs with detected extended emission. We
detected temporal breaks in the decay of the extended emission of three bright GRBs: GRB\,090510,
GRB\,090902B and GRB\,090926A.
In the top panel of Fig.~\ref{fig_temporalBreaksLuminosity} we show their
luminosities as functions of rest-frame time, as well
as the best fitting broken power-law models.
The later points in the light curves
are very important to constrain the break, but they also would be the most
affected by any unaccounted-for systematic uncertainties arising, for example, from the background estimation or the exposure calculation. In the bottom panel of Fig.~\ref{fig_temporalBreaksLuminosity}
we again report the luminosity as a function of the rest-frame time, but for all
the GRBs in the subsample. In Table~\ref{tab_extended} we report the results of this analysis.
For the three GRBs with temporal breaks we report the decay index starting from the peak flux and
before the break $\alpha_{1}$, the decay index after the break $\alpha_{2}$, and the break time $t_{b}$.
For all other GRBs, we report the decay index for the whole extended
emission starting from the peak flux, and the decay index for the light curve
starting from the end of the low-energy (GBM) emission.

Referring to Table~\ref{tab_extended}, we also define the ``late-time decay index''
$\alpha_{L}$, which corresponds to the decay index measured after the GBM T$_{95}$
($\alpha_{L}=\alpha$) for all GRBs except the three for which we detect
temporal breaks, for which it corresponds to the decay index after the break
($\alpha_{L}=\alpha_{2}$). In the third panel of Fig.~\ref{fig_ExtraComponent}
we report $\alpha_{L}$ for all of the GRBs of the subsample.

\subsection{Joint GBM-LAT Spectral Fits}
\label{sub_joint_fits_results}

For each GRB detected with the LAT we performed joint GBM-LAT spectral analyses in
two time intervals, following the procedure described in
\S\ref{subsec_lat_gbm_spectral_analysis}. We started by analyzing
data taken in the ``GBM'' time window for all detected GRBs. The
results of this analysis are presented in Tables \ref{tab_bandCrisis} and \ref{tab_JointFitGBMT90}.
Since the emission at energies $>$100~MeV is delayed with
respect to that at lower energies, we also performed
a spectral analysis in the time interval between the first Transient class $\gamma$-ray detected by
the LAT within the energy-dependent ROI of the GRB and the GBM T$_{95}$, in order to maximize the signal-to-noise ratio at high energies ($E > 100$~MeV).
We report the results of this analysis for all the bursts detected by the LAT in Table \ref{tab_JointFitLATPROMPT}.
The first table (Table \ref{tab_bandCrisis}) summarizes the model that best fits the data for each GRB, ordered by fluence. We also report the off-axis angle, which is a proxy for the detection efficiency of the LAT for equal exposure time (high off-axis angle means low efficiency).
Tables \ref{tab_JointFitGBMT90} and \ref{tab_JointFitLATPROMPT} contain
three sets of columns: the main component section, the additional
component section, and two columns with the total fluence (in the
10~keV--10~GeV band) and the value of PG-stat (see \S\ref{subsub_joint_spectral_fit}) with the number of
degrees of freedom. Each spectrum is modeled by one main
component (either a Band model or a Comptonized model or a logarithmic parabola) and one or two additional components (power-law and/or exponential cutoff) when needed (see again \S\ref{sub_sub_spectral_models}). 
The SBPL model does not provide the best fit for any GRB in our sample, so we do not include it in either Table \ref{tab_JointFitGBMT90} or ~\ref{tab_JointFitLATPROMPT}. Only the columns corresponding to the
parameters of the components used in the best fitting model are entered. When a spectrum requires additional components, we report separately the fluence corresponding to the main component and the fluence corresponding to the additional components.

To elaborate on the table entries, consider the results of the time integrated analysis reported in Table~\ref{tab_JointFitGBMT90}: the first entry refers to the spectrum of GRB\,080825C, which is best described by a Band model, thus only the columns referring to the parameters of the Band model are filled, and only the total fluence is reported. On the other hand, the spectrum of GRB\,090926A is described by a Band model plus a power law with an exponential cutoff. Correspondingly, all columns for the parameters of those
components are filled, as well as the columns for the total fluence and the fluences for the first component (Band) and the second component (power law with an exponential cutoff), respectively. The spectrum of GRB\,100724B is
instead described by a Band model with an exponential cutoff, so all of
the corresponding columns are filled. Note that there are no
partial fluences reported in this case, since the exponential cutoff
is a multiplicative term. In the case of GRB\,110731A, we reported in Table~\ref{tab_JointFitGBMT90} both the Band-only fit and the Band plus power law fit, even if the extra component is not significant according to our criteria, since the power law is clearly detected in the other time interval as reported in Table~\ref{tab_JointFitLATPROMPT} and thus Band plus power law is arguably a more accurate model for the ``GBM'' time window as well.

Some bursts have been detected only by the LLE photon counting analysis since they were outside the nominal LAT FoV ($\theta>70$ deg, see Table \ref{tab_GRBs}) at the time of the trigger. These include GRB\,090227B, 100826A, 101123A, and 110625A.
GRB\,101014A was detected too close to the Earth's Limb at the time of the trigger, resulting in a very low exposure for the LAT due to the zenith-angle cut (see \S~\ref{subsection_cuts}). For these LLE-only detections, it is not possible to obtain a spectrum from LAT standard data, and so we use only GBM data.

\subsubsection{Extra components}
\label{subsub_extracomponents}
We found that four GRBs clearly require a power-law added to the Band spectrum in both time intervals that we studied. 
Two cases, GRB\,090510 and GRB\,090902B, are already known \citep{GRB090510:ApJ,090902B_PAPER}. 
The two additional cases are GRB\,080916C and GRB\,110731A. During the ``GBM'' interval for GRB\,080916C, we obtain  a value of PG-stat $S=519$ (with 356 d.o.f.) with the Band model alone, while we obtain $S=485$ (with 354 d.o.f.) adding an extra power-law. The value $\Delta S = 34$ is well above our detection threshold of 25 (see \S~\ref{subsub_modelselection}). It corresponds to a chance probability of $\simeq 1\times10^{-5}$ or possibly lower (see Fig.~\ref{fig_deltaS}).
The possibility for an extra component was already considered in our first publication on this GRB \citep{GRB080916C:Science}, but the significance of the power-law was not high enough to claim a firm detection. Now, thanks to a better understanding of the background in the LAT with the use of the BKGE, and a better calibration of the GBM instrument, we obtained convincing evidence for such a claim. We also detect an extra component in GRB\,110731A, as published in \cite{GRB110731A:Fermi}.
In the ``GBM'' time interval, the significance of this component is below our threshold, but in the LAT time interval, with a better signal-to-noise ratio, we obtain $\Delta S=42$. This result is fully compatible with what we already published.

\begin{figure}[!ht]
\begin{center}
\includegraphics[width=\columnwidth]{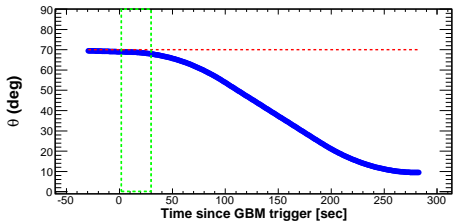}
\caption{Position of GRB\,100414A in the FoV as a function of the time since the GBM trigger. The y-axis is the off-axis angle. The green box is the GBM T$_{90}$ while the red dashed line represents the edge of the FoV.}
\label{fig:100414A-pointing}
\end{center}
\end{figure}

We also find an extra component in GRB\,100414A, but in this case we highlight some possible problems with the analysis.
We refer to Fig.~\ref{fig:100414A-pointing} that shows the off-axis angle of the GRB as a function of the time since the GBM trigger.
During the GRB prompt emission, this GRB was at the edge of the FoV of the LAT, where the effective area is small.
In addition, the ARR maneuver was particularly fast in terms of angular speed for this GRB and happened during the GBM T$_{90}$, resulting in rapidly changing backgrounds and effective area at the source location, which could create large and difficult to evaluate systematic uncertainties. Indeed, in the ``GBM'' time interval the spectrum is better described by a Comptonized model with an additional power-law, while in the LAT time interval the statistically preferred model is a Band function. In this case we cannot significantly claim the detection of the extra power-law component.

We confirm the detection of a cutoff around 1.5~GeV in the extra component of GRB\,090926A as previously published by \citet{GRB090926A:Fermi}, and we also significantly detect a new cutoff at lower energies in GRB\,100724B.
For the latter, considering again the ``GBM'' time interval, we find $S=977$ with 469 d.o.f. using the Band model, while adding an exponential cutoff we find $S=734$ with 468 d.o.f. The value $\Delta S=243$ is well above our threshold $\Delta S=28$. Discussion of the physical implication of these findings is outside the scope of the present paper. \citet{GRB110731A:Fermi} found a hint for another cutoff at high energy with a significance of $\sim 4 \sigma$ in the time interval starting from the LAT T$_{05}$ and ending at the GBM T$_{95}$. We refer the reader to that paper for details.


\section{Discussion}
\label{sec_discussion}

In this section, we describe the emergent properties of LAT-detected GRBs revealed by this study.

\subsection{Broadband spectroscopy}
\label{subsec_Broadband_spectroscopy}
\subsubsection{A Band model crisis?}
Before the launch of \Fermi, GRBs were mainly studied in the energy range from a few~keV to a few~MeV, with the catalog of BATSE \citep{Kaneko:06,2008ApJ...677.1168K}, constituting the largest sample available to date.
Several spectroscopic analyses have been performed on that sample, showing that most of the GRB spectra are well described by a Band model, a Comptonized model, or a smoothly broken power-law (SBPL) model \citep{Preece:00}.
LAT-detected GRBs are bright in the GBM energy band, which is very similar to the BATSE band, and thus we can compare our detection statistics with those found in the bright BATSE sample by \citet{Kaneko:06}.
In Table~\ref{tab_bandCrisis} we report all LAT-detected GRBs, ordered by fluence, and the model that best describes the spectrum over the GBM time interval. For convenience we also report their off-axis angles $\theta$ at the trigger times.
We exclude GRBs outside of the nominal FoV ($\theta>70$\de). We also exclude GRB\,101014A which was too close to the Earth limb to allow a spectroscopic study.

\citet{Kaneko:06} found that the spectra of $\sim$85~\% of the brightest 350 BATSE GRBs are well described by a Band function,  while we find that 70\% of LAT-detected GRBs are well described by either a Band model or a Comptonized model, which is similar to a Band model with a very soft value of $\beta$.
Given the small size of our sample, the two fractions are very similar. Additionally, \citet{Kaneko:06}  found that 5\% of BATSE GRBs require the more complex SBPL model, while no LAT-detected GRB requires it. Again, this is very likely to be due just to the small size of our sample.

On the other hand, Table~\ref{tab_bandCrisis} shows that the spectra of all of the brightest bursts inside the LAT FoV present significant deviations from a Band function, requiring additional components. Other GRBs, observed with low $\theta$ angle and correspondingly high effective area, show deviations as well. 
The phenomenological Band model, implemented for BATSE GRB observations up to a few~MeV, does not seem to describe bright or well-observed LAT-detected GRBs sufficiently.

For each GRB with a very high signal-to-noise ratio in the LAT data, we find that the Band model needs to be supplemented with additional components or modified with a cutoff. There is no common recipe to fit all \Fermi~GRBs:
for the bright GRBs~090510 and 090902B, an additional power-law component, extending from low to high energies is required; for GRB\,100724B a cut-off in the high energy spectrum is needed in order to explain the rapid drop-off of the flux at high energies; the case of GRB\,090926A is even more complex, with both a power-law and a exponential cut-off required to describe the spectrum. Other works \citep{Guiriec2011,ZhangBing2011} use a thermal component added to the Band function.
This difficulty arises thanks to the greatly broadened energy coverage provided by  \Fermi with respect to BATSE, and accurate GRB spectroscopy in the \Fermi era requires improved broad band modeling.

\subsection{Energetics}
\label{subsec_energetics}

\citet{Cenko+11} and \citet{Racusin+11} have studied the energetics of the afterglows of LAT-detected GRBs and concluded that they are among the most luminous afterglows observed by {\it Swift}. We start our analysis by examining the properties of LAT-detected GRBs in the context of the prompt emission and compare the high-energy properties measured by the LAT to the low-energy properties measured by the GBM.

\subsubsection{Prompt Phase Energetics}
We first study the fluence, and then continue with the subsample of GRBs that have a measured redshift and examine intrinsic GRB quantities. Even though intrinsic properties are, by far, more interesting for understanding the physics, properties measured in the observer's frame (such as the fluence or the peak flux) are sometimes more instructive from the experimental point of view, as they can reveal observational biases and selection effects.

\begin{figure}[ht!]
\begin{center}
\includegraphics[width=1.0\columnwidth,trim=10 5 40 40,clip=true]{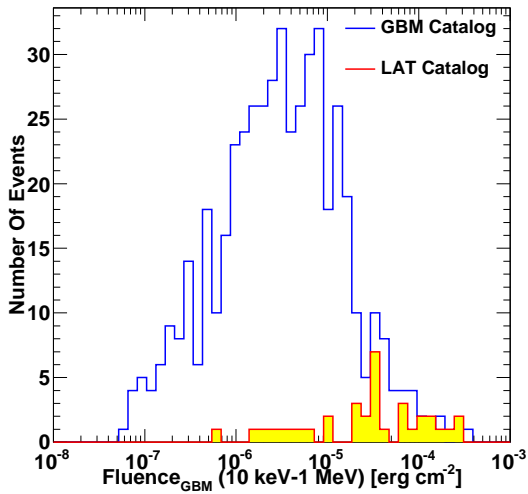}
\caption{Distribution of the energy fluences in the 10~keV--1~MeV energy range for the bursts detected by the LAT compared with the fluences in the same energy band for the entire sample of GRBs in the GBM spectral catalog \citep{Goldstein+12}.}
\label{fig_Fluence_Catalogs}
\end{center}
\end{figure}

In Fig.~\ref{fig_Fluence_Catalogs}, we compare the fluence measured by the GBM in the 10~keV--1~MeV energy band for the full GBM spectral catalog \citep{Goldstein+12} to the 10~keV--1~MeV fluence of LAT-detected GRBs. 
Since the LAT observations are photon-limited, the detection efficiency is directly related to the source fluence~\citep{2009ApJ...701.1673B}. This is in contrast to the GBM data, which are background dominated and the peak flux is a better proxy for the detection efficiency.

In general, LAT detected GRBs are among the brightest detected by the GBM, populating the right-hand side of the fluence distribution.
The brightest GRB in the GBM catalog is GRB\,090618 \citep{2009GCN..9535....1M}, also detected by AGILE (MINICAL and Super-AGILE) \citep{2009GCN..9524....1L} and {\it Swift}-BAT \citep{2009GCN..9512....1S}, but not detected by the LAT because it occurred outside its FoV ($\theta$=132\de). The second brightest GRB in the GBM catalog is the LAT-detected GRB\,090902B. More interestingly, there are a few cases of bursts that were not particularly bright in the GBM, yet were detected by the LAT, namely short GRBs 081024 and 090531, which have a relatively small fluences compared to the rest of the GBM-catalog bursts, mainly because of their short durations ($<$20\% and $<$30\% quantile of the distribution). The former was detected by the LAT up to $\sim$GeV energies \citep{LAT081024B}, while the latter was detected only at low energies by the LLE analysis. Note however that the published GBM catalog includes bursts only up to the beginning of 2010 July. Thus, it does not contain a significant part of our sample, and
in
particular GRB\,100724B, which has the highest fluence in the GBM energy range in our sample (see Table~\ref{tab_bandCrisis}).

\begin{figure}[ht!]
\begin{center}
\includegraphics[width=1.0\columnwidth,trim=10 5 40 40,clip=true]{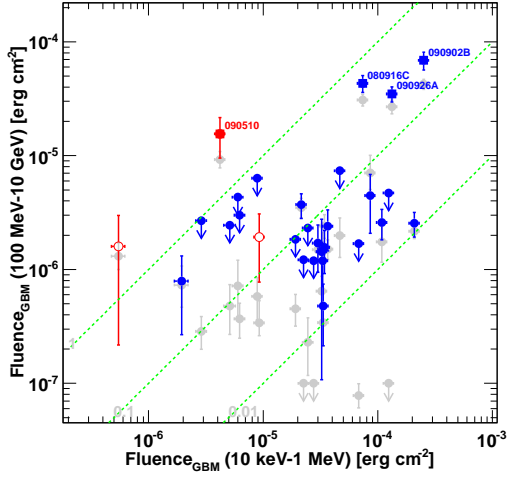}
 \includegraphics[width=1.0\columnwidth,trim=10 10 40 0,clip=true]{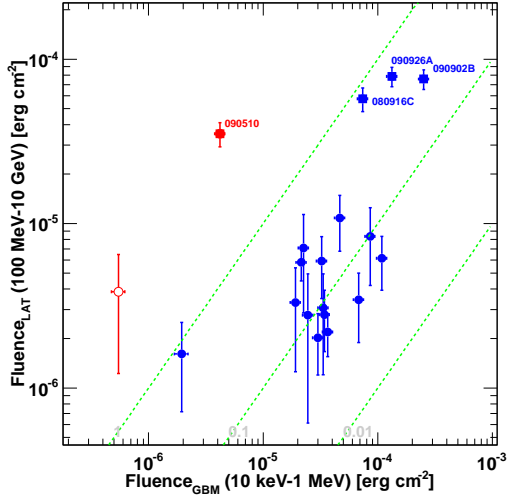}
\caption{Fluence measured by the LAT versus the fluence measured by the GBM in the ``GBM'' time window (top panel) and in the ``LAT'' time window (bottom panel). The three dashed lines denote the 100\%, 10\% and 1\% fluence ratios.  Colored symbols follow the convention of Fig.~\ref{fig_Onset}. Additionally, we also use gray circles for joint-fit results.}
\label{fig_FluenceFluence}
\end{center}
\end{figure}

\begin{figure}[ht!]
\begin{center}
\includegraphics[width=1.0\columnwidth]{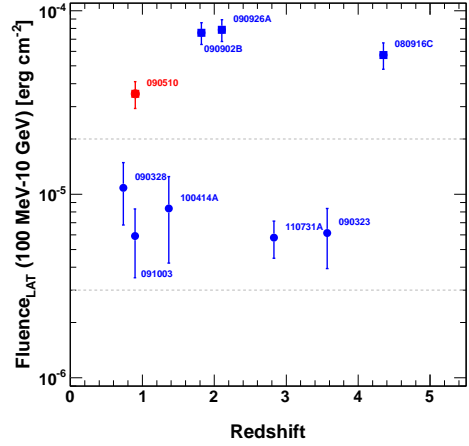}
\caption{Fluence measured by the LAT during the ``LAT'' time interval versus the redshift. The two dashed lines in this figure denote a fluence of 3$\times$10$^{-6}$\,erg\,cm$^{-2}$ and 2$\times$10$^{-5}$\,erg\,cm$^{-2}$, with the first number corresponding to an approximate empirical LAT detection threshold and the second simply denoting a minimum fluence for the four hyper-fluent bursts. The symbol convention is the same as in Fig.~\ref{fig_Onset}.}
\label{fig_redshift1}
\end{center}
\end{figure}

The top panel of Fig.~\ref{fig_FluenceFluence} shows the fluence measured by the LAT versus the fluence measured by the GBM in the ``GBM'' time window. The plotted GBM fluences were produced by the joint GBM-LAT spectral analysis in this study, in accordance with the best-fit spectral model described in Table~\ref{tab_JointFitGBMT90}.
LAT fluences calculated from the LAT-only maximum-likelihood analysis and from the joint GBM-LAT spectral fits are both shown in the figure.
Generally speaking, the agreement is good, however, for bright bursts the two methods produce results that are in slight disagreement. This arises because we use a two-component model in joint GBM-LAT spectral fits, with the low-energy component (a Band model or a Comptonized model) having a non-negligible contribution at high energy. Thus, both the photon index and the normalization for the power-law component are different with respect to the maximum-likelihood analysis, which uses a power law only.

The bulk of the LAT GRB population, primarily composed of long GRBs, has a ratio of high- (100~MeV--10~GeV) to low-energy (10~keV--1~MeV) fluence $\lta$20\%.
It is interesting to note that the three short LAT-detected bursts (red symbols in Fig.~\ref{fig_FluenceFluence}) have a greater ratio of high- to low-energy fluence than the bulk of the long-GRB population (blue symbols).
Two short bursts GRBs\,080825C and 090510 have the two highest ratios (over 100\%), and the short burst GRB\,090227B also has a relatively high ratio ($\sim$10\%). This reflects the well-known fact that short GRBs have harder spectra than do long duration bursts. 
On the other hand, since the high-energy emission typically lasts longer than the low-energy emission, and since in this plot the integration time is the same (the GBM T$_{90}$) for both axes, only part of the emission at high energies is included in the calculation of the fluence.
For this reason, we also integrate the fluence between 100~MeV and 10~GeV over the full LAT T$_{90}$ time window
and in the bottom panel we compare this quantity with the fluence as measured by the GBM during the GBM time window.  In this way, we better account for the energetics in the LAT energy range.
The LAT measurements in this panel were all derived from the likelihood analysis of this study. Because we were not able to measure durations in the LAT energy range for all bursts, this panel has fewer entries than in the top panel. Similarly to the above, short GRBs appear considerably more efficient at radiating at high energies than at low energies.

In both panels of Fig.~\ref{fig_FluenceFluence}, we can see the four hyper-fluent LAT bursts, GRBs 080916C, 090510, 090902B, and 090926A, having evidently greater emission in the LAT energy range compared to the rest of the GRB population. The discrepancy increases when comparing the high-energy emission measured in the generally-longer LAT time window to the low-energy emission measured in the GBM time window, a result of the bright extended high-energy emissions of these four bursts.

It is worth examining whether the four brightest LAT bursts appear bright because they are systematically closer to us compared to the rest of the GRB population. As can be seen in Fig.~\ref{fig_redshift1}, which shows the fluence in the LAT energy range and the LAT time window versus the redshift, this is not the case. In the figure we denote an empirical LAT-detection threshold, for which we caution the reader that since the minimum fluence at which the LAT can detect a GRB depends on the position of the GRB in the LAT FoV, as well as on the intrinsic properties of the GRB (photon index, duration, etc.), this threshold is just a crude estimate for reference.

To quantify the energy release at the source in some source-frame energy range E$_{1}$--E$_{2}$, we compute the isotropic equivalent energy E$_\mathrm{iso}$ as:
\begin{equation}
\mathrm{E_{iso}}= 4\,\pi\,d_L(z)^{2}\,\frac{S(E_1,E_2,z)}{1+z},
\label{eiso1}
\end{equation}
where $d_L(z)$ is the luminosity distance of a source at redshift $z$, and $S(E_1,E_2,z)$ is the fluence of the source integrated in the source frame energy range E$_{1}$ and E$_{2}$:
\begin{equation}
S(E_1,E_2,z)=\int_{E_{1}/(1+z)}^{E_{2}/(1+z)} E\,\frac{dN(E)}{dE} dE,
\label{eiso2}
\end{equation}
with $\frac{dN(E)}{dE}$ describing the spectral model. The choice of the energy band used to compute the isotropic energy is important and requires some discussion. In order to calculate the bolometric isotropic energy, the energy band must be as broad as possible. On the other hand, the calculation in principle should include only the portion of the spectrum that has been directly measured (i.e., constrained by the data) or a potentially-inaccurate extrapolation would be required.
Considering the spectral coverage of the two instruments onboard  \Fermi, we chose to integrate between the E$_{1}$=1~keV and E$_{2}$=10~GeV source-frame energies. We start at 1~keV source-frame, which corresponds to a few~keV observer-frame and is slightly outside of the GBM energy band, to make comparisons with some studies already in the literature.
In addition, we compute the isotropic equivalent energy in a narrower band (1~keV--10~MeV), covering mainly the energy range of the GBM detectors. The latter choice allows us to directly compare our results with those of previous works, namely \citet{Amati+02,Racusin+11} who adopted a source-frame range between 10~keV and 10~MeV, \citet{Amati+06,Butler+07} who adopted a slightly broader source-frame range extending from 1~keV to 10~MeV, and \citet{Cenko+11} who used an observer-frame range between 1~keV and 10~MeV.


\begin{figure}[ht!]
\begin{center}
\includegraphics[width=1.0\columnwidth,trim=10 10 30 0,clip=true]{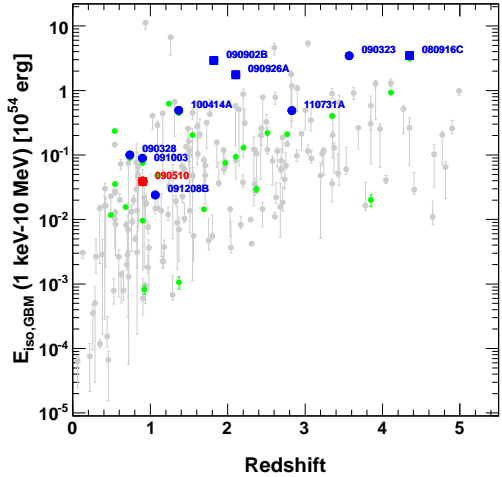}
\caption{Isotropic energy in the 1~keV--10~MeV energy range of LAT-detected GRBs (blue/red symbols) compared with Swift GRBs \citep{Butler+07} (grey symbols) and GBM GRBs \citep{Goldstein+12} (green). (Blue/red) squares denote the LAT-detected GRBs (with a measured redshift).}
\label{fig_eiso1}
\end{center}
\end{figure}

In Fig.~\ref{fig_eiso1} we plot E$_\mathrm{iso}$ in the 1~keV--10~MeV energy range versus the redshift in the prompt (``GBM'') time interval.
The energy range matches that of previous works (\citet{Butler+07} for Swift bursts and  \citet{Goldstein+12} for GBM bursts), allowing direct comparisons of E$_\mathrm{iso}$. For a given redshift, LAT-detected GRBs are generally brighter than the average burst in agreement with the findings from other works \citep{Cenko+11,Racusin+11}. We note that although GRBs\,110731A and 090510 have a moderate 1~keV--10~MeV E$_\mathrm{iso}$, they have been detected by the LAT. For these two bursts, the observational conditions were very favorable for detection, since they were nearly on-axis for the LAT at the times of the GBM triggers (13\ded6 for GRB\,090510 and 3\ded4 for GRB\,110731A off-axis angles).

Before proceeding, we would like to make an important point concerning the definition of ``bolometric'' luminosity of the prompt phase for GRBs. Before  \Fermi, the properties of prompt spectra of GRBs were known up to $\sim$~MeV energies, and there was no way to account for the higher-energy portion of the spectrum ($>$10~MeV) in the total energy budget. This is reasonable as long as the high-energy emission does not constitute a significant part of the total emitted energy.
Using LAT detections of GRBs, it has been discovered that extra power-law components are more common in GRBs compared to what was previously thought. More importantly, even if the high-energy emission can last longer than the usual~keV-to-MeV emission, in some cases (GRBs 090510, 090902B, 090926A) it contributes significantly during the prompt phase. These two considerations suggest that the total energy budget at high energies can be an important fraction of the total energy reservoir.

\begin{figure}[ht!]
\begin{center}
\includegraphics[width=1.0\columnwidth,trim=10 10 30 0,clip=true]{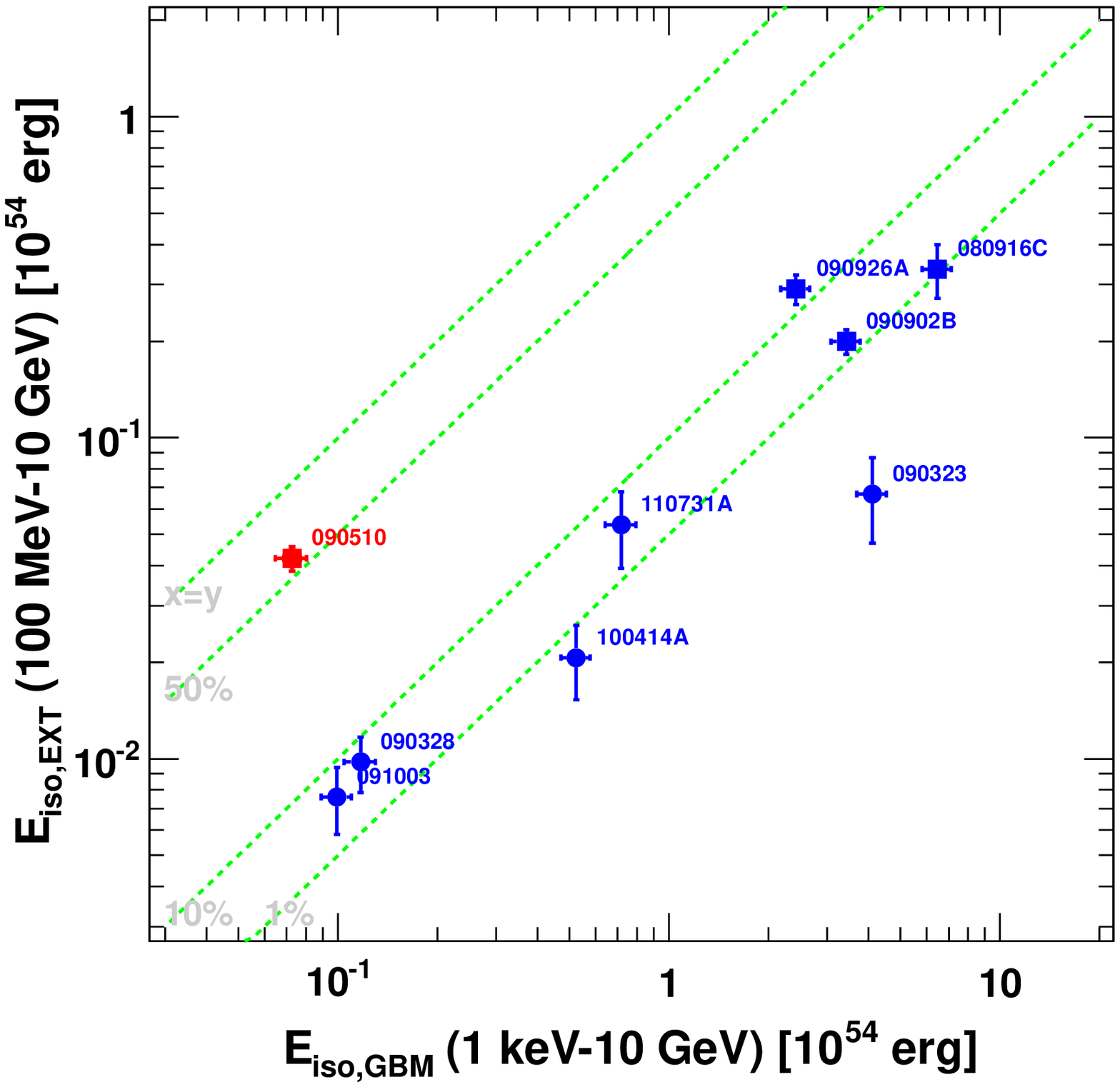}
\includegraphics[width=1.0\columnwidth,trim=10 10 30 0,clip=true]{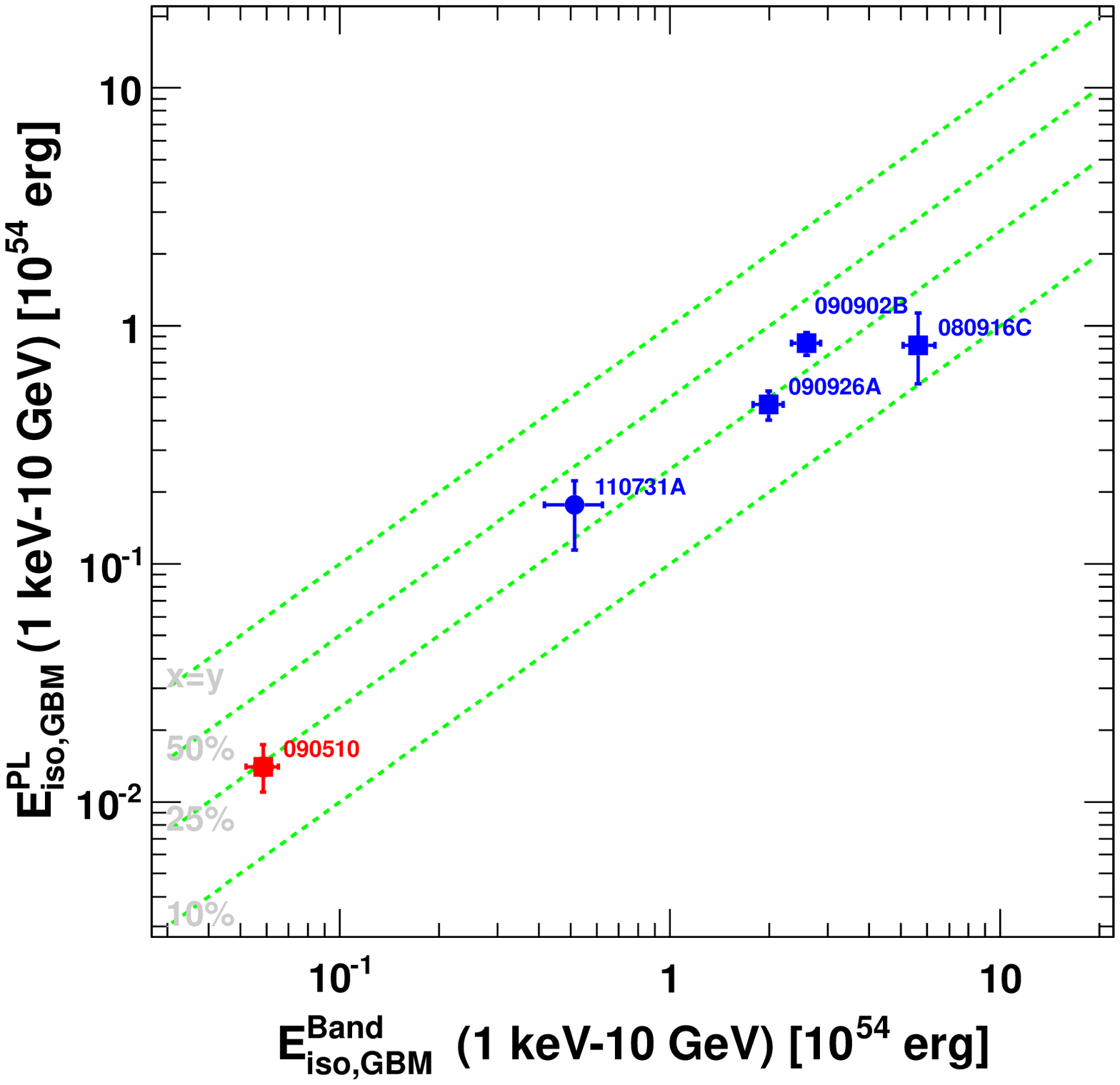}
\caption{Top: Isotropic equivalent energy in the 100~MeV--10~GeV versus the 1~keV--10~GeV energy range.
Bottom: Radiated energy corresponding to the power-law spectral component versus that corresponding to the Band component.
 The symbol convention is the same as in Fig.~\ref{fig_Onset}.}
\label{fig_eiso2}
\end{center}
\end{figure}

In Fig.~\ref{fig_eiso2} (top panel) we try to address this issue by plotting the amount of energy radiated by the source between 100~MeV and 10~GeV during the temporal extended emission compared to that radiated in the wider 1~keV--10~GeV energy range in the ``GBM'' time interval.
As can be seen, the fraction of energy radiated in the form of high-energy $\gamma$ rays during the temporal extended phase is typical $\lta$10\% of the total energy radiated during the prompt phase.
The short GRB\,090510 has an especially high fraction of $\sim$50\%. 

For the few bursts for which we can significantly separate the contributions from the extra component (power law) and the main component (the ``Band'' model), we can calculate the fraction of the energy during the prompt emission that is associated to each of these two spectral components.
In the bottom panel of Fig.~\ref{fig_eiso2} we show the emitted energy corresponding to each component for the ``GBM'' time interval.
As shown, the energy radiated during the prompt emission by the power-law component is between 10\% and 50\% of the energy radiated by the Band component.
The numerical results of this analysis can be found in Table~\ref{tab_eiso_components}.

\subsubsection{Highest Energy Photons}
Events with source-frame-corrected energy up to 50-100~GeV have been measured in GRBs by the LAT, including from high-redshift GRBs (up to z=4.35 from GRB\,080916C \citealt{Greiner+09}).
In order to produce $\gamma$ rays of such high energies within the first few seconds of the burst, particle acceleration must be efficient in a GRB. Internal-opacity constraints also indicate that these high-energy-photon detections require large bulk Lorentz factors for the jet.
Moreover, high-energy $\gamma$ rays from high-redshift GRBs offer a valuable tool for measuring the opacity of the Universe due to interaction of $>$10~GeV $\gamma$ rays with optical and UV photons of the Extragalactic Background Light~\citep{2010ApJ...723.1082A}.
Finally, the short time delay observed in LAT GRBs between low and high energy events can be used to place tight constraints on any energy dependence of the speed of light in vacuum as postulated by some quantum gravity theories~\citep{GRB090510:Nature}.

\begin{figure}[ht!]
\begin{center}
\includegraphics[width=1.0\columnwidth,trim=10 20 50 40,clip=true]{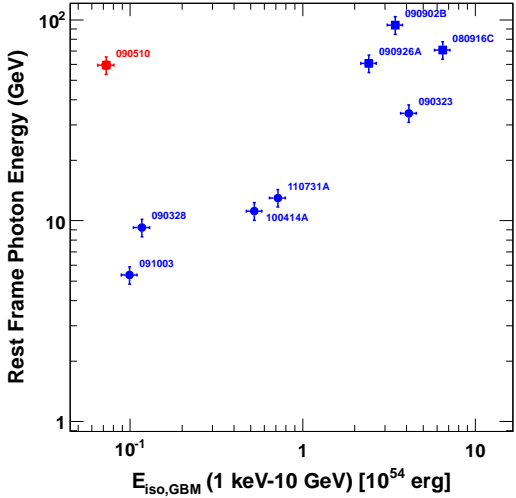}
\caption{Rest-frame-corrected energy of the highest-energy event recorded during the time resolved analysis versus E$_{\rm{iso}}$. Data points are from Table~\ref{tab_energymax_all}. The symbol convention is the same as in Fig.~\ref{fig_Onset}.}
\label{fig_Eiso_Emax}
\end{center}
\end{figure}

Figure~\ref{fig_Eiso_Emax} shows the source-frame-corrected energy of the highest-energy events
with a high ($>$0.9) probability of being associated with the GRB, detected in the time-resolved likelihood analysis, versus E$_{\rm{iso}}$. For long bursts, the most energetic photons appear in the brightest GRBs.
Interestingly, our only short GRB with a measured redshift, GRB\,090510, does not follow the correlation pattern followed by LAT detected long bursts.
More statistics are needed to determine whether this pattern is significant.

\subsubsection{Extended Phase Energetics}
\begin{figure}[ht!]
\begin{center}
 \includegraphics[width=1.0\columnwidth,trim=10 0 40 40,clip=true]{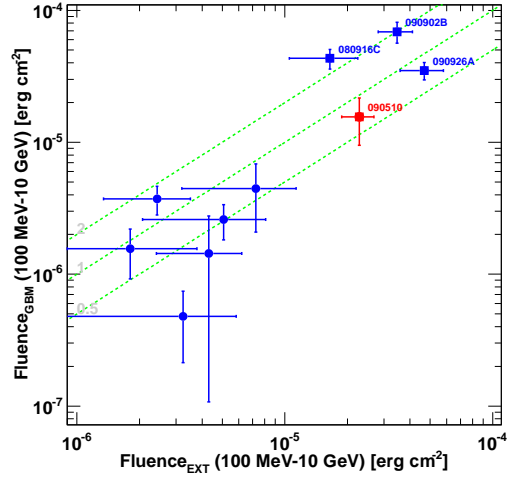}
\caption{Fluence in the 100~MeV--10~GeV energy range measured in the  ``GBM'' versus the ``EXT'' time intervals. The dashed lines correspond to ratios of 0.5, 1, and 2.  The symbol convention is the same as in Fig.~\ref{fig_Onset}.}
\label{fig_PromptExt}
\end{center}
\end{figure}

\begin{figure}[ht!]
\begin{center}
  \includegraphics[width=1.0\textwidth,trim=10 20 0 0,clip=true,angle=-90]{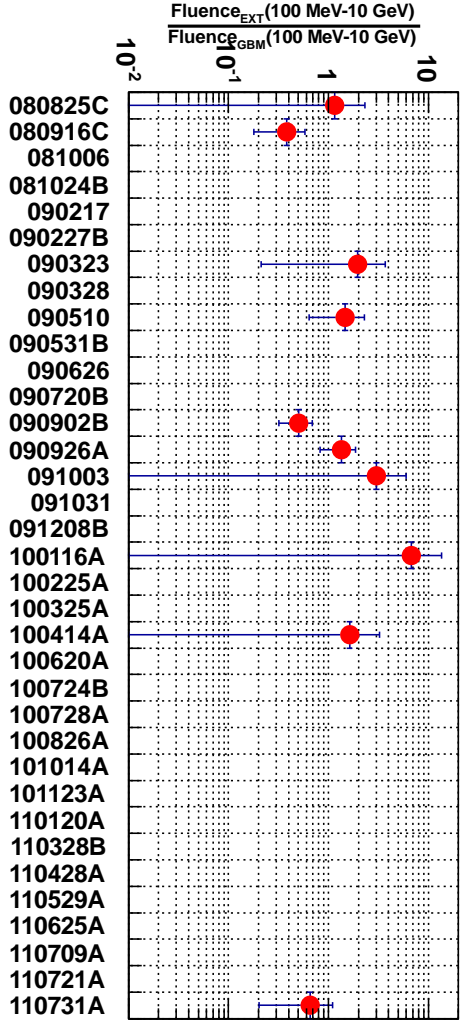}
\caption{Ratio of the 100~MeV--10~GeV fluence measured in the ``EXT'' over that measured in the ``GBM'' time intervals plotted for each burst that has significant extended emission in the LAT data.}
\label{fig_PromptExt2}
\end{center}
\end{figure}

We have explored the energy budget of the highly energetic GRBs during the prompt phase. Now we focus on the temporally extended phase.
First, we compare the energy radiated above 100~MeV during the prompt and temporally extended phases. Since we are comparing energies in the same band, we increase the statistics of our sample by comparing fluences, a quantity that does not require knowing the redshift.
Figure~\ref{fig_PromptExt} shows the 100~MeV--10~GeV fluence measured during the ``GBM'' time interval versus the fluence measured in the ``EXT'' time interval, and Fig.~\ref{fig_PromptExt2} shows the ratio of these quantities for all GRBs with a LAT detection in both time intervals.
We note that most of the ratios are compatible with unity. This implies that above 100~MeV the energy released during the prompt emission is similar to the energy released during the temporally extended emission.


\begin{figure}
\begin{center}
\includegraphics[width=1.0\columnwidth]{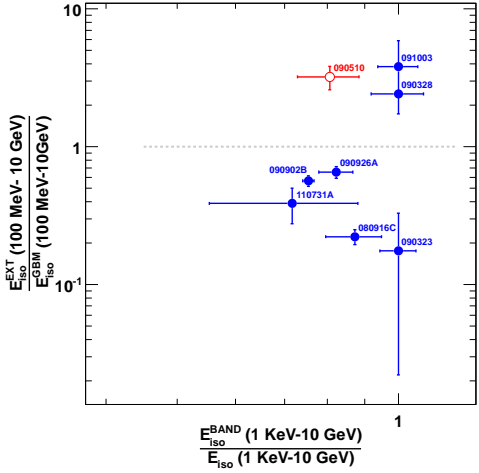}
\caption{The $\gamma$-ray efficiency of the temporally extended emission phase versus the efficiency of the prompt Band component.
The y-axis shows the ratio between the energy released during the temporally extended emission phase and the energy released during the prompt GBM phase, and the x-axis shows the ratio between the bolometric isotropic equivalent energy radiated by the Band component over the total radiated energy during the prompt emission.  
The symbol convention is the same as in Fig.~\ref{fig_Onset}.}
\label{fig_energetics_4}
\end{center}
\end{figure}

To study the relative efficiencies of the Band and extra power-law components during the prompt and temporally extended emission phases
we calculate the ratio of the source-frame isotropic equivalent energy, as measured by the LAT above 100~MeV in the temporally extended phase (the ``EXT'' time window), to the same quantity measured during the GBM time window. 
This is what we display in the y-axis of Fig.~\ref{fig_energetics_4}.
We now know that high-energy emission can be produced during both the prompt and the temporally extended phases, and the y-axis shows the relative importance of these two phases. 
The GRBs in the plot occupy two regions: ``$\gamma$-ray-afterglow dominated" GRBs, with  $E^{EXT}_{iso}>E^{GBM}_{iso}$ like (GRB\,090510, 091003 and 090328) and ``prompt-$\gamma$-ray dominated" GRBs, for which $E^{EXT}_{iso}<E^{GBM}_{iso}$.

The ``$\gamma$-ray-afterglow dominated" GRBs in our sample (GRB\,090510, 091003 and 090328) do not necessarily have a dominant power-law component in the prompt phase.
This could imply that the energy radiated by the extra component during the prompt phase can be dominated by the energy radiated by the main prompt component described by a Band function. 
Note that the LAT sensitivity to GRB\,090328 at the time of the GBM trigger was not optimal, and part of the emission may not have been detected.
This is certainly true for long bursts, such as GRBs 091003 and 090328 while it is not true for GRB\,090510, for which the power law component has been detected.
The majority of LAT-detected bursts radiate more efficiently at high energies during the prompt GBM phase (GRBs below the horizontal line).
We define such bursts as ``prompt-$\gamma$-ray dominated" GRBs.  The five such bursts follow an expected trend: the more important the power-law component in the prompt emission phase, the brighter the late-time emission becomes compared to the prompt high-energy $\gamma$-ray emission.
As already noted, each of the four hyper-fluent GRBs has evidence of an extra component, as does GRB\,110731A.

\subsection{High-Energy Spectral Properties}
\begin{figure}[ht!]
\begin{center}
\includegraphics[width=1.0\textwidth,trim=10 20 0 0,clip=true,angle=-90]{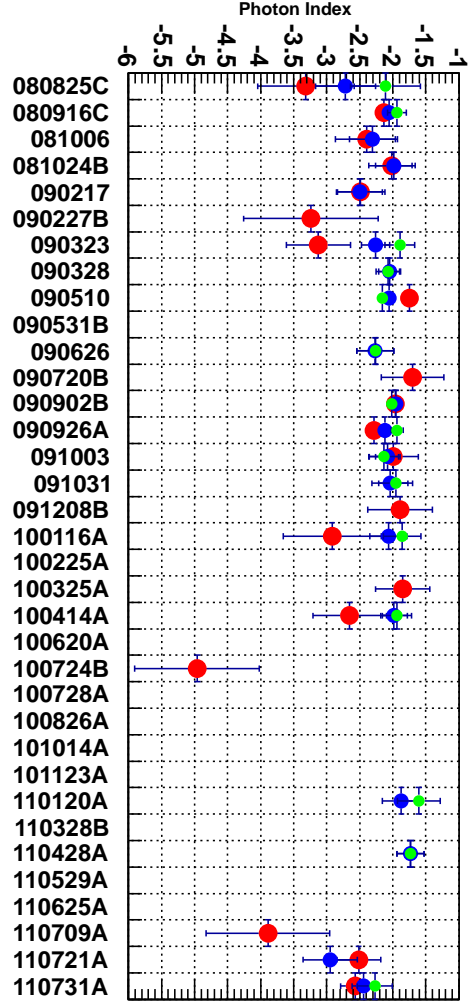}
\caption{Photon index $\Gamma$ of the likelihood-detected bursts as measured in three time windows: ``GBM'' (red), ``LAT'' (blue), and ``EXT'' (green).}
\label{fig_Index}
\end{center}
\end{figure}

In the previous section we discussed the energetics of \Fermi-LAT GRBs, and we now consider their spectral properties.
Since our primary interest is reporting observations related to \Fermi-LAT data, we focus on the spectral properties at high energies, with special emphasis on the role of the extra component. We start from the LAT-only analysis.
Figure~\ref{fig_Index} shows the photon indices of all GRBs detected by the likelihood analysis as measured in three different time windows.
Almost all photon index values are compatible with a value of $-$2 for all three time windows; using the estimated errors as weights, we obtain the average values $<\gamma_{GBM}>$ = $-$2.08$\pm$0.04 in the ``GBM'' time window, $<\gamma_{LAT}>$ = $-$2.05$\pm$0.03 in the ``LAT'' time window, and $<\gamma_{EXT}>$ = $-$2.00$\pm$0.04, in the ``EXT'' time window. There is a selection effect such that any bursts with a photon index considerably softer than $\sim$$-$2 are less detectable by the LAT.
Interestingly, GRB\,100724B, which has the steepest photon index during the ``GBM''  has the second largest GBM-measured duration, while the GRB with the shortest duration, GRB\,090510, has one of the hardest photon indices.

\begin{figure}[ht!]
\begin{center}
\includegraphics[width=1.0\columnwidth,trim=10 20 50 0]{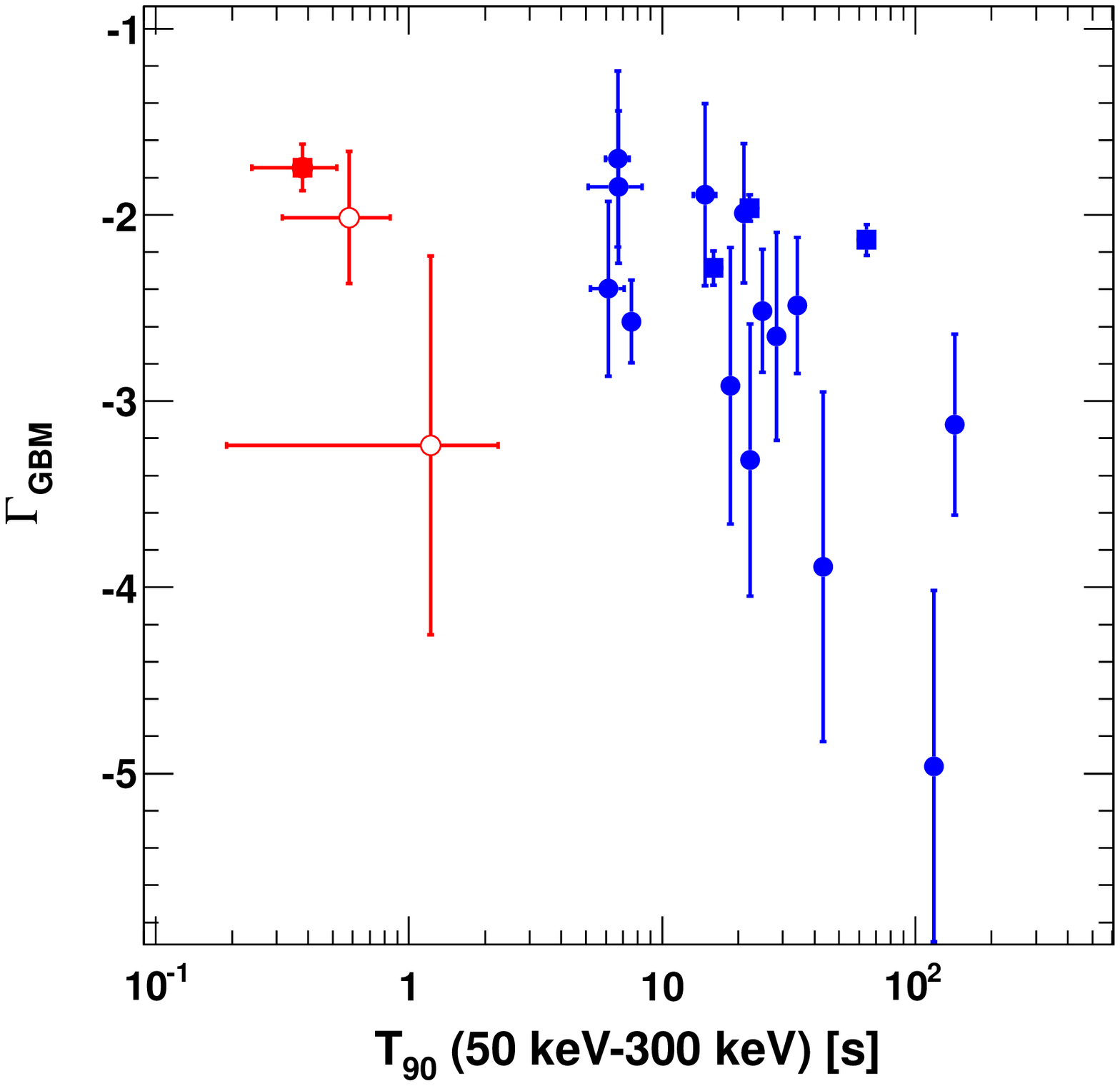}
\includegraphics[width=1.0\columnwidth,trim=10 20 50 0]{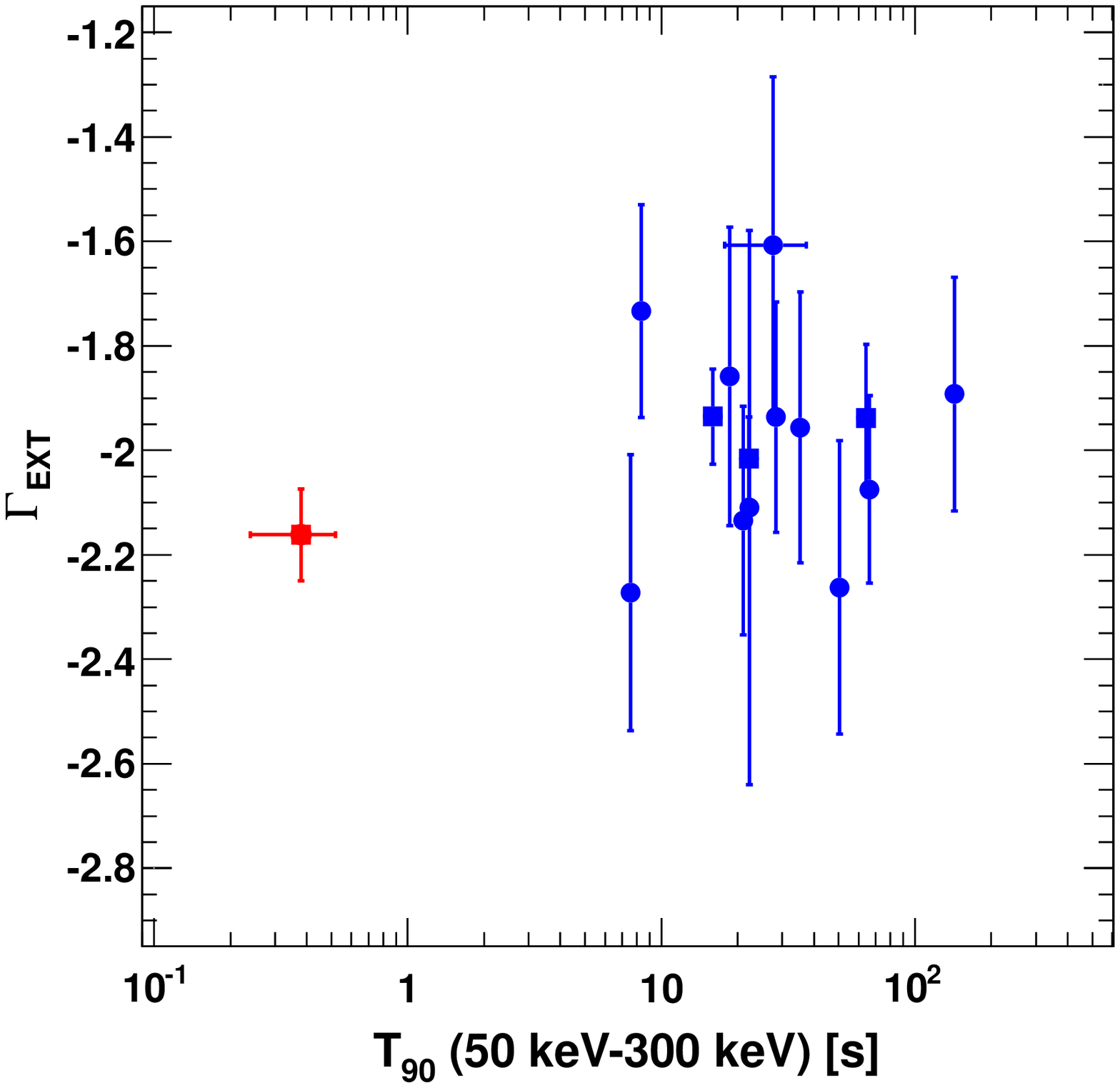}
\caption{Top: Power-law photon index measured in the GBM time window and (Bottom) in the EXT time window.  The symbol convention is the same as in Fig.~\ref{fig_Onset}.}
\label{fig_Index_Duration}
\end{center}
\end{figure}

To further explore whether the photon indices depend on the duration, we plot in Fig.~\ref{fig_Index_Duration} the value of the photon index of the extra power-law as measured in the ``GBM'' time window $\Gamma_{GBM}$ (top panel) and in the ``EXT'' time window $\Gamma_\mathrm{EXT}$ (bottom panel) versus the GBM T$_\mathrm{90}$.
The photon index has a mild inverse correlation with the duration of the burst (top panel), in agreement with our results above and previous findings that the spectra of short duration GRBs tend to be harder \citep{2004RvMP...76.1143P}. On the other hand, when the spectral analysis is performed during the ``EXT'' time window (bottom panel), during which the signal from the GRB is no longer detected by the GBM but is still bright in the LAT energy window, this mild correlation disappears.
Note that some of the GRBs (like GRB\,100724B) do not have detected extended emission and are reported only in the top panel.

\begin{figure}[ht!]
\begin{center}
\includegraphics[width=1.0\columnwidth,trim=10 20 50 0,clip=true]{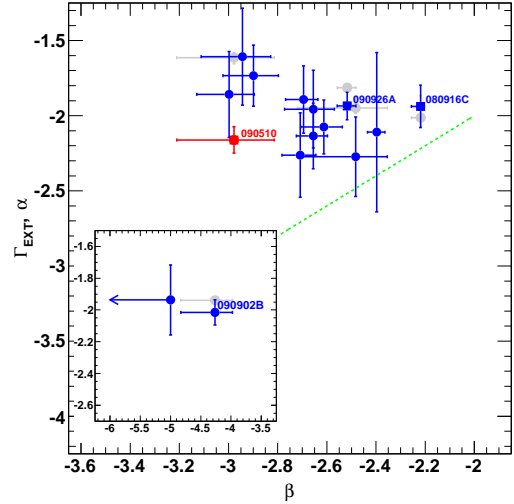}
\caption{Red/Blue symbols: photon index $\Gamma_\mathrm{EXT}$ of the power-law spectrum as measured by the LAT during the ``EXT'' time interval versus the value of the $\beta$ parameter of the Band function. Grey symbols: photon index $\alpha$ of the extra power-law component obtained by our joint GBM-LAT fits as measured in the ``GBM'' time window versus $\beta$.  The symbol convention is the same as in Fig.~\ref{fig_Onset}.}
\label{fig_BetaGamma}
\end{center}
\end{figure}

To further investigate this, we show in Fig.~\ref{fig_BetaGamma} the power-law photon index $\Gamma_\mathrm{EXT}$ of the GRB emission in the LAT energy range as measured during the ``EXT'' time interval versus the value of the high-energy power-law index $\beta$ of the Band function as measured in the prompt ``GBM'' time interval. The value $\Gamma_\mathrm{EXT}$ was obtained by our LAT-only likelihood analysis and the $\beta$ value was obtained by our joint GBM-LAT spectral fits.
We measured $\beta$ using either a Band-only or a Band-plus-power-law spectral model. For the cases where the more complex Band-plus-power-law spectral model also provided a good fit (i.e., when all the parameters were constrained and the fit converged), we selected the $\beta$ value found for the more complex model. 
For those cases, in addition to $\Gamma_\mathrm{EXT}$ we also plot the fitted values of the extra power-law component photon index $\alpha$ versus $\beta$.
Table~\ref{tab_JointFitGBMT90} summarizes the numerical values of the parameters of the model that best fits the LAT-GBM data.
An important selection effect must be kept in mind:  distinguishing an extra power-law spectral component is difficult when it is softer than the high-energy component of the Band function.
As can be seen in the figure, the power-law component described by $\Gamma_\mathrm{EXT}$ is typically harder than the high-energy emission measured during the prompt phase by GBM, described by $\beta$.
Furthermore, the two quantities do not seem to be correlated.
The most extreme case, GRB\,090902B, is shown in the inset of that figure, together with GRB\,100414A for which we detect the temporally extended emission, while the $\beta$ index of the Band function is only an upper limit. In fact for these bursts the best fit model found by our procedure were the Comptonized plus power law and the Comptonized alone, respectively.   Therefore it is very reasonable that when we replace the Comptonized model with a Band function, the resulting $\beta$ parameter is very steep, and not constrained toward lower values.

In two cases, GRBs 090510 and 090926A, the extra power-law component that is significantly detected during the prompt emission is harder than the power-law of the extended emission.
For the first case, this is probably caused by the hard-to-soft spectral evolution of the extra component, as demonstrated by the results of the time-resolved likelihood analysis shown in Fig.~\ref{like_090510016}. For the case of GRB\,090926A the extra power-law component during the prompt emission is significantly attenuated at high energies and the model that best fits the emission during the ``GBM'' time window consists of a Band function plus a {\rm Comptonized} model and has a very high peak energy. The (exponential) spectral cutoff of GRB\,090926A is not significantly detected at later times. Overall the temporal evolution of the extra power-law component of this GRB can be described as very soft/weak at the start, progressively becoming harder but also demonstrating a roll-off at around 10~GeV, and then becoming softer again with an index of $\Gamma_{EXT}\sim-2$.

In the other three cases for which we significantly detect the extra power-law component during the prompt phase (GRBs 080916C, 090902B, and 110731A) (see $\S$\ref{subsub_modelselection}), the photon index of the extra-power law in the prompt ``GBM'' time interval $\gamma$ is compatible with the index of the power-law in the LAT energy range measured during the temporally extended emission $\Gamma_\mathrm{EXT}$.

The picture emerging from the analyses described in this subsection suggests that the high-energy ($>$GeV) emission is dominated by a single long-lasting component, well described by a power-law function of a photon index typically near $-$2, independent of burst properties such as the duration, the brightness, or the spectral properties of the lower-energy prompt emission.


\subsection{Extended Emission Temporal Decay}
\label{subsub_extEmission_ch5}
%


\begin{figure}[ht!]
\begin{center}
\includegraphics[width=1\columnwidth,trim=10 20 50 0,clip=true]{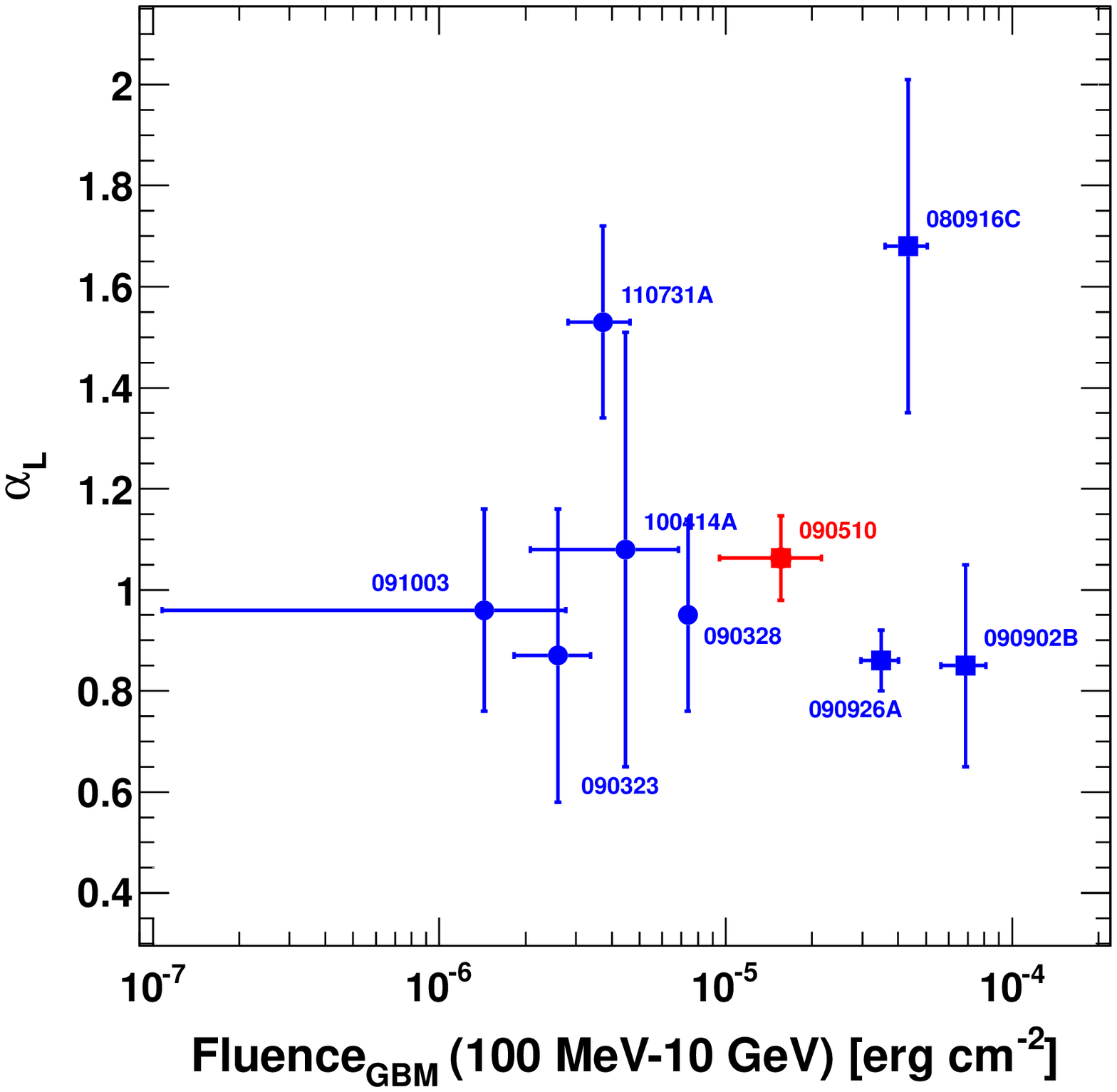}
\includegraphics[width=1\columnwidth,trim=10 20 50 0,clip=true]{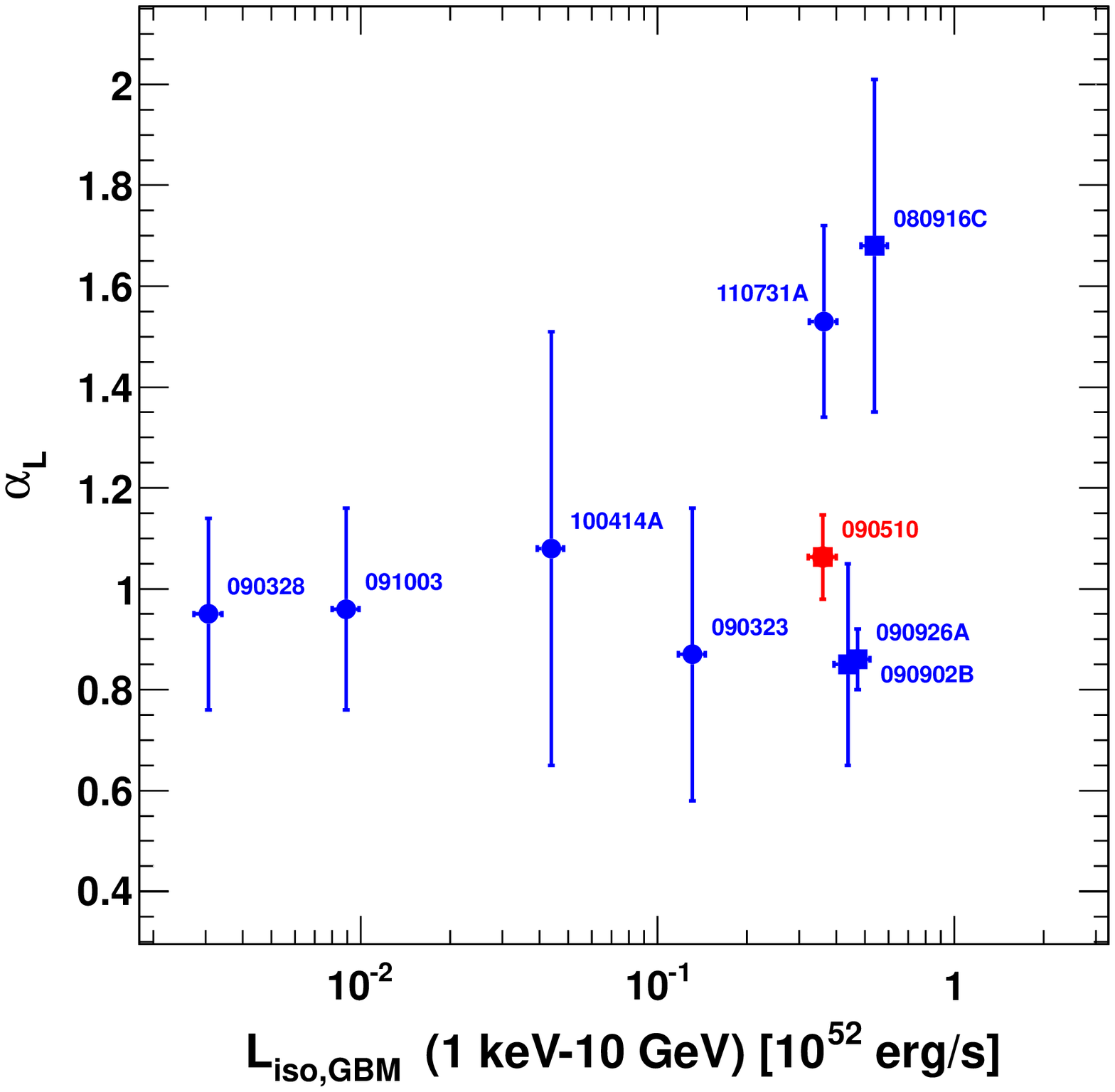}
\caption{Value of the ``late-time decay index'' as a function of the
fluence between 100~MeV and 10~GeV in the ``GBM'' time interval (Top) and of the
isotropic luminosity between 1~keV--10~GeV, source frame.  The value of $\alpha_{L}$ is
$\sim$1, except for GRB\,080916C and GRB\,110731A, which notably have the shortest
durations when measured in the source frame (see text).  The symbol convention is the same as in Fig.~\ref{fig_Onset}.}
\label{fig_agpower}
\end{center}
\end{figure}

\begin{figure}[ht!]
\begin{center}
\includegraphics[width=1\columnwidth,trim=10 20 50 0,clip=true]{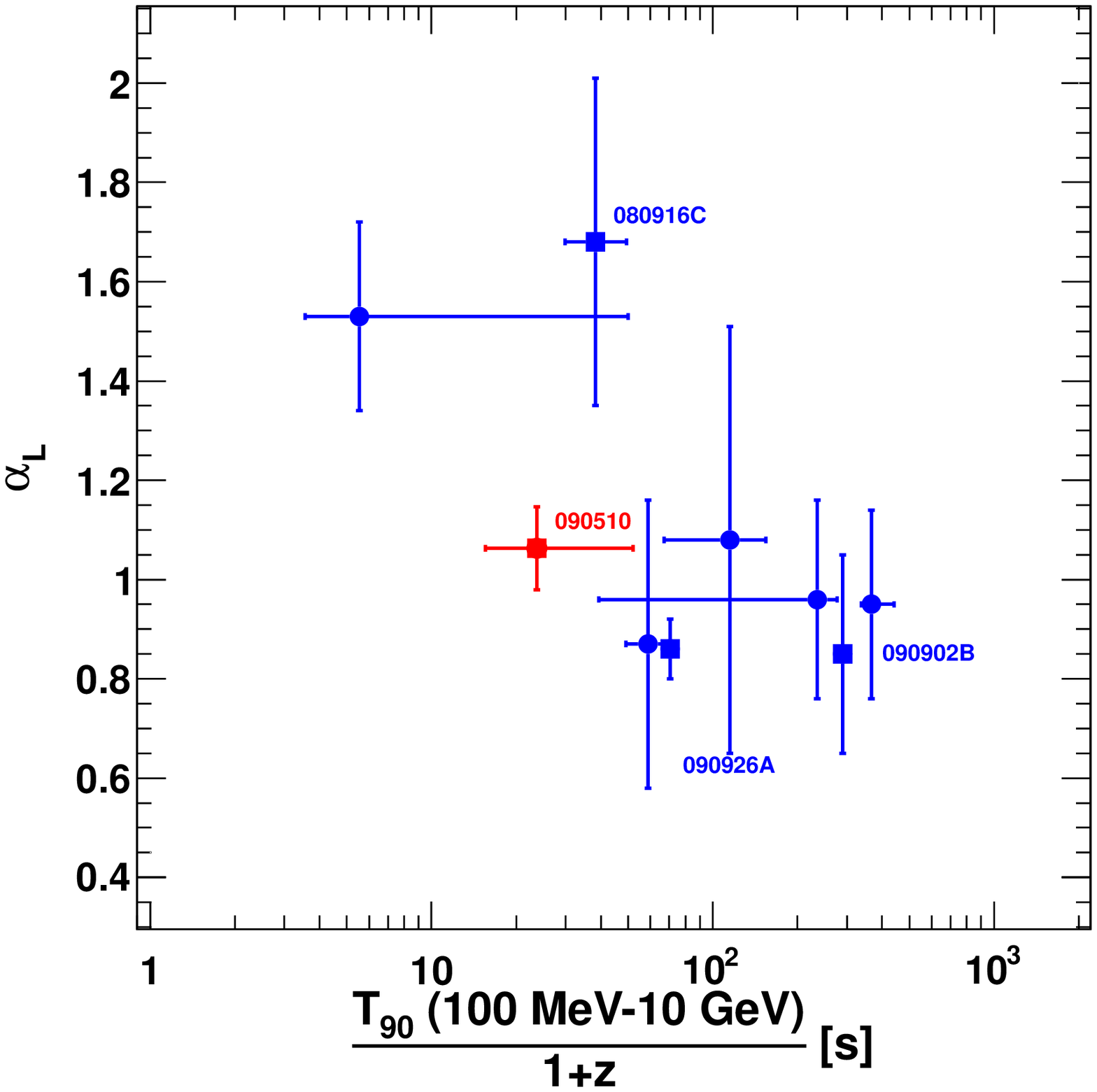}
\caption{Value of the ``late-time decay index'' $\alpha_{L}$ as a function of the LAT T$_{90}$ in the source reference frame.  The symbol convention is the same as in Fig.~\ref{fig_Onset}.}
\label{fig_decay_duration}
\end{center}
\end{figure}

In Fig.~\ref{fig_agpower} we report the ``late-time decay index'' $\alpha_{L}$
as a function of the fluence measured by the LAT in the GBM interval (top panel) and of the luminosity in the ``GBM'' time interval (lower panel). The
values of $\alpha_{L}$ seem to cluster around 1, which in the context of the
fireball model indicates an adiabatic expansion of the fireball (see
$\S$\ref{sub_interpretation_extended}). There are two exceptions: GRB\,080916C
and GRB\,110731A.
To investigate this a little further, we plot in  Fig.~\ref{fig_decay_duration} the value of $\alpha_L$ as a function of the intrinsic duration of the GRB at high energy.
Both GRB\,080916C and GRB\,110731A have the shortest intrinsic LAT T$_{90}$ among long GRBs. This suggests that we have probably observed only the first steep part of the
decay after the prompt phase and that we cannot exclude the existence of a flattening or a break at later times that would reconcile them with the other bursts.

\subsection{LAT Detection Rate}
\label{subsec_lat_rate}
\begin{figure}[ht!]
\begin{center}
\includegraphics[width=1.0\columnwidth,trim=10 0 50 50]{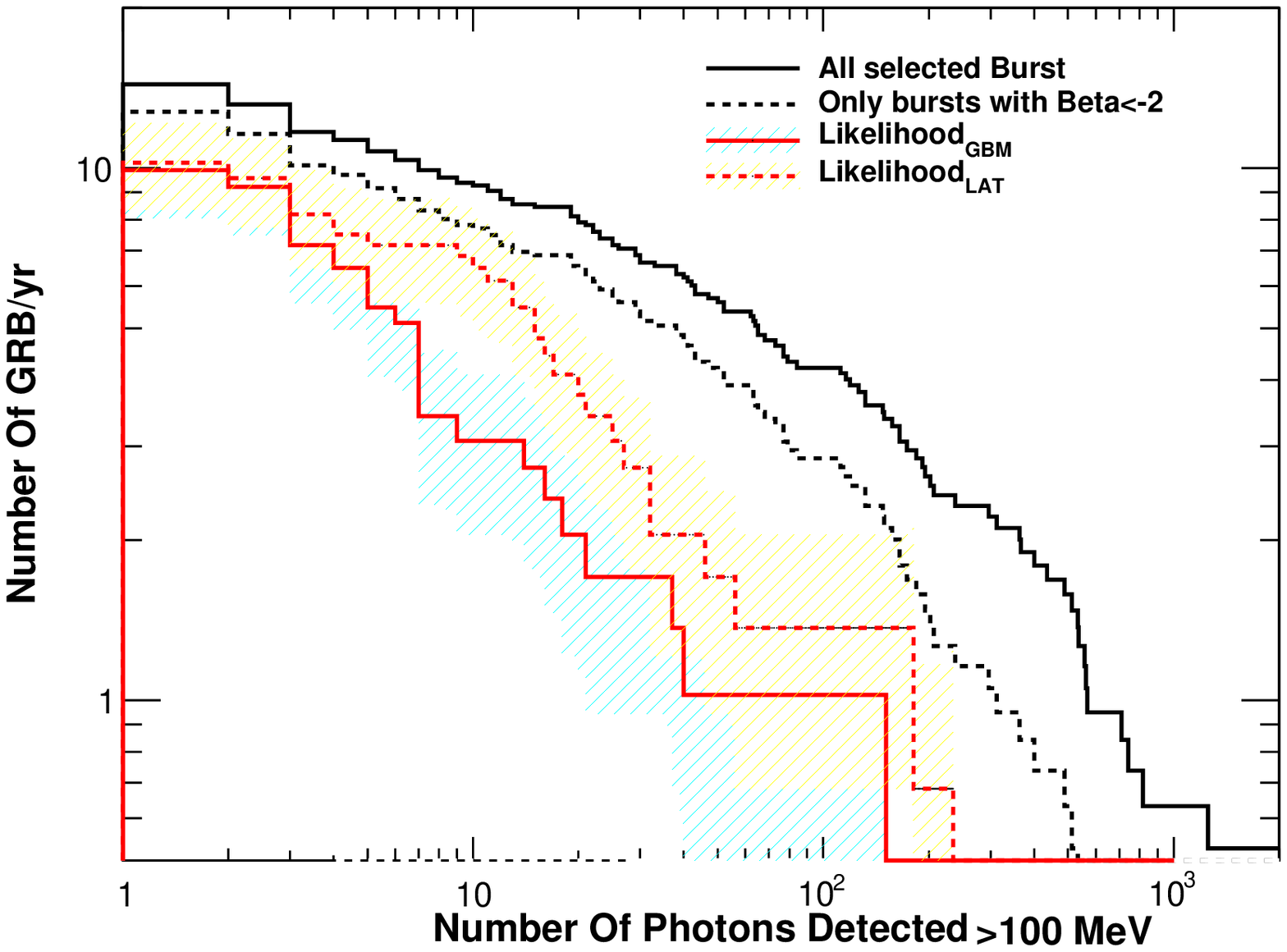}
\caption{Comparison between the observed yearly rate of LAT GRB detections to the pre-launch expectations.
Black lines are taken from \citet{2009ApJ...701.1673B} for an energy threshold of 100~MeV, using the bright BATSE GRB sample from  \citet{Kaneko:08} as input.
The dashed black line corresponds to an input distribution from which hard bursts with $\beta\ge-2$ have been removed.
The red lines indicate the observed number of GRBs as a
    function of the number of events predicted by the best-fit model.
    The hatched regions correspond to the statistical uncertainties
    assuming Poisson statistics.}
\label{fig_GRBYield}
\end{center}
\end{figure}
\citet{2009ApJ...701.1673B} have reported the number of expected GRBs per year detectable by
the LAT as a function of the number of excess events. This rate was estimated with Monte Carlo simulations
using the predicted pointing history for the first year of observations.
This calculation was performed using a standard survey profile without any pointed-mode observations (due to
a positive response to ARR or planned Target Of Opportunity).
The spectral model was a simple Band function, with parameters distributed according
to the sample of bright BATSE GRBs \citep{Kaneko:08}.
The all-sky burst rate was assumed to be 50 GRB yr$^{-1}$ full sky (above the peak flux in 256~ms of 10 ph s$^{-1}$ cm$^{-2}$ in the 50--300~keV band or with an energy flux greater than 2$\times$10$^{-5}$ erg cm$^{-2}$) in the 20--2000~keV band, derived from the BATSE catalog of bright bursts.
\citet{2009ApJ...701.1673B} calculated the number of expected $\gamma$-rays using the bright BATSE GRB sample and also repeated the calculation with the hardest-spectrum (index $\beta>$-2)  GRBs removed as the numbers of $\gamma$-rays at high energies would have been unphysically large.

In addition, \citet{2009ApJ...701.1673B} used simplified detection criteria, based entirely on the numbers of detected photons assuming a negligible contribution from background, or using a semi-analytical model to compute the value of the Test Statistic. For the latter, an isotropic background was assumed, but no additional sources were added to the simulation, including the bright Earth limb.
The results of these simulations, taken from \citet{2009ApJ...701.1673B}, are shown in Fig.~\ref{fig_GRBYield}.
We compare these results with the numbers of events above 100~MeV predicted by the best-fit model, including all bursts from Table~\ref{tab_likelihoods}.
In this comparison we use both the values obtained by integrating the spectrum in the GBM time window and in the LAT time window.
Several interesting features are evident from this plot.
First of all, the number of detected GRBs is somewhat less than expected.
Additionally, the differences between the predicted and observed numbers of GRBs increase for bursts with many $\gamma$-rays in the LAT data.
The absence of very bright bursts (with several hundreds of $\gamma$-rays detected above 100~MeV) could
be due to the systematic uncertainties that are propagated in the simulation when extrapolating the Band function
fits to high energies over a very-large lever arm.
Especially when the high-energy photon index is close to $-$2, a small change of the flux value could create large uncertainties on the number of
detected events at high energies, when extrapolated.
This has been specifically tested using bright GBM bursts that were not detected by the LAT, and the bias introduced by fitting GBM-only data for bursts has been estimated by adding LAT upper limits in the spectral fit \citep{2012ApJ...754..121T}.
On the other hand, intrinsic deviations from a pure Band function, such as spectral cut offs, spectral breaks, or curvature in the spectra could influence the number of predicted LAT detected GRBs.

%

\subsection{Detectability of GBM bursts}
\label{subsec:gbm_detectability}
%

Although many observed properties may be considered in
      classifying the detectability of GBM GRBs by the LAT, we limit
      the current analysis to the competing effects that the effective
      area decreases with increasing off-axis angle $\theta$ while the
      solid angle increases with $\theta$.
      
\begin{figure}[ht!]
\begin{center}
\includegraphics[width=1.0\columnwidth]{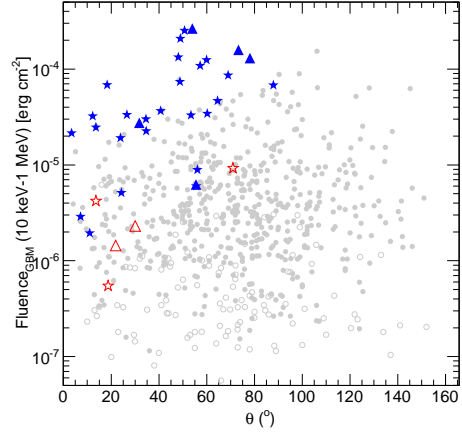}
\caption{Sensitivity plot for GBM GRBs showing the fluence in the 10~keV-1~MeV energy band as a function of the LAT off-axis position ($\theta$). 
Filled symbols indicate long duration bursts while empty
    symbols denote short bursts. Gray circles denote GBM bursts that were
    not detected by the LAT, stars denote LAT bursts detected using the 
    standard likelihood analysis, and triangles denote bursts detected by
    the LLE analysis only.  For clarity, long duration LAT-detected GRBs
    are plotted in blue, while short duration LAT-detected GRBs are plotted in 
    red.}
\label{fig_FluenceTheta}
\end{center}
\end{figure}

It follows that there are more GRBs at large $\theta$, although the LAT can detect only the brightest. Fig.~\ref{fig_FluenceTheta} shows the fluence in the GBM energy band as a function of $\theta$.
Using the  sample of GRBs through August 2011 that is available at the HEASARC web site\footnote{The GBM Burst catalog: \url{http://heasarc.gsfc.nasa.gov/W3Browse/fermi/fermigbrst.html}}, 
we display both the LAT and LLE detected GRBs. For LAT detections we use the fluence computed by our analysis, while for GBM-detected GRBs we use the value obtained from the GBM Burst catalog.
Generally speaking, the LAT-detected GRBs are among the brightest GBM GRBs occurring in the LAT FoV. On the other hand, there are some exceptions where GRBs with a modest energy fluence or with a suboptimal viewing angle have still been detected by the LAT. These cases highlight the importance of secondary considerations other than $\theta$ or the fluence. In terms of GBM fluence, short bursts
are easier to detect.
Also, we note that the location in the FoV of the GRB at the time of the GBM trigger is not always representative of the quality of the exposure obtained during the burst. For example GRB\,110625A was far off-axis at the time of the trigger (87\de), but the high-energy emission was detected by the likelihood analysis a few hundred seconds after the GBM trigger when the GRB was well inside the FoV of the LAT. LLE bursts (triangles) occur typically at larger incidence angles, indicating that the FoV of the LAT is larger for LLE data sample than when using standard event classification. There is also one case of a relatively bright GBM burst (GRB\,110328B), where the off-axis angle was relatively small ($\sim$32\de) but the GRB was detected only using LLE analysis. This is explained by the results of the combined spectral analysis (summarized in Table\ref{tab_JointFitGBMT90}), which show that the best fit spectral  model is a Comptonized model cutting-off approximately at 1.2~MeV, implying
suppression of high-energy emission.

\clearpage
\newpage
\section{Interpretation}
\label{sec_interpretation}

In this study, we have characterized the high-energy emission observed from 35 GRBs detected by the LAT. While the number of LAT GRBs is a small fraction of the number detected by the GBM~\citep{Paciesas+12,Goldstein+12}, there are a few emission features that show up only at high energies and distinguish the LAT GRBs:
\begin{itemize}
\item high fluence and energy release,
\item temporally extended emission lasting longer than the GBM-detected emission,
\item delayed onset with respect to the GBM-detected emission, and
\item presence of an extra power-law component in the spectrum.
\end{itemize}
Here we discuss plausible interpretations of the emission properties observed with the LAT, salient features of these models, and possible issues.

\subsection{Fluence and Energetics of LAT Bursts}

The distribution of fluences of LAT GRBs (see Fig.~\ref{fig_FluenceFluence}) provides hints of two classes: a hyper-fluent class currently comprising four members, GRBs~080916C~\citep{GRB080916C:Science}, 090510~\citep{GRB090510:Nature,GRB090510:ApJ,GRB090510:Agile}, 090902B~\citep{GRB090902B:Fermi}, and 090926A~\citep{GRB090926A:Fermi}, and which have a typical 100~MeV--10~GeV fluence of $\sim$(3--8)$\times 10^{-5}$~erg\,cm$^{-2}$; and a larger class with a lower typical fluence of $\sim$(2--10)$\times 10^{-6}$~erg\,cm$^{-2}$.  The GBM fluences for the hyper-fluent class are also higher, $\sim$1.3~times the LAT fluence for the short burst GRB\,090510 and $\sim$3--10 times the LAT fluences for the 3 long bursts (see Fig.~\ref{fig_FluenceFluence}, bottom panel).  For comparison, we note that the typical fluence for the GBM long bursts is $\sim 2\times 10^{-6}$~erg\,cm$^{-2}$ in the 8~keV--1~MeV range and $\sim 10^{-5}$~erg\,cm$^{-2}$ in the 8~keV--40~MeV range, based on Band function fits to the spectra ~\citep{Goldstein+12}.  
It is evident that most of the LAT bursts do seem to be very bright in the GBM, especially when comparing their 10~keV--1~MeV fluences (see Figs.~\ref{fig_FluenceTheta} and \ref{fig_Fluence_Catalogs}) to the 8~keV--1~MeV fluence of the typical GBM bursts~\citep{Goldstein+12}.

In the cases of 9 LAT bursts for which the redshift information is available, the isotropic equivalent energy $E_{\rm iso}$ in the LAT energy range (100~MeV--10~GeV) is also higher for the three hyper-fluent long bursts (see Figs.~\ref{fig_redshift1} and \ref{fig_eiso2} top panel).  The ratio of the $E_{\rm iso}$ (100~MeV--10~GeV) to the total $\gamma$-ray energy $E_{\rm iso}$ (1~keV--10~GeV) for the long bursts is $\sim$(5--25)\%.  Interestingly in the case of GRB\,090510, the only short LAT burst with known redshift, this ratio is $\sim 70\%$ and is clearly distinct from the long bursts.  The bottom panel of Figure~\ref{fig_eiso2} shows that for the bright bursts, including GRB\,090510 with its additional PL spectral component, the ratio of isotropic equivalent energies in the PL and Band components is concentrated at $\sim 25\%$.  Thus the high ($\sim 70\%$) LAT-to-GBM $E_{\rm iso}$ ratio for GRB\,090510 is a combination of high Band $E_{\rm pk}$, typical for short hard class, and a very hard ($-1.61$) PL photon index which must cut off at high energies.  
The bursts with $\lesssim 10\%$ LAT-to-GBM $E_{\rm iso}$ do not allow for the detection of an additional PL spectral component, though it could still be present.  The additional PL spectral component is most likely responsible for the high fluence detected by the LAT, as also indicated in Fig.~\ref{fig_energetics_4} for five of the eight brightest bursts.

The isotropic-equivalent energies of the LAT bursts calculated here are largely consistent with the energies calculated by~\citet{Cenko+11} and show that LAT bursts possibly compose the most energetic sub-sample of GRBs (see Fig.~\ref{fig_eiso1}).  The range of $E_{\rm iso}$ for short bursts in the pre-\Fermi era was (0.0033--10.2)$\times 10^{52}$~erg~\citep{Ghirlanda+09}.
GRB 090510 is clearly at the high end of that range with $E_{\rm iso} \simeq 7\times 10^{52}$~erg.
Although the sample is rather small, the detected redshifts of LAT bursts do not show a concentration of bursts at any particular range (see Fig.~\ref{fig_eiso1}).  \cite{Racusin+11} showed that the redshift distributions are statistically consistent for {\it Swift}-BAT detected GRBs, those detected by both GBM and BAT, and the small sample of LAT-detected bursts with measured redshifts.  
The only redshifts available for the GBM sample are for those bursts that also triggered BAT or LAT. Therefore, whether LAT-detected GRBs follow the redshift distribution of the rest of GBM-detected bursts is still an open question.

Another interesting feature of the LAT emission is that the 100~MeV--10~GeV fluences in the ``GBM'' and ``EXT'' time intervals are within a factor $\sim 2$ of each other for a handful of bursts with high-significance detections (see Fig.~\ref{fig_PromptExt}).  This may indicate an approximately equal efficiency of the GRB fireball to produce high-energy emission during the coasting (prompt) and deceleration (afterglow) phases, in the context of the early-afterglow model as the origin of LAT emission.


\subsection{Temporally Extended Emissions}
\label{sub_interpretation_extended}

The flux of LAT-detected emission at late times decays rather smoothly and can generally be fitted with a power law $F_{\nu} \propto t^{-\alpha_{L}}$ (see \S~\ref{subsubsec_temporallyextended_ch4}, and Figs.~\ref{fig_temporalBreaksLuminosity} and \ref{fig_ExtraComponent}). Such behavior also is typically observed in X-ray, UV, and optical wavelengths after the prompt $\gamma$-ray emission and is attributed to the afterglow emission.  The apparent non-variation of the photon index for individual bursts (see Fig.~\ref{fig_Index}) in the ``EXT'' time interval as compared to the ``LAT'' time interval also suggests that the temporally extended LAT emission resembles afterglow rather than prompt emission, for which the photon index is likely to vary with time. The burst-averaged values for the photon index in these two intervals: $\Gamma_{\rm EXT} = -2.00\pm 0.04$ and $\Gamma_{\rm LAT} = -2.05 \pm 0.03$ are also very similar.
The slightly larger values for the burst-averaged photon index $\Gamma_{\rm GBM} = -2.08 \pm 0.04$ in the
earlier ``GBM'' time interval could be due to a plausible contamination by the prompt emission in the LAT.  Indeed, the high-energy photon index of the Band function, $\beta_{\rm Band}$, is systematically softer than $\Gamma_{\rm EXT}$ in the joint fit to the GBM and LAT data (Fig.~\ref{fig_BetaGamma}), suggesting that the hard spectral component becomes dominant at late times.

Remarkably, the ``late-time decay index'' is always close to $\alpha_{L} = 1$ (see Fig.~\ref{fig_ExtraComponent} and Fig.~\ref{fig_agpower}), except in two cases, GRBs~080916C and 110731A, which could be affected by an observational bias (see \S\ref{subsub_extEmission_ch5}).  The clustering around $\alpha_{L}=1$ suggests a common emission mechanism, even though our limited sample does not allow firm conclusions.  In the context of afterglow emission, the bolometric flux decays as $\propto t^{-\alpha}$, with $\alpha=1$ and $\alpha=10/7$ for an adiabatic fireball and a radiative fireball in a constant density environment~\citep{Sari97,Katz+97,Ghisellini+10}, respectively. The flux decay in a particular energy band is more complicated, and depends on the fast- or slow-cooling spectral models~\citep{Sari+98} as well as on the surrounding environment (i.e., whether it is uniform density interstellar medium (ISM) or with wind-type density profile~\citep{Sari+98,Chevalier+00,Panaitescu+00}).  In particular,
the relation
between the flux-decay slope $\alpha$ and spectral index $\beta$ for the flux density $F_\nu (t) \propto t^{-\alpha} \nu^{-\beta}$ varies between different parts of the spectrum. LAT-detected $\gtrsim 100$~MeV emission is likely to be from the fast-cooling part of the spectrum for which $\alpha = (12\beta -2)/7$ for a radiative fireball and $\alpha = (3\beta -1)/2$ for an adiabatic fireball, both for the ISM and wind environments~\citep{Sari+98,Granot+02}.  In the LAT data, $\beta = -\Gamma_{\rm EXT}-1 = 1.00\pm 0.04$, and $\alpha_{\rm adiabatic} = 1$ and $\alpha_{\rm radiative} = 10/7$, both of which are equal to their respective bolometric flux-decay indices.  Thus, a simple interpretation of $\alpha_{L} \approx 1$ flux-decay index for most LAT bursts indicates that the $\gtrsim$100~MeV emission is more likely from an adiabatic fireball~\citep{Kumar+09,DePasquale+10,Razzaque+10} rather than from a radiative fireball, as~\citet{Ghisellini+10} had suggested.

For three bright bursts (GRBs~090510, 090902B and 090926A), a broken power law fits the LAT data better than a single power law (see \S~\ref{subsubsec_temporallyextended_ch4}). After the time of peak flux, the initial flux decay is much steeper than the later decay. The initial steep-decay phase is likely due to a transition from the prompt to afterglow emission.  An additional short-lived emission component, such as the high-latitude emission from the fireball which decays quickly~\citep{Kumar+00} and dominates the underlying afterglow emission, may in principle explain the initial steep decay.

\onecolumn
\subsection{Delayed Onset of LAT-detected Emission}
For most bursts, the onset of the LAT-detected emission, as measured by LAT T$_{05}$ (100~MeV--10~GeV), is delayed with respect to the onset of the GBM-detected emission, measured by GBM T$_{05}$ (50~keV--300~keV) (see Fig.~\ref{fig_Onset}).  Delays of up to 40~s have been detected in long bursts, with a few seconds being the typical value.  The delay is $\sim 0.5$~s for GRB\,090510 and $\gtrsim 0.05$~s for GRB\,081024B, both of which are short bursts.  The origin of the delayed onset of the LAT emission is poorly understood.

One interpretation of the delayed LAT onset is based on the early afterglow model for the temporally-extended LAT emission~\citep{Kumar+09,Ghisellini+10,DePasquale+10,Razzaque+10}.  The bolometric flux from a coasting fireball increases as $\propto t^2$~\citep{Sari97}, both for an adiabatic and a radiative fireball, before it decelerates and enters a self-similar phase~\citep{Blandford+76,Rees+94}.  The time required for the flux to increase and be detected by the LAT corresponds to the delayed onset of the LAT emission in this scenario.  It also implies that the peak-flux time of the LAT is of the order of the fireball deceleration time.  The corresponding jet bulk Lorentz factor can be estimated for an ISM of constant density $n=$1\,cm$^{-3}$ as~\citep{Blandford+76,Sari+98,Ghisellini+10}, 


\begin{equation}
\Gamma_0 = \left[ \frac{3E_{k,\rm iso} (1+z)^3}
{32\pi n m_pc^5 t^3_{\rm peak}} \right]^{1/8}
\times \begin{cases} a^{-1/8} ; ~a=4 & (adiabatic)
\cr a^{-5/32}; ~a=7 & (radiative),\end{cases}
\label{Gamma_0}
\end{equation}
where $E_{k,\rm iso}$ is the isotropic-equivalent jet kinetic energy immediately before deceleration.

%

In the case of a wind environment, with the wind parameter $A = 3.02\times 10^{35} A_\star$~cm$^{-1}$ for a $10^{-5} M_\odot$~yr$^{-1}$ mass-loss rate in the wind of velocity $10^3$~km\,s$^{-1}$ and ${A_\star\sim1}	$~\cite{Chevalier+00}, the jet bulk Lorentz factor can be estimated as~\citep{Chevalier+00,Panaitescu+00}:
\begin{equation}
\Gamma_0 = \left[  \frac{E_{k,\rm iso} (1+z)}
{16\pi A m_pc^3 t_{\rm dec}} \right]^{1/4},
\label{Gamma_02}
\end{equation}
where $t_{\rm peak} \approx t_{\rm dec}$ for the adiabatic and radiative fireballs.

\begin{figure}[t]
\begin{centering}
\includegraphics[width=1.0\columnwidth,trim=10 20 50 30,clip=true]{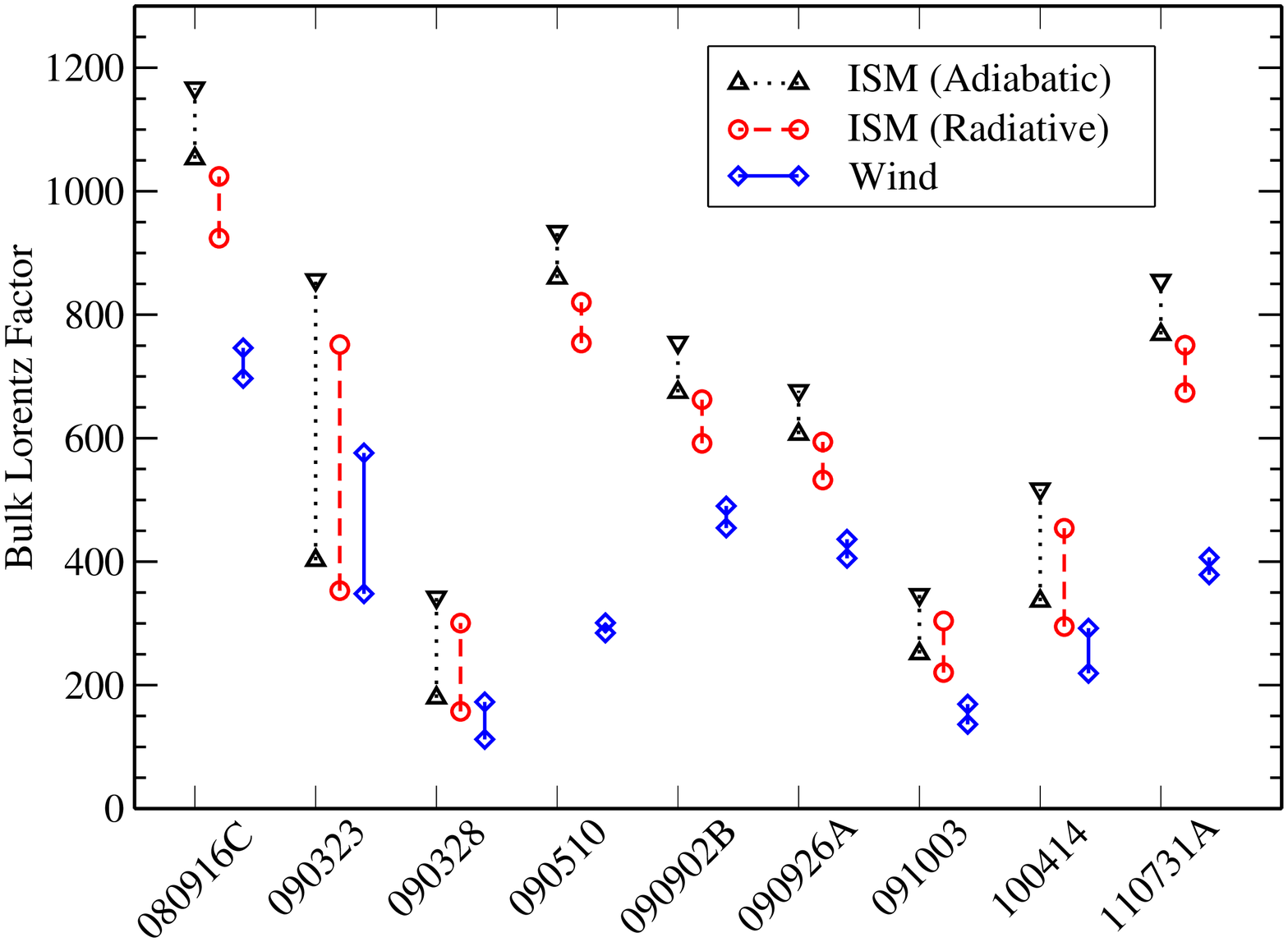}
\caption{
Bulk Lorentz factors of the LAT bursts derived on the assumption that the peak flux time in the LAT (Fig.~\ref{fig_ExtraComponent}) represents the fireball-deceleration time through Eqs.~(\ref{Gamma_0}) and (\ref{Gamma_02}).  We also assumed a constant ISM density of $n=1$~cm$^{-3}$, a wind parameter with $A_\star = 0.1$ and a kinetic energy four times the $\gamma$-ray energy, $E_{k,\rm iso} = 4\times E_{\gamma,\rm iso}$, for this illustrative plot.  The range of $\Gamma_0$ in each case represents the $1 \sigma$ error on $t_{\rm peak}$.}
\label{fig:bulkLorentz}
\end{centering}
\end{figure}

Figure \ref{fig:bulkLorentz} illustrates the range of the bulk Lorentz factors calculated using Eqs.~\ref{Gamma_0} and \ref{Gamma_02} for the nine LAT bursts with known redshifts.
The range depends on the uncertainty of the measurement of the peak flux time in the LAT (see Fig.~\ref{fig_ExtraComponent}).  We assumed $n=1$\,cm$^{-3}$, $A_\star = 0.1$ and $E_{k,\rm iso}$ is four times larger than the isotropic-equivalent $\gamma$-ray energy $E_{\gamma,\rm iso}$ in the Band or Comptonized (in the case of GRB\,100414A) component.  The dependence of $\Gamma_0$ on the ISM density ($\propto n^{-1/8}$) is rather mild.  Thus, the dominant uncertainty of $\Gamma_0$ in the ISM environment comes from the peak-flux time.  Note that $\Gamma_0$ needs to be large in order to explain the delayed onset and peak of the LAT emission as results of early afterglow.  These estimates of $\Gamma_0$ are similar to $\Gamma_{\rm min}$ values calculated from $\gamma\gamma$ pair production opacities for the four brightest LAT
bursts~\citep{
GRB080916C:Science,GRB090510:ApJ,GRB090902B:Fermi,GRB090926A:Fermi}. For GRB\,110731A, detailed multiwavelength modeling suggests a wind environment. In the case of a wind environment, $\Gamma_0$ is usually smaller with milder $t_{dec}^{-1/4}$ dependence.


The temporal variability of $>$100~MeV emission in GRBs~090902B \citep{GRB090902B:Fermi} and 090926A \citep{GRB090926A:Fermi} argues against a simple forward shock interpretation in the prompt phase, since such variability is characteristic of internal shocks.  However, an energy-dependent transition between the prompt and afterglow contributions in the LAT flux is possible.


In the context of the internal shock scenario, the delayed onset of the LAT-detected emission could arise from late internal shocks produced via inverse Compton (IC) with plausible evolution of the microphysical parameters from the early internal shocks~\citep{Wang+09,Bosnjak+09,Toma+11}.  Hadronic emission such as proton/ion synchrotron radiation and/or photopion-induced cascade radiation could also account for this delay through the time required for proton/ion acceleration and cooling as well as to form cascades~\citep{Asano+09,Razzaque+10b,Wang+09}.  However, a challenge for the internal shocks scenario is explaining the temporally extended LAT-detected emission often lasting $\sim 10^2$--$10^3$~s (see Fig.~\ref{fig_Duration}) without associated detectable~keV--MeV emission.
\twocolumn

\subsection{Spectral Models of LAT-detected Emissions}

A power-law spectral component that dominates LAT-detected emission has been detected in the brightest LAT bursts: GRBs 080916C, 090510, 090902B, 090926A, and 110731A.  This component is in addition to the Band function or the Comptonized model that typically describes the~keV--MeV emission.  The top panel of Fig.~\ref{fig_Index_Duration} shows that this power-law component is hard in the prompt phase ($\Gamma_{\rm GBM} \sim -2$), allowing for a high-significance detection.  In other bursts it can be softer, and consequently not easily detectable.  In the ``EXT'' time window, however, the power-law component is hard (Fig.~\ref{fig_Index_Duration}, bottom panel) without any contamination from the~keV--MeV photons.  Whether or not the same hard power-law component in the prompt phase evolves into the power law in the ``EXT'' time window is a central issue in GRB science.

Early afterglow models for the temporally extended LAT-detected emission \citep{Kumar+09,Ghisellini+10,DePasquale+10,Razzaque+10} imply that a power-law component from the forward shock that propagates into the external medium surrounding the GRB \citep{Meszaros+97,Sari+98} arises early in the prompt phase when the fireball is still coasting.  A high jet bulk Lorentz factor seems to be required in this scenario as mentioned earlier.  IC scattering of soft target photons, either synchrotron or photospheric, by relativistic electrons can also produce an additional power-law component~\citep{Wang+09,Bosnjak+09,GRB090510:ApJ,Toma+11}.  The IC component contributes most significantly in the $\gg 1$~GeV range. Hadronic emission models, either proton/ion synchrotron radiation or photopion-induced cascade radiation, can produce an additional spectral component as well, able to dominate the LAT-detected emission in the prompt phase~\citep{Asano+09,Razzaque+10b,Wang+09}.  However,
these models require a much
larger total energy budget than the leptonic models, especially if the jet bulk Lorentz factor is high, which seems to be the case for LAT bursts.

Finally, significant cutoffs in the additional power-law component have been detected in the time-integrated prompt spectra of GRBs~090926A and 110731A.  Electron-positron pair production by high-energy photons with~keV--MeV photons is a plausible origin of these multi-GeV cutoffs~\citep{Krolik+91,Fenimore+93,Baring+97,Lithwick+01}.  Detection of such cutoffs in some future LAT bursts will be helpful in determining the bulk Lorentz factors of the jets, as well as in answering whether $\gamma\gamma$ opacity plays a role in the observed low detection rate of LAT bursts.

\subsection{Summary and conclusion}
\label{sec_conclusion}

We have compiled a catalog of all GRBs significantly detected by the \Fermi-LAT. For each of these bursts we have examined the spectral and temporal behavior of their high-energy emission. In this ensemble of bursts we have searched for common patterns in flux behavior in order to obtain an unbiased view of high-energy emission from GRBs. We have also compared the LAT-detected emission with the lower-energy emission detected by the GBM from a much greater number of bursts, and sought theoretical interpretations of the LAT observations.

In general LAT bursts are also among the brightest bursts seen by GBM.
They are also the most energetic when redshift measurements allow determination of the total luminosity.
Although based on only 4 bursts, there seems to be an emergent class of hyper-fluent LAT GRBs.

A common characteristic of the LAT-detected emission is that it is delayed with respect to the GBM emission.  
This delay is longer for long bursts, with some indications that the onset time increases with the energy. 
LAT bursts also generally have longer durations in the LAT energy range than in the GBM energy range for the same bursts.

LAT GRBs exhibit a temporally extended phase during which the LAT flux decays following a single or broken power law.  
The photon index in this phase is also distributed in a relatively narrow range.  
The index of power-law flux decay (later index in case of broken power-law fits) is typically close to $F_{\nu}\propto t^{-1}$ with only a few exceptions.

The temporally extended LAT-detected emission is consistent with that expected from afterglow (forward shock) emission from a relativistic blast wave.  
An adiabatic fireball model is favored over a radiative fireball model by the measured $\propto t^{-1}$ LAT flux-decay behavior in the majority of bursts.

The spectra of LAT GRBs are typically well described by a power-law with a fairly narrow distribution of indices, centered at $-2.0$ although deviations (spectral cutoffs) from a pure power law have been detected in GRBs~090926A and 110731A in the~GeV range. Joint GBM-LAT spectral fits require an additional power-law component in all bright LAT bursts, indicating that the Band function alone is inadequate to fit the spectra of these bursts.

Several models exist in the literature for the delayed onset of LAT-detected emission and the additional power-law component. The early afterglow model for temporally extended LAT-detected emission can explain both the delayed onset and the additional component, but other models involving internal shocks cannot be ruled out.
The detection of additional bright LAT bursts will help to characterize and explain cutoffs in the power-law spectra, determine the bulk Lorentz factors, and constrain GRB energetics.

\clearpage
\newpage
\section{Tables}
\label{sec_tables}
In the following section, we present the results of our catalog in tabular form. Additionally, we provide all the numbers shown here in an electronic FITS\footnote{\url{http://fits.gsfc.nasa.gov/}} file format.

Table~\ref{tab_intervals_fit} summarizes the intervals in which we performed the time-integrated spectral analysis described in \S\ref{subsec_analysis_sequence} on page \pageref{subsec_analysis_sequence}.

Table~\ref{tab_GRBs} contains the list of LAT-detected GRBs, including the trigger time and position information used as input to our analysis pipeline. Each GRB was detected using the standard likelihood analysis (denoted by ``Like=1'' in the table) and/or the LAT Low Energy analysis (denoted using ``LLE=1''). We also list the redshift (errors are omitted) and the reference number of the LAT GCN circular, if one was issued.
Since the LAT localizations are obtained iteratively, we report only the final localization.

Table~\ref{tab_durations} shows a comparison between the various duration estimates obtained using the standard LAT data and the LLE analysis. We also report the duration of the bursts as reported in the GBM catalog \citep{Paciesas+12} indicating whether the burst is classified as short (S) or long (L). The final two columns report the maximum significance of the source in the likelihood analysis (Max TS) and the post-trials detection significance obtained by the LLE analysis.

Likelihood analysis results are summarized in Table~\ref{tab_likelihoods}, where we report for each interval and for each GRB, the number of events actually detected inside the ROI, the predicted number of events from the source, the detection significance, and the values of the measured photon flux, energy fluence, and isotropic equivalent energy (if a redshift is available).
For the cases where the significance is below our detection threshold, we report upper limits. 
Three bursts detected by the LLE analysis are included in this table, the other 4 bursts (GRBs\,090531B, 101014A, 101123A and 110529A) had too few events to even compute an upper limit during the ``GBM'' time interval.

Table~\ref{tab_localizations} shows our best reconstructed direction with associated errors.

The highest-energy events associated with each GRB are summarized in Tables \ref{tab_energymax_gbm}, \ref{tab_energymax_ext}, and \ref{tab_energymax_all}. 
In these tables we used different time-window to perform the analysis, and we indicate the number of events associated with the GRB, the energy, and the arrival time of the highest-energy event. 
We also report the probability of the event being associated with the GRB computed as described in \S\ref{susubbsec_event_probability}.

The temporally extended high-energy emission is systematically studied in this paper, and the relative quantities are summarized in Table~\ref{tab_extended}.
We report the results obtained by fitting the photon flux light curves with simple power laws starting from the position of the peak flux and from the position of the GBM T$_{95}$. When the statistics allow, we also perform a broken power-law fit. A font in bold letters indicates the parameters that best reproduce the late time decay of the $\gamma$-ray flux.

Next we summarize the results of our joint spectral-fit analyses. In
Table~\ref{tab_bandCrisis} we report the spectral model that best fits
the data during the ``GBM'' time interval. Then we present the whole
range of results from the joint-fit spectral analyses as obtained in
the ``GBM'' time interval (Table~\ref{tab_JointFitGBMT90}) and in the
interval extending from the first detection of a GRB photon by the LAT
up to the GBM T$_{95}$ (Table~\ref{tab_JointFitLATPROMPT}).
Only bursts detected by the LAT ($TS > 20$, see $\S$~\ref{subsec_analysis_sequence}) in the ``GBM'' time interval are included in Table~\ref{tab_JointFitLATPROMPT}.
In these two tables, we display the parameters of the main component and the
parameters of any additional components required to describe the
spectrum. For the cases that more than one component is needed, we
compute the energy fluence for each spectral component separately. In
Table\ref{tab_eiso_components} we report the isotropic equivalent
energy in aggregate and also per spectral component for the best-fit
spectral model.

Finally, we address the systematic uncertainties of our results by using a different set of data-selection cuts and we compare our standard results obtained with the Pass 6 event selection to the results obtained with the new Pass 7 data selection. This is summarized in Table~\ref{tab_systematics} and described in Appendix~\ref{sec_systematics}.

\begin{deluxetable}{lll}
\tablecolumns{3}
\tablewidth{0pt}
\tablecaption{Definitions of intervals used in time-integrated spectral analysis}
\tablehead{\colhead{Name}&\colhead{Definition}&\colhead{Description}}
\startdata
GBM&\makebox[30pt][l]{GBM}T$_{05}$ -- GBM T$_{95}$& Bulk of the GBM-detected
emission\\
LAT&\makebox[30pt][l]{LAT}T$_{05}$ -- LAT T$_{95}$& Bulk of the LAT-detected
emission\\
PRE&\makebox[30pt][l]{GBM}T$_{05}$ -- LAT T$_{05}$& Interval between GBM and LAT emission onsets\\
JOINT&\makebox[30pt][l]{LAT}T$_{05}$ -- GBM T$_{95}$& Interval when both the GBM
and LAT significantly detect emission\\
EXT&\makebox[30pt][l]{GBM}T$_{95}$ -- LAT T$_{95}$& Interval between end of
GBM-detected and LAT-detected emission\\
LATTE && Interval between the start of the first and the end of the last\\
& & bins with TS $>$ 16, as found by the time-resolved likelihood analysis\\
\enddata
\label{tab_intervals_fit}
\end{deluxetable}

\begin{deluxetable}{lrrrrrrrrrr}
\tabletypesize{\tiny}
\tablecolumns{11}
\tablewidth{0pt}
\tablecaption{Sample of \Fermi-LAT GRBs, from August 2008 to August 2011. 
\label{tab_GRBs}}
\tablehead{
\colhead{GRB NAME} & \colhead{DATE} & \colhead{GBM Trigger Time} & \colhead{R.A.} & \colhead{Dec.} & \colhead{$\theta$} & \colhead{Loc. Err.$^{b}$} & \colhead{Like.} & \colhead{LLE} & \colhead{Redshift} & \colhead{LAT GCN Number} \\ 
\colhead{ } & \colhead{ } & \colhead{(MET$^a$)} & \colhead{Deg., J2000} & \colhead{Deg., J2000} & \colhead{Deg.} & \colhead{} & \colhead{} & \colhead{} & \colhead{} & \colhead{}} 
\startdata
 080825C  &  2008-08-25 14:13:48.1  &    241366429.105 &      233.9  &       $-$4.5  &  60.3  & 0.75\de$^{\gamma}$ & 1 & 0 & - & 8183 \\ 
 080916C  &  2008-09-16 00:12:45.6  &    243216766.614 &     119.85  &     $-$56.64  &  48.8  & 0.36''$^{\star}$ & 1 & 1 & 4.35 & 8246 \\ 
  081006  &  2008-10-06 14:29:34.1  &    244996175.173 &     136.32  &     $-$62.05  &  11.0  & 0.52\de$^{\gamma}$ & 1 & 0 & - &  \\ 
 081024B  &  2008-10-24 21:22:40.8  &    246576161.864 &     322.95  &       21.2  &  18.7  & 0.22\de$^{\gamma}$ & 1 & 1 & - & 8407 \\ 
  090217  &  2009-02-17 04:56:42.5  &     256539404.560 &     204.83  &      $-$8.42  &  34.5  & 0.35\de$^{\gamma}$ & 1 & 1 & - & 8903 \\ 
 090227B  &  2009-02-27 18:31:01.4  &     257452263.410 &      10.48  &      29.24  &  71.0  & 1.00\de$^{\bigtriangleup}$ & 1 & 1 & - &  \\ 
  090323  &  2009-03-23 00:02:42.6  &     259459364.630 &     190.71  &    17.053  &  57.2  & 0.36'' $^{\star}$ & 1 & 1 & 3.57 & 9021 \\ 
  090328  &  2009-03-28 09:36:46.5  &     259925808.510 &      90.67  &   $-$41.715  &  64.6  & 0.72'' $^{\star}$ & 1 & 1 & 0.74 & 9044,9077 \\
\enddata
\tablenotetext{a}{Mission Elapsed Time: seconds since 2001-1-1 00:00:00 UTC.}
\tablenotetext{b}{Uncertainties on the localizations from from: $^{\gamma}$ \Fermi-LAT, $^{\dagger}$ \Fermi-GBM, $^{\star}$ Swift-XRT/Swift-UVOT, $^{\bigtriangleup}$ IPN.} 
\end{deluxetable}

\begin{deluxetable}{lrrrrrrrrr}
\tabletypesize{\tiny}
\tablecolumns{10}
\tablewidth{0pt}
\tablecaption{Comparisons between duration estimators\label{tab_durations}}
\tablehead{
\colhead{GRB NAME} & \colhead{CLASS$^{a}$} & \colhead{GBM T$_{05}$} & \colhead{GBM T$_{95}$} & \colhead{LAT T$_{05}$} & \colhead{LAT T$_{95}$} & \colhead{LLE T$_{05}$} & \colhead{LLE T$_{95}$} & \colhead{Max TS} & \colhead{LLE Significance} \\ 
\colhead{ } & \colhead{} & \colhead{s } & \colhead{s } & \colhead{s } & \colhead{s } & \colhead{s } & \colhead{s } & \colhead{} & \colhead{$\sigma$, Post Trials}} 
\startdata
080825C & L & 1.22 & 23.4 $\pm$ 0.2 & 3.3$_{-0.1}^{+0.2}$ & 29$_{-3}^{+17}$ & - & - & 57 & 3.4 \\ 
080916C & L & 1.28 & 65.5 $\pm$ 0.8 & 5.0$_{-0.3}^{+0.5}$ & 210$_{-50}^{+60}$ & 4.1$_{-0.1}^{+0.2}$ & 80$_{-20}^{+30}$ & 1450 & 26.1 \\ 
081006 & L & $-$0.26 & 5.9 $\pm$ 0.9 & $>$0.7 & $>$100 & - & - & 72 & 1.1 \\ 
081024B & S & $-$0.06 & 0.5 $\pm$ 0.3 & $>$0.05 & $>$200 & -0.1$_{-0.3}^{+0.1}$ & 2.1$_{-0.3}^{+0.2}$ & 111 & 4.5 \\ 
090217 & L & 0.83 & 34.9 $\pm$ 0.7 & 6.2$_{-5.0}^{+0.5}$ & 70$_{-40}^{+110}$ & 0$_{-8}^{+2}$ & 14.0$_{-0.8}^{+7.3}$ & 105 & 10.9 \\ 
090227B & S & $-$0.06 & 1 $\pm$ 1 & - & - & $-$0.01 $\pm$ 0.01 & 1.6$_{-0.8}^{+0.3}$ & 30 & 20.5 \\ 
090323 & L & 8.19 & 152 $\pm$ 1 & 16$_{-5}^{+47}$ & 290$_{-30}^{+50}$ & 6.9$_{-2.1}^{+1.0}$ & 185$_{-6}^{+13}$ & 136 & 14.4 \\ 
090328 & L & 4.35 & 70 $\pm$ 2 & 19$_{-4}^{+33}$ & 650$_{-40}^{+130}$ & 9 $\pm$ 1 & 90$_{-50}^{+10}$ & 107 & 14.2 \\
\enddata
\tablenotetext{a}{In accordance with convention, we define as Short (S) those GRBs with GBM T$_{90}<$2 s, and Long (L) those with T$_{90}>$2 s. Durations of GBM bursts are from \citep{Paciesas+12}.}
\end{deluxetable}


\newgeometry{left=1cm,right=1cm,bottom=0.1cm,top=7.0cm}
\begin{landscape}
\thispagestyle{empty}
\begin{deluxetable}{lrrrrrrrr}
\tabletypesize{\tiny}
\tablecolumns{9}
\tablewidth{0pt}
\tablecaption{Results from Likelihood Analysis\label{tab_likelihoods}}
\tablehead{
\colhead{GRB NAME} & \colhead{Interval (t$_{0}$-t$_{1}$)} & \colhead{Trans. Ev.} & \colhead{Trans. Ev.} & \colhead{Test Statistic} & \colhead{Spectral Index} & \colhead{Flux} & \colhead{Fluence} & \colhead{$\rm E_{iso}$ (100~MeV--10~GeV)} \\ 
\colhead{ } & \colhead{s} & \colhead{in the ROI} & \colhead{Predicted} & \colhead{(TS)} & \colhead{ } & \colhead{cm$^{-2}$ s$^{-1}$ ($\times 10^{-5}$)} & \colhead{erg cm$^{-2}$ ($\times 10^{-5}$)} & \colhead{erg ($\times 10^{52}$)}} 
\startdata
080825C &  GBM (1.2--23.4) & 7 & 6.8 & 36 & $-$3.3 $\pm$ 0.7 & 20 $\pm$ 10 & 0.16 $\pm$ 0.06 & - \\ 
 &  LAT (3.2--29.4) & 11 & 10.1 & 57 & $-$2.7 $\pm$ 0.5 & 28 $\pm$ 10 & 0.3 $\pm$ 0.1 & - \\ 
 &  JOINT (3.2--23.4) & 7 & 6.8 & 38 & $-$3.3 $\pm$ 0.7 & 30 $\pm$ 10 & 0.16 $\pm$ 0.06 & - \\ 
 &  EXT (23.4--29.4) & 4 & 3.5 & 25 & $-$2.1 $\pm$ 0.5 & 40 $\pm$ 20 & 0.2 $\pm$ 0.2 & - \\ 
 &  LATTE (3.2--56.2) & 14 & 11.5 & 50 & $-$2.7 $\pm$ 0.4 & 16 $\pm$ 5 & 0.3 $\pm$ 0.1 & - \\ 
\hline
080916C &  GBM (1.3--65.5) & 156 & 150.3 & 1450 & $-$2.13 $\pm$ 0.08 & 82 $\pm$ 7 & 4.3 $\pm$ 0.7 & 150 $\pm$ 20 \\ 
 &  LAT (5.0--209.8) & 201 & 180.0 & 1382 & $-$2.05 $\pm$ 0.07 & 29 $\pm$ 2 & 5.7 $\pm$ 0.9 & 160 $\pm$ 20 \\ 
 &  JOINT (5.0--65.5) & 146 & 140.2 & 1338 & $-$2.10 $\pm$ 0.08 & 81 $\pm$ 7 & 4.3 $\pm$ 0.8 & 140 $\pm$ 10 \\
\enddata
\end{deluxetable}

\end{landscape}
\restoregeometry

\begin{deluxetable}{lrrrrr}
\tabletypesize{\tiny}
\tablecolumns{6}
\tablewidth{0pt}
\tablecaption{\Fermi-LAT Localizations\label{tab_localizations}}
\tablehead{
\colhead{GRB NAME} & \colhead{R.A.} & \colhead{Dec.} & \colhead{68\%} & \colhead{90\%} & \colhead{95\%} \\ 
\colhead{ } & \colhead{Deg., J2000} & \colhead{Deg., J2000} & \colhead{Deg.} & \colhead{Deg.} & \colhead{Deg.}} 
\startdata
 080825C  &  233.95  &  $-$4.55  &  0.77  &  1.24  &  1.55  \\ 
 080916C  &  119.87  &  $-$56.58  &  0.07  &  0.10  &  0.12  \\ 
  081006  &  136.43  &  $-$62.10  &  0.51  &  0.76  &  0.89  \\ 
 081024B  &  322.94  &  21.05  &  0.29  &  0.46  &  0.56  \\ 
 090217  &  204.79  &  $-$8.41  &  0.35  & 0.51 & 0.59 \\ 
  090323  &  190.64  &  17.03  &  0.10  &  0.16  &  0.20  \\ 
  090328  &  90.54  &  $-$42.01  &  0.13  &  0.17  &  0.19  \\ 
  090510  &  333.50  &  $-$26.53  &  0.04  &  0.06  &  0.07  \\
\enddata
\end{deluxetable}


\begin{deluxetable}{lrrrr}
\tabletypesize{\tiny}
\tablecolumns{5}
\tablewidth{0pt}
\tablecaption{Highest energy events for \Fermi-LAT GRBs: GBM Durations\label{tab_energymax_gbm}}
\tablehead{
\colhead{GRB NAME} & \colhead{Number of events} & \colhead{Energy} & \colhead{Arrival time} & \colhead{Probability } \\ 
\colhead{ } & \colhead{($>$ 100 MeV, P$>$0.9)} & \colhead{GeV} & \colhead{s} & \colhead{}} 
\startdata
080825C & 7 & 0.29 & 3.25 & 0.9854 \\ 
080916C & 143 & 13.22 & 16.54 & 0.9999 \\ 
081006 & 7 & 0.65 & 1.80 & 0.9997 \\ 
081024B & 7 & 3.07 & 0.49 & 1.0000 \\ 
090217 & 11 & 0.87 & 14.83 & 0.9960 \\ 
090227B & 3 & 0.24 & 0.48 & 0.9996 \\ 
090323 & 12 & 0.48 & 92.74 & 0.9682 \\ 
090510 & 36 & 31.31 & 0.83 & 1.0000 \\ 
090720B & 2 & 1.45 & 0.22 & 0.9997 \\ 
090902B & 155 & 11.16 & 11.67 & 0.9999 \\ 
090926A & 149 & 3.19 & 9.48 & 0.9990 \\ 
091003 & 6 & 2.83 & 6.47 & 0.9997 \\ 
091208B & 3 & 1.18 & 3.41 & 0.9958 \\ 
100116A & 4 & 0.86 & 101.30 & 0.9973 \\ 
100325A & 4 & 0.84 & 0.35 & 0.9990 \\ 
100414A & 4 & 0.64 & 19.89 & 0.9442 \\ 
100620A & 3 & 0.27 & 3.77 & 0.9886 \\ 
100724B & 16 & 0.22 & 61.75 & 0.9805 \\ 
110120A & 4 & 0.46 & 0.87 & 0.9570 \\ 
110709A & 3 & 0.17 & 30.63 & 0.9596 \\ 
110721A & 15 & 0.86 & 0.86 & 0.9937 \\ 
110731A & 38 & 0.88 & 5.52 & 0.9974 \\ 
\enddata
\end{deluxetable}

\begin{deluxetable}{lrrrr}
\tabletypesize{\tiny}
\tablecolumns{5}
\tablewidth{0pt}
\tablecaption{Highest energy events for \Fermi-LAT GRBs: EXT Durations\label{tab_energymax_ext}}
\tablehead{
\colhead{GRB NAME} & \colhead{Number of events} & \colhead{Energy} & \colhead{Arrival time} & \colhead{Probability } \\ 
\colhead{ } & \colhead{($>$ 100 MeV, P$>$0.9)} & \colhead{GeV} & \colhead{s} & \colhead{}} 
\startdata
080825C & 3 & 0.57 & 28.29 & 0.999 \\ 
080916C & 33 & 1.46 & 124.16 & 0.998 \\ 
090323 & 11 & 7.50 & 195.42 & 1.000 \\ 
090328 & 14 & 3.83 & 264.42 & 0.956 \\ 
090510 & 141 & 3.90 & 1.55 & 1.000 \\ 
090626 & 9 & 2.09 & 111.63 & 0.998 \\ 
090902B & 108 & 12.54 & 45.61 & 0.999 \\ 
090926A & 80 & 19.56 & 24.83 & 1.000 \\ 
091003 & 11 & 1.79 & 76.78 & 0.993 \\ 
091031 & 5 & 1.19 & 79.75 & 0.996 \\ 
100116A & 7 & 2.20 & 105.71 & 1.000 \\ 
100414A & 11 & 4.72 & 288.26 & 0.999 \\ 
110120A & 2 & 1.82 & 72.46 & 0.998 \\ 
110428A & 5 & 2.62 & 14.79 & 0.999 \\ 
110731A & 18 & 1.90 & 8.27 & 1.000 \\ 
\enddata
\end{deluxetable}

\begin{deluxetable}{lrrrr}
\tabletypesize{\tiny}
\tablecolumns{5}
\tablewidth{0pt}
\tablecaption{Highest energy events for \Fermi-LAT GRBs: Time resolved analysis\label{tab_energymax_all}}
\tablehead{
\colhead{GRB NAME} & \colhead{Number of events} & \colhead{Energy} & \colhead{Arrival time} & \colhead{Probability } \\ 
\colhead{ } & \colhead{($>$ 100 MeV, P$>$0.9)} & \colhead{GeV} & \colhead{s} & \colhead{}} 
\startdata
080825C & 10 & 0.57 & 28.29 & 0.997 \\ 
080916C & 181 & 13.22 & 16.54 & 1.000 \\ 
081006 & 10 & 0.79 & 12.08 & 0.955 \\ 
081024B & 11 & 3.07 & 0.49 & 1.000 \\ 
090217 & 16 & 1.23 & 179.08 & 0.907 \\ 
090323 & 28 & 7.50 & 195.42 & 1.000 \\ 
090328 & 23 & 5.32 & 697.80 & 0.926 \\ 
090510 & 186 & 31.31 & 0.83 & 1.000 \\ 
090626 & 15 & 2.09 & 111.63 & 0.999 \\ 
090720B & 2 & 1.45 & 0.22 & 0.997 \\ 
090902B & 276 & 33.39 & 81.75 & 0.949 \\ 
090926A & 239 & 19.56 & 24.83 & 1.000 \\ 
091003 & 20 & 2.83 & 6.47 & 1.000 \\ 
091031 & 7 & 1.19 & 79.75 & 0.999 \\ 
091208B & 4 & 1.18 & 3.41 & 0.956 \\ 
100116A & 14 & 13.12 & 296.43 & 0.993 \\ 
100325A & 5 & 0.84 & 0.35 & 0.990 \\ 
100414A & 19 & 4.72 & 288.26 & 1.000 \\ 
100620A & 6 & 0.27 & 3.77 & 0.994 \\ 
100724B & 16 & 0.22 & 61.75 & 0.988 \\ 
100728A & 5 & 13.54 & 5461.08 & 0.987 \\ 
110120A & 6 & 1.82 & 72.46 & 0.999 \\ 
110428A & 6 & 2.62 & 14.79 & 1.000 \\ 
110625A & 6 & 2.42 & 272.44 & 0.986 \\ 
110709A & 5 & 0.42 & 41.75 & 0.921 \\ 
110721A & 22 & 1.73 & 0.74 & 0.998 \\ 
110731A & 64 & 3.39 & 435.96 & 0.998 \\ 
\enddata
\end{deluxetable}

\newgeometry{left=1cm,right=1cm,bottom=0.1cm,top=7.0cm}
\begin{landscape}
\thispagestyle{empty}
\begin{deluxetable}{lrrrrrrrr}
\tabletypesize{\tiny}
\tablecolumns{9}
\tablewidth{0pt}
\tablecaption{Temporally extended high-energy emission\label{tab_extended}}
\tablehead{
\colhead{GRB NAME} & \colhead{Peak Flux $F_{p}$} & \colhead{Peak-Flux Time $t_{p}$} &
\colhead{$\alpha$ (SPL)}  & \colhead{$\alpha$(SPL)} & \colhead{$\alpha_1$ (BPL)}
& \colhead{$\alpha_2$ (BPL)} & \colhead{Break Time (BPL) $t_{b}$} \\
\colhead{ } & \colhead{cm$^{-2}$ s$^{-1}$ ($\times 10^{-5}$)} & \colhead{s}
& \colhead{from the peak flux} & \colhead{from GBM T$_{95}$} & \colhead{ } &
\colhead{} & \colhead{s} }
\startdata
080916C & 500 $\pm$ 100 & 6.6 $\pm$ 0.9   & 1.37 $\pm$ 0.07 & \textbf{1.8 $\pm$ 0.3}  & - & - & - \\
090323 & 6 $\pm$ 3 & 40 $\pm$ 30 & 1.0 $\pm$ 0.3 & \textbf{0.9 $\pm$ 0.3}  & - & - & - \\
090328 & 9 $\pm$ 4 & 40 $\pm$ 30 & 1.0 $\pm$ 0.3 & \textbf{0.9 $\pm$ 0.2}  & - & - & - \\
090510 & 3900 $\pm$ 600 & 0.9 $\pm$ 0.1 & 1.8 $\pm$ 0.2 & 1.82 $\pm$ 0.2  &   $2.2 \pm  0.1$ & \textbf{1.1 $\pm$ 0.1} &  $7 \pm   1$ \\
090902B & 600 $\pm$ 100 & 9 $\pm$ 1  & 1.56 $\pm$ 0.06 & 1.4 $\pm$ 0.1  &  $1.7 \pm  0.2$ & \textbf{0.8 $\pm$ 0.2} & $130 \pm    50$ \\
091003 & 8 $\pm$ 3 & 22 $\pm$ 9 & 1.0 $\pm$ 0.2  & 1.6 $\pm$ 0.3  & $2.7 \pm 0.09$ & \textbf{0.86 $\pm$  0.07} & $40 \pm   5$\\
091003    &     8 $\pm$ 3    & 22.5  $\pm$ 9.1  & 1.0 $\pm$ 0.2  & \textbf{1.0 $\pm$ 0.2}  & - & - & - \\
100414A & 70 $\pm$ 30 & 20 $\pm$ 10 & 1.7 $\pm$ 0.3 & \textbf{1.1 $\pm$ 0.4}  & - & - & - \\
110731A & 220 $\pm$ 60 & 4.9 $\pm$ 0.7  & 1.8 $\pm$ 0.2 & \textbf{1.5 $\pm$ 0.2}  & - & - & - \\
\enddata
\tablecomments{Using numbers from this table we also define the ``late-time
decay index'' $\alpha_{L}$, which is equal to $\alpha$ from GBM T$_{95}$ for all GRBs,
except the 3 for which we detect the break time, for which
$\alpha_{L}=\alpha_{2}$. The corresponding value is also marked with the bold font.}
\end{deluxetable}

%
\end{landscape}
\restoregeometry

\begin{deluxetable}{lrrrrrrrrrrrr}
\tabletypesize{\tiny}
\tablecolumns{13}
\tablewidth{0pt}
\tablecaption{The best spectral model for the GRB during the GBM interval, ordered by fluence\label{tab_bandCrisis}}
\tablehead{
\colhead{ } & \colhead{Fluence} & \colhead{Best model} & \colhead{$\theta$} \\
\colhead{ } & \colhead{10~keV - 10~GeV} & \colhead{ } & \colhead{deg} \\
\colhead{ } & \colhead{($10^{-7}$ erg/cm$^2$)} & \colhead{ } & \colhead{ }}
\startdata
100724B & $ 4665^{-76}_{+78} $   & Band with exponential cutoff &  48.9 \\
090902B & $ 4058^{-24}_{+25} $   & Comptonized + Power law & 50.8 \\
090926A & $ 2225^{-48}_{+50} $    & Band + Power law with exponential cutoff & 48.1 \\
080916C & $ 1795^{-39}_{+41} $   & Band + Power law & 48.8 \\
090323 & $ 1528^{-44}_{+44} $    & Band & 57.2 \\
100728A & $ 1293^{-27}_{+28} $   & Comptonized & 59.9 \\
100414A & $ 1098^{-27}_{+35} $   & Comptonized + Power law & 69.0 \\
090626 & $ 927^{-16}_{+17} $     & Logarithmic parabola & 18.3 \\
\enddata
\tablecomments{We exclude from this table all GRBs outside the nominal LAT FOV (with $\theta>$70\de) and GRB\,101014A, which was detected too close to the Earth limb.}
\end{deluxetable}


\newgeometry{left=1cm,right=1cm,bottom=0.1cm,top=7.0cm}
\thispagestyle{empty}
\begin{landscape}
\begin{deluxetable}{@{\extracolsep{4pt}}lrrrrrrrr@{\extracolsep{20pt}}r@{\extracolsep{4pt}}rr@{\extracolsep{20pt}}r@{\extracolsep{4pt}}r}
\tabletypesize{\tiny}
\tablecolumns{14}
\tablewidth{0pt}
\tablecaption{Results for the joint fit over the interval GBM T05-GBM T95\label{tab_JointFitGBMT90}}
\tablehead{
\colhead{} & \multicolumn{8}{c}{Main component} & \multicolumn{3}{c}{Additional components} & \multicolumn{1}{c}{} & \multicolumn{1}{c}{}\\[3pt]
\cline{2-9}\cline{10-12}\\[5pt]

\colhead{GRB} & \multicolumn{3}{c}{Band} & \multicolumn{2}{c}{Comptonized} & \multicolumn{2}{c}{Log. Parabola} & \multicolumn{1}{c}{Fluence} &
\multicolumn{1}{c}{Power law} & \multicolumn{1}{c}{Cut Off} & \multicolumn{1}{c}{Fluence} &
\multicolumn{1}{c}{Total Fluence} & \multicolumn{1}{c}{Statistic} \\
\cline{2-4} \cline{5-6} \cline{7-8} \cline{9-9}\cline{10-10}\cline{11-11}\cline{12-12}\cline{13-13}\cline{14-14}\\

& \colhead{E$_0$} & \colhead{$\alpha$} & \colhead{$\beta$} & 
 \colhead{E$_0$} & \colhead{$\alpha$} &  
 \colhead{b} & \colhead{E$_p$} & 
 \colhead{10~keV - 10~GeV} & 
 \colhead{$\alpha$}  &
 \colhead{E$_{c}$} &
 \colhead{10~keV - 10~GeV} &
 \colhead{10~keV - 10~GeV} &
 \colhead{Stat./Dof} \\ 

& \colhead{(keV)} & & & 
 \colhead{(keV)} &  
 & & \colhead{(keV)} & 
 \colhead{($10^{-7}$ erg/cm$^2$)} &
 & 
 \colhead{(MeV)} & 
 \colhead{($10^{-7}$ erg/cm$^2$)} &
 \colhead{($10^{-7}$ erg/cm$^2$)} & \\[5pt]
}
\startdata
080825C &$ 141^{-5}_{+5} $ & $ -0.65^{-0.02}_{+0.02} $ & $ -2.40^{-0.04}_{+0.03} $ &                          &                          &                          &                          &                          &                          &                          &                          & $ 517^{-20}_{+21} $ & 1002.2/821 \\ 
\noalign{\smallskip}\hline\noalign{\smallskip}
080916C &$ 269^{-19}_{+21} $ & $ -0.65^{-0.06}_{+0.05} $ & $ -2.22^{-0.04}_{+0.02} $ &                          &                          &                          &                          & $ 1614^{-12}_{+12} $ & $ 2.01^{-0.07}_{+0.15} $ &                          & $ 181^{-102}_{+114} $ & $ 1795^{-39}_{+41} $ & 485.1/354 \\ 
\noalign{\smallskip}\hline\noalign{\smallskip}
081006 &$ 496^{-197}_{+394} $ & $ -0.48^{-0.26}_{+0.34} $ & $ -2.30^{-0.10}_{+0.08} $ &                          &                          &                          &                          &                          &                          &                          &                          & $ 56^{-9}_{+10} $ & 477.3/478 \\ 
\noalign{\smallskip}\hline\noalign{\smallskip}
081024B &$ 1313^{-580}_{+1196} $ & $ -0.93^{-0.13}_{+0.16} $ & $ -2.12^{-0.13}_{+0.10} $ &                          &                          &                          &                          &                          &                          &                          &                          & $ 30^{-5}_{+6} $ & 354.3/357 \\ 
\noalign{\smallskip}\hline\noalign{\smallskip}
090217 &$ 504^{-27}_{+30} $ & $ -0.86^{-0.02}_{+0.02} $ & $ -2.56^{-0.05}_{+0.05} $ &                          &                          &                          &                          &                          &                          &                          &                          & $ 512^{-15}_{+16} $ & 495.7/358 \\
\enddata
\tablenotetext{a}{These GRBs have such a large off-axis angle that the corresponding effective area for the LAT (Transient class) is negligible. Accordingly, only GBM data have been used during the spectral analysis, and the fluence has been computed extrapolating the best fit model up to the LAT energy range.}
\tablecomments{Each GRB is modeled by one main component, and eventually
one or more additional components. So for example, the spectrum of GRB\,080825C is well described by a Band model, thus only the
corresponding columns are filled. The spectrum of GRB\,090926A is instead modeled by a Band model plus a Power law times an Exponential 
cutoff (see main text).}
\end{deluxetable}
\end{landscape}

\restoregeometry
\clearpage
\newgeometry{left=1cm,right=1cm,bottom=0.1cm,top=7.0cm}
\begin{landscape}
\thispagestyle{empty}
\begin{landscape}
\begin{deluxetable}{@{\extracolsep{4pt}}lrrrrrrrr@{\extracolsep{20pt}}r@{\extracolsep{4pt}}rr@{\extracolsep{20pt}}r@{\extracolsep{4pt}}r}
\tabletypesize{\tiny}
\tablecolumns{14}
\tablewidth{0pt}
\tablecaption{Results for the joint fit over the interval between the first photon detected by the LAT inside the energy-dependent ROI and the GBM T95 \label{tab_JointFitLATPROMPT}}
\tablehead{
\colhead{} & \multicolumn{8}{c}{Main component} & \multicolumn{3}{c}{Additional components} & \multicolumn{1}{c}{} & \multicolumn{1}{c}{}\\[3pt]
\cline{2-9}\cline{10-12}\\[5pt]

\colhead{GRB} & \multicolumn{3}{c}{Band} & \multicolumn{2}{c}{Comptonized} & \multicolumn{2}{c}{Log. Parabola} & \multicolumn{1}{c}{Fluence} &
\multicolumn{1}{c}{Power law} & \multicolumn{1}{c}{Cut Off} & \multicolumn{1}{c}{Fluence} &
\multicolumn{1}{c}{Total Fluence} & \multicolumn{1}{c}{Statistic} \\
\cline{2-4} \cline{5-6} \cline{7-8} \cline{9-9}\cline{10-10}\cline{11-11}\cline{12-12}\cline{13-13}\cline{14-14}\\

& \colhead{E$_0$} & \colhead{$\alpha$} & \colhead{$\beta$} & 
 \colhead{E$_0$} & \colhead{$\alpha$} &  
 \colhead{b} & \colhead{E$_p$} & 
 \colhead{10~keV - 10~GeV} & 
 \colhead{$\alpha$}  &
 \colhead{E$_{c}$} &
 \colhead{10~keV - 10~GeV} &
 \colhead{10~keV - 10~GeV} &
 \colhead{Stat./Dof} \\ 

& \colhead{(keV)} & & & 
 \colhead{(keV)} &  
 & & \colhead{(keV)} & 
 \colhead{($10^{-7}$ erg/cm$^2$)} &
 & 
 \colhead{(MeV)} & 
 \colhead{($10^{-7}$ erg/cm$^2$)} &
 \colhead{($10^{-7}$ erg/cm$^2$)} & \\[5pt]
}
\startdata
080825C &$ 126^{-5}_{+5} $ & $ -0.65^{-0.02}_{+0.02} $ & $ -2.43^{-0.05}_{+0.04} $ &                          &                          &                          &                          &                          &                          &                          &                          & $ 358^{-17}_{+18} $ & 975.9/821 \\ 
\noalign{\smallskip}\hline\noalign{\smallskip}
080916C &$ 260^{-17}_{+22} $ & $ -0.65^{-0.06}_{+0.05} $ & $ -2.20^{-0.04}_{+0.02} $ &                          &                          &                          &                          & $ 1498^{-11}_{+12} $ & $ 2.00^{-0.06}_{+0.12} $ &                          & $ 191^{-98}_{+107} $ & $ 1689^{-36}_{+38} $ & 474.6/354 \\ 
\noalign{\smallskip}\hline\noalign{\smallskip}
081006 &                         &                          &                          &                          &                          & $ 0.25^{-0.05}_{+0.05} $ & $ 6765^{-1598}_{+2309} $ &                          &                          &                          &                          & $ 46^{-9}_{+11} $ & 461.7/479 \\ 
\noalign{\smallskip}\hline\noalign{\smallskip}
081024B &                         &                          &                          &                          &                          & $ 0.13^{-0.03}_{+0.03} $ & $ 46287^{-22078}_{+69759} $ &                          &                          &                          &                          & $ 22^{-4}_{+5} $ & 338.9/358 \\ 
\noalign{\smallskip}\hline\noalign{\smallskip}
090217 &$ 526^{-29}_{+32} $ & $ -0.88^{-0.02}_{+0.02} $ & $ -2.57^{-0.05}_{+0.05} $ &                          &                          &                          &                          &                          &                          &                          &                          & $ 500^{-15}_{+16} $ & 497.7/358 \\
\enddata
\tablecomments{Each GRB is modeled by one main component, and eventually
one or more additional components. So for example, the spectrum of GRB\,080825C is well described by a Band model, thus only the
corresponding columns are filled. The spectrum of GRB\,090926A is instead modeled by a Band model plus a Power law times an Exponential 
cutoff (see main text).}
\end{deluxetable}
\end{landscape}


\end{landscape}
\restoregeometry

\begin{deluxetable}{lrrrr}
\tabletypesize{\tiny}
\tablecolumns{5}
\tablewidth{0pt}
\tablecaption{Isotropic equivalent energy by component.}
\tablehead{\colhead{GRB NAME} & \colhead{Best Model} & \colhead{E$_{iso}$} & \colhead{E$^{Band}_{iso}$} & \colhead{E$^{PL}_{iso}$} \\
\colhead{ } & \colhead{ } & \colhead{} & \colhead{} & \colhead{} \\
\colhead{ } & \colhead{ } & \colhead{10$^{52}$ erg} & \colhead{10$^{52}$ erg} & \colhead{10$^{52}$ erg}}
\startdata
GRB080916C & BP & 647.2$^{+12.8}_{-12.3}$ & 564.6$^{+38.1}_{-4.1}$ & 82.69$^{+29.42}_{-24.24}$ \\
GRB090323 & B & 411.7$^{+11.7}_{-11.7}$ & 411.7$^{+11.7}_{-11.7}$ & -- \\
GRB090328 & B & 11.7$^{+0.5}_{-0.5}$ & 11.7$^{+0.5}_{-0.5}$ & -- \\
GRB090510 & BP & 7.3$^{+0.3}_{-0.3}$ & 5.9$^{+0.3}_{-0.3}$ & 1.41$^{+0.30}_{-0.27}$ \\
GRB090902B & BP & 343.6$^{+2.6}_{-2.6}$ & 259.2$^{+2.6}_{-2.6}$ & 84.46$^{+3.90}_{-3.75}$ \\
GRB090926A & BC & 242.0$^{+5.1}_{-5.0}$ & 199.1$^{+6.5}_{-6.1}$ & 46.74$^{+4.48}_{-4.61}$ \\
GRB091003 & B & 9.9$^{+0.3}_{-0.3}$ & 9.9$^{+0.3}_{-0.3}$ & -- \\
GRB091208B & B & 3.0$^{+0.4}_{-0.3}$ & 3.0$^{+0.4}_{-0.3}$ & -- \\
GRB100414A & CP & 52.5$^{+1.2}_{-1.0}$ & -- & -- \\
GRB110731A & B BP & 71.7$^{+2.8}_{-2.7}$ & 51.4$^{+9.8}_{-8.3}$ & 17.72$^{+4.25}_{-6.03}$ \\
\enddata
\label{tab_eiso_components}
\end{deluxetable}

\newgeometry{left=2cm,right=1cm,bottom=1cm,top=7.0cm}
\begin{landscape}
\begin{center}
\begin{deluxetable}{lrrrrrrrrrr}
\tabletypesize{\tiny}
\tablecolumns{5}
\tablewidth{0pt}
\tablecaption{Systematic Uncertainties\label{tab_systematics}}
\tablehead{\colhead{} & \colhead{} & \colhead{A: CATALOG} & \colhead{B: Pass7} & \colhead{$\frac{A-B}{\sqrt{\sigma_{A}^2+\sigma_{B}^2}}$}& \colhead{C: DIFF} & \colhead{$\frac{A-C}{\sqrt{\sigma_{A}^2+\sigma_{C}^2}}$}& \colhead{D: DIFF-F} & \colhead{$\frac{A-D}{\sqrt{\sigma_{A}^2+\sigma_{D}^2}}$}& \colhead{E: DIFF-B} & \colhead{$\frac{A-E}{\sqrt{\sigma_{A}^2+\sigma_{E}^2}}$}}
\startdata
   080825C & Flux ($\times10^{-5}$~m$^{-2}$s$^{-1}$)\dotfill & 20 $\pm$ 10& 27 $\pm$ 10&  -0.5& $<$20& \nodata& $<$70& \nodata& \nodata& \nodata\\            &   Spectral idx\dotfill & $-$3.3 $\pm$ 0.7& $-$3.1 $\pm$ 0.6&  -0.2& \nodata& \nodata& \nodata& \nodata& \nodata& \nodata\\            &   Temporal idx\dotfill & \nodata& \nodata& \nodata& \nodata& \nodata& \nodata& \nodata& \nodata& \nodata\\            & Loc. err. (deg.)\dotfill &  0.77&  0.85& -0.08&  0.77&  0.01&  0.85& -0.08&  1.65& -0.87\\    080916C & Flux ($\times10^{-5}$~m$^{-2}$s$^{-1}$)\dotfill & 82 $\pm$ 7& 84 $\pm$ 7&  -0.2& 100 $\pm$ 10&  -1.5& 100 $\pm$ 20&  -0.8& 90 $\pm$ 20&  -0.4\\            &   Spectral idx\dotfill & $-$2.13 $\pm$ 0.08& $-$2.21 $\pm$ 0.08&   0.7& $-$2.2 $\pm$ 0.1&   0.5& $-$2.2 $\pm$ 0.2&   0.3& $-$2.2 $\pm$ 0.1&   0.5\\            &   Temporal idx\dotfill & 1.8 $\pm$ 0.3& 1.26 $\pm$ 0.07&   1.8& 1.28 $\pm$ 0.08&   1.7& 1.4 $\pm$ 0.1&   1.3& 1.3 $\pm$ 0.1&   1.6\\            & Loc. err. (deg.)\dotfill &  0.07&  0.07& -0.00&  0.07& -0.00&  0.10& -0.03&  0.08& -0.01\\
\enddata
\end{deluxetable}
\end{center}

\end{landscape}
\restoregeometry

\section{Acknowledgments}
The \Fermi LAT Collaboration acknowledges generous ongoing support
from a number of agencies and institutes that have supported both the
development and the operation of the LAT as well as scientific data analysis.
These include the National Aeronautics and Space Administration and the
Department of Energy in the United States, the Commissariat \`a l'Energie Atomique
and the Centre National de la Recherche Scientifique / Institut National de Physique
Nucl\'eaire et de Physique des Particules in France, the Agenzia Spaziale Italiana
and the Istituto Nazionale di Fisica Nucleare in Italy, the Ministry of Education,
Culture, Sports, Science and Technology (MEXT), High Energy Accelerator Research
Organization (KEK) and Japan Aerospace Exploration Agency (JAXA) in Japan, and
the K.~A.~Wallenberg Foundation, the Swedish Research Council and the
Swedish National Space Board in Sweden.

Additional support for science analysis during the operations phase is gratefully
acknowledged from the Istituto Nazionale di Astrofisica in Italy and the Centre National d'\'Etudes Spatiales in France.

\bibliography{mnemonic,GLAST_GRB,refs}
\bibliographystyle{aa}

\appendix
\twocolumn
\section{Systematic Errors}
\label{sec_systematics}

In this appendix we report possible sources of systematic uncertainties in our results and how we estimate or ameliorate them.

The most important source of systematic errors arises from potentially inaccurate descriptions of the responses of the GBM and the LAT. The parametrization of the response of the LAT to incident $\gamma$ rays is tabulated in instrument response functions (IRFs), produced using Monte Carlo simulations and subsequently refined based on in-flight data. Even though the results of these simulations have been verified extensively against flight data and also pre-launch using calibrated sources \cite{LATperform}, any imperfections in the simulation model or in the simulation procedure can propagate in the IRFs affecting all our results.

Additionally, any relative calibration errors between the GBM and the LAT and any errors in the description of the response of the GBM can affect joint spectral fits, manifesting as distortions in the spectral shapes and biases in the measured parameters.

Finally, the results of joint spectral fits also can be affected by the motion of the GRB in the instruments' fields of view which creates variations of their responses over time. These effects are minimized by producing response matrices that accurately describe the response of the instruments at any instant of the observation (see \S~\ref{subsec_lat_gbm_spectral_analysis}).

Another source of systematic uncertainty is the background estimates. For Transient-class events, background estimation is performed using a procedure that has an estimated systematic uncertainty of 10--15\% and negligible statistical errors (as described in \S~\ref{subsubsec_LAT_bkg}). For LLE and GBM data the backgrounds are estimated using interpolations of the event rate before and after the burst, a procedure the uncertainty of which primarily arises from limited statistics and is estimated to be $\sim$10\% for LLE and less for the GBM data. For observations involving large variations of the instrument's pointing (e.g., in ARRs) or observations of locations near the Earth's limb, the systematic errors can increase possibly up to the magnitude of the statistical errors. Any
mis-estimations of the LAT backgrounds can affect the final results, especially those for longer time scales such as duration estimates. The maximum likelihood analyses are not particularly sensitive to errors in the background estimates since the background level
is a loosely-constrained parameter in the fitting; thus any systematic errors are partially ``fit out''.


In order to evaluate the impact of the above uncertainties on the maximum likelihood analysis results, we have repeated the analysis using different sets of cuts. The magnitude of the difference between the results obtained with these alternative data sets and the standard one can be used as an order-of-magnitude estimate of the systematic uncertainties in our (standard) results.

First, we have repeated the maximum-likelihood analysis using Diffuse-class events (``Pass 6 V7 Diffuse Class''), adopting the standard isotropic template available at the FSSC site\footnote{\url{fermi.gsfc.nasa.gov/ssc/data/access/lat/BackgroundModels.html}} as the representation of the non-rejected charged particle background. Because the Diffuse class has significantly less background contamination than Transient class, any uncertainties in the background estimates are minimal.  Thus a comparison against this set of results can reveal the uncertainties arising from any inaccuracy in the background estimates for our standard set of results. Furthermore, because the two analyses employ different sets of IRFs this test is also sensitive to systematics of the IRFs in general.

\begin{figure*}[ht!]
\begin{center}
\includegraphics[width=0.5\textwidth]{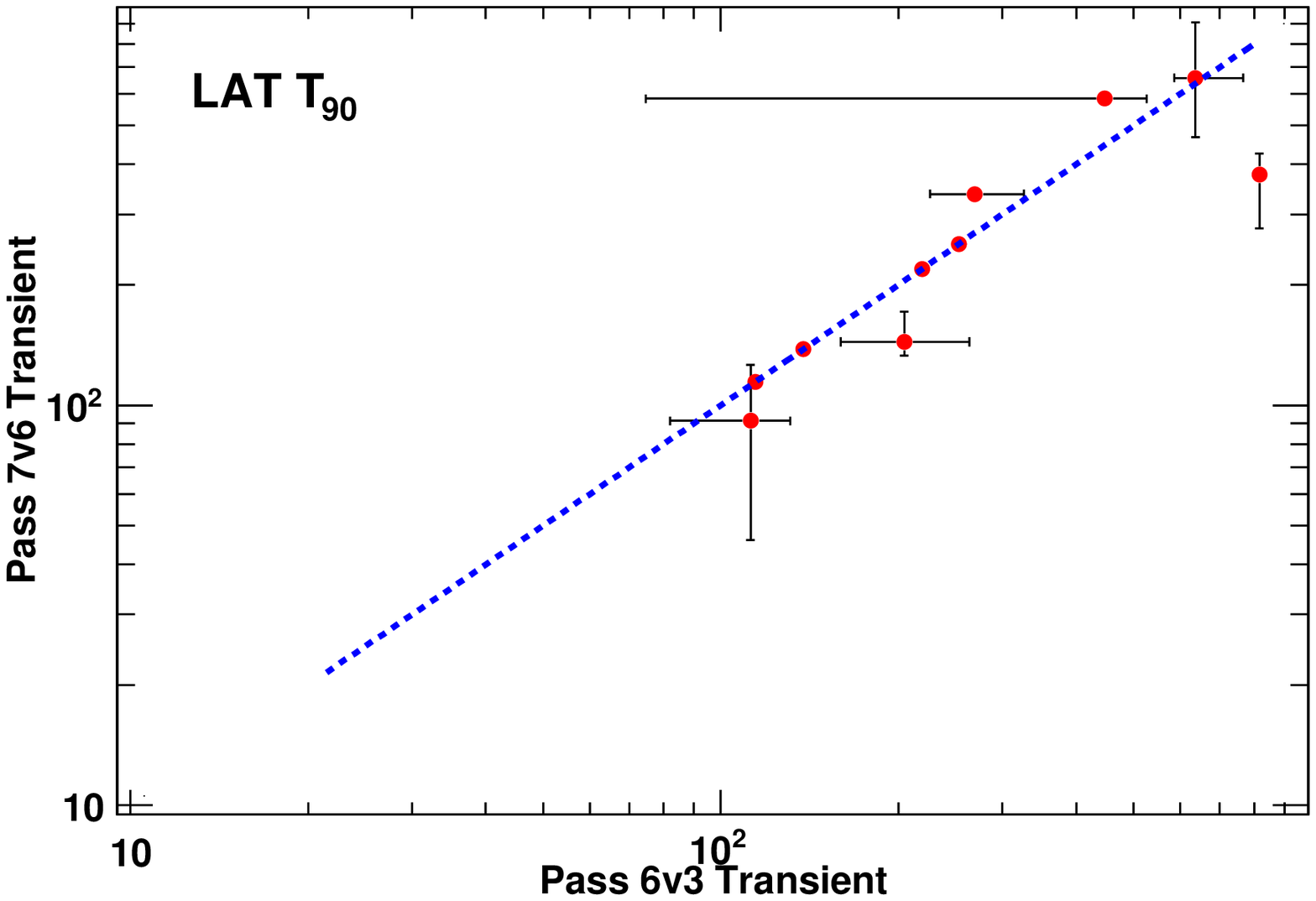}\includegraphics[width=0.5\textwidth]{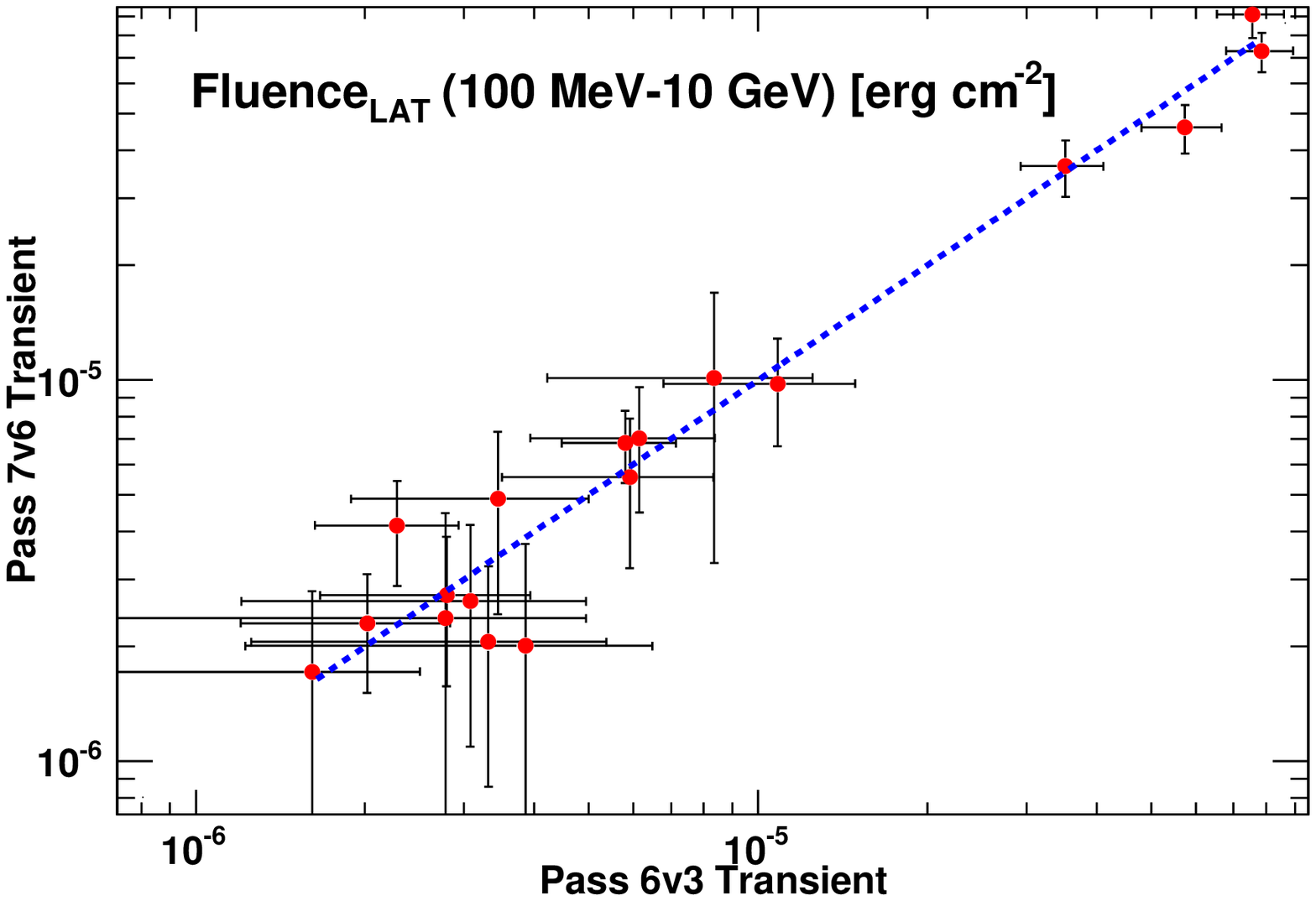}
\includegraphics[width=0.5\textwidth]{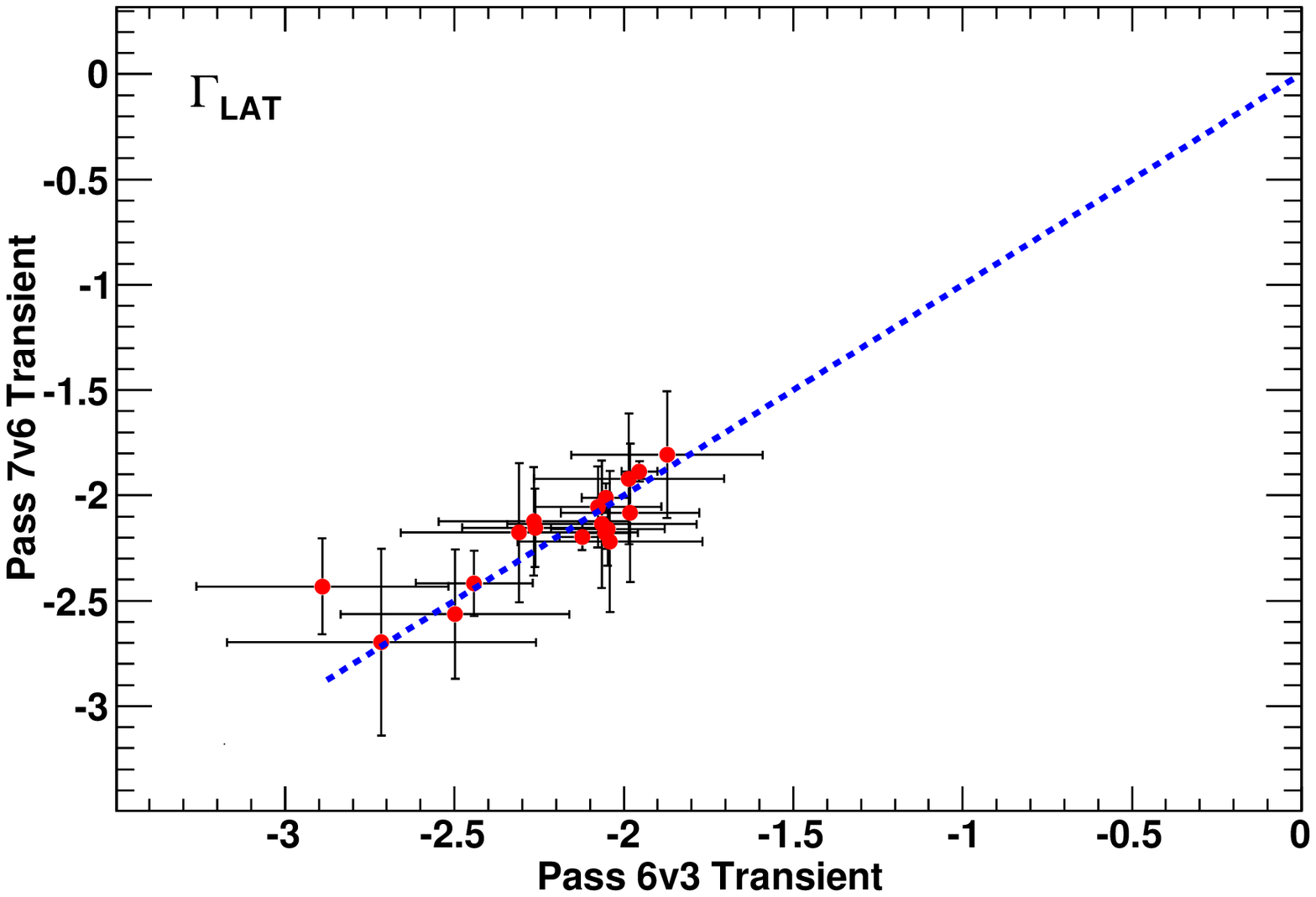}\includegraphics[width=0.5\textwidth]{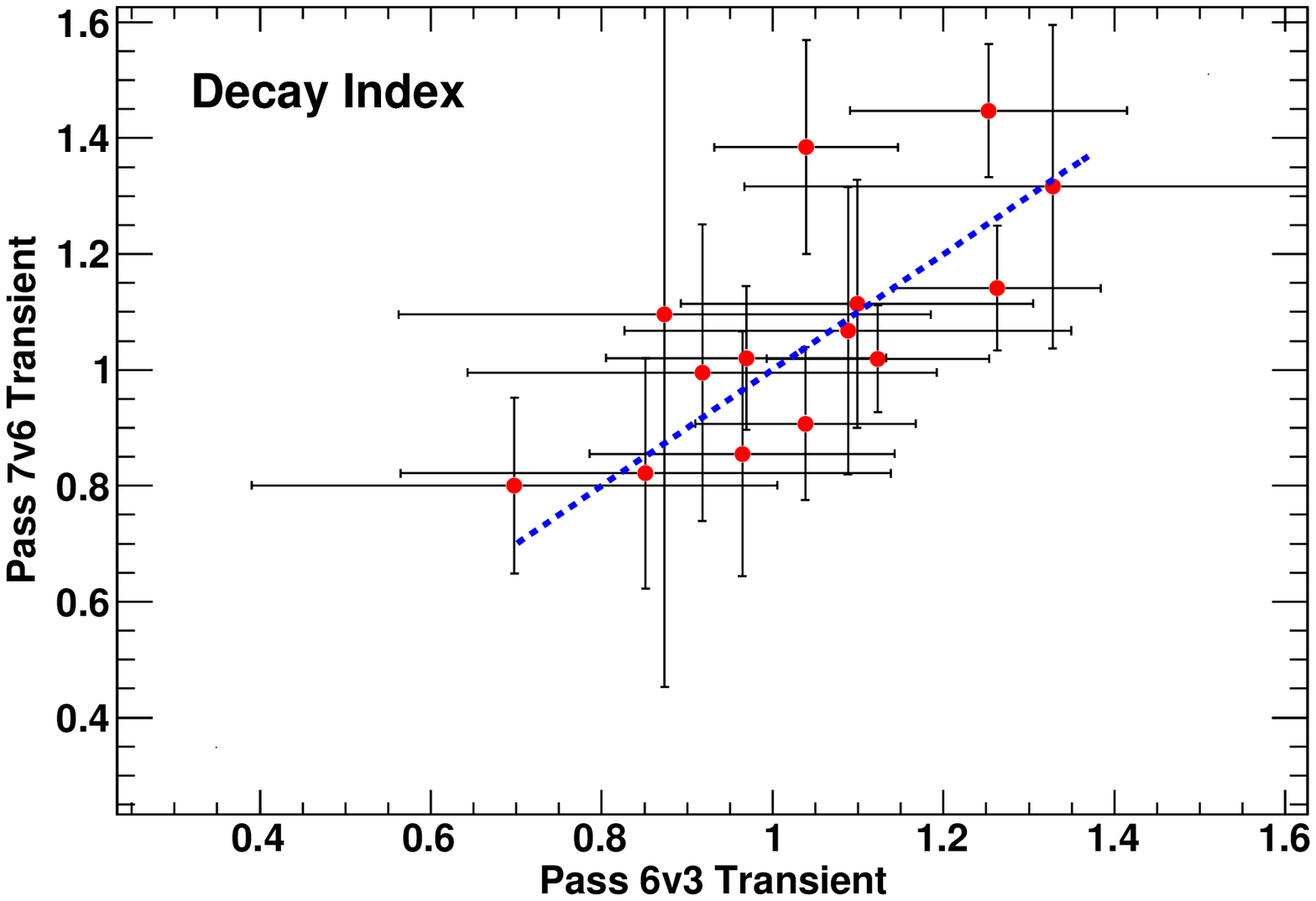}

\caption{\label{fig:systematic_plots}Comparison between our standard results produced with the ``Pass6 V3 Transient'' set of cuts, versus the results obtained with the more recent ``Pass7 V6 Transient'' set of cuts. From top left and clockwise: LAT T$_{90}$, Fluence in the 100~MeV--10~GeV energy range during the ``LAT'' time interval, index of the power law temporal decay, and photon index of the emission detected by the LAT.}
\end{center}
\end{figure*}

We continued by splitting the Diffuse-class data sample into two independent data sets depending on which portion of the tracker each event was converted (front versus back). Events produced by photons converting in the first 12 layers of the tracker (``front'') suffer on-average less multiple scattering than those converting at the next 4 layers of the tracker (``back'') since the front layers have thinner converter foils (see \S \ref{sec_introduction} for a description of the instrument). The decreased magnitude of multiple scattering for front-converting events provides significantly better angular resolution. In addition, the front-converting events have a significantly smaller fraction of their energy measured by the calorimeter than back-converting events, which results to lower-energy ($<$ few~GeV) front-converting events being reconstructed with a worse energy resolution than back-converting events. A comparison against this sample can be sensitive to systematics of the IRFs associated to the particular properties of front- versus back-converted events.

Finally, we repeated the analysis using a more recent iteration of the set of event selection cuts for the LAT data, specifically the ``Pass 7 Transient V6'' selection, which benefits from more robust and accurate classification algorithms and increased refinement using flight data. Again, a comparison of our results from this data set can reveal differences affecting any parts of the IRFs.

We refer to the standard configuration as ``CATALOG'', to the Diffuse class as ``DIFF'', to the front and back as ``DIFF:F'' and ``DIFF:B'' respectively, and to the ``Pass 7 Transient V6''  as ``Pass7''. Table \ref{tab_systematics} summarizes the results of the above tests, quoting for each analysis the photon flux, spectral index, index of the temporal-decay power law along with their statistical errors, and the estimated localization error. We also report the absolute difference between the CATALOG and each of the test configurations. In Fig. \ref{fig:systematic_plots} we compare the results between the Pass7 and CATALOG results. The quantities compared (from top left and clockwise) are the LAT T$_{90}$, the Fluence in the 100~MeV--10~GeV energy range during the ``LAT'' time interval, the index of the power-law temporal decay, and the photon index of the emission detected by the LAT. As can be seen, there are no discernible differences within errors.

\begin{figure}[ht!]
\begin{center}
\includegraphics[width=1.0\columnwidth,trim=20 10 40 30, clip=true]{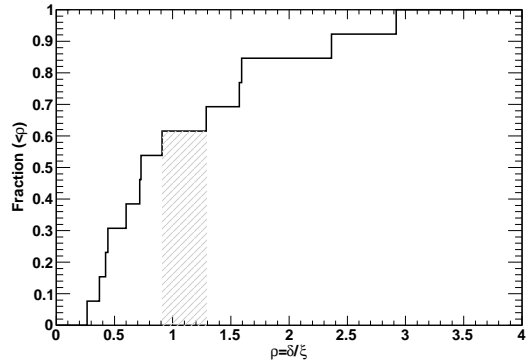}
\caption{Cumulative distribution of the number of GRBs (with good X-ray localization) over some $\rho$, defined as the ratio between the angular separation of the best LAT localization and the X-ray position over the 1$\sigma$ statistical error of the LAT localization. The vertical band highlights the range where the cumulative distribution reaches the 68\% level.}
\label{fig_cum_loc_error}
\end{center}
\end{figure}

We also estimated the error in the localizations obtained with the LAT. For 13 of the GRBs localized by the LAT, a Swift XRT position is also available. For those cases, we calculated the quantity $\rho=\delta/\epsilon$, which is the ratio between the angular separation ($\delta$) between the LAT and the XRT position over the estimated LAT 1$\sigma$ localization error $\epsilon$. In Fig.\ref{fig_cum_loc_error}, we plot the cumulative distribution of the number of GRBs with $\rho$. The number of GRBs in this sample is very limited, and thus we cannot draw any firm conclusions, but we note that, as expected, the 68\% quantile of the distribution is consistent with the 68\% (or 1$\sigma$) estimated error.

To estimate the effects arising from relative mis-calibrations between the GBM and the LAT in the joint-spectral fit results, we introduced a flux normalization factor for each detector, letting all but one such factor be free to vary during the fit. This is basically equivalent to a rigid effective area correction across the whole bandpass of each instrument, relative to one detector chosen as reference (we chose the LAT). This procedure could give spurious results if the model used for the fit contains localized features or components, which is not the case for the models we used. We introduced these factors for the brightest GRBs of our sample: GRBs
080916C, 090323, 090328, 090510, 090902B, 090926A, 100724B, 100826A, 100414A, and 110731A. For all other GRBs, the factors were effectively unconstrained by the fit, because the inter-calibration systematic errors were small compared to the statistical errors or because the systematic errors were dominated by other components. The resulting correction was less than 5\% for NaI detectors, and less than 15\% for BGO detectors. According to these initial tests, relative inter-calibration uncertainties are important only in the case of bright GRBs, for which statistical errors are small.

We also tested our GBM background estimation procedure. We first considered real spectra from time intervals well outside any GRB emission. For each of these intervals $I_{fake}$, the actually observed spectrum was compared with the spectrum predicted by a background model obtained from the fit of two intervals surrounding $I_{fake}$, obtained with the procedure described in \S~\ref{subsubsec_GBM_BKG}. We selected a couple of GRBs, and we defined different background models by selecting different time intervals around the GRB times. These validation studies showed that the procedure has, under normal circumstances, a systematic error of $\sim 3$\%, which we have added to all of our predicted background spectra.

\onecolumn
\section{Fermi LAT Gamma-Ray Bursts}
\label{sec_fermi_lat_grb}
In this appendix we give detailed information on individual LAT-detected GRBs. We summarize the information previously published in refereed papers
and GCN circulars. We also include figures showing the GBM/LAT composite light curves as well as, when possible, the results of the LAT time-resolved spectral analyses.\\
\subsection{Conventions and Styles for Figures}
\label{sec_fermi_lat_grb_intro}

Each composite light curve consists of either 4 or 5 panels, showing the emission (in counts) recorded by the GBM NaI's (first two panels from the top), by
  the GBM BGO (third panel), by the LAT within the LLE event selection (fourth panel) and, if any, the selected LAT Transient-class events above 100~MeV
  (bottom  panel).
 
\begin{itemize}
\item The GBM NaI light curves were obtained by summing all the NaI detectors (typically 2 or 3) for which the GRB position was within 50\de
    from the detector normal pointing axis. We also selected the BGO detector that faces the burst. We used GBM TTE data and selected the channels corresponding to
    the energy ranges of 8--20~keV and 20--250~keV for the NaI detectors, and 0.2--5~MeV for the BGO detector.
\item The LLE light curve corresponds to the selection cuts discussed in $\S$~\ref{subsection_cuts}, which were applied to LAT events
    with energies above 10~MeV. As the gamma-ray signal in the LAT is proportional to the LAT effective area, it depends strongly on the GRB
    off-axis angle $\theta$ (and spectrum) at any time. In order to reflect the amplitude of this modulation, the grey curve displayed in the LLE
    panel shows the $\cos[\theta(t)]$ function (ranging from 0 to 1 over the full extent of the panel).
\item In the last light curve, we selected the LAT Transient-class events in a 12\de ROI which have a reconstructed
energy above 100~MeV. We represent, as filled circles, the events which also have a probability $>$0.9 of being associated with the GRB (see
$\S$~\ref{susubbsec_event_probability}).
\item In each panel, vertical dashed lines indicate the GBM trigger time (in red, at T=0), the GBM T$_{05}$ and the GBM T$_{95}$
    (both in green). Other lines indicate the time of the LAT highest-energy event associated with the GRB within the GBM T$_{90}$ (in magenta, from
    Table~\ref{tab_energymax_gbm}) and during the LAT emission (in blue, from Table~\ref{tab_energymax_all}). If the two events are identical, then only
    the blue line is displayed.
\end{itemize}

When possible, we add a figure for the $>$100~MeV flux light curve, showing how the temporally extended emission develops and then decays as a
  function of time $F(t)$. 
  \begin{itemize}
  \item The GBM T$_{95}$ is indicated by a vertical red dashed line.
  \item For each time bin where the GRB was significantly detected (i.e. TS$>$16, see step 2 in $\S$~\ref{subsec_analysis_sequence}), we
  also show the value of the photon index (we use here the convention $N(E)\propto E^\beta$ where $N$ is the fitted photon flux and $\beta$ is
  typically negative). 
    \item For the bins with no detection, we fixed the power-law index to $\beta$=-2.0 and then report the value of the flux upper
  limit. 
    \item When the statistics are large enough, we give the decay indices from the fit $F(t)\propto t^{-\alpha}$ of a power-law 
    (starting from the latest time between the peak flux time $t_p$ and the time of the GBM T$_{95}$) and of a broken power law (starting from $t_p$).
 If a significant break is found in the latter fit, the broken power law is displayed as a filled grey line and the power law as a dashed grey
  line. The line styles are reversed in the opposite case.
\end{itemize}

In two cases (GRBs\,090323 and 090328) where the ARR maneuver caused a particularly bright increase of the background during the GBM prompt emission
we also show the LLE light curve and the relative background estimation.
  

\FloatBarrier \subsection{GRB\,080825C}
\begin{figure}[ht!]
\begin{center}
\includegraphics[width=5.0in]{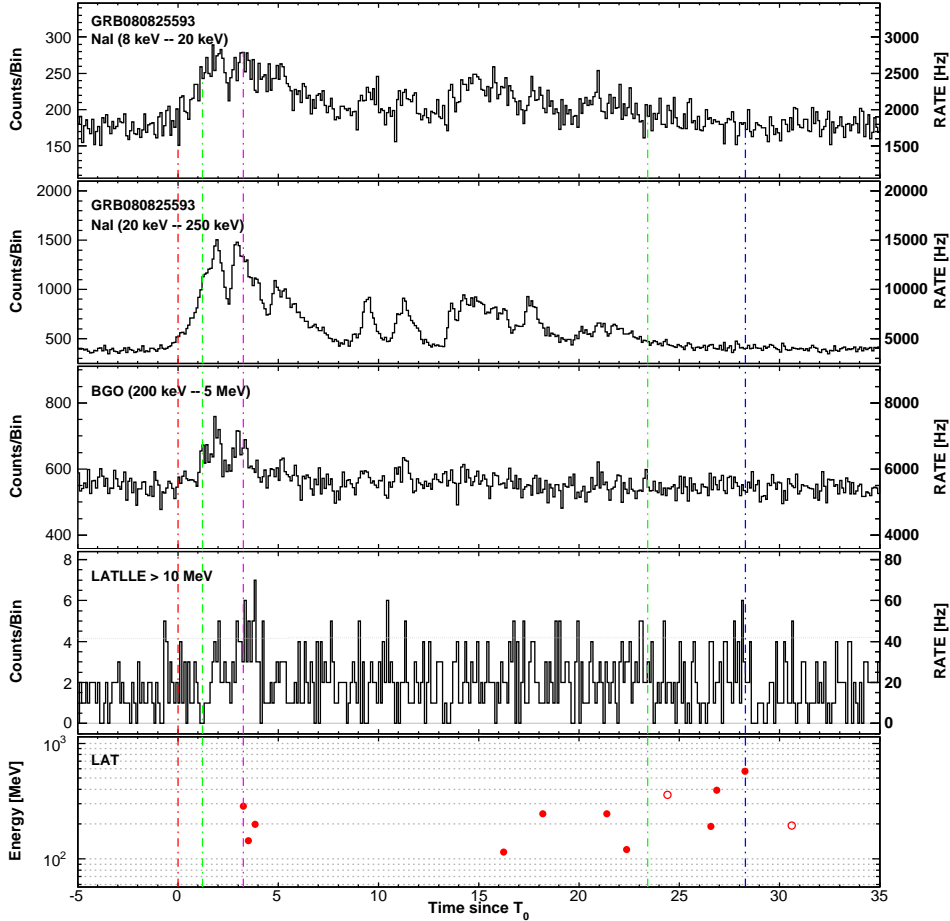}
\caption{Composite light curve for GRB\,080825C: summed GBM/NaI detectors (first two panels), GBM/BGO (third panel), LLE (fourth panel) and LAT Transient-class events above 100~MeV within a 12\de ROI  (last panel). See $\S$~\ref{sec_fermi_lat_grb_intro} for more information on lines and symbols in the LAT panels.}
\label{compo_080825C}
\end{center}
\end{figure}

\begin{figure}[ht!]
\begin{center}
\includegraphics[width=2.2in]{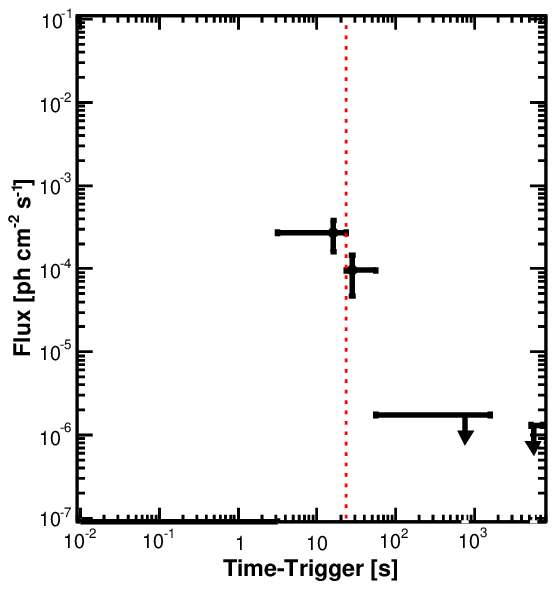}
\includegraphics[width=2.2in]{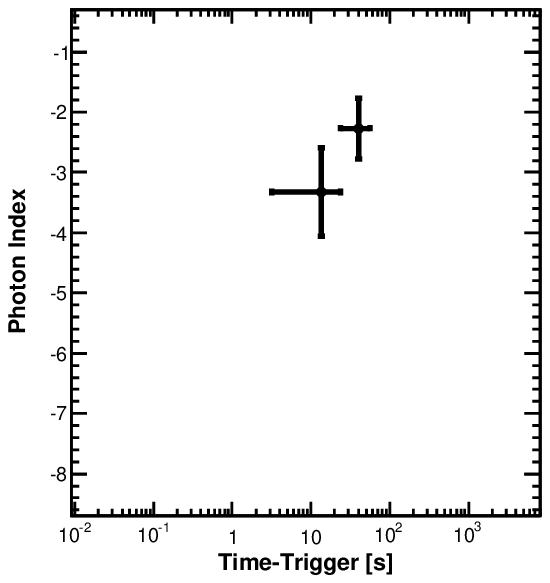}
\caption{Likelihood light curve for GRB\,080825C (flux above 100~MeV on the left, photon index on the right). See $\S$~\ref{sec_fermi_lat_grb_intro} for more information on lines and symbols.}
\label{like_080825C}
\end{center}
\end{figure}

The long GRB\,080825C triggered the GBM flight software at T$_{0}$=14:13:48 UT on 25 August 2008 \citep[trigger 241366429,][]{2008GCN..8141....1V}.
Although this faint burst had an off-axis angle of 60\ded3 at the trigger time, where the effective area is a factor $\sim$3 less than on axis,
the LAT detected it significantly and the LAT preliminary localization was delivered via GCN \citep{2008GCN..8183....1B}, with a statistical error of 0\ded95.
A detailed analysis was published by the \Fermi LAT collaboration in \cite{GRB080825C_LATpaper}.
The composite light curve (Fig.~\ref{compo_080825C}) shows a multi-peak structure in the GBM signal, while the number of counts is not
large enough at high energy to study the temporal profile in details.
The LAT emission, especially above 100~MeV, seems to coincide with the second bright pulse in the GBM.
The high-energy emission is also clearly visible at later times, and the highest-energy event (0.57~GeV) is detected at
T$_{0}$+28.29~s, i.e. after the end of the GBM emission.
However, as the temporally extended high-energy emission is faint, the LAT time-resolved likelihood analysis returned a significant flux in two time bins
only (Fig.~\ref{like_080825C}).

Note that an LLE light curve of GRB\,080825C was reported in the paper on GRB\,090217 published by the \Fermi LAT
collaboration \citep{LAT_090217}, which indicated a $\sim$5$\sigma$ signal after integration over the first $\sim$4~s, slowly increasing to
$\sim$9$\sigma$ after $\sim$30~s.
We could not confirm this signal excess in LLE data as our analysis is based on a different detection algorithm, which is not tuned to slowly accumulating
signals. This algorithm is mostly sensitive to the short variability time scales as it looks for the highest-significant
excess among all considered time bins in the LLE light curve (see $\S$~\ref{subsubsect_detection_LLE}).
A 3.2$\sigma$ post-trial significance (4.2$\sigma$ pre-trial) was found, thus no LLE results are reported for this burst in the catalog.

\FloatBarrier \subsection{GRB\,080916C}
The long, bright GRB\,080916C triggered the GBM at T$_{0}$=00:12:46 UT on 16 September 2008 \citep[trigger 243216766,][]{2008GCN..8245....1G}.
This burst would have been bright enough to trigger an ARR of the \Fermi spacecraft, but the repointing capability of the spacecraft was enabled only a few
weeks later, on 8 October 2008.
GRB\,080916C was easily detected by the LAT, which delivered a localization via GCN \citep{2008GCN..8246....1T}, with a statistical error of 0\ded09.
It had an off-axis angle of 48\ded8 at the trigger time and it exited the FoV of the LAT after $\sim$3000~s.
Swift Target of Opportunity (TOO) observations started $\sim$17~hours after the trigger time \citep{2008GCNR..166....1S}. A possible X-ray counterpart was
found by Swift-XRT 3.1~arcmin away from the LAT position \citep{2008GCN..8253....1K}, and further observations confirmed the existence of a fading
source \citep{2008GCN..8261....1P}.
Follow-up observations with the Gamma-Ray burst Optical/Near-infrared Detector (GROND) yielded a high photometric redshift of
z=4.35$\pm$0.15 \citep{Greiner+09} which, combined with its brightness, makes GRB\,080916C the most energetic burst ever detected, with an isotropic equivalent
energy $\mathrm{E_{iso}}\simeq6.5\times 10^{54}$~erg (1~keV--10~GeV, within the GBM T$_{90}$).\\

\begin{figure}[ht!]
\begin{center}
\includegraphics[width=5.0in]{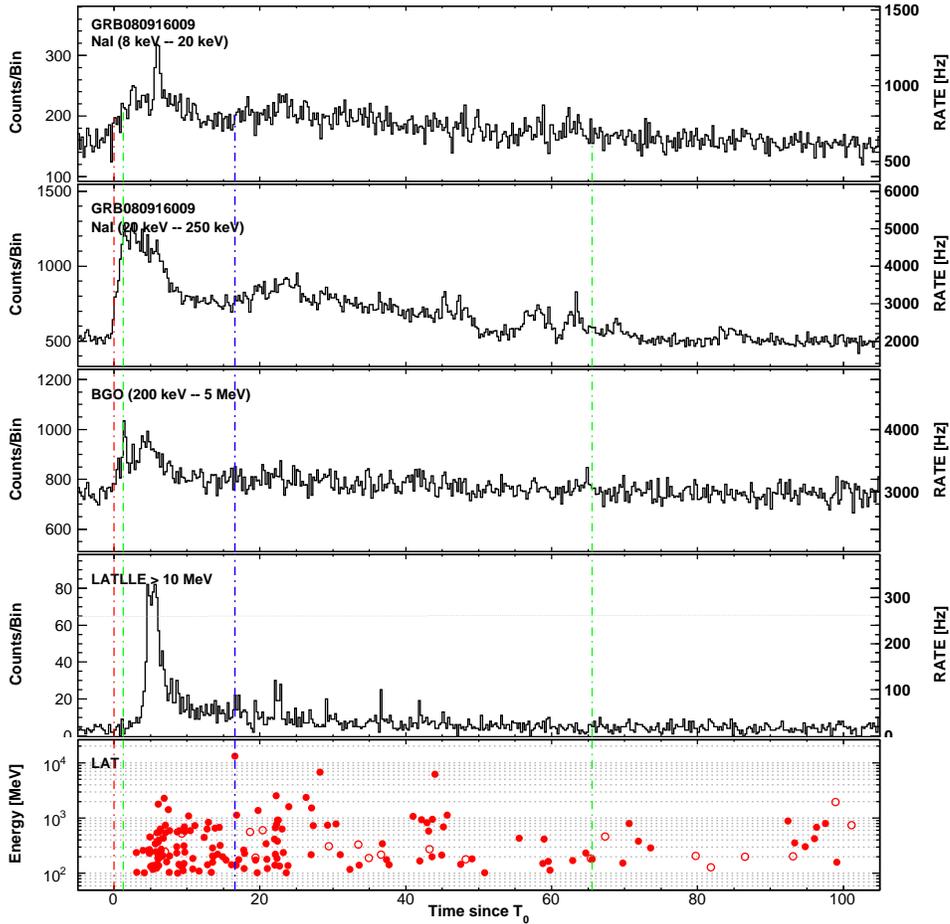}
\caption{Composite light curve for GRB\,080916C: summed GBM/NaI detectors (first two panels), GBM/BGO (third panel), LLE (fourth panel) and LAT Transient-class events above 100~MeV within a 12\de ROI  (last panel). See $\S$~\ref{sec_fermi_lat_grb_intro} for more information on lines and symbols in the LAT panels.}
\label{compo_080916C}
\end{center}
\end{figure}

The LAT emission peaked $\sim$5~s after the trigger time, coinciding with the second GBM bright pulse (Fig.~\ref{compo_080916C}).
Approximately 180 Transient-class events are recorded above 100~MeV within the LAT T$_{90}$$\sim$210~s, including many GeV events.
The highest-energy event (13.22~GeV), which is detected at T$_{0}$+16.54~s, does not coincide with any noticeable feature in the GBM light curve.
In the first paper published by the \Fermi LAT collaboration \citep{GRB080916C:Science}, the prompt emission spectrum of GRB\,080916C was fitted over
six decades in energy by the empirical Band function.
This previous analysis also searched for possible deviations from the Band function, and did not provide any evidence for a deficit or
  a signal excess at the highest energies in the LAT. In particular, the significance for an additional power-law component was found to be small,
  $\sim$2$\sigma$. We repeated the analysis and found that an additional power law is actually required (4-5$\sigma$, see
  $\S$~\ref{subsub_extracomponents}). It is worth stressing the improvements which have been brought to the analysis procedure since the first
  post-launch GRB studies and which support this new result. First of all, we now use the Background Estimator tool (see $\S$~\ref{subsubsec_LAT_bkg}) which provides a much more accurate
  description of the backgrounds in the spectral fits. In addition, we benefit from a better calibration of both the GBM and the LAT
  instruments. Finally, we base our assessment of the significance of any new spectral feature on dedicated and extended Monte-Carlo simulations.
 These improvements, along with a new choice of the time intervals (based on our estimates for the durations of the emission in the GBM and the
  LAT) as well as a different spectral shape, also explain the differences in our results (Tables ~\ref{tab_JointFitGBMT90}
  and~\ref{tab_JointFitLATPROMPT}) with respect to the original publication.

\begin{figure}[ht!]
\begin{center}
\includegraphics[width=2.2in]{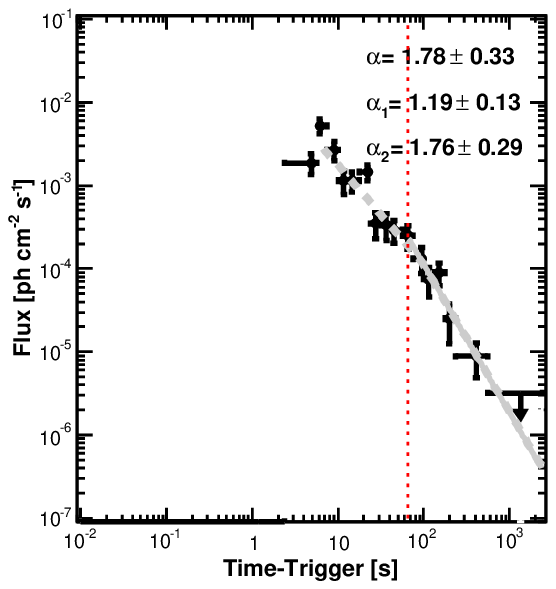}
\includegraphics[width=2.2in]{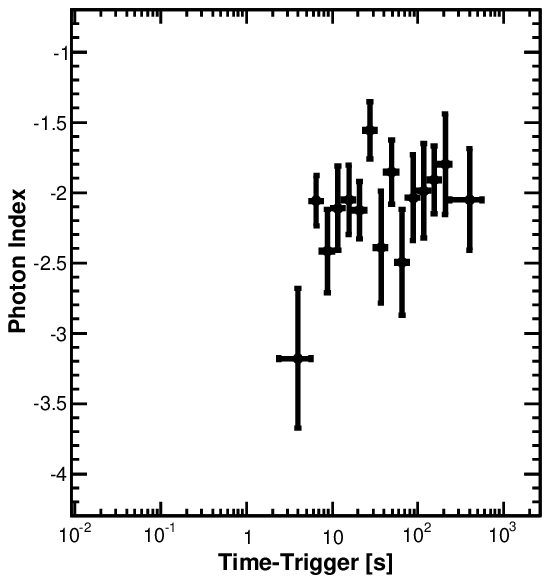}
\caption{Likelihood light curve for GRB\,080916C (flux above 100~MeV on the left, photon index on the right).See $\S$~\ref{sec_fermi_lat_grb_intro} for more information on lines and symbols.}
\label{like_080916C}
\end{center}
\end{figure}

The high-energy emission of GRB\,080916C lasts much longer than the GBM estimated duration.
The LAT time-resolved likelihood analysis resulted in a well sampled light curve of the high-energy flux up to $\sim$560~s (Fig.~\ref{like_080916C}).
Its first point suggests that the spectrum is significantly softer than the LAT emission at later times, where the photon
index fluctuates consistently around $\beta$=-2.
The decay of the flux as a function of time follows a simple power law starting from the GBM T$_{95}$, with a decay index
  $\alpha$=1.78$\pm$0.33.
This steep decay is similar to the first part of the decay observed in GRBs~090510, 090902B and 090926A (Table~\ref{tab_extended}) for which a
  significant break was found in the flux light curve. This suggests that GRB\,080916C was observed during the transition from the prompt phase to the
  afterglow phase as discussed in $\S$~\ref{sub_interpretation_extended}.

\FloatBarrier \subsection{GRB\,081006}
\begin{figure}[ht!]
\begin{center}
\includegraphics[width=5.0in]{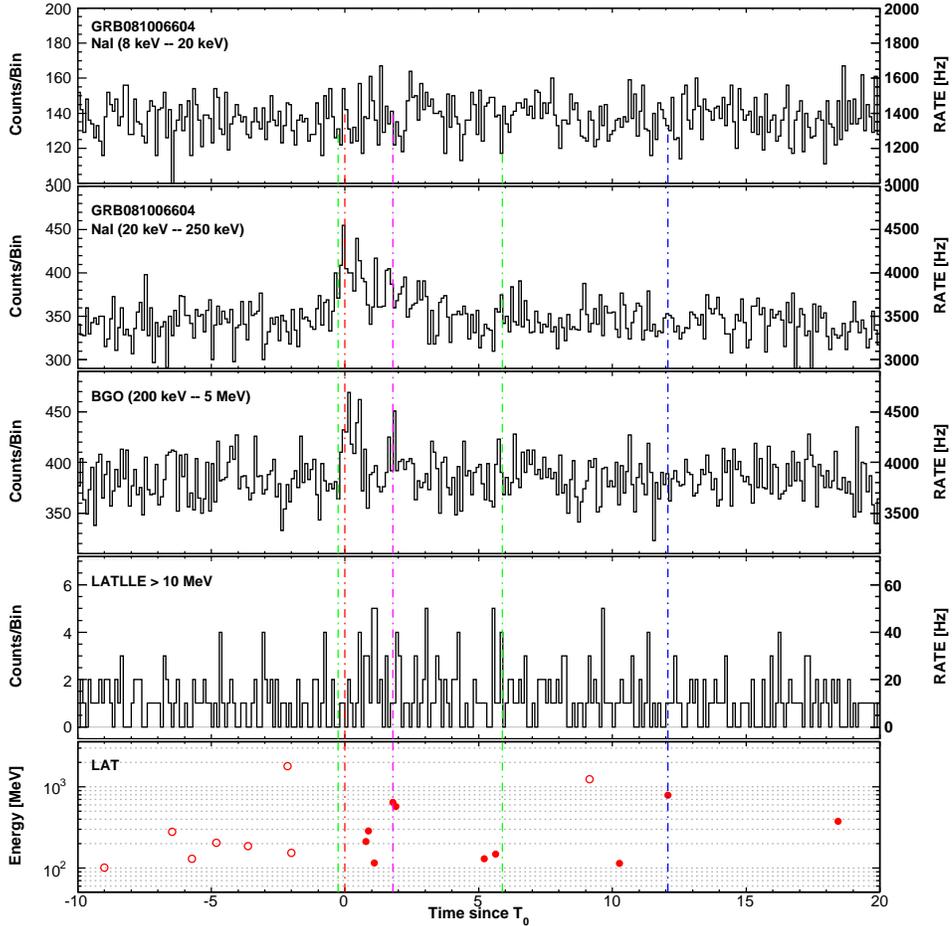}
\caption{Composite light curve for GRB\,081006: summed GBM/NaI detectors (first two panels), GBM/BGO (third panel), LLE (fourth panel) and LAT Transient-class events above 100~MeV within a 12\de ROI  (last panel). See $\S$~\ref{sec_fermi_lat_grb_intro} for more information on lines and symbols in the LAT panels.}
\label{compo_081006}
\end{center}
\end{figure}

\begin{figure}[ht!]
\begin{center}
\includegraphics[width=2.2in]{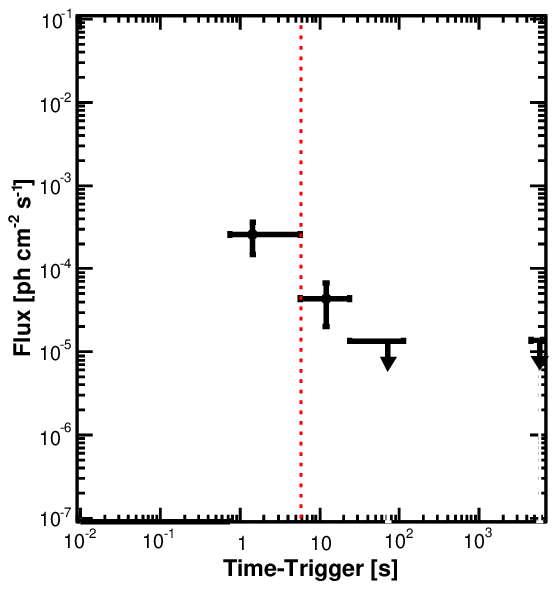}
\includegraphics[width=2.2in]{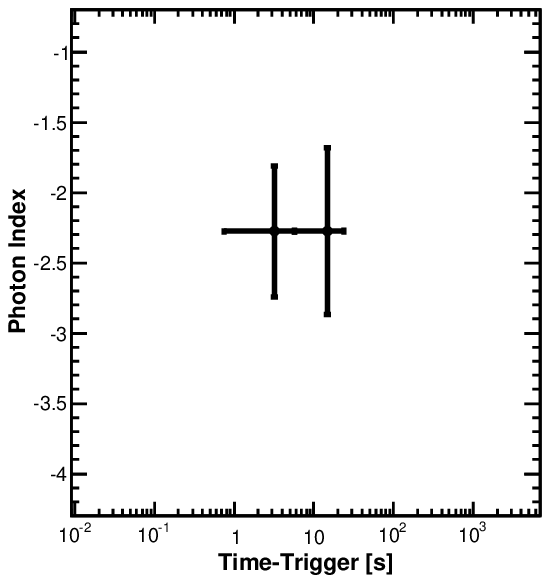}
\caption{Likelihood light curve for GRB\,081006 (flux above 100~MeV on the left, photon index on the right). See $\S$~\ref{sec_fermi_lat_grb_intro} for more information on lines and symbols.}
\label{like_081006}
\end{center}
\end{figure}

The long GRB\,081006 triggered the GBM at T$_{0}$=14:29:34 UT on 6 October 2008 \citep[trigger 244996175,][]{GBMGCN_081006}.
It was a faint burst, both in the GBM and in the LAT (despite an initial off-axis angle of 11\de).
No significant emission was detected in the LLE light curve (Fig.~\ref{compo_081006}) despite a 2.7$\sigma$ fluctuation observed shortly after the
trigger time.
More interestingly, this burst was detected and localized by the LAT likelihood analysis using Transient-class events above 100~MeV, with a maximum
TS$\sim$72 (see Table~\ref{tab_likelihoods}).
Taking into account uncertainties in the calculation of the LAT T$_{90}$, the high-energy emission could be simultaneous with the low-energy emission
(i.e. happening on very similar time scales) or it could last much longer as a significant signal excess is detected above the estimated background up to
$\sim$T$_{0}$+115~s.
This time corresponds to the entrance of the LAT in the South Atlantic Anomaly, and was thus reported as a lower limit to the duration in
Table~\ref{tab_durations}.
In spite of this hint for a temporally extended high-energy emission, the LAT likelihood analysis did not find any significant signal in the ``EXT''
time interval, and could not provide good time-resolved spectral measurements (Fig.~\ref{like_081006}).

\FloatBarrier \subsection{GRB\,081024B}
\begin{figure}[ht!]
\begin{center}
\includegraphics[width=5.0in]{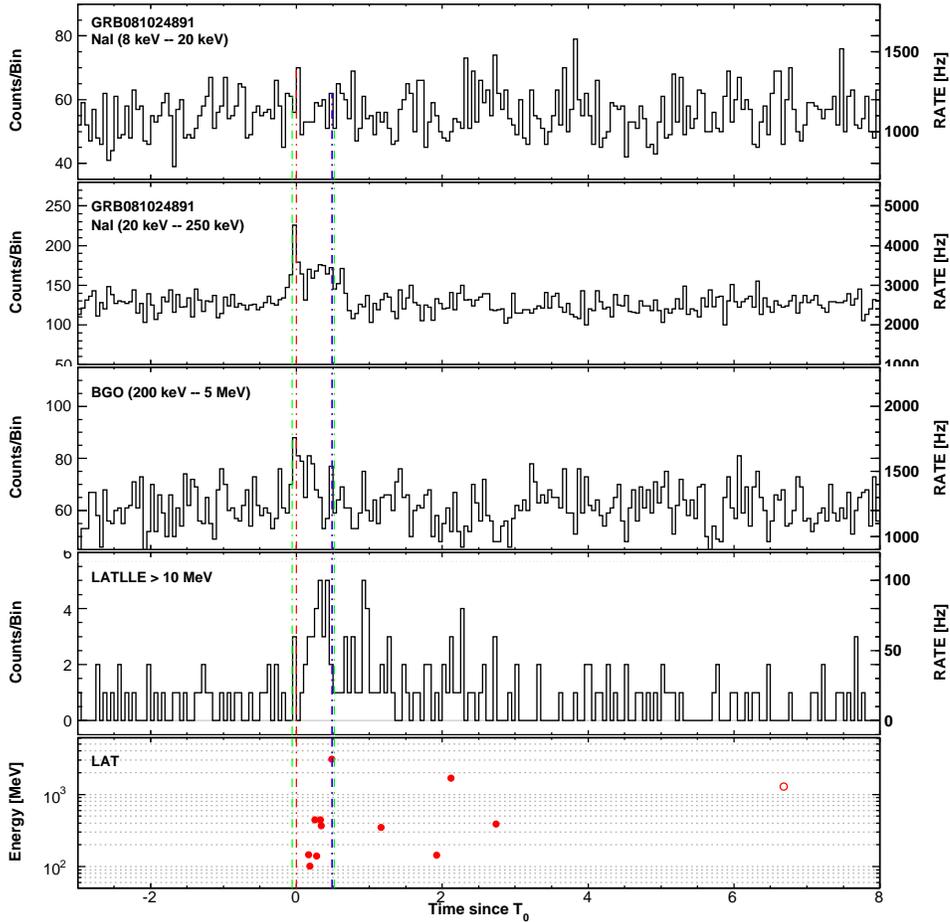}
\caption{Composite light curve for GRB\,081024B: summed GBM/NaI detectors (first two panels), GBM/BGO (third panel), LLE (fourth panel) and LAT Transient-class events above 100~MeV within a 12\de ROI  (last panel). See $\S$~\ref{sec_fermi_lat_grb_intro} for more information on lines and symbols in the LAT panels.}
\label{compo_081024B}
\end{center}
\end{figure}

\begin{figure}[ht!]
\begin{center}
\includegraphics[width=2.2in]{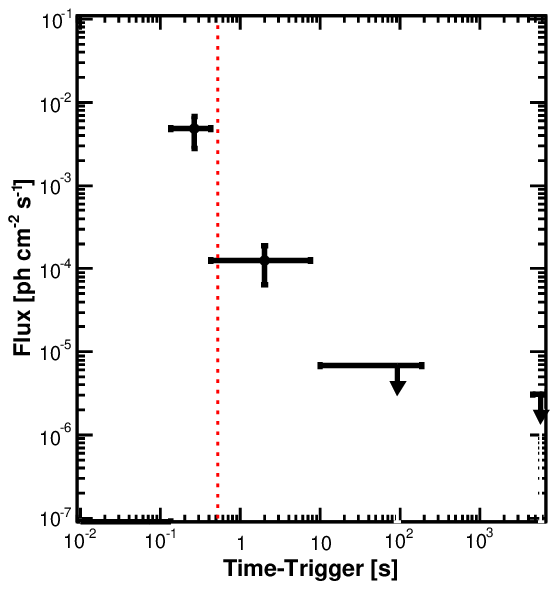}
\includegraphics[width=2.2in]{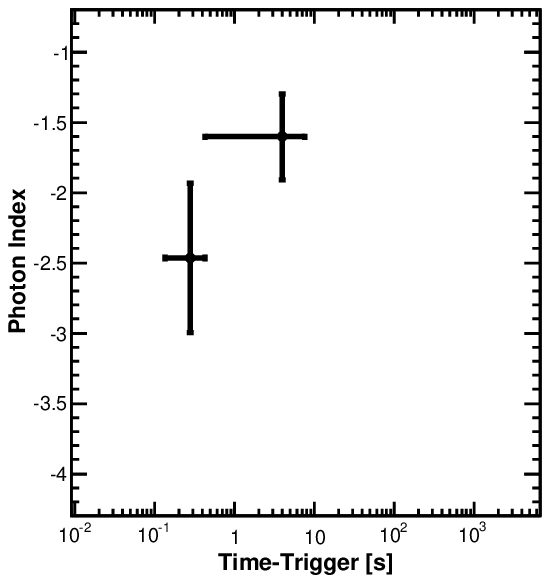}
\caption{Likelihood light curve for GRB\,081024B (flux above 100~MeV on the left, photon index on the right). See $\S$~\ref{sec_fermi_lat_grb_intro} for more information on lines and symbols.}
\label{like_081024B}
\end{center}
\end{figure}

GRB\,081024B triggered the GBM at T$_{0}$=21:22:41 UT on 24 October 2008 \citep[trigger 246576161,][]{2008GCN..8408....1C} and it was
 the first LAT detection of a short burst.
The LAT preliminary localization was delivered via GCN \citep{2008GCN..8407....1O}, with a statistical error of 0\ded16.
Follow-up observations by Swift and ground-based telescopes did not find any conclusive evidence for an afterglow
  counterpart \citep{2008GCNR..178....1G}.
Historically, GRB\,081024B represents the first clear detection of a temporally extended emission from a short GRB at~GeV
energies \citep{LAT081024B,Corsi081024B}.
Whereas the low-energy emission observed by the GBM lasts $\sim$0.5~s, the high-energy emission is visible up to $\sim$3~s after the trigger
time (Fig.~\ref{compo_081024B}).
The highest-energy event (3.07~GeV) is detected at T$_{0}$+0.49~s, i.e. very close in time to the end of the GBM emission.
A LAT T$_{90}$ could not be derived due to the small number of Transient-class events above 100~MeV, however the LLE duration ($\sim$2.3~s) indicates
a significantly longer duration of the LAT emission at tens-of-MeV energies.
Due to the low photon statistics, the LAT likelihood analysis did not find any significant signal in the ``EXT''
interval, and could not provide good time-resolved spectral measurements (Fig.~\ref{like_081024B}).

\FloatBarrier \subsection{GRB\,090217}
\begin{figure}[ht!]
\begin{center}
\includegraphics[width=5.0in]{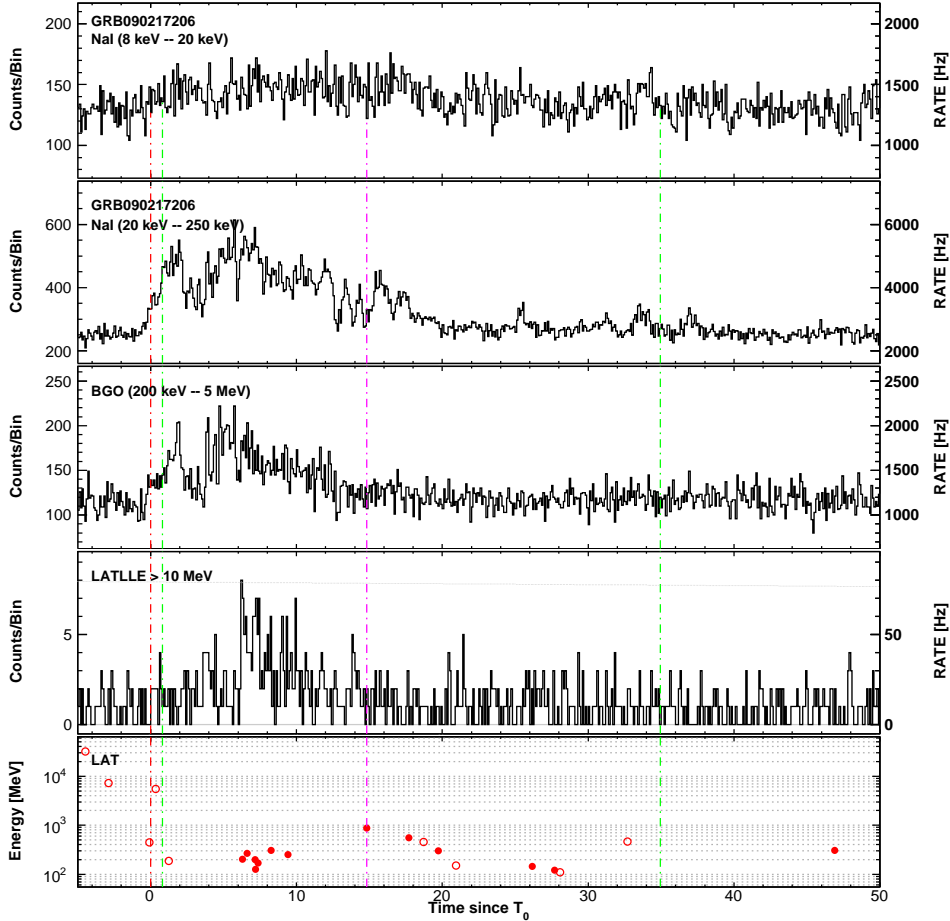}
\caption{Composite light curve for GRB\,090217: summed GBM/NaI detectors (first two panels), GBM/BGO (third panel), LLE (fourth panel) and LAT Transient-class events above 100~MeV within a 12\de ROI  (last panel). See $\S$~\ref{sec_fermi_lat_grb_intro} for more information on lines and symbols in the LAT panels.}
\label{compo_090217}
\end{center}
\end{figure}

\begin{figure}[ht!]
\begin{center}
\includegraphics[width=2.2in]{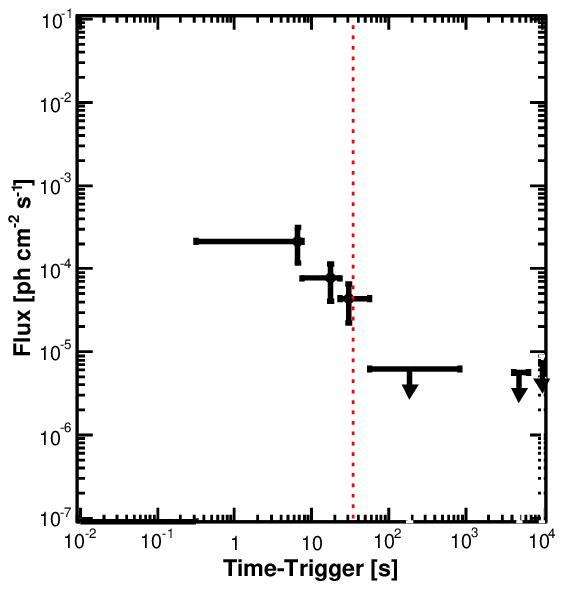}
\includegraphics[width=2.2in]{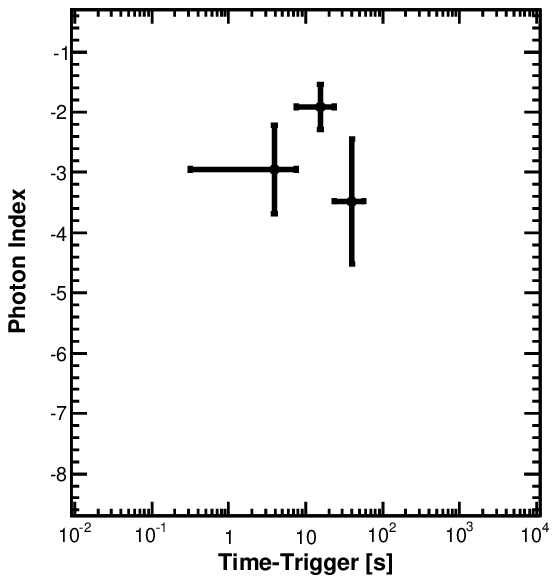}
\caption{Likelihood light curve for GRB\,090217 (flux above 100~MeV on the left, photon index on the right). See $\S$~\ref{sec_fermi_lat_grb_intro} for more information on lines and symbols.}
\label{like_090217}
\end{center}
\end{figure}

The long GRB\,090217 triggered the GBM at T$_{0}$=04:56:42.56 UT on 17 February 2009 \citep[trigger 256539404,][]{GBMGCN_090217}.
The LAT preliminary localization was delivered via GCN \citep{GCN_090217}, with a statistical error of 0\ded36.
A detailed analysis was published by the \Fermi LAT collaboration in \cite{LAT_090217}.
No X-ray counterpart was found in \Swift TOO observations of the LAT preliminary localization that covered only the center of the LAT error circle \citep{XRTGCN_090217}, and therefore no redshift is
available for this burst.
GRB\,090217 is a bright burst both in LLE and in LAT Transient-class data above 100~MeV.
The LLE light curve shows a series of pulses coincident with the GBM emission (Fig.~\ref{compo_090217}).
The highest-energy event (0.87~GeV) during this prompt emission is detected at T$_{0}$+14.83~s and is not associated with any noticeable structure of the GBM light curve.
The LAT T$_{95}$=68$^{+109}_{-40}$~s is not accurate enough to conclude if the high-energy emission extends later than the low-energy emission
(GBM T$_{95}\sim$35~s).
The off-axis angle of GRB\,090217 remained below 60\de until T$_{0}$+500~s, but no additional signal was found and upper limits are reported up to 10~ks
(Fig.~\ref{like_090217}).


\FloatBarrier \subsection{GRB\,090227B}
\begin{figure}[ht!]
\begin{center}
\includegraphics[width=5.0in]{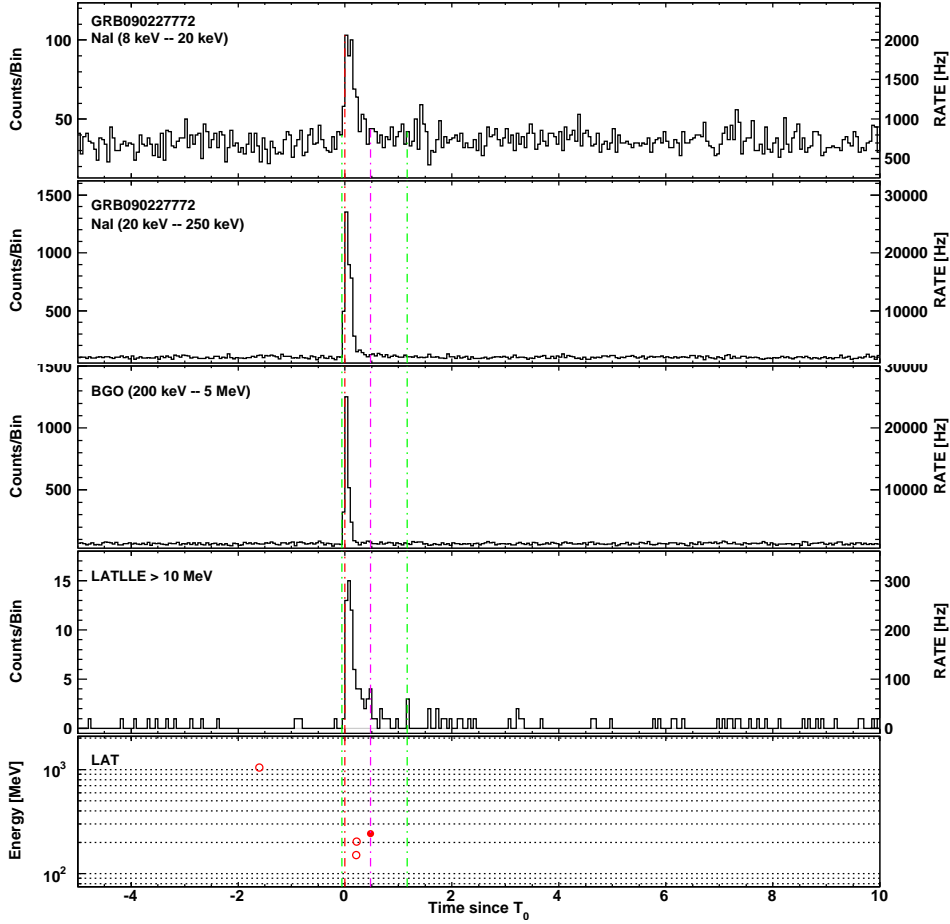}
\caption{Composite light curve for GRB\,090227B: summed GBM/NaI detectors (first two panels), GBM/BGO (third panel), LLE (fourth panel) and LAT Transient-class events above 100~MeV within a 12\de ROI  (last panel). See $\S$~\ref{sec_fermi_lat_grb_intro} for more information on lines and symbols in the LAT panels.}
\label{compo_090227B}
\end{center}
\end{figure}

The short GRB\,090227B triggered the GBM at T$_{0}$=18:31:01.41 UT on 27 February 2009 \citep[trigger 257452263,][]{2009GCN..8921....1G}.
GRB\,090227B had an initial off-axis angle of 71\de from the LAT boresight and the ARR triggered by the GBM brought it down to $\sim$20\de after $\sim$300~s.
The triangulation of the burst by the Interplanetary Network (IPN) provided a position with a 3$\sigma$ error box area of 1.5 square
degrees \citep{2009GCN..8925....1G} which we used in our analysis.
The GBM light curve of GRB\,090227B consists of one single pulse which was also significantly detected in the LLE data, with comparable durations
(Fig.~\ref{compo_090227B}).
A TS$\sim$30 was obtained by the LAT likelihood analysis based on the 3 Transient-class events recorded above 100~MeV during the GBM time
window, thus the burst is included in the catalog.
However, due to the position of the burst in the LAT FoV during the main emission, no LAT T$_{90}$ could be derived due to the paucity of events.
We could also not improve upon the IPN localization as no reliable TS map could be obtained.
For the same reason, no time-resolved likelihood analysis could be performed with the LAT.

\FloatBarrier \subsection{GRB\,090323}
The long, bright GRB\,090323 triggered the GBM at T$_{0}$=00:02:42.63 UT on 23 March 2009 \citep[trigger 259459364,][]{2009GCN..9021....1O}.
It had an initial off-axis angle of 57\ded2, where the LAT effective area is low, but it triggered an ARR of the \Fermi spacecraft which allowed the LAT
to detect its late emission phase and to localize it with a statistical error of 0\ded09 \citep{2009GCN..9021....1O}.
Specifically, GRB\,090323 was detected by the LAT on-gound Automated Science Processing (ASP) which searches for LAT counterparts to known GRBs.
\Swift TOO observations started $\sim$19.5~hours after the trigger time.
A possible X-ray counterpart was found by \Swift-XRT 1.9~arcmin away from the LAT
position \citep{2009GCN..9024....1K}, and further observations confirmed the existence of a fading source \citep{2009GCN..9031....1P}.
Follow-up observations of the X-ray afterglow with GROND in 7 bands started $\sim$27~hours after the trigger time, providing a preliminary
photometric redshift of z=4.0$\pm$0.3 \citep{2009GCN..9026....1U}.
\citet{2009GCN..9028....1C} reported a spectroscopic redshift of z=3.57 based on observations of the optical afterglow using the Gemini Multi-Object
Spectrograph (GMOS) mounted on the Gemini South Telescope.
Combined with its brightness, this makes GRB\,090323 the second most energetic LAT-detected burst after GRB\,080916C, with an isotropic equivalent
energy $\mathrm{E_{iso}}\simeq4.1\times 10^{54}$~erg (1~keV--10~GeV, within the GBM T$_{90}$).
The burst was also detected in the radio band \citep{2009GCN..9043....1H,2009GCN..9047....1V}.
A dedicated analysis of the near-infrared and optical follow-up observations of GRB\,090323 is presented in \citet{McBreen+10}.

\begin{figure}[ht!]
\begin{center}
\includegraphics[width=5.0in]{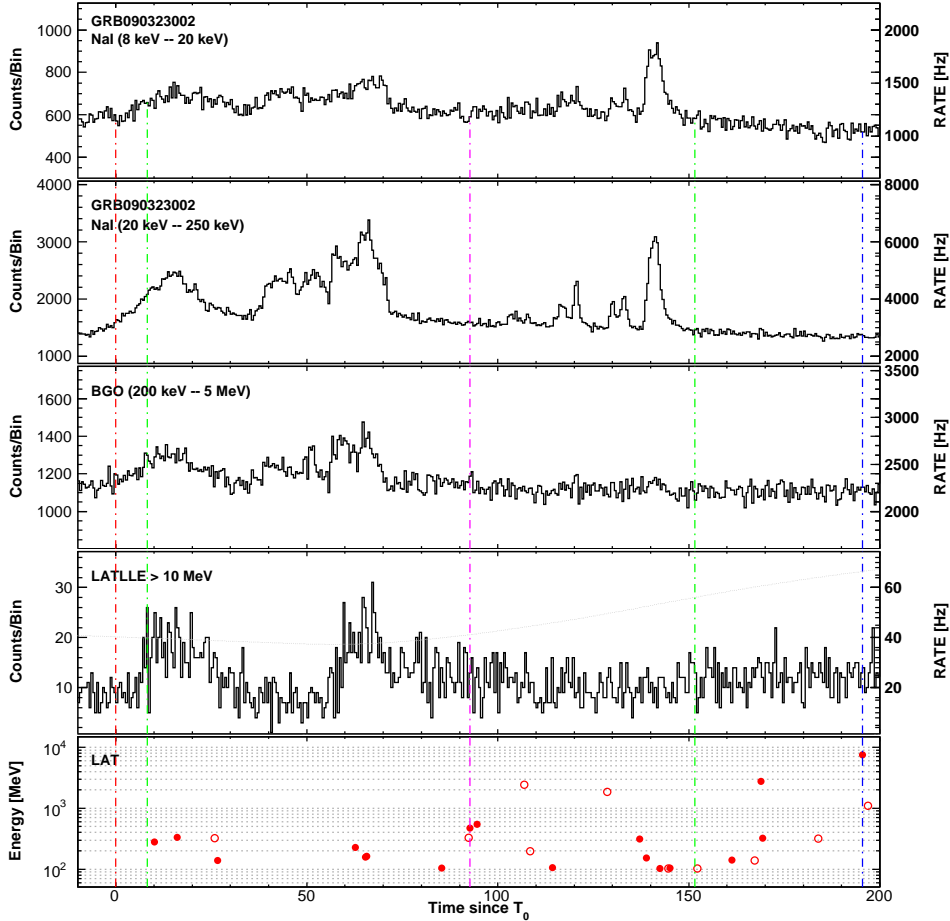}
\caption{Composite light curve for GRB\,090323: summed GBM/NaI detectors (first two panels), GBM/BGO (third panel), LLE (fourth panel) and LAT Transient-class events above 100~MeV within a 12\de ROI  (last panel). See $\S$~\ref{sec_fermi_lat_grb_intro} for more information on lines and symbols in the LAT panels.}
\label{compo_090323}
\end{center}
\end{figure}

\begin{figure}[ht!]
\begin{center}
\includegraphics[width=5.0in]{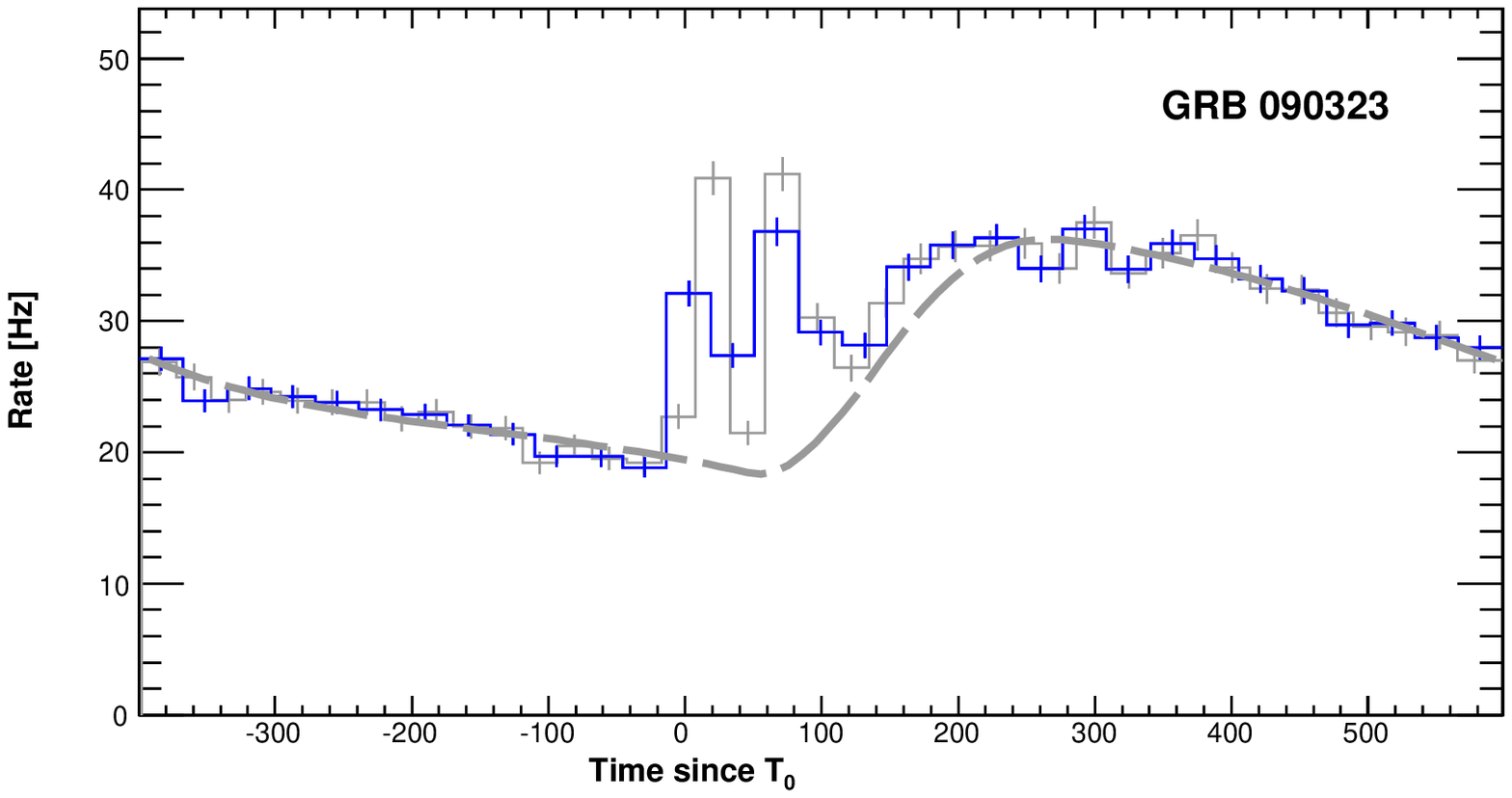}
\caption{Background estimation in the LLE light curve of GRB\,090323. Gray and blue histograms are the count rates in the LAT detector for two different time binning. The dashed gray line is the best fit background.}
\label{LLE090323002}
\end{center}
\end{figure}

\begin{figure}[ht!]
\begin{center}
\includegraphics[width=2.2in]{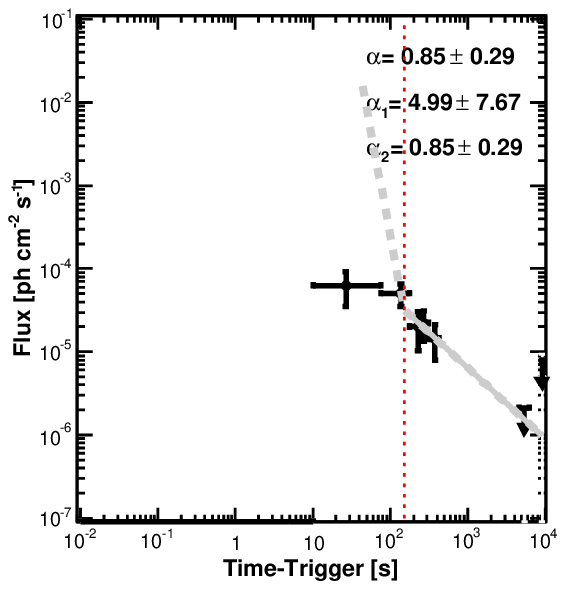}
\includegraphics[width=2.2in]{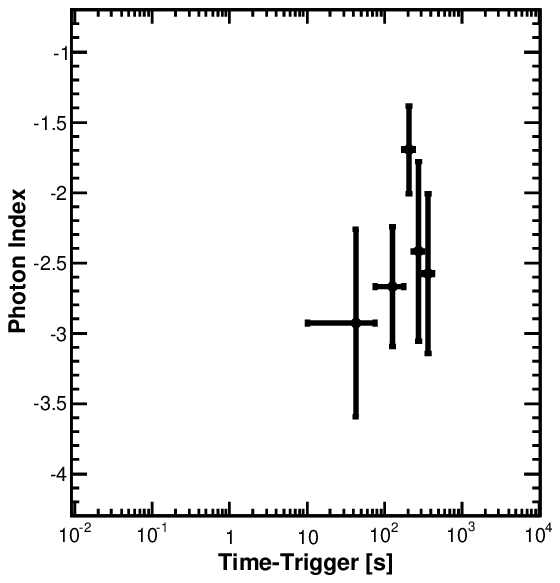}
\caption{Likelihood light curve for GRB\,090323 (flux above 100~MeV on the left, photon index on the right). See $\S$~\ref{sec_fermi_lat_grb_intro} for more information on lines and symbols.}
\label{like_090323}
\end{center}
\end{figure}

The GBM light curve of GRB\,090323 consists of several pulses and lasts $\sim$150~s (Fig.~\ref{compo_090323}).
The LLE light curve shows two bright long pulses which somehow coincide with two broad pulses observed in the GBM light curve.
The ARR caused the GBM and LAT orientations to change very rapidly with time, requiring a careful evaluation of the
instruments' responses and backgrounds as the spacecraft is slewing.
In particular, the burst Zenith angle increased from 67\de at T$_{0}$ to 84\de at T$_{0}$+300~s, causing a rise in the LAT count rate due to the
entrance of the Earth's limb in the instrument's FoV.
As illustrated in Fig.~\ref{LLE090323002}, the analysis of LLE data accounts for this effect, following the background estimation method discussed
in~$\S$~\ref{subsec_backgrounds}.
In the LAT likelihood analysis, we reduced the contamination from the Earth's limb by simply rejecting the time intervals in which the burst Zenith
angle was larger than 105\de.
Indications of long-lasting high-energy emission are seen in the LAT Transient-class data where multi-GeV events were recorded well after the
GBM emission, similarly to the 7.50~GeV event detected at T$_{0}$+195.42~s.
The LAT T$_{95}$=294$^{+55}_{-25}$~s confirms the temporal extension of the high-energy emission, and the LAT time-resolved likelihood analysis returned a
significant signal up to T$_{0}$+422~s, with a temporal decay index $\alpha$=0.85$\pm$0.29 (Fig.~\ref{like_090323}).
GRB\,090323 became occulted after $\sim$570~s and, in the next orbit, the spacecraft entered the SAA only $\sim$50~s after the burst
exited occultation, thus only upper limits are reported at later times, up to $\sim$10~ks.

\FloatBarrier \subsection{GRB\,090328}
The long, bright GRB\,090328 triggered the GBM at T$_{0}$=09:36.46 UT on 28 March 2009 \citep[trigger 259925808, ][]{2009GCN..9044....1M}.
GRB\,090328 had an initial off-axis angle of 64\ded6 in the LAT and the ARR triggered by the GBM brought it down to $\sim$10\de after $\sim$300~s.
The LAT preliminary localization was delivered via GCN \citep{2009GCN..9044....1M}, with a statistical error of 0\ded11.
\Swift TOO observations started $\sim$16~hours after the trigger time \citep{2009GCNR..207....1M}.
A possible X-ray counterpart was found by \Swift-XRT $\sim$10~arcmin away from the LAT position \citep{2009GCN..9045....1K}, and further observations
confirmed the existence of a fading source \citep{2009GCN..9052....1R}.
Observations of a candidate optical afterglow were also reported by \citet{2009GCN..9046....1K} and \citet{2009GCN..9048....1O}.
More observations of the afterglow were conducted in the optical \citep{2009GCN..9058....1A}, in the optical/NIR with GROND
\citep{2009GCN..9054....1U}, and in the radio band \citep{2009GCN..9060....1F}.
\citet{2009GCN..9053....1C} reported a spectroscopic redshift of z=0.736 based on observations of the optical afterglow using the GMOS
  spectrograph mounted on the Gemini South Telescope.
A dedicated analysis of the near-infrared and optical follow-up observations of GRB\,090328 is presented in \citet{McBreen+10}.\\

\begin{figure}[ht!]
\begin{center}
\includegraphics[width=5.0in]{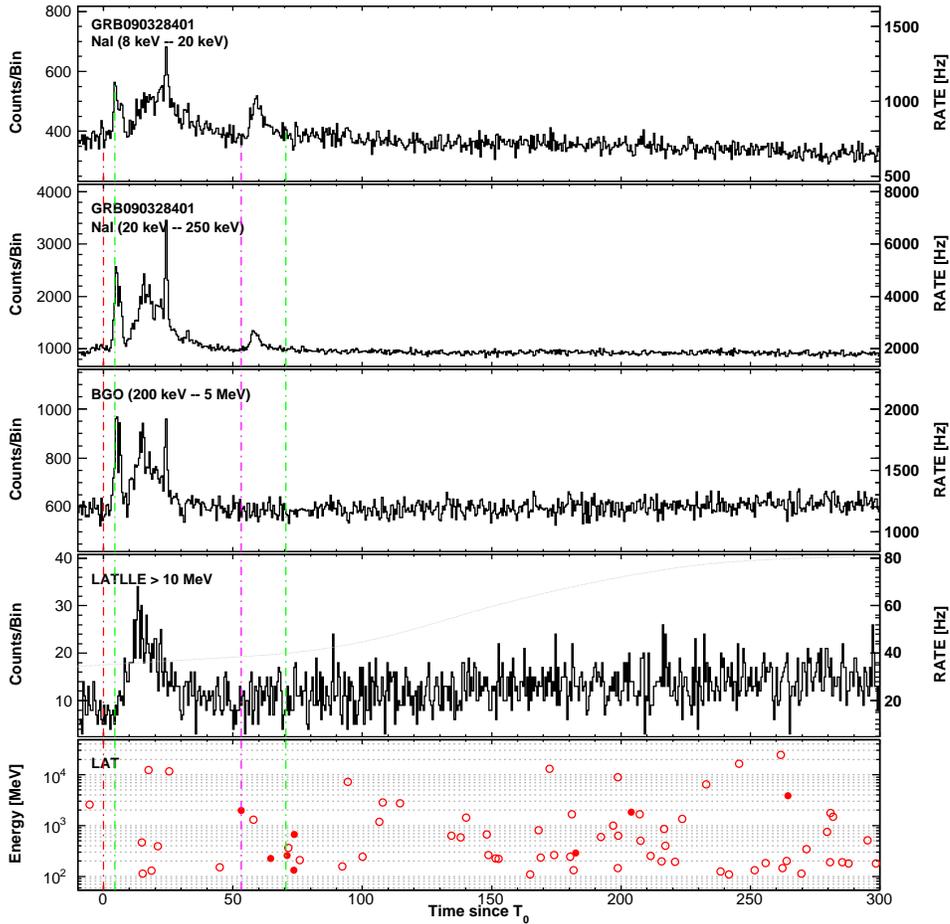}
\caption{Composite light curve for GRB\,090328: summed GBM/NaI detectors (first two panels), GBM/BGO (third panel), LLE (fourth panel) and LAT Transient-class events above 100~MeV within a 12\de ROI  (last panel). See $\S$~\ref{sec_fermi_lat_grb_intro} for more information on lines and symbols in the LAT panels.}
\label{compo_090328}
\end{center}
\end{figure}

\begin{figure}[ht!]
\begin{center}
\includegraphics[width=5.0in]{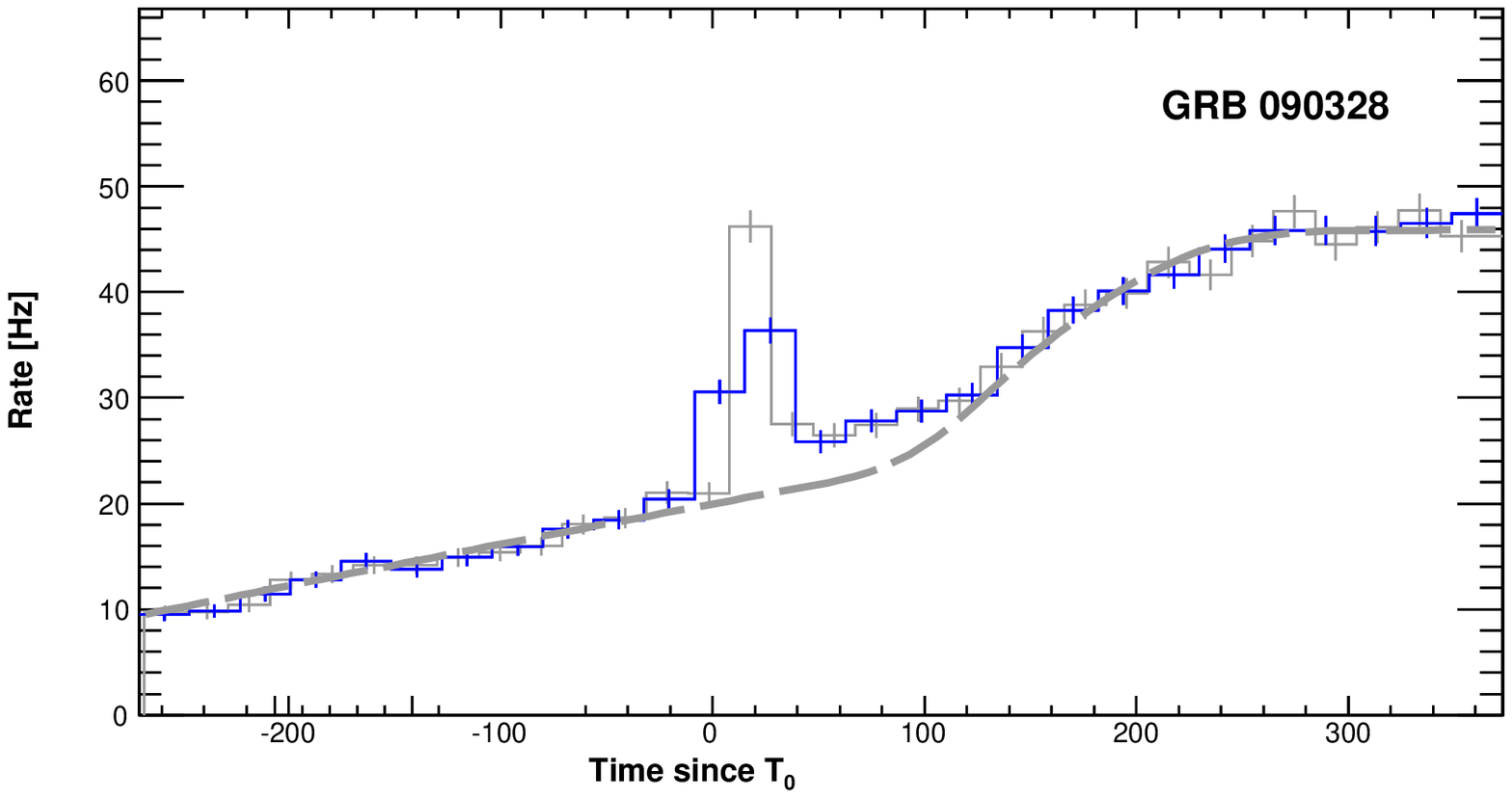}
\caption{Background estimation in the LLE light curve of  GRB\,090328401. Gray and blue histograms are the count rates in the LAT detector for two different time binning. The dashed gray line is the best fit background.}
\label{LLE090328401}
\end{center}
\end{figure}

The GBM light curve of GRB\,090328 consists of several pulses and lasts $\sim$70~s (Fig.~\ref{compo_090328}).
The LLE light curve shows one single, long bright pulse which coincides with the second broad pulse observed in the GBM light curve.
In addition, the first narrow spike in the GBM light curve has no LLE counterpart, indicating an initially soft spectrum.
The ARR caused an increase in the background rate in the LLE light curve as the burst off-axis angle was decreasing (third panel of
Fig.~\ref{compo_090328}). Fig.~\ref{LLE090328401} shows the results of the background estimation in the analysis of LLE data.

\begin{figure}[ht!]
\begin{center}
\includegraphics[width=2.2in]{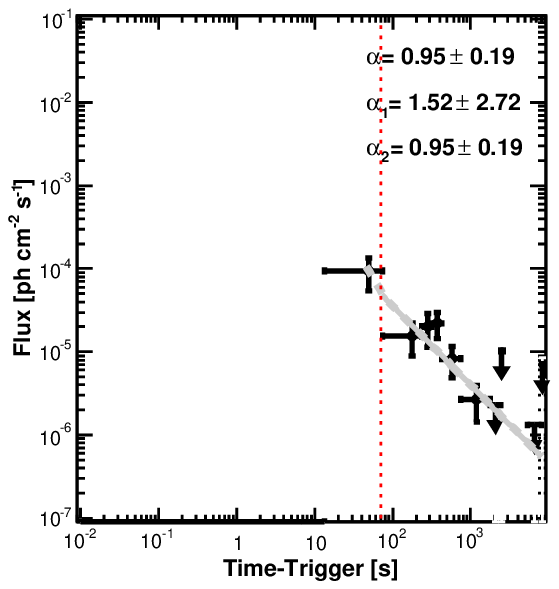}
\includegraphics[width=2.2in]{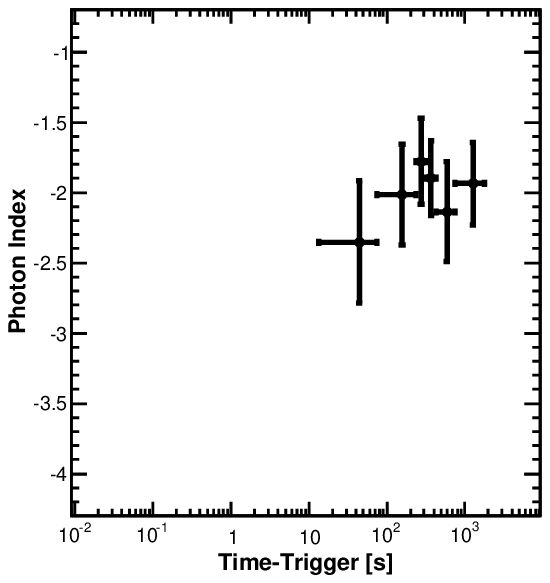}
\caption{Likelihood light curve for GRB\,090328 (flux above 100~MeV on the left, photon index on the right). See $\S$~\ref{sec_fermi_lat_grb_intro} for more information on lines and symbols.}
\label{like_090328}
\end{center}
\end{figure}

In the preliminary analysis of LAT data, \citet{2009GCN..9077....1C} reported that GRB\,090328 high-energy emission lasted until $\sim$900~s
  post trigger.
Our analysis of the LAT Transient-class data above 100~MeV provided a LAT T$_{95}$=653$^{+134}_{-45}$~s which confirms the temporal extension of the
burst emission in the LAT. We could also confirm that the highest-energy events detected by the LAT which are spatially coincident with the burst
position arrived hundreds of seconds after the trigger time. Multi-GeV events were recorded well after the GBM emission, in particular two
3.83~GeV and 5.32~GeV events detected at T$_{0}$+264.42~s and T$_{0}$+697.80~s, respectively.
Unlike GRB\,090323, the ARR for GRB\,090328 was excellent and started just after the burst exited occultation.
During the next two orbits, observations were only interrupted by occultations, with no passage through the SAA.
As a result, the LAT time-resolved likelihood analysis detected a significant signal up to T$_{0}$+1.78~ks, with a temporal decay index
$\alpha$=0.95$\pm$0.19 (Fig.~\ref{like_090328}).

\FloatBarrier \subsection{GRB\,090510}
The short, bright GRB\,090510 is the only burst detected so far by the LAT onboard flight software (trigger
  263607783). The LAT onboard localization was delivered via GCN \citep{2009GCN..9334....1O}, with a statistical error of 7~arcmin.
Combined with an initial off-axis angle of 13\ded6, GRB\,090510 caused an exceptionally bright emission in the LAT as reported in the
follow-up analysis by \citet{2009GCN..9350....1O}, and it triggered an ARR of the \Fermi spacecraft.
GRB\,090510 was also significantly ($>$5$\sigma$) detected by the AGILE-GRID above 100~MeV \citep{2009GCN..9343....1L,GRB090510:Agile}.
At lower energies, GRB\,090510 triggered both the \Swift-BAT \citep{2009GCN..9331....1H,2009GCNR..218....1H} and the GBM \citep[trigger 263607781, at
T$_{0}$=00:22:59.97 UT on 10 May 2009, ][]{2009GCN..9336....1G} instruments. Both the \Swift-XRT and GBM positions were consistent with the LAT onboard
localization.
Follow-up observations of the candidate optical afterglow found by \Swift-UVOT \citep{2009GCN..9332....1M} were conducted with the Nordic Optical Telescope
\citep{2009GCN..9340....1M} and in the optical/NIR with GROND \citep{2009GCN..9352....1O}.
\citet{2009GCN..9353....1R} reported a spectroscopic redshift of z=0.903$\pm$0.003 based on observations with the VLT/FORS2 instrument.
A dedicated analysis of the near-infrared and optical follow-up observations of GRB\,090510 is presented in \citet{McBreen+10}, 
and analysis of the broadband observations including gamma-ray, X-ray, and optical are presented in \citet{DePasquale+10}.

\begin{figure}[ht!]
\begin{center}
\includegraphics[width=5.0in]{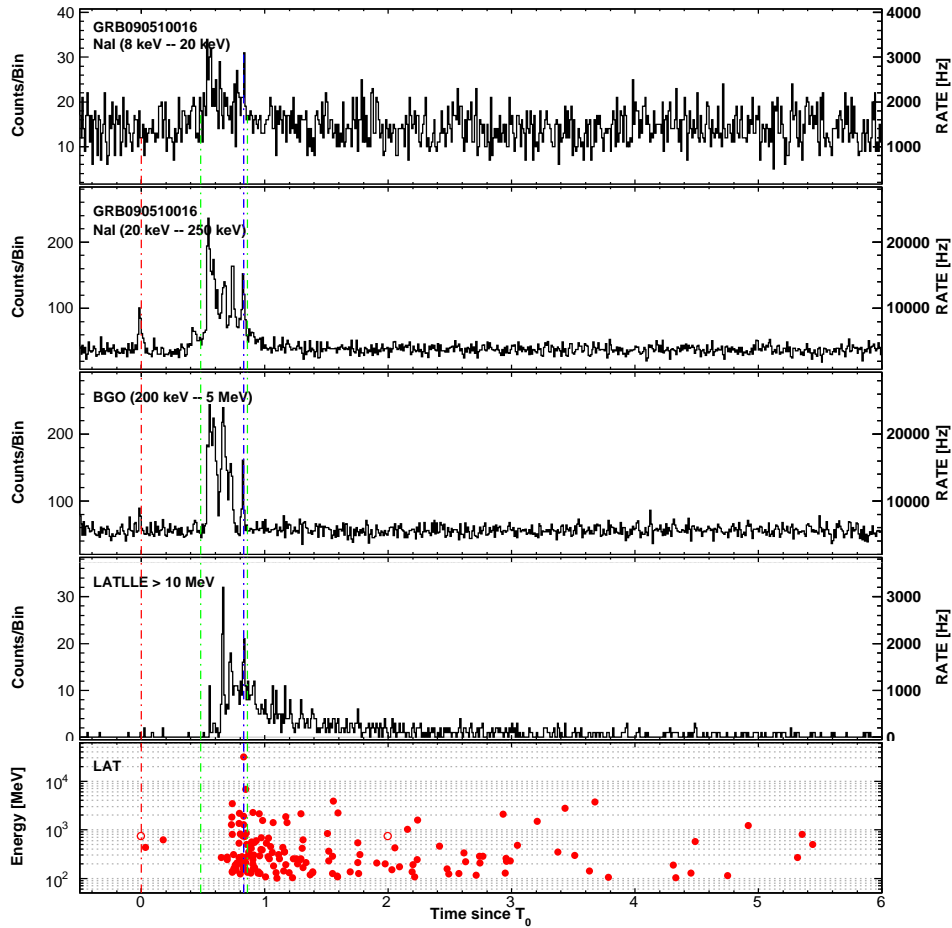}
\caption{Composite light curve for GRB\,090510: summed GBM/NaI detectors (first two panels), GBM/BGO (third panel), LLE (fourth panel) and LAT Transient-class events above 100~MeV within a 12\de ROI  (last panel). See $\S$~\ref{sec_fermi_lat_grb_intro} for more information on lines and symbols in the LAT panels.}
\label{compo_090510016}
\end{center}
\end{figure}

As shown in Fig.~\ref{compo_090510016}, the GBM triggered on a precursor in GRB\,090510 light curve.
The main emission in the GBM consists of several pulses, with a maximum at T$_{0}$+0.6~s and a duration of $\sim$0.6~s.
The temporal structure of the LAT emission shows fast variability on timescales as short as 20~ms.
The LLE light curve shows a series of short spikes coinciding with the GBM pulses and appearing on top of a smoother and longer single pulse.
Two of the three LAT Transient-class events recorded above 100~MeV at the time of the precursor (between T$_{0}$ to T$_{0}$+0.2~s) have high
probabilities to be associated with the burst.
The main emission in the LAT starts at T$_{0}$+0.6~s and lasts much longer than the GBM estimated duration, with 180 Transient-class events recorded
above 100~MeV within the LAT T$_{90}$$\sim$45~s (see Table~\ref{tab_likelihoods}).
many GeV events are recorded during and well after the GBM emission, similarly to the 31.31~GeV event detected a
T$_{0}$+0.83~s in coincidence with a short bright spike in the GBM light curve.
This photon candidate has been used by the \Fermi LAT collaboration to set the best lower limit on the energy scale at which postulated quantum-gravity
effects create violations of Lorentz invariance, disfavoring models which predict a linear variation of the speed of light with photon energy below
the Planck energy scale $E_\mathrm{Planck}$=1.22$\times10^{19}$~GeV \citep{GRB090510:Nature}.

In the time-resolved spectral analysis published by the \Fermi LAT collaboration \citep{GRB090510:ApJ}, the prompt emission spectrum of GRB\,090510 was
fitted over more than six decades in energy by the combination of the empirical Band function with a high-energy power law.
The hard power law is detected from the onset of the main emission in the LAT, and it dominates the Band
function not only at high energy but also below $\sim$20~keV.
Our GBM-LAT joint spectral analysis in the GBM time window confirms these results, yielding a peak energy $E_{p}$$\sim$3.6~MeV for the Band function
and a spectral slope of 1.60$\pm$0.04 for the additional power-law component. The total isotropic equivalent energy is ($7.3\pm0.3)\times 10^{52}$~erg
(1~keV--10~GeV, within the GBM T$_{90}$).

\begin{figure}[ht!]
\begin{center}
\includegraphics[width=2.2in]{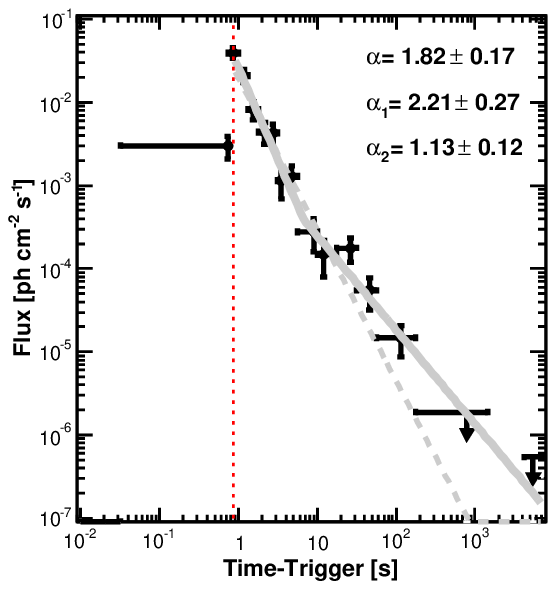}
\includegraphics[width=2.2in]{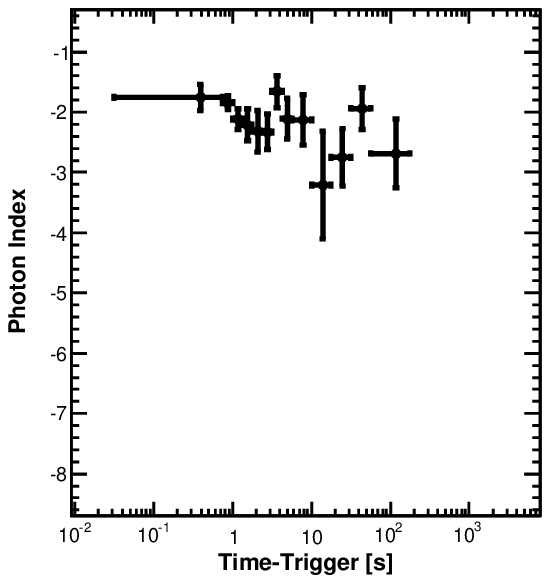}
\caption{Likelihood light curve for GRB\,090510 (flux above 100~MeV on the left, photon index on the right). See $\S$~\ref{sec_fermi_lat_grb_intro} for more information on lines and symbols.}
\label{like_090510016}
\end{center}
\end{figure}

The LAT time-resolved likelihood analysis resulted in a well sampled light curve of the high-energy flux up to T$_{0}$+178~s (Fig.~\ref{like_090510016}).
No significant spectral evolution was detected.
The decay of the flux as a function of time can be fitted with a simple power law starting from the GBM T$_{95}$, with a decay index
$\alpha$=1.82$\pm$0.17 somewhat steeper than the index of 1.38$\pm$0.07 reported in \cite{DePasquale+10}.
However, the fit of the flux light curve with a broken power law from the peak flux time $t_p$=T$_{0}$+0.9~s up to T$_{0}$+$\sim$8~ks (including flux
upper limits after T$_{0}$+178~s) returned a significant break at $t_b$=7.0$\pm$1.5~s, along with a steeper initial decay ($\alpha_1$=2.21$\pm$0.27)
and a smoother decay ($\alpha_2$=1.13$\pm$0.12) at later times.

\FloatBarrier \subsection{GRB\,090531B}
\begin{figure}[ht!]
\begin{center}
\includegraphics[width=5.0in]{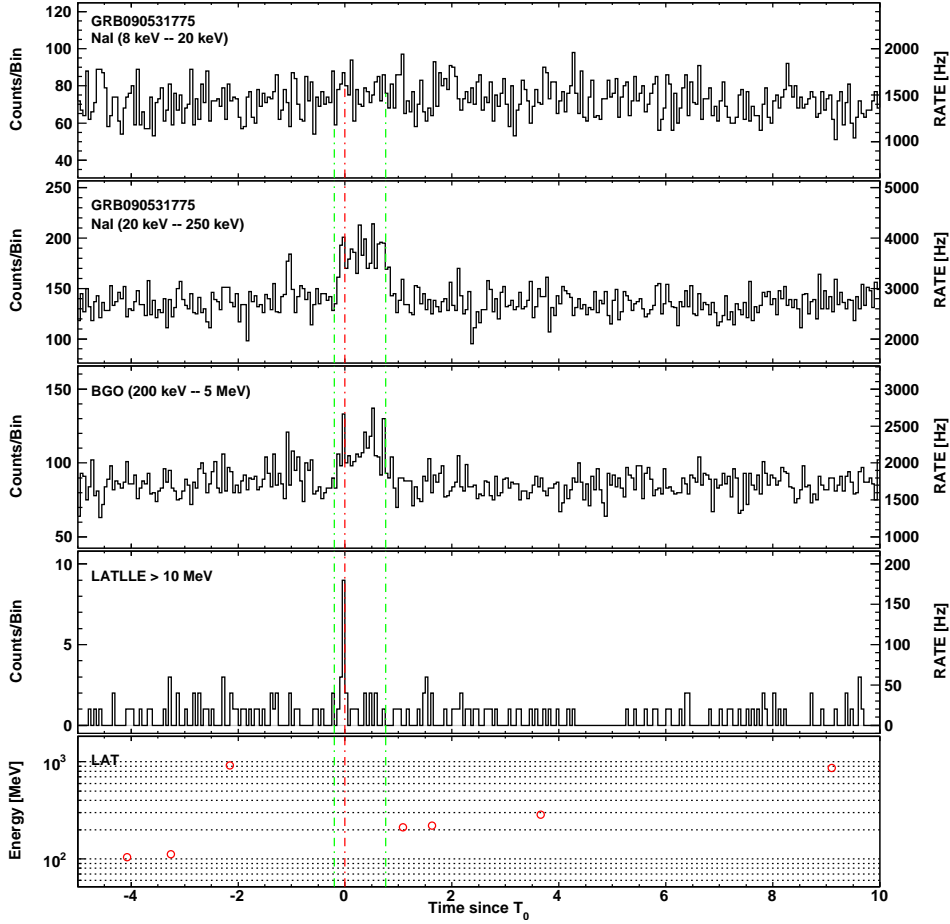}
\caption{Composite light curve for GRB\,090531B: summed GBM/NaI detectors (first two panels), GBM/BGO (third panel), LLE (fourth panel) and LAT Transient-class events above 100~MeV within a 12\de ROI  (last panel). See $\S$~\ref{sec_fermi_lat_grb_intro} for more information on lines and symbols in the LAT panels.}
\label{compo_090531B}
\end{center}
\end{figure}

The short GRB\,090531B triggered the GBM at T$_{0}$=18:35:56.49 UT on 31 May 2009 \citep[trigger 265487758,][]{2009GCN..9501....1G},
and it was also detected by the \Swift-BAT \citep{2009GCN..9461....1C} and \Swift-XRT \citep{2009GCN..9463....1S} instruments.
It is a relatively faint burst, both in the GBM and in the LAT (despite an initial off-axis angle of 21\ded9).
Only a few LAT Transient-class events above 100~MeV are compatible with the \Swift localization, therefore no significant emission was found in the
likelihood analysis. 
GRB\,090531B was detected in the LLE data only, and the LLE light curve shows a significant signal excess which is temporally coincident with the first
pulse detected by the NaI and BGO detectors (Fig.~\ref{compo_090531B}).

\FloatBarrier \subsection{GRB\,090626}
\begin{figure}[ht!]
\begin{center}
\includegraphics[width=5.0in]{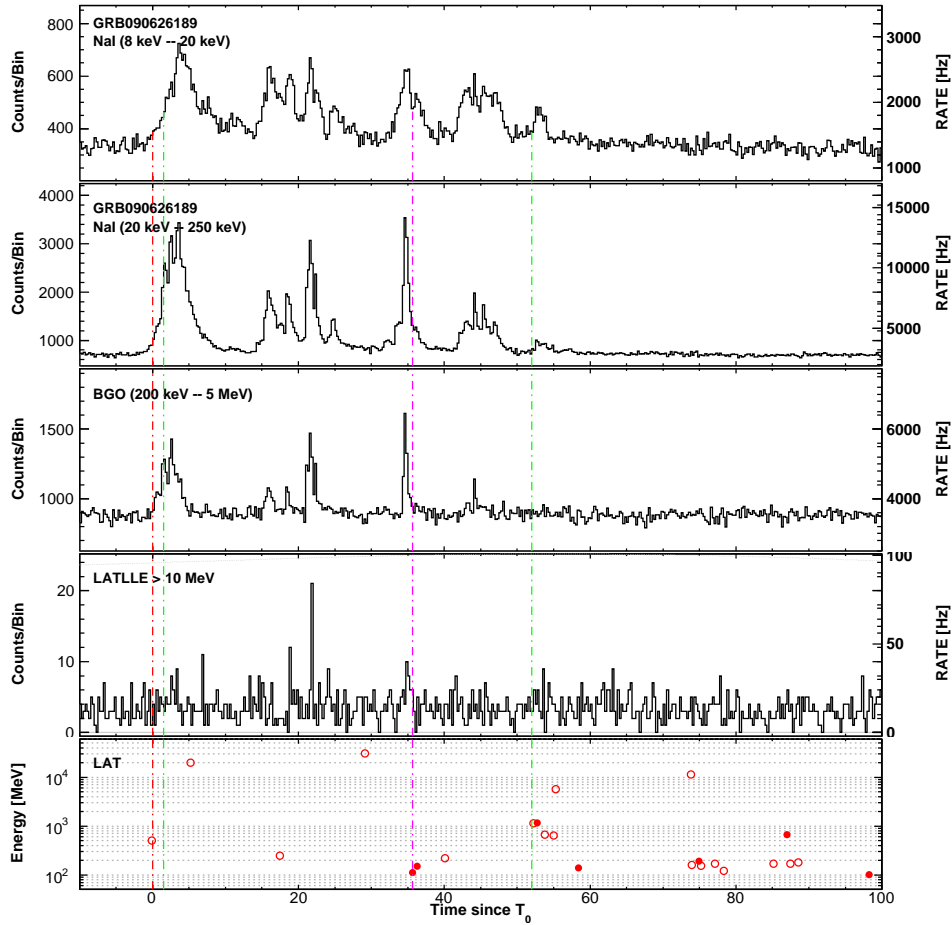}
\caption{Composite light curve for GRB\,090626: summed GBM/NaI detectors (first two panels), GBM/BGO (third panel), LLE (fourth panel) and LAT Transient-class events above 100~MeV within a 12\de ROI  (last panel). See $\S$~\ref{sec_fermi_lat_grb_intro} for more information on lines and symbols in the LAT panels.}
\label{compo_090626}
\end{center}
\end{figure}

\begin{figure}[ht!]
\begin{center}
\includegraphics[width=2.2in]{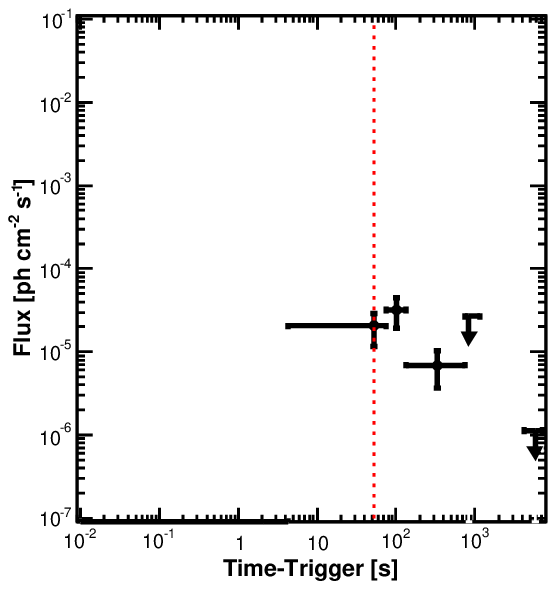}
\includegraphics[width=2.2in]{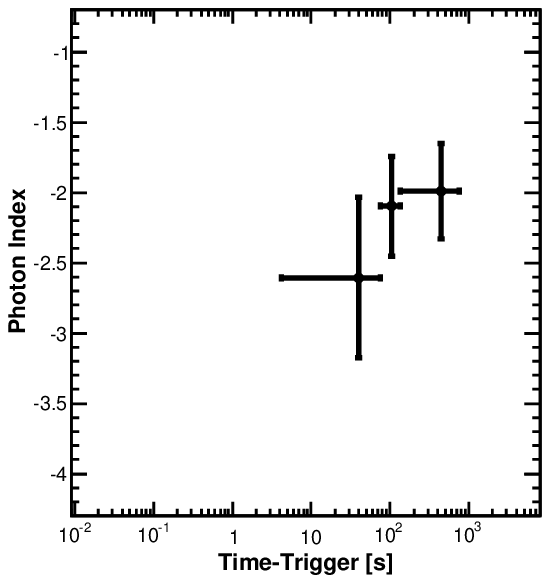}
\caption{Likelihood light curve for GRB\,090626 (flux above 100~MeV on the left, photon index on the right). See $\S$~\ref{sec_fermi_lat_grb_intro} for more information on lines and symbols.}
\label{like_090626}
\end{center}
\end{figure}

The long GRB\,090626 triggered the GBM at T$_{0}$=04:32:08.88 UT on 26 June 2009 \citep[trigger 267683530,][]{2009GCN..9579....1V}.
It was also detected by the LAT on-gound ASP which searches for LAT counterparts to known GRBs, and the
LAT preliminary localization was delivered via GCN \citep{090626_LAT}, with a statistical error of 0\ded32 (95\% confidence level).
The GBM light curve of GRB\,090626 consists of several bright pulses and lasts $\sim$55~s (Fig.~\ref{compo_090626}).
The LLE light curve shows one single, faint short pulse which coincides with the second bright pulse observed in the BGO light curve.
However, this signal excess was not significant enough to claim an LLE detection (see Table~\ref{tab_GRBs}).
In the preliminary analysis of LAT data, \citet{090626_LAT} reported that GRB\,090626 high-energy emission lasted until T$_{0}$+$\sim$250~s.
Our analysis of the LAT Transient-class data above 100~MeV provided a LAT T$_{95}$=300$^{+338}_{-53}$~s which confirms the temporal extension of the
burst emission in the LAT. In addition, a 2.09~GeV event is recorded at T$_{0}$+111.63~s.
The LAT time-resolved likelihood analysis returned a significant flux in three time bins only up to T$_{0}$+750~s (Fig.~\ref{like_090626}).

\FloatBarrier \subsection{GRB\,090720B}
\begin{figure}[ht!]
\begin{center}
\includegraphics[width=5.0in]{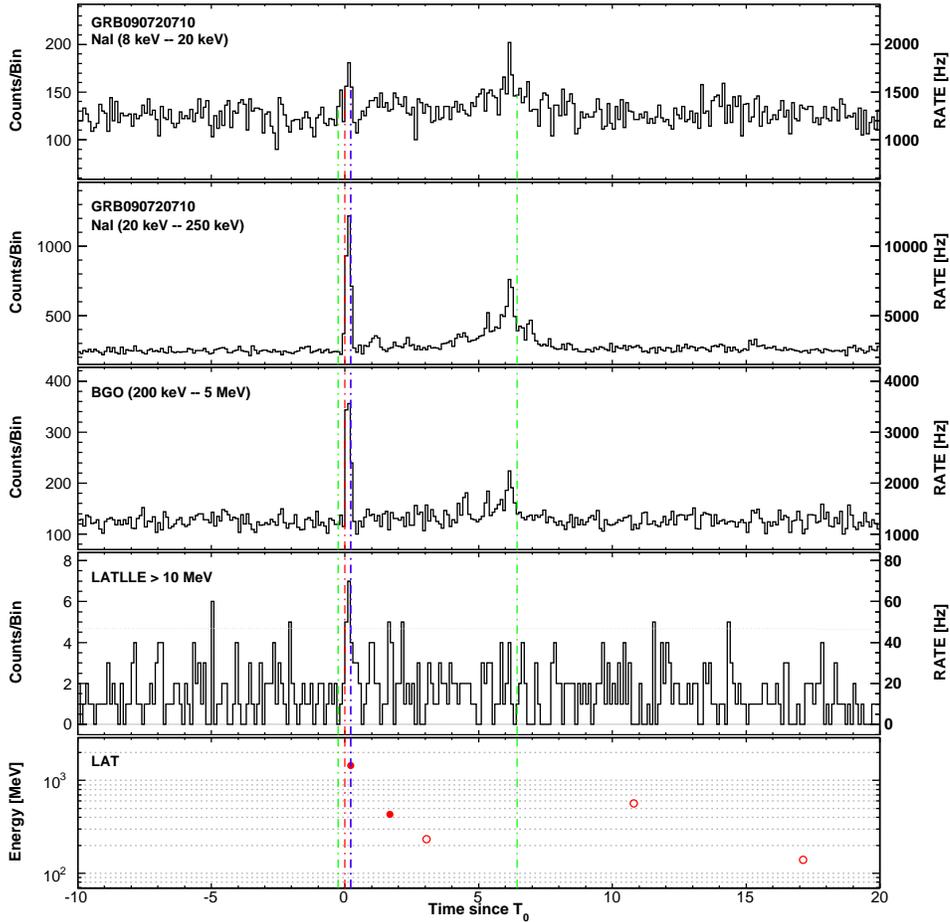}
\caption{Composite light curve for GRB\,090720B: summed GBM/NaI detectors (first two panels), GBM/BGO (third panel), LLE (fourth panel) and LAT Transient-class events above 100~MeV within a 12\de ROI  (last panel). See $\S$~\ref{sec_fermi_lat_grb_intro} for more information on lines and symbols in the LAT panels.}
\label{compo_090720B}
\end{center}
\end{figure}

\begin{figure}[ht!]
\begin{center}
\includegraphics[width=2.2in]{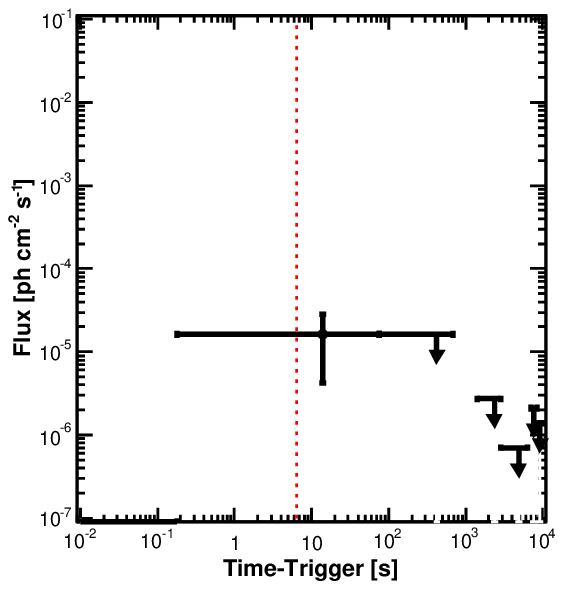}
\includegraphics[width=2.2in]{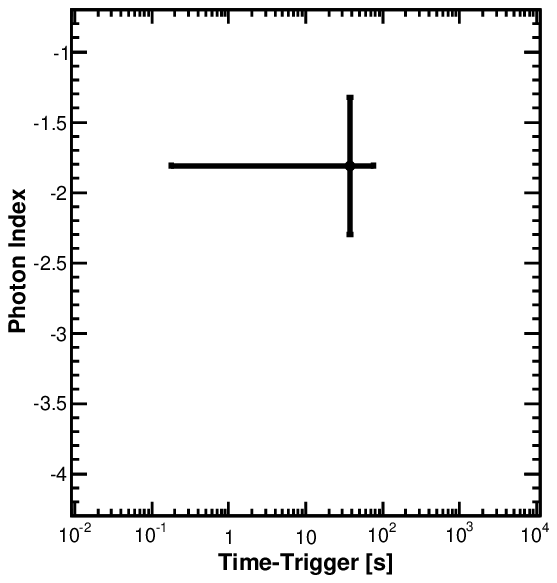}
\caption{Likelihood light curve for GRB\,090720B (flux above 100~MeV on the left, photon index on the right). See $\S$~\ref{sec_fermi_lat_grb_intro} for more information on lines and symbols.}
\label{like_090720B}
\end{center}
\end{figure}

The long GRB\,090720B triggered the GBM at T$_{0}$=17:02:56.91 UT on 20 July 2009 \citep[trigger 269802178,][]{2009GCN..9698....1B}.
The GBM light curve consists of one short hard pulse followed by a wider pulse (Fig.~\ref{compo_090720B}).
GRB\,090720B had an off-axis angle of 56\ded1 in the LAT at the trigger time, where the effective area is a factor $\sim$3 less than on axis.
The burst was not significantly detected in the LLE data and the LAT likelihood analysis returned a TS$\sim$25 based on the 3
Transient-class events recorded above 100~MeV during the GBM time window, including a 1.45~GeV event at T$_{0}$+0.22~s.
No LAT T$_{90}$ could be derived due to the large Zenith angle of the burst.
The LAT time-resolved likelihood analysis returned a marginal detection in one time bin only, ending at T$_{0}$+75~s (Fig.~\ref{like_090720B}).

\FloatBarrier \subsection{GRB\,090902B}
The long, bright GRB\,090902B triggered the GBM at T$_{0}$=11:05:08.31 UT on 2 September 2009 \citep[trigger 273582310,][]{Betta_090902B}.
In spite of an initial off-axis angle of 50\ded8, GRB\,090902B caused exceptionally bright emission in the LAT and it triggered an ARR of the \Fermi
spacecraft.
The LAT preliminary localization was delivered via GCN \citep{LAT_090902B}, with a statistical error of 0\ded04.
\Swift TOO observations started $\sim$12.5~hours after the trigger time \citep{2009GCNR..249....1S}.
A possible X-ray counterpart was found by \Swift-XRT
3.2~arcmin away from the LAT position \citep{XRT_090902B}, and further observations confirmed the existence of a fading source \citep{XRT2_090902B}.
Follow-up detections in the optical were reported by the \Swift-UVOT team \citep{UVOT_090902B,UVOT2_090902B} and by several observers operating ground-based
telescopes \citep{Perley_090902B,Guidorzi_090902B,Pandey_090902B}. GRB\,090902B was also detected in the optical/NIR \citep{Olivares_090902B} and in
the radio band \citep{Alexander_090902B,Chandra_090902B}.
\citet{Z_090902B} reported a spectroscopic redshift of z=1.822 based on observations of the optical afterglow using the GMOS spectrograph
mounted on the Gemini South Telescope.
Combined with its brightness, this makes GRB\,090902B the third most energetic LAT-detected burst after GRB\,080916C and GRB\,090323, with an isotropic
equivalent energy $\mathrm{E_{iso}}\simeq3.4\times 10^{54}$~erg (1~keV--10~GeV, within the GBM T$_{90}$).
A dedicated analysis of the near-infrared and optical follow-up observations of GRB\,090902B is presented in \citet{McBreen+10}.

\begin{figure}[ht!]
\begin{center}
\includegraphics[width=5.0in]{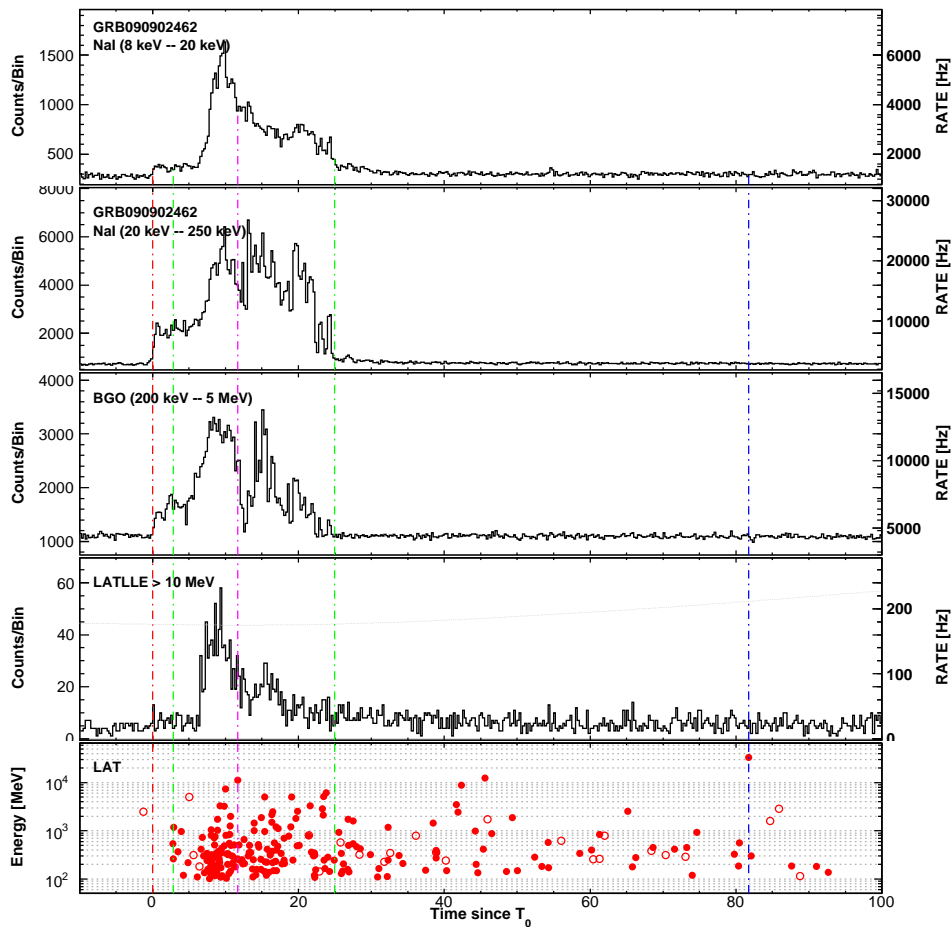}
\caption{Composite light curve for GRB\,090902B: summed GBM/NaI detectors (first two panels), GBM/BGO (third panel), LLE (fourth panel) and LAT Transient-class events above 100~MeV within a 12\de ROI  (last panel). See $\S$~\ref{sec_fermi_lat_grb_intro} for more information on lines and symbols in the LAT panels.}
\label{compo_090902B}
\end{center}
\end{figure}

As shown in Fig.~\ref{compo_090902B}, the GBM light curve of GRB\,090902B is complex both in the NaI and BGO detectors, probably resulting from the
  overlap of many small pulses.
After a plateau phase of $\sim$6~s similar to what is observed at lower energies, the LLE light curve shows a series of short spikes on top of two
broad and partially overlapping pulses, which seem to coincide with two distinct emission episodes visible in both the NaI and BGO light curves.
The temporal structure of the LAT emission shows fast variability on timescales as short as $\sim$100~ms.
In the first paper published by the \Fermi LAT collaboration \citep{GRB090902B:Fermi}, the prompt emission spectrum of GRB\,090902B was fitted over
more than six decades in energy by the combination of the empirical Band function with a high-energy power law.
The hard power law is detected from the trigger time, and it dominates the Band function not only at high
energies but also below $\sim$50~keV as already reported in the preliminary joint analysis of GBM and LAT data \citep{LAT2_090902B}.
Our GBM-LAT joint spectral analysis in the GBM time window confirms these results, yielding similar parameters for the Band function and a spectral slope of
1.94$\pm$0.01 for the additional power-law component.
Note that alternative spectral models have been studied in details \citep{ryde09,2011ApJ...730....1L}, and that the peculiar spectrum of GRB\,090902B
has also been used to constrain several theoretical models \citep{2011MNRAS.417.1584B,2012MNRAS.420..468P}.

\begin{figure}[ht!]
\begin{center}
\includegraphics[width=2.2in]{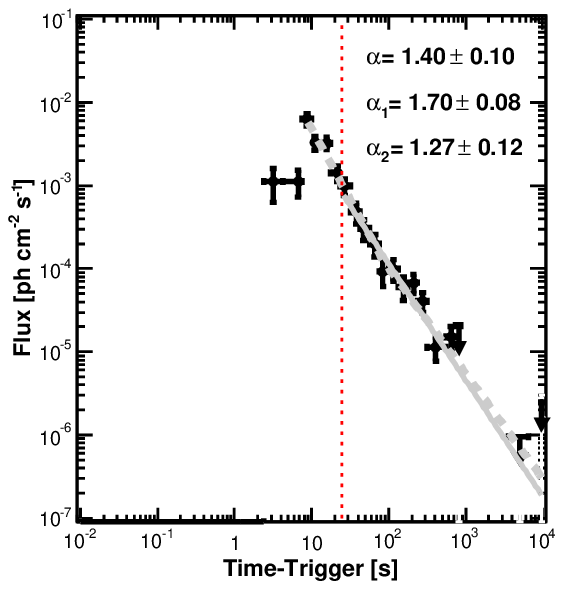}
\includegraphics[width=2.2in]{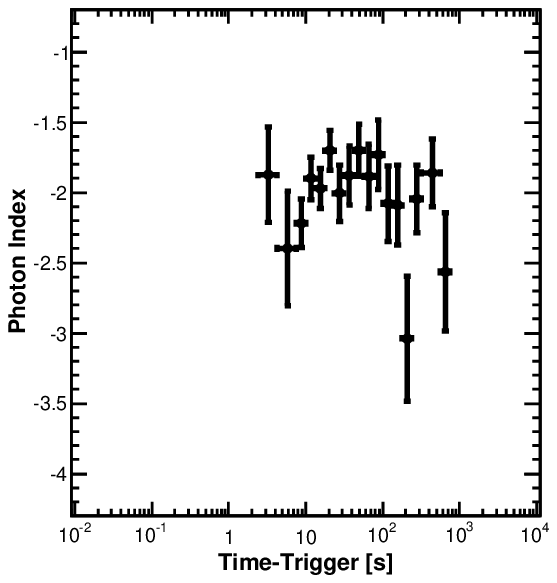}
\caption{Likelihood light curve for GRB\,090902B (flux above 100~MeV on the left, photon index on the right). See $\S$~\ref{sec_fermi_lat_grb_intro} for more information on lines and symbols.}
\label{like_090902B}
\end{center}
\end{figure}

The LAT emission contains many GeV events during and well after the GBM emission, similarly to the 33.39~GeV event detected at T$_{0}$+81.75~s.
This photon candidate has the highest energy ever observed from a burst and it has been used by the \Fermi LAT collaboration to probe the Extragalactic
Background Light as a function of redshift in the optical-UV range \citep{2010ApJ...723.1082A}.
The temporally extended high-energy emission reaches at least the end of the first GTI (LAT T$_{95}$$>$825~s) and $\sim$300 Transient-class events
are recorded above 100~MeV until this time (see Table~\ref{tab_likelihoods}).
The LAT time-resolved likelihood analysis resulted in a well sampled light curve of the high-energy flux up to T$_{0}$+750~s (Fig.~\ref{like_090902B}).
No significant spectral evolution was detected.
The decay of the flux as a function of time can be fitted with a simple power law starting from the GBM T$_{95}$, with a decay index
$\alpha$=1.40$\pm$0.10, in agreement with the result reported by \citet{GRB090902B:Fermi}.
Similarly to GRB\,090510, however, the fit of the flux light curve with a broken power law from the peak flux time $t_p$=T$_{0}$+8.7~s up to
T$_{0}$+$\sim$8~ks (including flux upper limits after T$_{0}$+750~s) returned a significant break at $t_b$=130$\pm$50~s, along with a steeper initial
decay ($\alpha_1$=1.70$\pm$0.19) and a smoother decay ($\alpha_2$=1.27$\pm$0.12) at later times.

\FloatBarrier \subsection{GRB\,090926A}
The long, bright GRB\,090926A triggered the GBM at T$_{0}$=04:20:26.99 UT on 26 September 2009 \citep[trigger 275631628,][]{GBM_090926}.
In spite of an initial off-axis angle of 48\ded1, GRB\,090926A caused exceptionally bright emission in the LAT and it triggered an ARR of the \Fermi
spacecraft.
The spacecraft initially remained in survey mode as long as the Earth avoidance angle condition was not satisfied, and GRB\,090926A became occulted by
the Earth $\sim$500~s after the trigger time.
At $\sim$T$_{0}$+3~ks, the LAT resumed observations and the spacecraft slewed to the burst position, keeping it in the LAT FoV
until 5~hours post trigger.
The LAT preliminary localization was delivered via GCN \citep{LAT_090926A}, with a statistical error of 0\ded04.

Swift TOO observations started $\sim$13~hours after the trigger time \citep{2009GCNR..254....1V,Swenson+10}. An X-ray counterpart was found by
\Swift-XRT 4~arcmin away from the LAT position \citep{2009GCN..9936....1G}, and further observations confirmed the existence of a fading source
with some flaring activity \citep{2009GCN..9961....1V}.
The optical afterglow of GRB\,090926A was discovered by the Skynet-PROMPT
telescopes \citep{2009GCN..9937....1H,2009GCN..9953....1H,2009GCN..9982....1H,2009GCN..9984....1H,2009GCN..10003...1H} and also detected by
\Swift-UVOT \citep{2009GCN..9938....1G,2009GCN..9948....1O}.
\citet{2009GCN..9942....1M} reported a spectroscopic redshift of z=2.1062 based on observations of the optical afterglow using the X-shooter
spectrograph mounted on the ESO-VLT UT2.
Combined with its brightness, this makes GRB\,090926A the fourth most energetic LAT-detected burst, with an isotropic
equivalent energy $\mathrm{E_{iso}}\simeq2.4\times 10^{54}$~erg (1~keV--10~GeV, within the GBM T$_{90}$).

\begin{figure}[ht!]
\begin{center}
\includegraphics[width=5.0in]{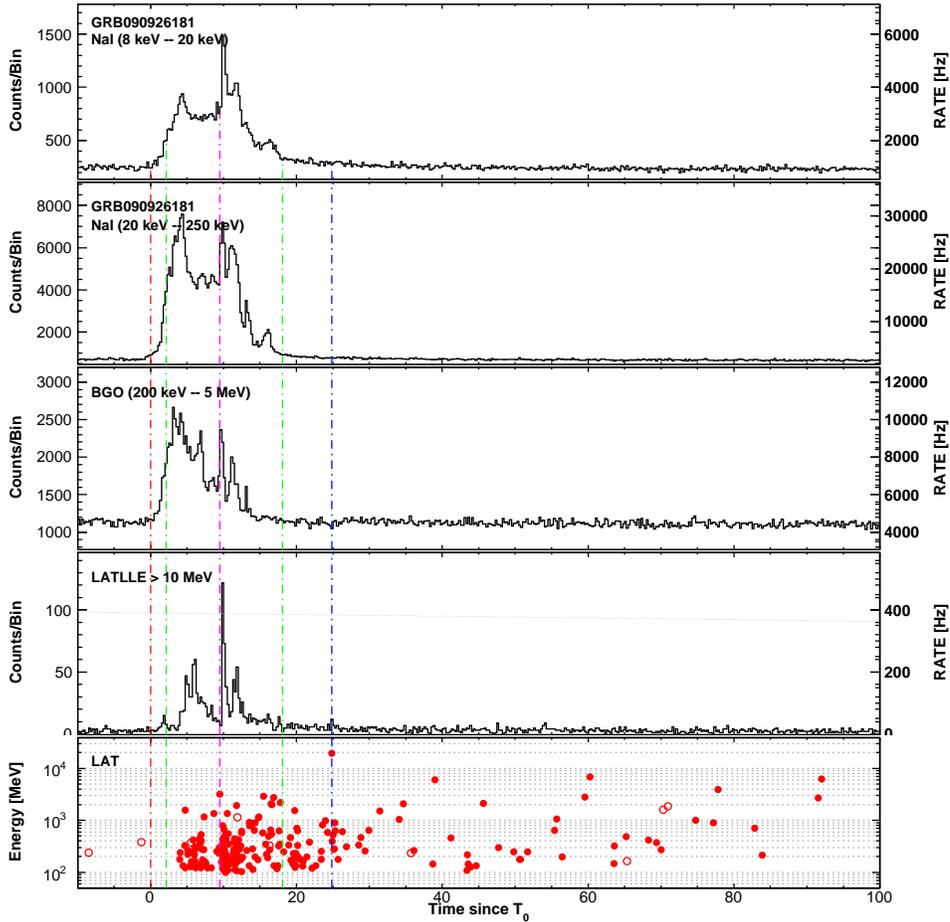}
\caption{Composite light curve for GRB\,090926A: summed GBM/NaI detectors (first two panels), GBM/BGO (third panel), LLE (fourth panel) and LAT Transient-class events above 100~MeV within a 12\de ROI  (last panel). See $\S$~\ref{sec_fermi_lat_grb_intro} for more information on lines and symbols in the LAT panels.}
\label{compo_090926A}
\end{center}
\end{figure}

As shown in Fig.~\ref{compo_090926A}, the light curve of GRB\,090926A exhibits a bright, short pulse at $\sim$T$_{0}$+10~s, in all energy bands covered
by the GBM and the LAT.
In the preliminary analysis of GBM and LAT data, \citet{2009GCN..9972....1B} fitted the emission spectrum of this pulse by the combination of the
empirical Band function with a high-energy power law.
In the time-resolved spectral analysis published by the \Fermi LAT collaboration \citep{2011ApJ...729..114A}, the high-energy power-law
component was found to start at the time of the bright pulse and to persist until $\sim$T$_{0}$+22~s. In this study, a spectral break was also found
at the highest energies, with a cutoff energy $\mathrm{E}_\mathrm{c}$$\sim$400~MeV during the bright pulse and $\mathrm{E}_\mathrm{c}$$\sim$1.4~GeV
for the time-integrated spectrum.
Our GBM-LAT joint spectral analysis in the GBM time window confirms these results, yielding $\mathrm{E}_\mathrm{c}$$\sim$1.5~GeV and a similar
spectral slope of 1.73$\pm$0.03 for the high-energy power-law component (Table~\ref{tab_JointFitGBMT90}).

\begin{figure}[ht!]
\begin{center}
\includegraphics[width=2.2in]{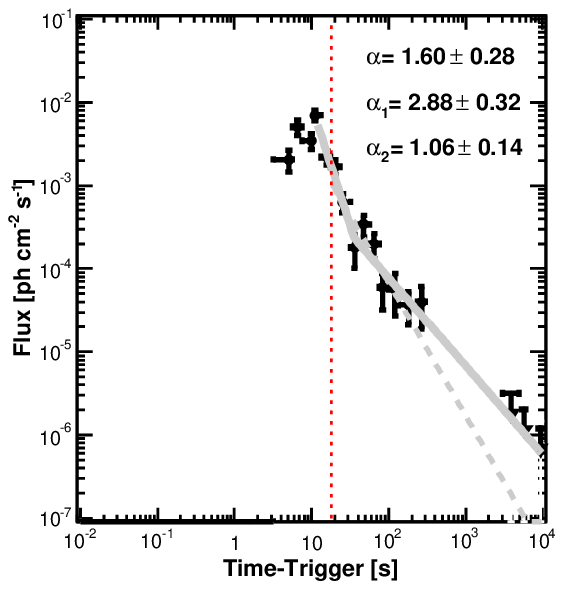}
\includegraphics[width=2.2in]{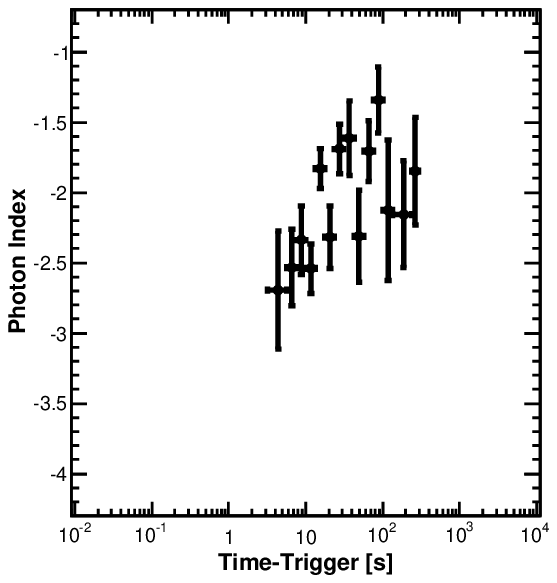}
\caption{Likelihood light curve for GRB\,090926A (flux above 100~MeV on the left, photon index on the right). See $\S$~\ref{sec_fermi_lat_grb_intro} for more information on lines and symbols.}
\label{like_090926A}
\end{center}
\end{figure}

The LAT emission contains many GeV events during and well after the GBM emission, similar to the 19.56~GeV event detected at T$_{0}$+24.83~s.
The temporally extended high-energy emission reaches at least T$_{0}$+225~s, and $\sim$230 Transient-class events
are recorded above 100~MeV until this time (see Table~\ref{tab_likelihoods}).
The LAT time-resolved likelihood analysis resulted in a well sampled light curve of the high-energy flux up to T$_{0}$+295~s (Fig.~\ref{like_090926A}).
The decay of the flux as a function of time can be fitted with a simple power law starting from the GBM T$_{95}$, with a decay index
$\alpha$=1.60$\pm$0.28, in agreement with the result reported by \citet{2011ApJ...729..114A}.
Similarly to GRB\,090510 and GRB\,090902B, however, the fit of the flux light curve with a broken power law from the peak flux time $t_p$=T$_{0}$+11.7~s up to
T$_{0}$+$\sim$8~ks (including flux upper limits after T$_{0}$+295~s) returned a significant break at $t_b$=40$\pm$5~s, along with a steeper initial
decay ($\alpha_1$=2.88$\pm$0.32) and a smoother decay ($\alpha_2$=1.06$\pm$0.14) at later times.
The right hand plot of Fig.~\ref{like_090926A} also suggests that the photon index in the first phase is steeper than the one in the final decay part.

\FloatBarrier \subsection{GRB\,091003}
The long GRB\,091003 triggered the GBM at T$_{0}$=04:35:45.5 UT on 3 October 2009 \citep[trigger 276237347,][]{2009GCN..9983....1R}.
The LAT preliminary localization was delivered via GCN \citep{2009GCN..9985....1M}, with a statistical error of 0\ded15.
\Swift TOO observations started $\sim$15.5~hours after the trigger time \citep{2009GCNR..253....1S}.
A fading source was detected in X-rays by \Swift-XRT \citep{2009GCN..9986....1S,2009GCN..9991....1P} and an UV/optical afterglow
candidate was found by \Swift-UVOT \citep{2009GCN..9987....1G,2009GCN..9990....1P}.
The optical afterglow was also detected by the William Herschel Telescope \citep{2009GCN..9995....1W} and a possible low redshift host galaxy was
found by the Lick Observatory \citep{2009GCN..9997....1P}. A spectroscopic redshift measurement of z=0.8969 was obtained using the GMOS spectrograph
mounted on the Gemini North Telescope \citep{2009GCN..10031...1C}.

\begin{figure}[ht!]
\begin{center}
\includegraphics[width=5.0in]{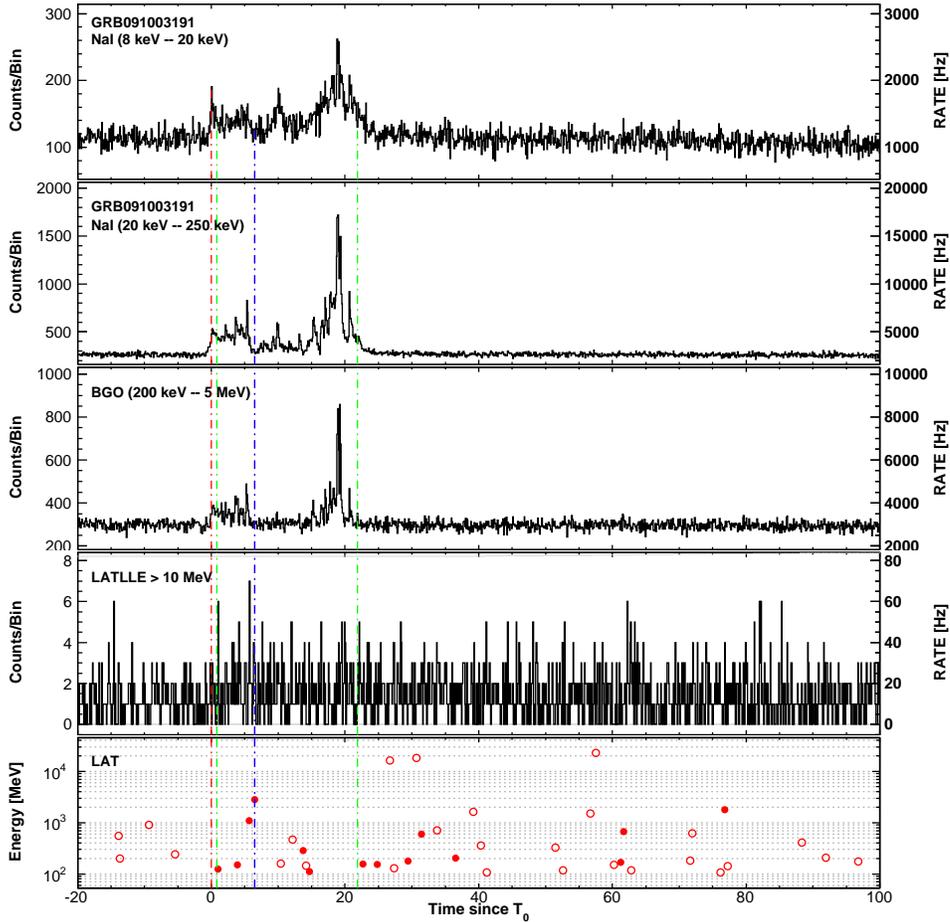}
\caption{Composite light curve for GRB\,091003: summed GBM/NaI detectors (first two panels), GBM/BGO (third panel), LLE (fourth panel) and LAT Transient-class events above 100~MeV within a 12\de ROI  (last panel). See $\S$~\ref{sec_fermi_lat_grb_intro} for more information on lines and symbols in the LAT panels.}
\label{compo_091003}
\end{center}
\end{figure}

\begin{figure}[ht!]
\begin{center}
\includegraphics[width=2.2in]{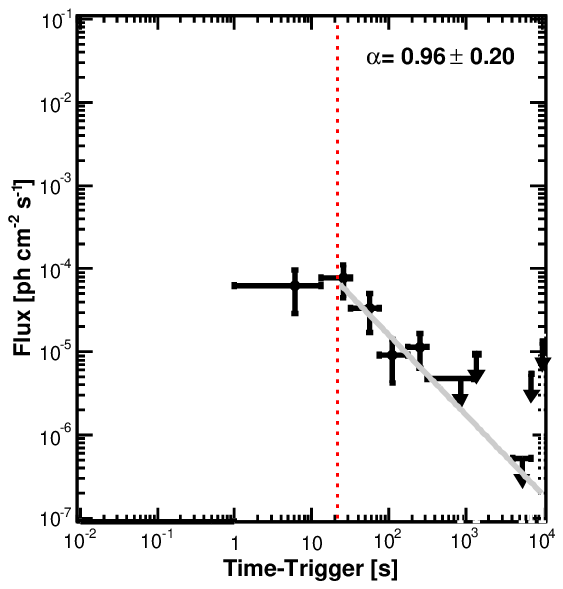}
\includegraphics[width=2.2in]{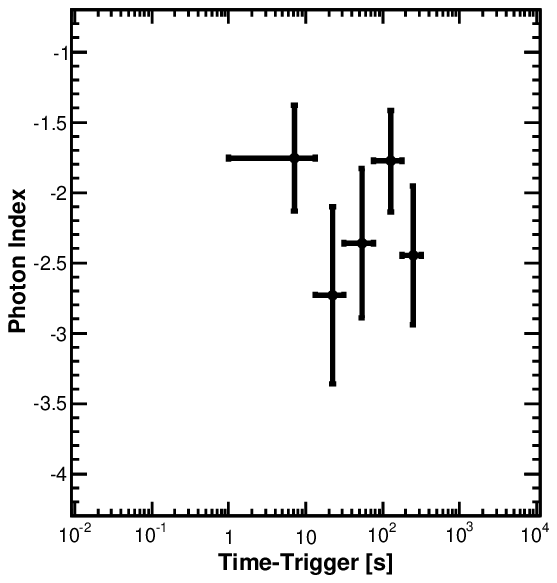}
\caption{Likelihood light curve for GRB\,091003 (flux above 100~MeV on the left, photon index on the right). See $\S$~\ref{sec_fermi_lat_grb_intro} for more information on lines and symbols.}
\label{like_091003}
\end{center}
\end{figure}

No significant emission was detected in the LLE light curve (Fig.~\ref{compo_091003}). The highest-energy event (2.8~GeV) is detected at
  T$_{0}$+6.47~s and does not coincide with any noticeable feature in the GBM light curve.
Although the LAT T$_{95}$=453$^{+86}_{-376}$~s suffers from a large uncertainty due to the relatively small statistics ($\sim$30 events),
the burst was detected up to this time with high significance by the LAT likelihood analysis of the Transient-class data above 100~MeV.
The LAT time-resolved likelihood analysis returned a significant flux up to T$_{0}$+316~s, with a temporal decay index $\alpha$=0.96$\pm$0.20
(Fig.~\ref{like_091003}).

\FloatBarrier \subsection{GRB\,091031}
\begin{figure}[ht!]
\begin{center}
\includegraphics[width=5.0in]{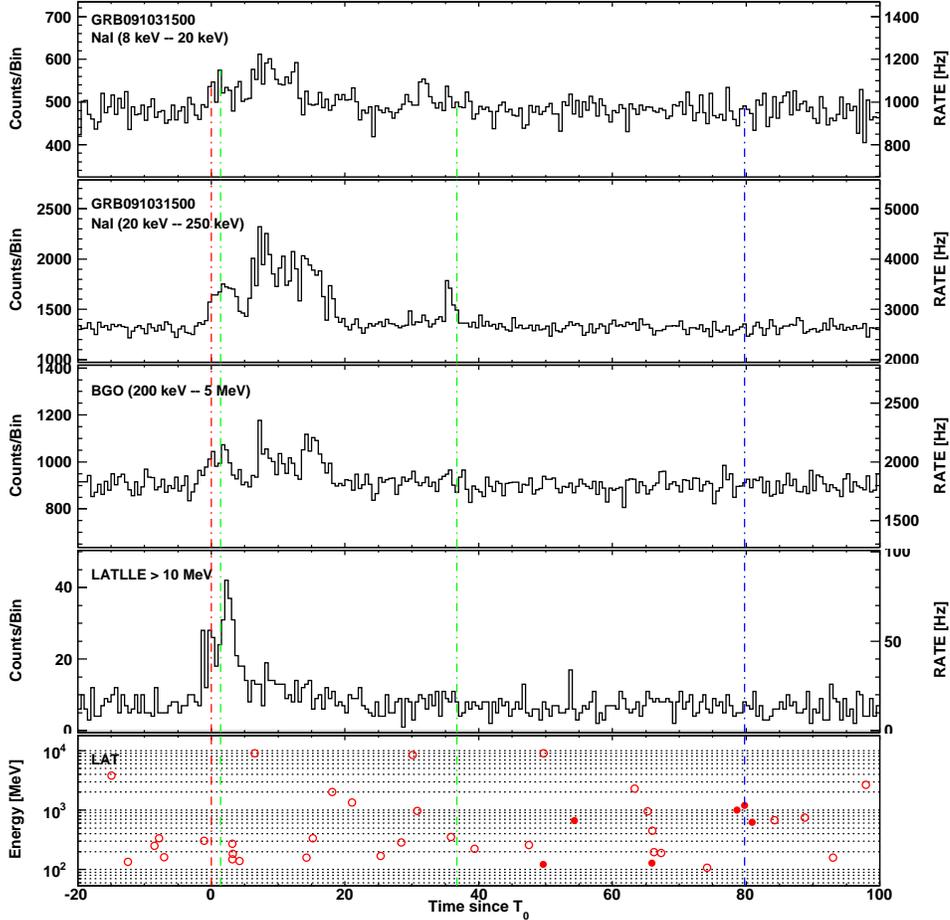}
\caption{Composite light curve for GRB\,091031: summed GBM/NaI detectors (first two panels), GBM/BGO (third panel), LLE (fourth panel) and LAT Transient-class events above 100~MeV within a 12\de ROI  (last panel). See $\S$~\ref{sec_fermi_lat_grb_intro} for more information on lines and symbols in the LAT panels.}
\label{compo_091031}
\end{center}
\end{figure}

\begin{figure}[ht!]
\begin{center}
\includegraphics[width=2.2in]{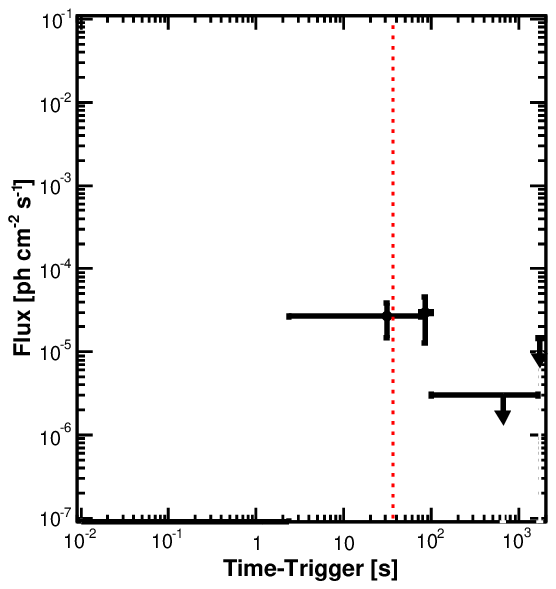}
\includegraphics[width=2.2in]{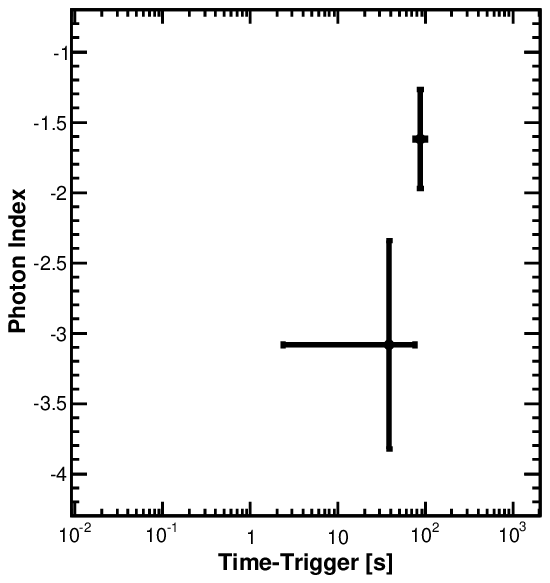}
\caption{Likelihood light curve for GRB\,091031 (flux above 100~MeV on the left, photon index on the right). See $\S$~\ref{sec_fermi_lat_grb_intro} for more information on lines and symbols.}
\label{like_091031}
\end{center}
\end{figure}

The long GRB\,091031 triggered the GBM at T$_{0}$=12:00:28.85 UT on 31 October 2009 \citep[trigger 278683230,][]{2009GCN..10115...1M}.
The LAT preliminary localization was delivered via GCN \citep{2009GCN..10163...1D}, with a statistical error of 0\ded2.
This burst was significantly detected in the LLE light curve (Fig.~\ref{compo_091031}) and above 100~MeV by the likelihood analysis up to the LAT
T$_{95}$=206$^{+12}_{-43}$~s.
The LAT time-resolved likelihood analysis returned a significant flux up to T$_{0}$+100~s (Fig.~\ref{like_091031}).
The highest-energy event (1.19~GeV) is detected with two other high-energy events at T$_{0}$+79.75~s, well after the end of the GBM emission.

\FloatBarrier \subsection{GRB\,091208B}
\begin{figure}[ht!]
\begin{center}
\includegraphics[width=5.0in]{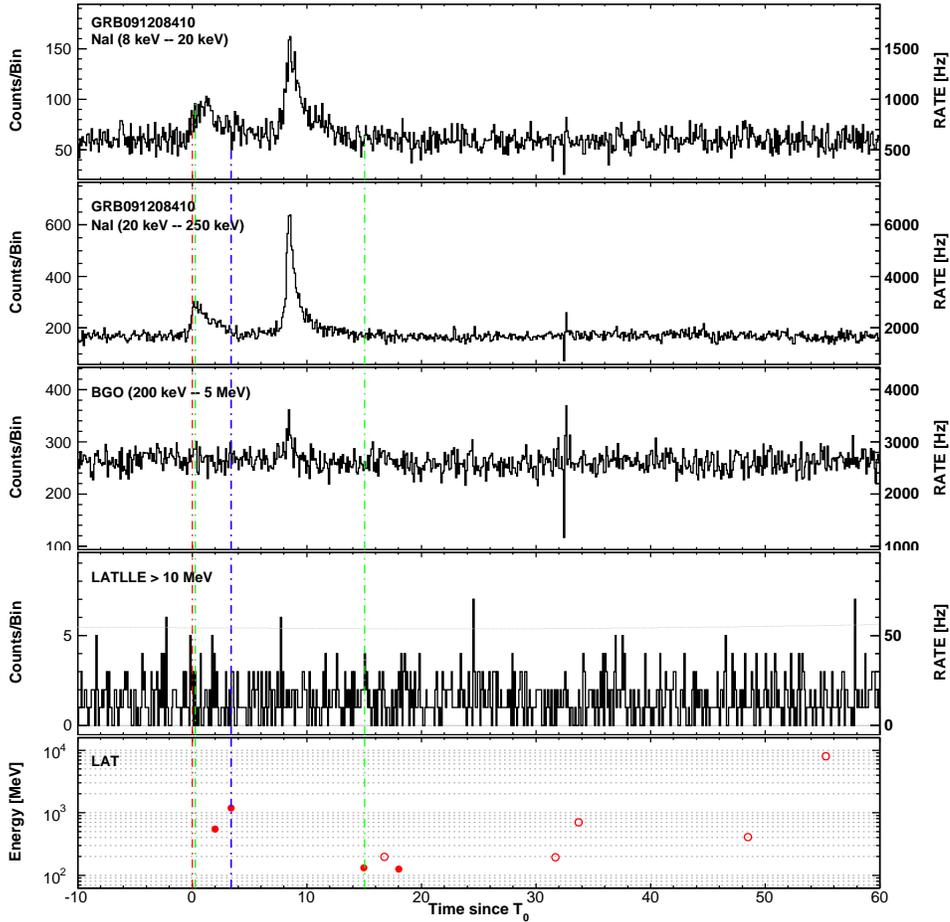}
\caption{Composite light curve for GRB\,091208B: summed GBM/NaI detectors (first two panels), GBM/BGO (third panel), LLE (fourth panel) and LAT Transient-class events above 100~MeV within a 12\de ROI  (last panel). See $\S$~\ref{sec_fermi_lat_grb_intro} for more information on lines and symbols in the LAT panels.}
\label{compo_091208410}
\end{center}
\end{figure}

\begin{figure}[ht!]
\begin{center}
\includegraphics[width=2.2in]{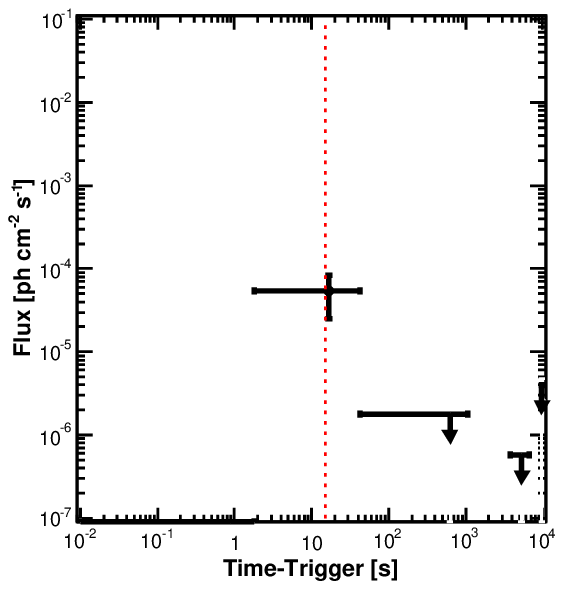}
\includegraphics[width=2.2in]{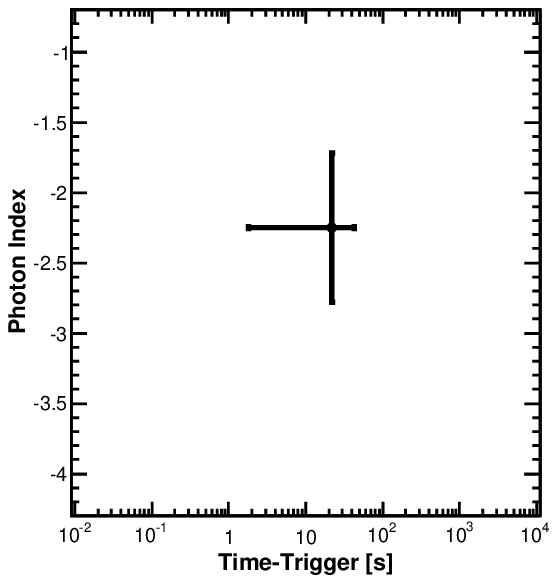}
\caption{Likelihood light curve for GRB\,091208B (flux above 100~MeV on the left, photon index on the right). See $\S$~\ref{sec_fermi_lat_grb_intro} for more information on lines and symbols.}
\label{like_091208410}
\end{center}
\end{figure}

The long GRB\,091208B triggered the GBM at T$_{0}$=09:49:57.96 UT on 8 December 2009 \citep[trigger 281958599,][]{2009GCN..10266...1M} and the
  \Swift-BAT at 09:49:57 UT \citep{2009GCN..10256...1P}.
\Swift-XRT observations started 115.2~s after the BAT trigger \citep{2010GCNR..266....1P}. A fading and uncataloged X-ray source was found and
\Swift-UVOT detected a bright afterglow candidate consistent with the XRT localization \citep{2009GCN..10267...1D,2009GCN..10256...1P}.
Several telescopes detected the bright optical transient
\citep{2009GCN..10279...1X,2009GCN..10275...1K,2009GCN..10273...1A,2009GCN..10271...1U,2009GCN..10269...1X,2009GCN..10262...1C,2009GCN..10260...1N,2009GCN..10258...1Y,2009GCN..10255...1D}.
A spectroscopic redshift measurement of z=1.063 was obtained using the GMOS spectrograph mounted on the Gemini North
Telescope \citep{2009GCN..10263...1W}, later confirmed by the HIRES-r spectrometer mounted on the 10~m Keck I telescope \citep{2009GCN..10272...1P}.
Using the XRT refined position \citep{2009GCN..10259...1O}, the LAT likelihood analysis found a marginal detection (TS=20) during the GBM T$_{90}$,
based on three Transient-class events associated to the burst.
The highest-energy event (1.18~GeV) is detected at T$_{0}$+3.41~s.
Due to the paucity of events (Fig.~\ref{compo_091208410}), no LAT T$_{90}$ could be derived and the LAT time-resolved likelihood analysis returned a
significant flux in one time bin only, ending at T$_{0}$+42~s (Fig.~\ref{like_091208410}).

\FloatBarrier \subsection{GRB\,100116A}
\begin{figure}[ht!]
\begin{center}
\includegraphics[width=5.0in]{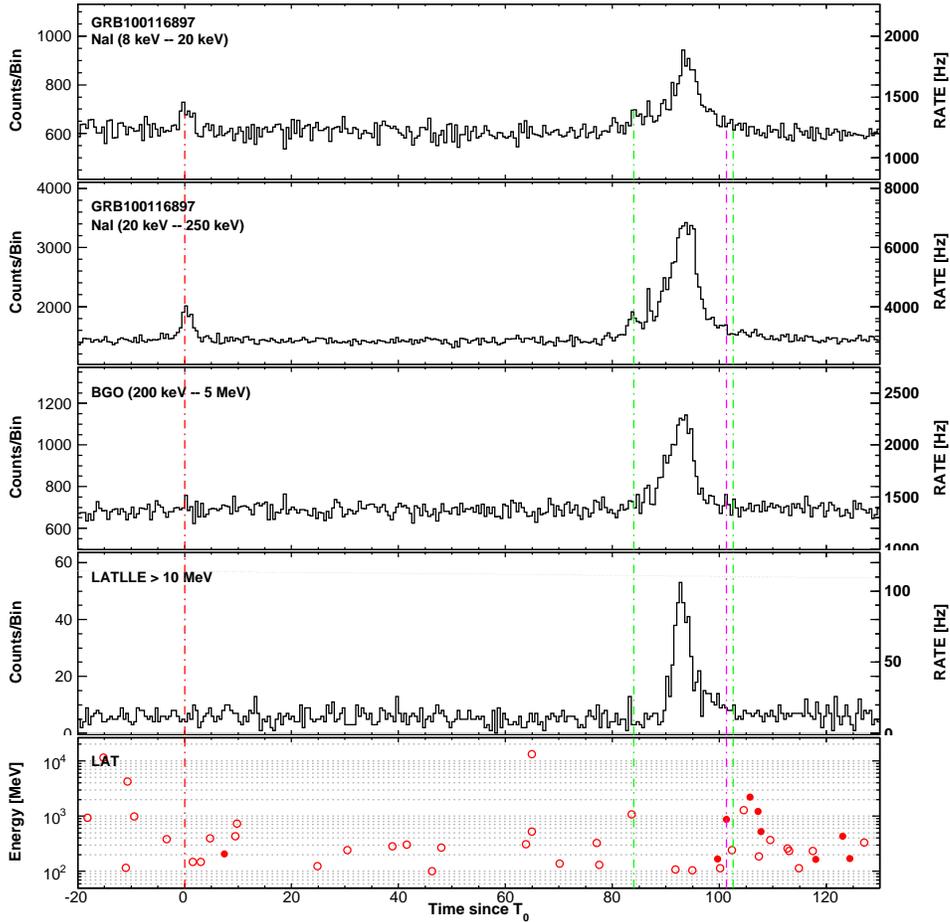}
\caption{Composite light curve for GRB\,100116A: summed GBM/NaI detectors (first two panels), GBM/BGO (third panel), LLE (fourth panel) and LAT Transient-class events above 100~MeV within a 12\de ROI  (last panel). See $\S$~\ref{sec_fermi_lat_grb_intro} for more information on lines and symbols in the LAT panels.}
\label{compo_100116A}
\end{center}
\end{figure}

\begin{figure}[ht!]
\begin{center}
\includegraphics[width=2.2in]{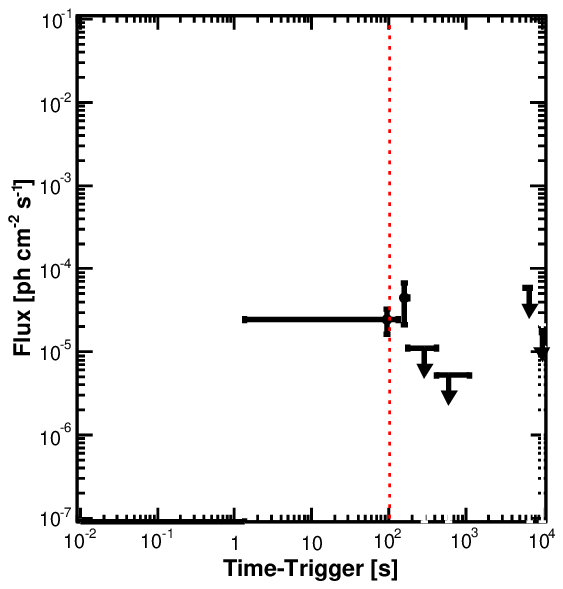}
\includegraphics[width=2.2in]{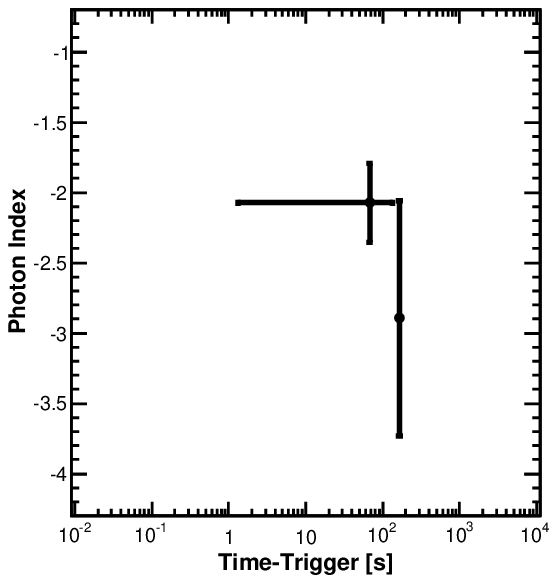}
\caption{Likelihood light curve for GRB\,100116A (flux above 100~MeV on the left, photon index on the right). See $\S$~\ref{sec_fermi_lat_grb_intro} for more information on lines and symbols.}
\label{like_100116A}
\end{center}
\end{figure}

The long GRB\,100116A triggered the GBM at T$_{0}$=21:31:00.24 UT on 16 January 2010 \citep[trigger 285370262,][]{2010GCN..10330...1B}.
The LAT preliminary localization was delivered via GCN \citep{2010GCN..10333...1M}, with a statistical error of 0\ded17.
As shown in Fig.~\ref{compo_100116A}, the GBM triggered on a precursor in GRB\,100116A light curve.
A very intense pulse is observed at $\sim$T$_{0}$+90~s, with a slow rise and a fast decay, probably due to the overlap of many smaller pulses during
the rising part. LAT low-energy events are recorded in temporal coincidence with this bright GBM pulse.
More interestingly, the Transient-class events above 100~MeV which are compatible with the burst position appear to be slightly delayed ($\sim$20~s)
with respect to both the LLE and GBM emission, and the highest-energy event (2.2~GeV) is detected at T$_{0}$+105.71~s, right at the end of the GBM
emission.
This temporally extended high-energy emission reaches at least the end of the first GTI (LAT T$_{95}$$>$141~s).
The LAT time-resolved likelihood analysis returned a significant flux up to T$_{0}$+178~s (Fig.~\ref{like_100116A}). 
A 13.12~GeV event with a probability higher than 99\% to be associated with the burst is detected at $\sim$T$_{0}$+296~s (see the discussion in
$\S$~\ref{subsubsec_res_HEevents}).

\FloatBarrier \subsection{GRB\,100225A}
\begin{figure}[ht!]
\begin{center}
\includegraphics[width=5.0in]{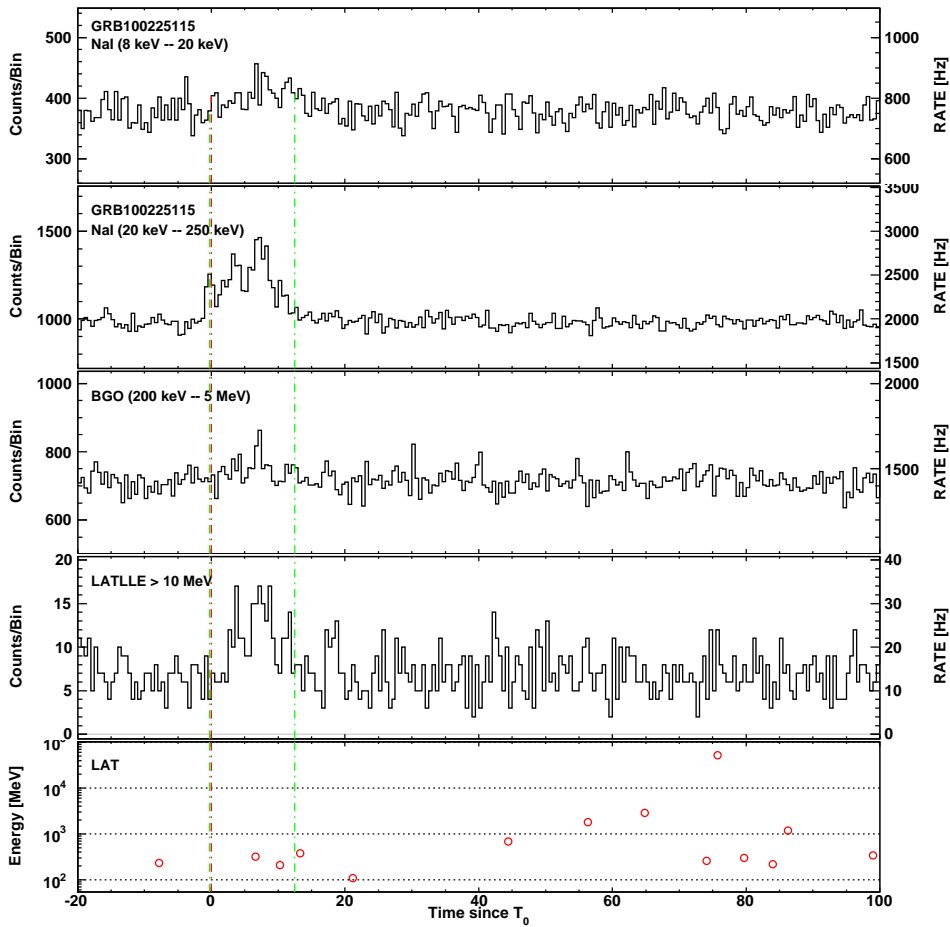}
\caption{Composite light curve for GRB\,100225A: summed GBM/NaI detectors (first two panels), GBM/BGO (third panel), LLE (fourth panel) and LAT Transient-class events above 100~MeV within a 12\de ROI  (last panel). See $\S$~\ref{sec_fermi_lat_grb_intro} for more information on lines and symbols in the LAT panels.}
\label{compo_100225A}
\end{center}
\end{figure}

The long GRB\,100225A triggered the GBM at T$_{0}$=02:45:31.15 UT on 25 February 2010 \citep[trigger 288758733,][]{2010GCN..10449...1F}.
This faint burst had an off-axis angle of 55\ded5 at the trigger time, where the LAT effective area is low.
A tentative localization with the LAT was delivered via GCN \citep{2010GCN..10450...1P}, with a statistical error of 0\ded9.
Only a few LAT Transient-class events above 100~MeV are actually compatible with the burst position, therefore no LAT T$_{90}$ could be derived and no
significant emission was found in the likelihood analysis. 
GRB\,100225A was detected in the LLE data only. The LLE light curve consists of a long duration pulse which mimics the light curve seen in the NaI
detectors (Fig.~\ref{compo_100225A}).

\FloatBarrier \subsection{GRB\,100325A}
\begin{figure}[ht!]
\begin{center}
\includegraphics[width=5.0in]{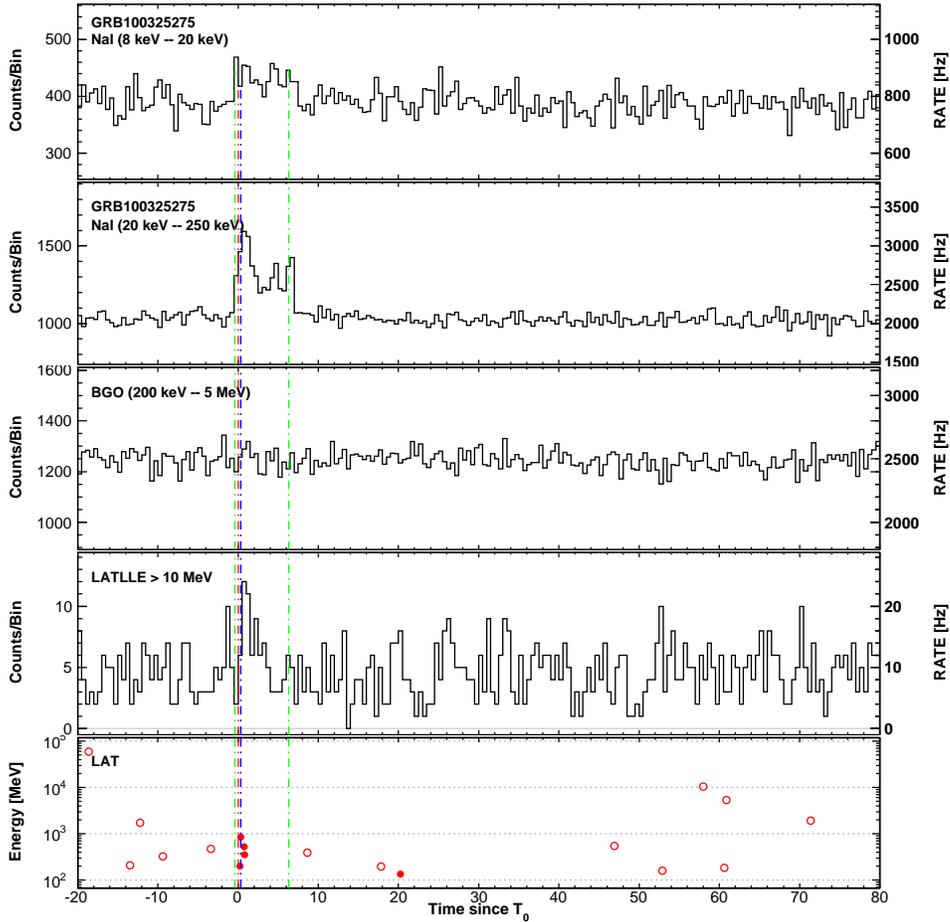}
\caption{Composite light curve for GRB\,100325A: summed GBM/NaI detectors (first two panels), GBM/BGO (third panel), LLE (fourth panel) and LAT Transient-class events above 100~MeV within a 12\de ROI  (last panel). See $\S$~\ref{sec_fermi_lat_grb_intro} for more information on lines and symbols in the LAT panels.}
\label{compo_100325A}
\end{center}
\end{figure}

\begin{figure}[ht!]
\begin{center}
\includegraphics[width=2.2in]{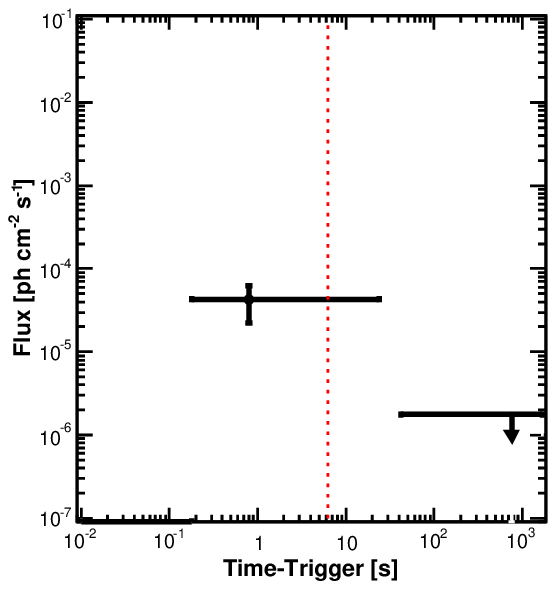}
\includegraphics[width=2.2in]{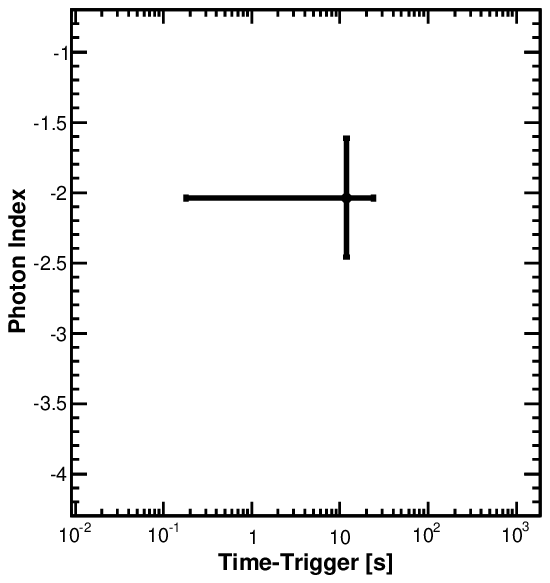}
\caption{Likelihood light curve for GRB\,100325A (flux above 100~MeV on the left, photon index on the right). See $\S$~\ref{sec_fermi_lat_grb_intro} for more information on lines and symbols.}
\label{like_100325A}
\end{center}
\end{figure}

The long GRB\,100325A triggered the GBM at T$_{0}$=06:36:08.02 UT on 25 March 2010 \citep[trigger 291191770,][]{2010GCN..10546...1V}.
The LAT preliminary localization was delivered via GCN \citep{2010GCN..10548...1D}, with a statistical error of 0\ded6.
The light curve in the NaI detectors consists of several overlapping pulses, whereas the burst is not visible in the BGO light curve and only marginally
detected in the LLE light curve (Fig.~\ref{compo_100325A}).
Due to the paucity of events, no LAT T$_{90}$ could be derived.
A cluster of four Transient-class events above 100~MeV are recorded within 0.57~s right after the trigger time, and the LAT time-resolved likelihood
analysis returned a significant flux up to T$_{0}$+23.7~s (Fig.~\ref{like_100325A}).
More interestingly, the time-integrated spectrum of GRB\,100325A during the GBM T$_{90}$ is best represented by a Band function, with a hard value for
the low-energy spectral slope $\alpha=-0.33\pm0.11$.

\FloatBarrier \subsection{GRB\,100414A}
The long GRB\,100414A triggered the GBM at T$_{0}$=02:20:21.99 UT on 14 April 2010 \citep[trigger 292904423,][]{2010GCN..10595...1F}.
It had an initial off-axis angle of 69\de in the LAT and the ARR triggered by the GBM brought it down to $\sim$10\de after $\sim$250~s.
The LAT preliminary localization was delivered via GCN \citep{2010GCN..10594...1T}, with a statistical error of 0\ded14.
\Swift TOO observations started $\sim$48~hours after the trigger time and a possible X-ray counterpart was found by \Swift-XRT
\citep{2010GCN..10601...1P,2010GCN..10605...1P}. Further observations confirmed the existence of a fading source \citep{2010GCN..10632...1P}.
Follow-up detections in the optical were reported by the \Swift-UVOT team \citep{2010GCN..10609...1L} and by other observers operating ground-based
telescopes \citep{2010GCN..10618...1M,2010GCN..10641...1U}. GRB\,100414A was also detected in the optical/NIR \citep{2010GCN..10607...1F} and in
the radio band \citep{2010GCN..10697...1K,2010GCN..10698...1F}.
\citet{2010GCN..10606...1C} reported a spectroscopic redshift of z=1.368 based on observations of the optical afterglow using the
GMOS spectrograph mounted on the Gemini North Telescope.

\begin{figure}[ht!]
\begin{center}
\includegraphics[width=5.0in]{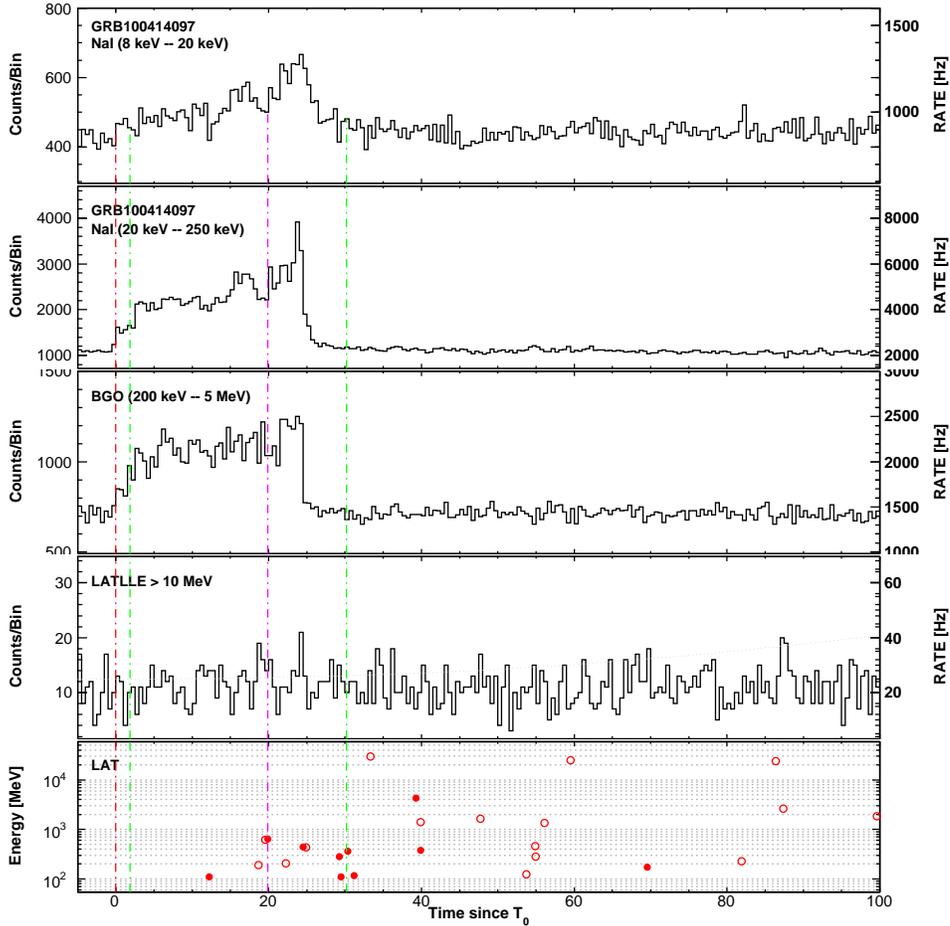}
\caption{Composite light curve for GRB\,100414A: summed GBM/NaI detectors (first two panels), GBM/BGO (third panel), LLE (fourth panel) and LAT Transient-class events above 100~MeV within a 12\de ROI  (last panel). See $\S$~\ref{sec_fermi_lat_grb_intro} for more information on lines and symbols in the LAT panels.}
\label{compo_100414A}
\end{center}
\end{figure}

\begin{figure}[ht!]
\begin{center}
\includegraphics[width=2.2in]{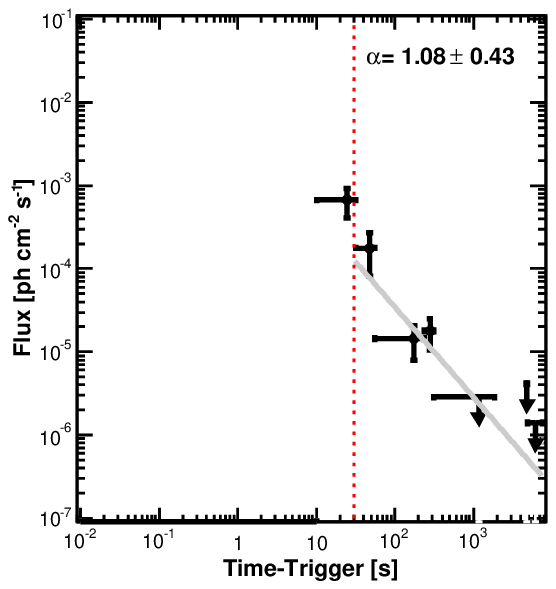}
\includegraphics[width=2.2in]{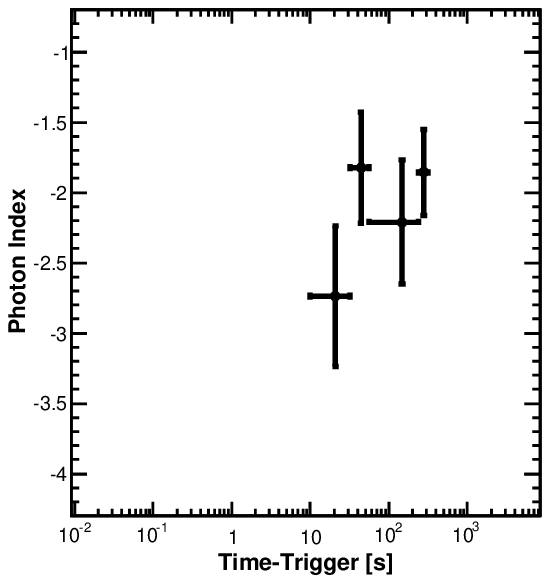}
\caption{Likelihood light curve for GRB\,100414A (flux above 100~MeV on the left, photon index on the right). See $\S$~\ref{sec_fermi_lat_grb_intro} for more information on lines and symbols.}
\label{like_100414A}
\end{center}
\end{figure}

The GBM light curve of GRB\,100414A consists of a single slowly rising pulse which ends abruptly after culminating at T$_{0}$+23~s
  (Fig.~\ref{compo_100414A}). No significant emission was detected in the LLE light curve.
Although the LAT T$_{95}$=289$^{+90}_{-111}$~s suffers from a large uncertainty due to the relatively small statistics ($\sim$20 events),
the burst was detected up to this time with high significance by the LAT likelihood analysis of the Transient-class data above 100~MeV.
The LAT time-resolved likelihood analysis returned a significant flux up to T$_{0}$+316~s, with a temporal decay index $\alpha$=1.08$\pm$0.43
(Fig.~\ref{like_100414A}).
More interestingly, the time-integrated spectrum of GRB\,100414A during the GBM T$_{90}$ is best represented by a Comptonized model with an
additional power-law component with a spectral slope of 1.75$\pm$0.07. However, as discussed in $\S$~\ref{subsub_extracomponents}, this additional
component is seen in the ``GBM'' time interval only (Tables~\ref{tab_JointFitGBMT90} and~\ref{tab_JointFitLATPROMPT}) and its existence is uncertain due to
possible systematic effects in the GBM-LAT joint spectral analysis during the ARR maneuver.

\FloatBarrier \subsection{GRB\,100620A}
\begin{figure}[ht!]
\begin{center}
\includegraphics[width=5.0in]{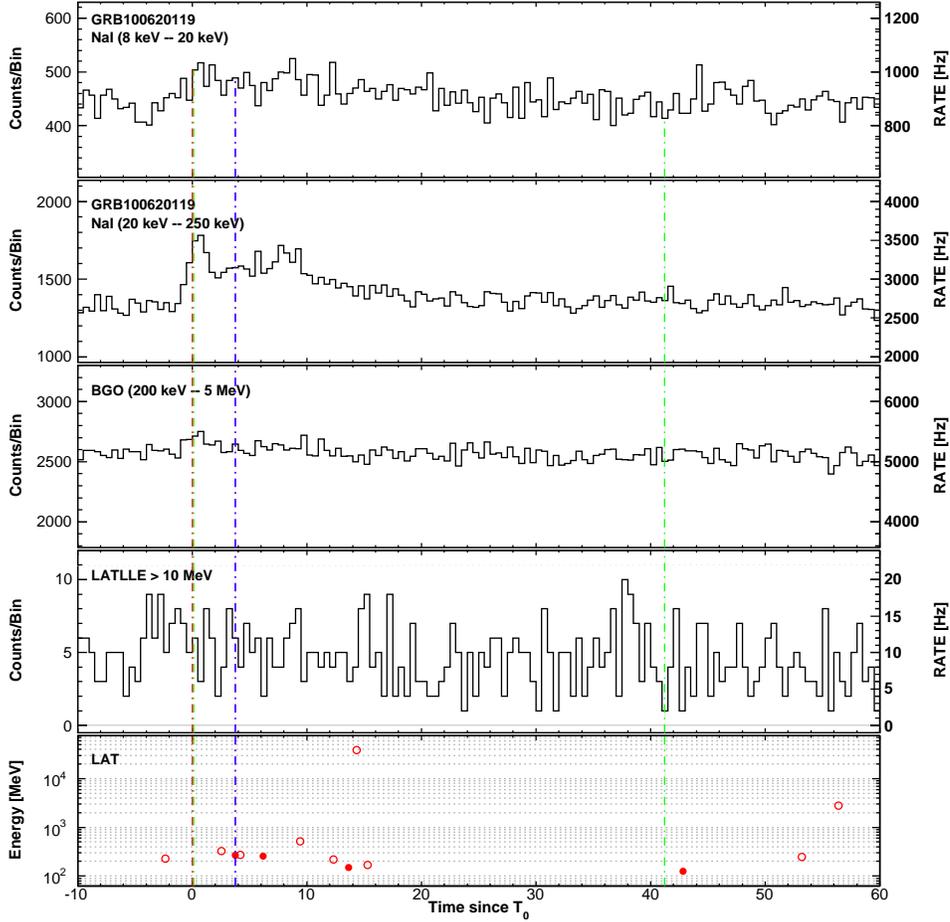}
\caption{Composite light curve for GRB\,100620A: summed GBM/NaI detectors (first two panels), GBM/BGO (third panel), LLE (fourth panel) and LAT Transient-class events above 100~MeV within a 12\de ROI  (last panel). See $\S$~\ref{sec_fermi_lat_grb_intro} for more information on lines and symbols in the LAT panels.}
\label{compo_100620A}
\end{center}
\end{figure}

\begin{figure}[ht!]
\begin{center}
\includegraphics[width=2.2in]{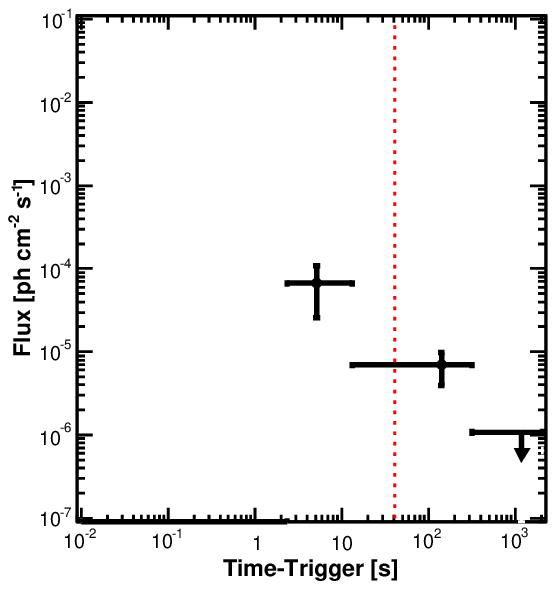}
\includegraphics[width=2.2in]{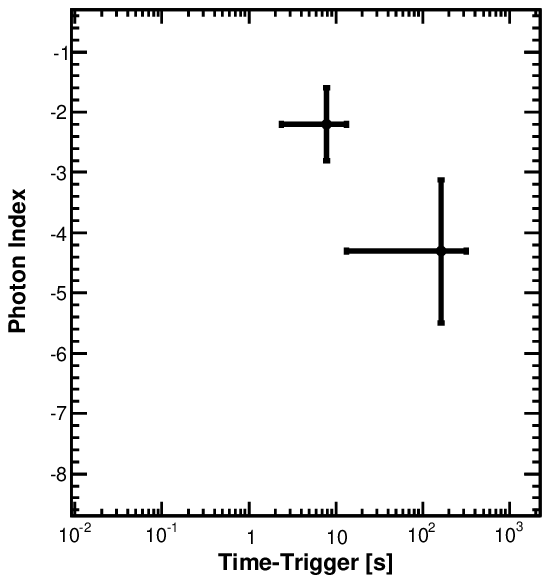}
\caption{Likelihood light curve for GRB\,100620A (flux above 100~MeV on the left, photon index on the right). See $\S$~\ref{sec_fermi_lat_grb_intro} for more information on lines and symbols.}
\label{like_100620A}
\end{center}
\end{figure}

The long GRB\,100620A triggered the GBM at T$_{0}$=02:51:29.1134 UT on 20 June 2010 (trigger 298695091). The best localization reported in the GBM
  catalog \citep{Paciesas+12} was used as an initial seed for our analysis.
Using LAT Transient-class events above 100~MeV, we could improve upon the GBM localization.  The LAT localization, which has a statistical error
of 0\ded71 (Table~\ref{tab_localizations}), is the final best position.
GRB\,100620A is a faint burst in the GBM, with no emission in the BGO detector nor in LLE data (Fig.~\ref{compo_100620A}).
No LAT T$_{90}$ could be derived due to the paucity of events, but accumulating signal in the LAT time-resolved likelihood analysis allowed us to
detect a significant flux up to T$_{0}$+316~s (Fig.~\ref{like_100620A}).

\FloatBarrier \subsection{GRB\,100724B}
The long GRB\,100724B triggered the GBM at T$_{0}$=00:42:05.98 UT on 24 July 2010 \citep[trigger 301624927,][]{2010GCN..10977...1B}.
Its off-axis angle in the LAT was 48\ded9 at the trigger time, and remained greater than 40\de for 2700~s despite the ARR triggered by the GBM, as
the \Fermi spacecraft remained in survey mode as long as the Earth avoidance angle condition was not satisfied.
The LAT preliminary localization was delivered via GCN \citep{2010GCN..10978...1T}, with a statistical error of 0\ded6.
The burst was also significantly detected by both the AGILE-GRID and the AGILE-MCAL~\citep{2010GCN..10994...1M,2010GCN..10996...1G}.

\begin{figure}[ht!]
\begin{center}
\includegraphics[width=5.0in]{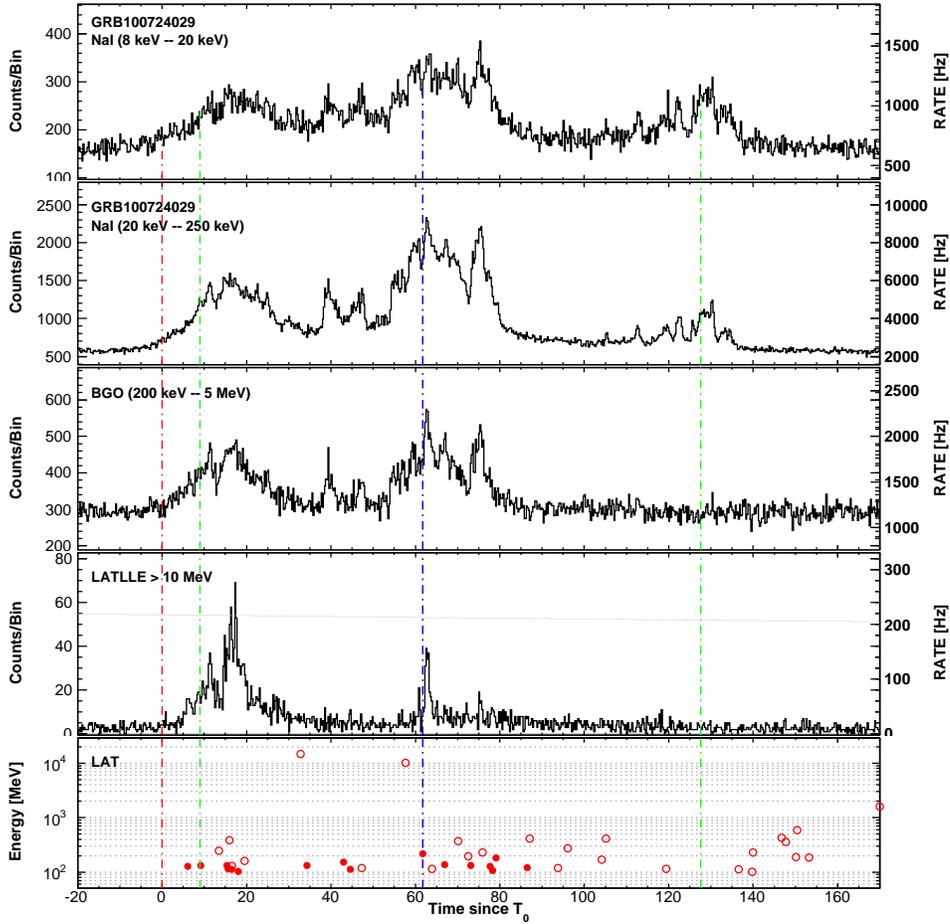}
\caption{Composite light curve for GRB\,100724B: summed GBM/NaI detectors (first two panels), GBM/BGO (third panel), LLE (fourth panel) and LAT Transient-class events above 100~MeV within a 12\de ROI  (last panel). See $\S$~\ref{sec_fermi_lat_grb_intro} for more information on lines and symbols in the LAT panels.}
\label{compo_100724B}
\end{center}
\end{figure}

\begin{figure}[ht!]
\begin{center}
\includegraphics[width=2.2in]{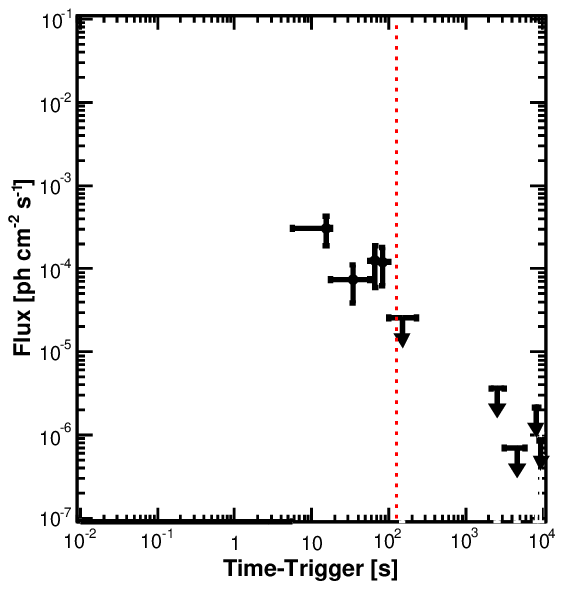}
\includegraphics[width=2.2in]{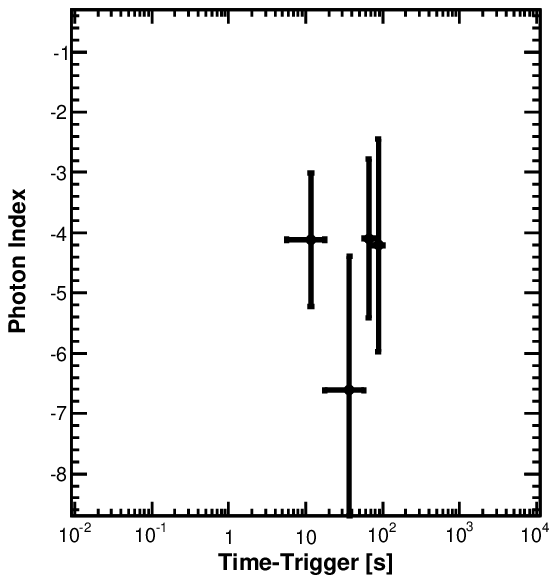}
\caption{Likelihood light curve for GRB\,100724B (flux above 100~MeV on the left, photon index on the right). See $\S$~\ref{sec_fermi_lat_grb_intro} for more information on lines and symbols.}
\label{like_100724B}
\end{center}
\end{figure}

GRB\,100724B is very bright in the GBM and in LLE data, with two main emission episodes (Fig.~\ref{compo_100724B}).
Surprisingly, it is not exceptionally bright in LAT Transient-class data above 100~MeV, and the highest-energy event (0.22~GeV) is detected at
T$_{0}$+61.75~s, during the second episode.
No LAT T$_{90}$ could be derived due to the large Zenith angle of the burst, but the burst was detected up to T$_{0}$+125~s with high significance by
the LAT likelihood analysis above 100~MeV.
This analysis actually revealed a fairly steep high-energy spectrum, with a photon index of $-4.96\pm0.94$ during the GBM T$_{90}$ and
$-4.85\pm0.92$ in the ``LATTE'' time interval. Similar indices were measured in the LAT time-resolved likelihood analysis (Fig.~\ref{like_100724B}).
More interestingly, the time-integrated spectrum of GRB\,100724B during the GBM T$_{90}$ is best
represented by a Band function with a hard value for the high-energy spectral slope $\beta=-2.00\pm0.01$ and with an exponential cutoff at
$\mathrm{E}_\mathrm{c}=40\pm3$~MeV (Table~\ref{tab_JointFitGBMT90} and Table~\ref{tab_JointFitLATPROMPT}).
The spectral analysis performed by \citet{Guiriec2011} was based on GBM data only and yielded similar results except for the spectral break whose
detection requires the addition of LAT data.
Conversely, our analysis is not in agreement with the results reported by \citet{2011A&A...535A.120D}, who found a single power-law spectral shape
extending up to 100~MeV with a photon index $\beta=-2.13^{+0.05}_{-0.04}$.
This difference could be explained by the larger effective area and the deeper calorimeter of the \Fermi-LAT \citep{Atwood:09}, which provides more sensitive
spectral measurements than the AGILE instruments.
Owing to its long duration ($\sim$120~s in the GBM) and despite the relatively low peak energy $E_{p}=263\pm4$~keV and the spectral break at
MeV energies, GRB\,100724B is the most fluent burst in the catalog, with a fluence of ($4.66\pm0.08)\times10^{-4}$~erg/cm$^2$ (10~keV--10~GeV, within the
GBM T$_{90}$).

\FloatBarrier \subsection{GRB\,100728A}
The long GRB\,100728A triggered the GBM at T$_{0}$=02:17:30.61 UT on 28 July 2010 \citep[trigger 301976252,][]{2010GCN..11006...1V} and the
  \Swift-BAT at 02:18:24 UT \citep{2010GCN..11004...1C}.
\Swift-XRT observations started 76.7~s after the BAT trigger and a bright, uncataloged X-ray source was immediately located \citep{2010GCNR..294....1C}.
The enhanced \Swift-XRT position \citep{2010GCN..11005...1B} enabled the detection of the optical/NIR afterglow
\citep{2010GCN..11007...1P,2010GCN..11008...1I,2010GCN..11020...1O}, but no redshift could be measured.

\begin{figure}[ht!]
\begin{center}
\includegraphics[width=5.0in]{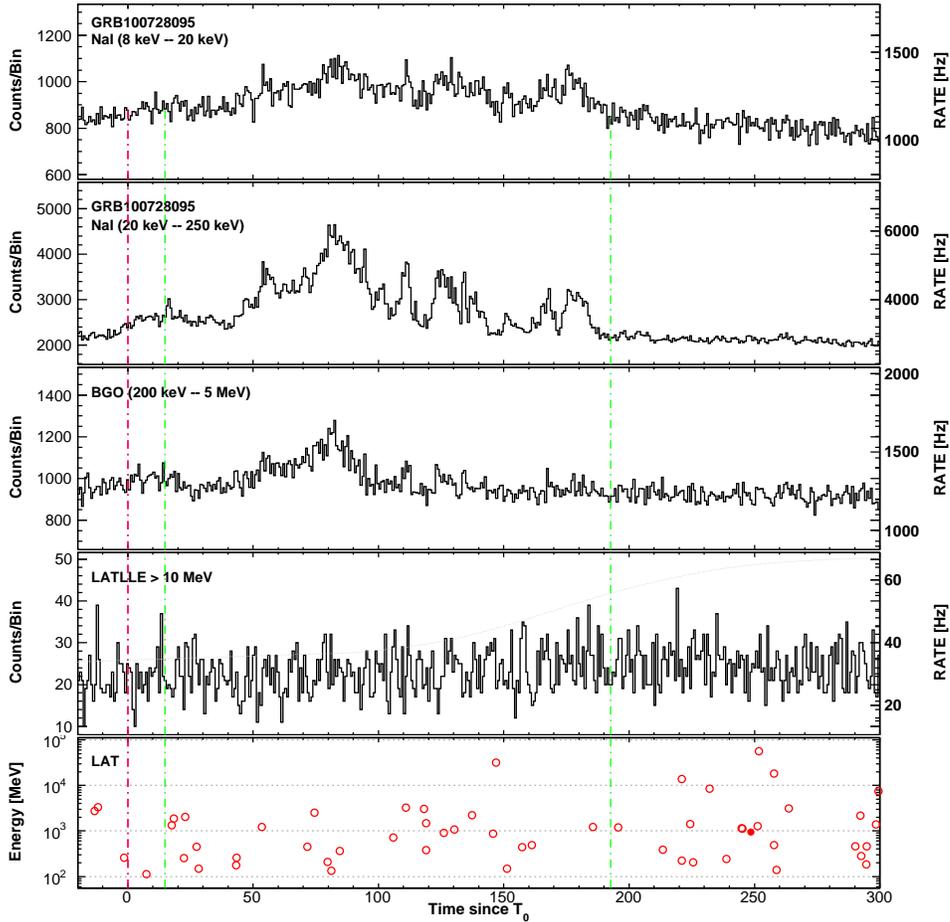}
\caption{Composite light curve for GRB\,100728A: summed GBM/NaI detectors (first two panels), GBM/BGO (third panel), LLE (fourth panel) and LAT Transient-class events above 100~MeV within a 12\de ROI  (last panel). See $\S$~\ref{sec_fermi_lat_grb_intro} for more information on lines and symbols in the LAT panels.}
\label{compo_100728A}
\end{center}
\end{figure}

\begin{figure}[ht!]
\begin{center}
\includegraphics[width=2.2in]{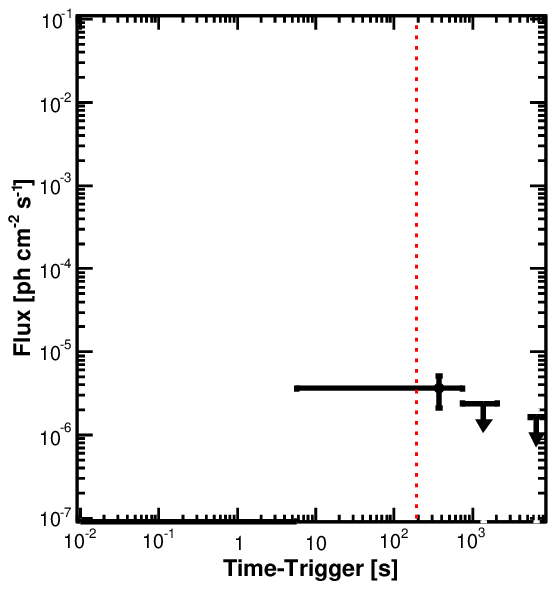}
\includegraphics[width=2.2in]{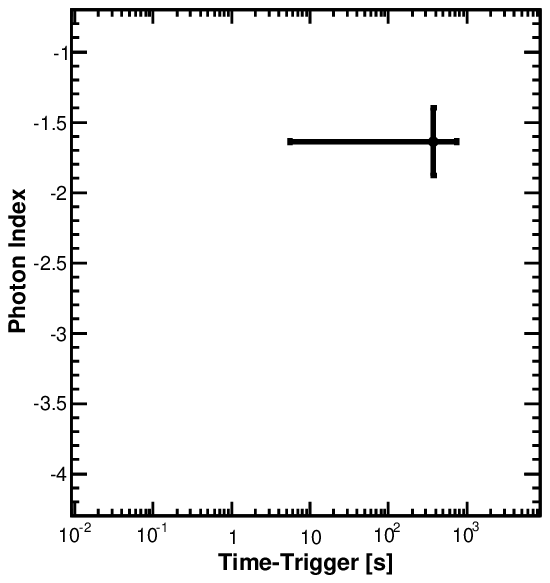}
\caption{Likelihood light curve for GRB\,100728A (flux above 100~MeV on the left, photon index on the right). See $\S$~\ref{sec_fermi_lat_grb_intro} for more information on lines and symbols.}
\label{like_100728A}
\end{center}
\end{figure}

The GBM light curve of GRB\,100728A shows a multi-peaked structure lasting approximately $\sim$190~s (Fig.~\ref{compo_100728A}). Most of the
  emission is detected at low energy, and the time-integrated spectrum of the burst during the GBM T$_{90}$ is best represented by a Comptonized model. 
GRB\,100728A had an initial off-axis angle of 59\ded9 in the LAT and the ARR triggered by the GBM brought it down to $\sim$10\de after $\sim$300~s.
Only a few LAT Transient-class events above 100~MeV are compatible with the burst position, therefore no LAT T$_{90}$ could be derived
and no significant emission was found in the likelihood analysis.
Accumulating signal in the LAT time-resolved likelihood analysis allowed us to detect a significant flux in one time bin, ending at T$_{0}$+750~s
(Fig.~\ref{like_100728A}). This detection confirms the temporal coincidence of the high-energy emission of GRB\,100728A with its flaring
activity in X-rays, as published in \cite{GRB100728A}.
The implications of the \Fermi-LAT observation and the possible connection between the gamma-ray emission and the X-ray activity of GRB\,100728A have also
been discussed in \citet{He:11} and \citet{Mao:12}.
A 13.54~GeV event with a probability higher than 98\% to be associated with the burst is detected $\sim$90 minutes after the trigger time (see the
discussion in $\S$~\ref{subsubsec_res_HEevents}).
This represents the only evidence in our catalog that high-energy events ($>$10~GeV) can arrive very late in time, confirming the results from \citet{1994Natur.372..652H} and suggesting that such events are rare.

\FloatBarrier \subsection{GRB\,100826A}
\begin{figure}[ht!]
\begin{center}
\includegraphics[width=5.0in]{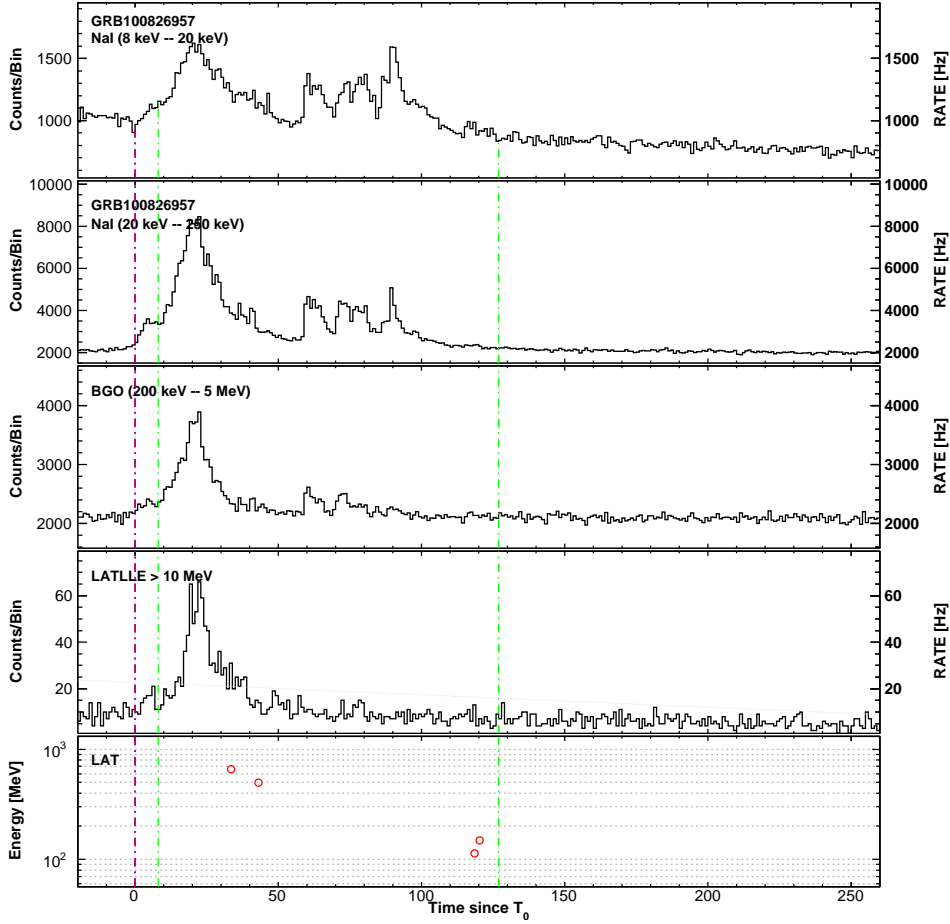}
\caption{Composite light curve for GRB\,100826A: summed GBM/NaI detectors (first two panels), GBM/BGO (third panel), LLE (fourth panel) and LAT Transient-class events above 100~MeV within a 12\de ROI  (last panel). See $\S$~\ref{sec_fermi_lat_grb_intro} for more information on lines and symbols in the LAT panels.}
\label{compo_100826A}
\end{center}
\end{figure}

The long GRB\,100826A triggered the GBM at T$_{0}$=22:58:22.89 UT on 26 August 2010 \citep[trigger 304556304,][]{2010GCN..11155...1M}.
The triangulation of the burst by the IPN provided a position with a 3$\sigma$ error box area of 1.5 square
degrees \citep{2010GCN..11156...1H} which we used in our analysis.
Only a few LAT Transient-class events above 100~MeV are compatible with the burst position, therefore no LAT T$_{90}$ could be derived and no
significant emission was found in the likelihood analysis.
GRB\,100826A was detected in the LLE data only \citep{2010GCN..11155...1M}. The LLE light curve has a very similar structure to
the GBM broad peak, with the maximum count rate occurring at $\sim$T$_{0}$+22~s (Fig.~\ref{compo_100826A}).
The burst is bright in the GBM and its time-integrated spectrum during the GBM T$_{90}$ is best represented by a Band function, with a hard value for
the high-energy spectral slope $\beta=-2.03\pm0.02$.

\FloatBarrier \subsection{GRB\,101014A}
\begin{figure}[ht!]
\begin{center}
\includegraphics[width=5.0in]{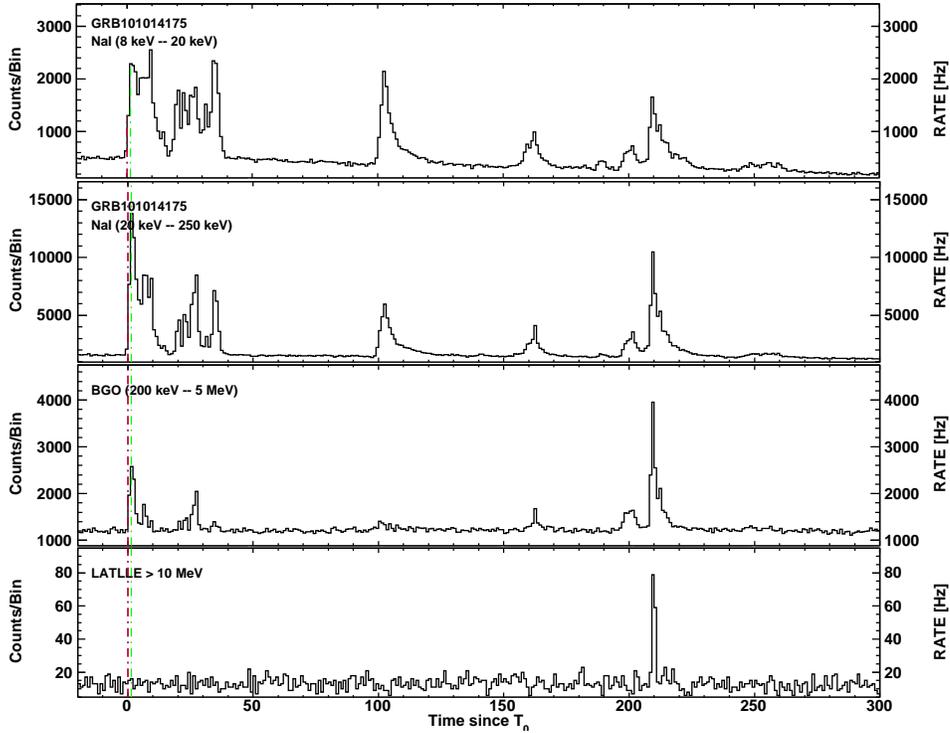}
\caption{Composite light curve for GRB\,101014A: summed GBM/NaI detectors (first two panels), GBM/BGO (third panel), LLE (fourth panel) and LAT Transient-class events above 100~MeV within a 12\de ROI  (last panel). See $\S$~\ref{sec_fermi_lat_grb_intro} for more information on lines and symbols in the LAT panels.}
\label{compo_101014A}
\end{center}
\end{figure}

The long GRB\,101014A triggered the GBM at T$_{0}$=04:11:52.62 UT on 14 October 2010 \citep[trigger 308722314,][]{2010GCN..11341...1T} and it has
  the longest GBM duration (T$_{90}$$\sim$450~s) in the catalog.
It had an initial off-axis angle of 54\de in the LAT and the ARR triggered by the GBM brought it down to $\sim$10\de after $\sim$200~s.
Because of the burst's proximity to the orbital pole, there was substantial contamination in the surrounding region owing to gamma-ray
emission from the Earth's limb \citep{2010GCN..11349...1T}. As a result, no LAT Transient-class events are left above 100~MeV after our selection
cuts ($\S$~\ref{subsection_cuts}).
We could thus not improve upon the GBM localization and no likelihood analysis was possible.
GRB\,101014A was detected in the LLE data only. Whereas the GBM light curve exhibits several emission episodes, the LLE light curve consists of 
a single, narrow pulse at $\sim$T$_{0}$+210~s (Fig.~\ref{compo_101014A}).

\FloatBarrier \subsection{GRB\,101123A}
\begin{figure}[ht!]
\begin{center}
\includegraphics[width=5.0in]{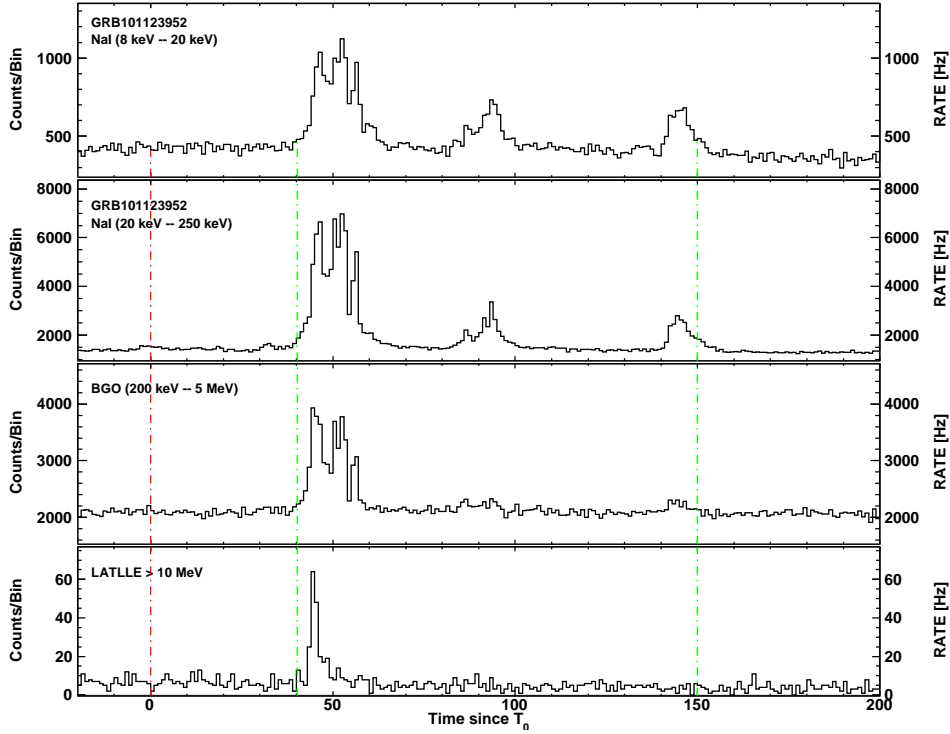}
\caption{Composite light curve for GRB\,101123A: summed GBM/NaI detectors (first two panels), GBM/BGO (third panel), LLE (fourth panel) and LAT Transient-class events above 100~MeV within a 12\de ROI  (last panel). See $\S$~\ref{sec_fermi_lat_grb_intro} for more information on lines and symbols in the LAT panels.}
\label{compo_101123A}
\end{center}
\end{figure}

The long GRB\,101123A triggered the GBM at T$_{0}$=22:51:34.97 UT on 23 November 2010 \citep[trigger 312245496,][]{2010GCN..11423...1G}.
It had an initial off-axis angle of 78\ded2 in the LAT and a large Zenith angle, thus no LAT Transient-class events are left above 100~MeV after our
selection cuts ($\S$~\ref{subsection_cuts}).
We could thus not improve upon the GBM localization and no likelihood analysis was possible.
GRB\,101123A was detected in the LLE data only. The LLE light curve consists of a single, narrow pulse at $\sim$T$_{0}$+45~s, in temporal coincidence
with the first pulse of the first bright emission episode observed in the GBM light curve (Fig.~\ref{compo_101123A})
The burst is relatively bright in the GBM and its time-integrated spectrum during the GBM T$_{90}$ is best represented by a Band function, with a hard
value for the high-energy spectral slope $\beta=-2.04\pm0.03$.

\FloatBarrier \subsection{GRB\,110120A}
\begin{figure}[ht!]
\begin{center}
\includegraphics[width=5.0in]{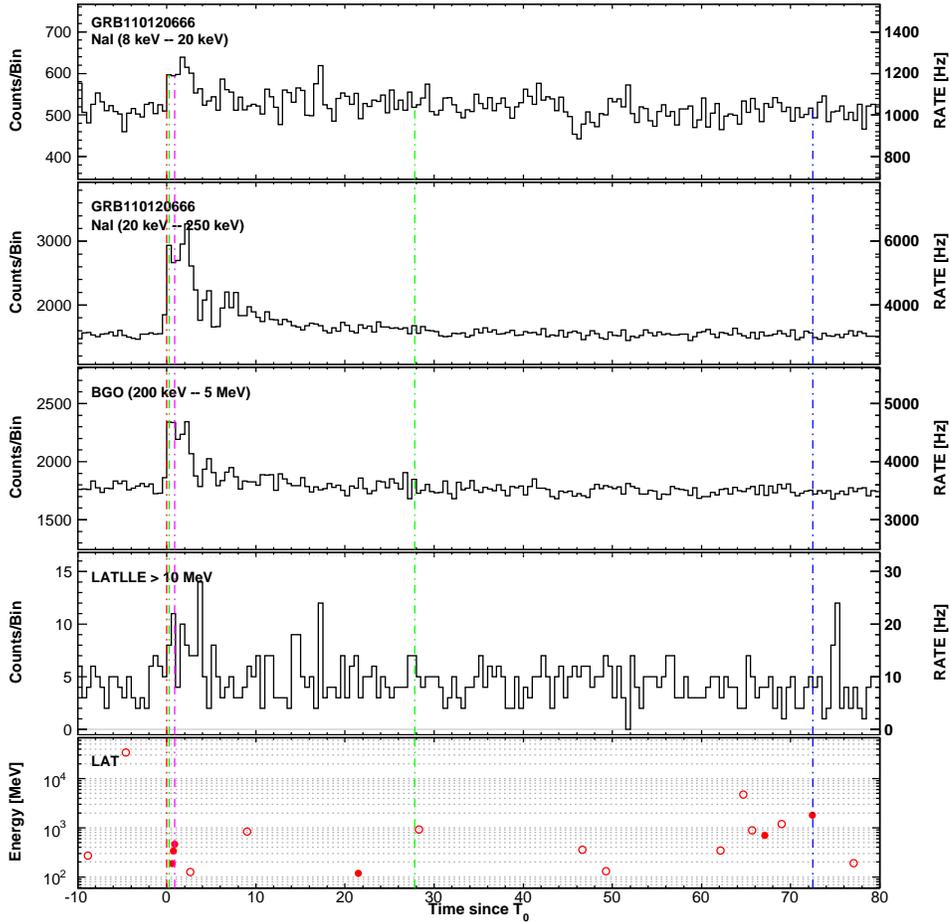}
\caption{Composite light curve for GRB\,110120A: summed GBM/NaI detectors (first two panels), GBM/BGO (third panel), LLE (fourth panel) and LAT Transient-class events above 100~MeV within a 12\de ROI  (last panel). See $\S$~\ref{sec_fermi_lat_grb_intro} for more information on lines and symbols in the LAT panels.}
\label{compo_110120A}
\end{center}
\end{figure}

\begin{figure}[ht!]
\begin{center}
\includegraphics[width=2.2in]{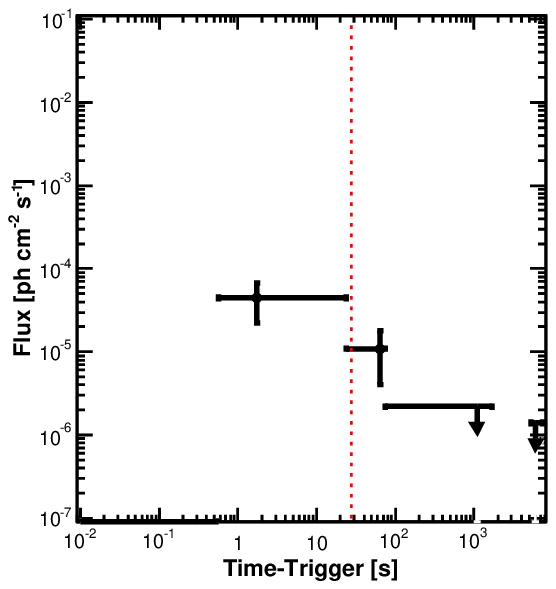}
\includegraphics[width=2.2in]{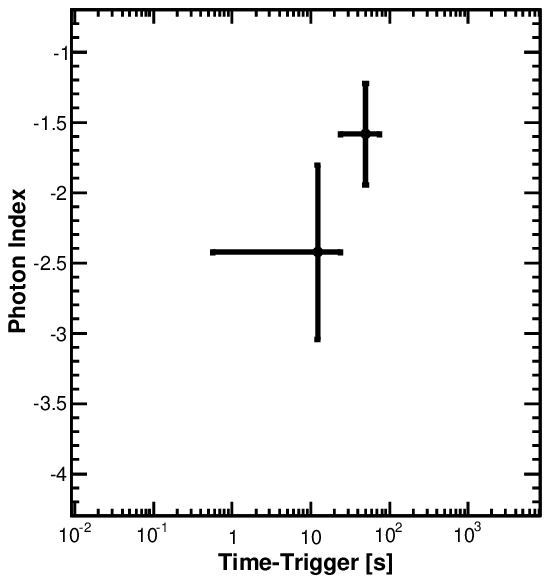}
\caption{Likelihood light curve for GRB\,110120A (flux above 100~MeV on the left, photon index on the right). See $\S$~\ref{sec_fermi_lat_grb_intro} for more information on lines and symbols.}
\label{like_110120A}
\end{center}
\end{figure}

The long GRB\,110120A triggered the GBM at T$_{0}$=15:59:39.23 UT on 20 January 2011 \citep[trigger 317231981,][]{2011GCN..11591...1L}.
In spite of an initial off-axis angle of 13\ded6, GRB\,110120A was relatively faint in the LAT.
The LAT preliminary localization was delivered via GCN \citep{2011GCN..11597...1O}, with a statistical error of 0\ded4.
The GBM light curve of GRB\,110120A consists of two overlapping pulses (Fig.~\ref{compo_110120A}).
The LLE light curve shows a small signal excess which coincides with the GBM emission, but this excess was not significant enough to claim an LLE detection
(see Table~\ref{tab_GRBs}).
Our analysis of the LAT Transient-class data above 100~MeV provided a LAT T$_{95}$=113$^{+21}_{-30}$~s which indicates the temporal extension of the
burst emission in the LAT. In addition, a 1.82~GeV event is recorded at T$_{0}$+72.46~s.
The LAT time-resolved likelihood analysis returned a significant flux in two time bins only, up to T$_{0}$+75~s (Fig.~\ref{like_110120A}).

\FloatBarrier \subsection{GRB\,110328B}
\begin{figure}[ht!]
\begin{center}
\includegraphics[width=5.0in]{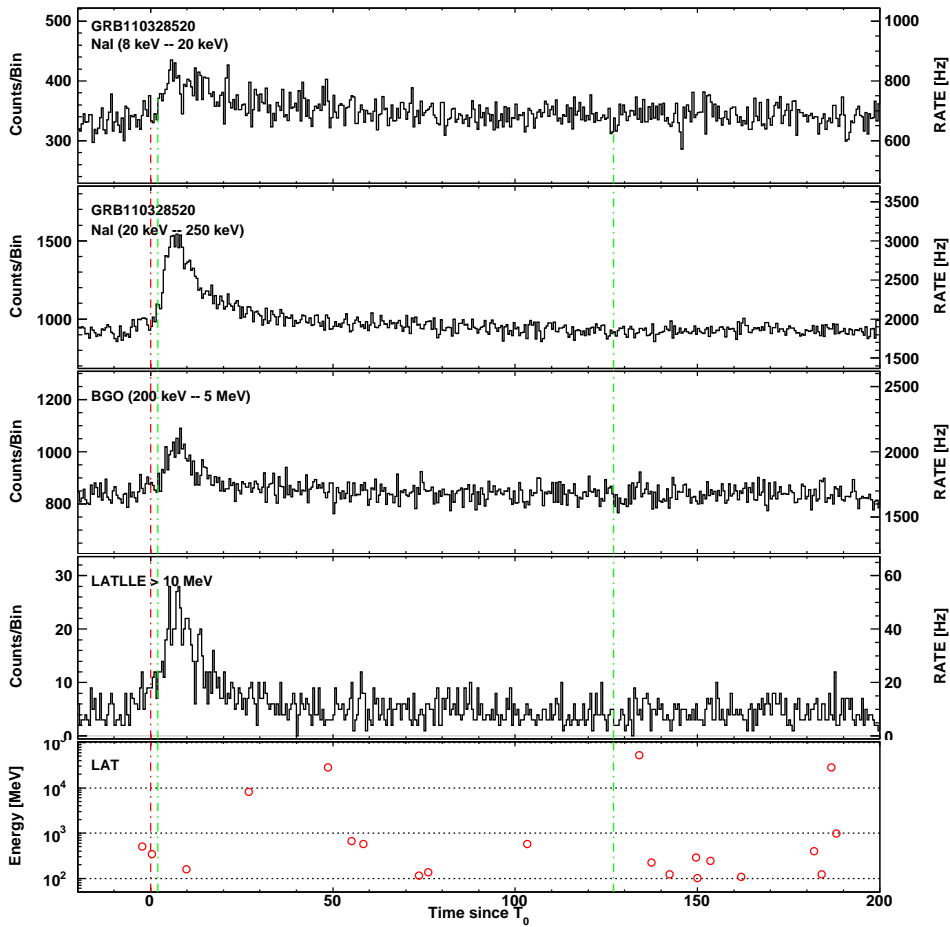}
\caption{Composite light curve for GRB\,110328B: summed GBM/NaI detectors (first two panels), GBM/BGO (third panel), LLE (fourth panel) and LAT Transient-class events above 100~MeV within a 12\de ROI  (last panel). See $\S$~\ref{sec_fermi_lat_grb_intro} for more information on lines and symbols in the LAT panels.}
\label{compo_110328B}
\end{center}
\end{figure}

The long GRB\,101123A triggered the GBM at T$_{0}$=12:29:19.19 UT on 28 March 2011 \citep[trigger 323008161,][]{GCN11831}.
Only a few LAT Transient-class events above 100~MeV are compatible with the burst position, therefore no LAT T$_{90}$ could be derived and no
significant emission was found in the likelihood analysis.
Using a lower energy threshold of 50~MeV, a tentative localization with the LAT was delivered via GCN \citep{GCN11835}, compatible with the GBM
localization and with a statistical error of 1\ded7.
GRB\,110328B was detected in the LLE data only.
The LLE light curve consists of a single pulse which starts approximately at the time of the
GBM trigger and which mimics the light curve seen in the NaI and BGO detectors (Fig.~\ref{compo_110328B}).

\FloatBarrier \subsection{GRB\,110428A}
The long GRB\,110428A triggered the GBM at T$_{0}$=09:18:30.41 UT on 28 April 2011 \citep[trigger 325675112,][]{2011GCN..12012...1T}.
It had an initial off-axis angle of 34\ded6 in the LAT and the ARR triggered by the GBM brought it down to $\sim$5\de after $\sim$200~s.
The LAT preliminary localization was delivered via GCN \citep{2011GCN..11982...1V}, with a statistical error of 0\ded15.
\Swift TOO observations started $\sim$55.6~ks after the trigger time and a possible X-ray counterpart was found by \Swift-XRT
\citep{2011GCN..11984...1M}. Further observations confirmed the existence of a fading source \citep{2011GCN..11989...1M}.

\begin{figure}[ht!]
\begin{center}
\includegraphics[width=5.0in]{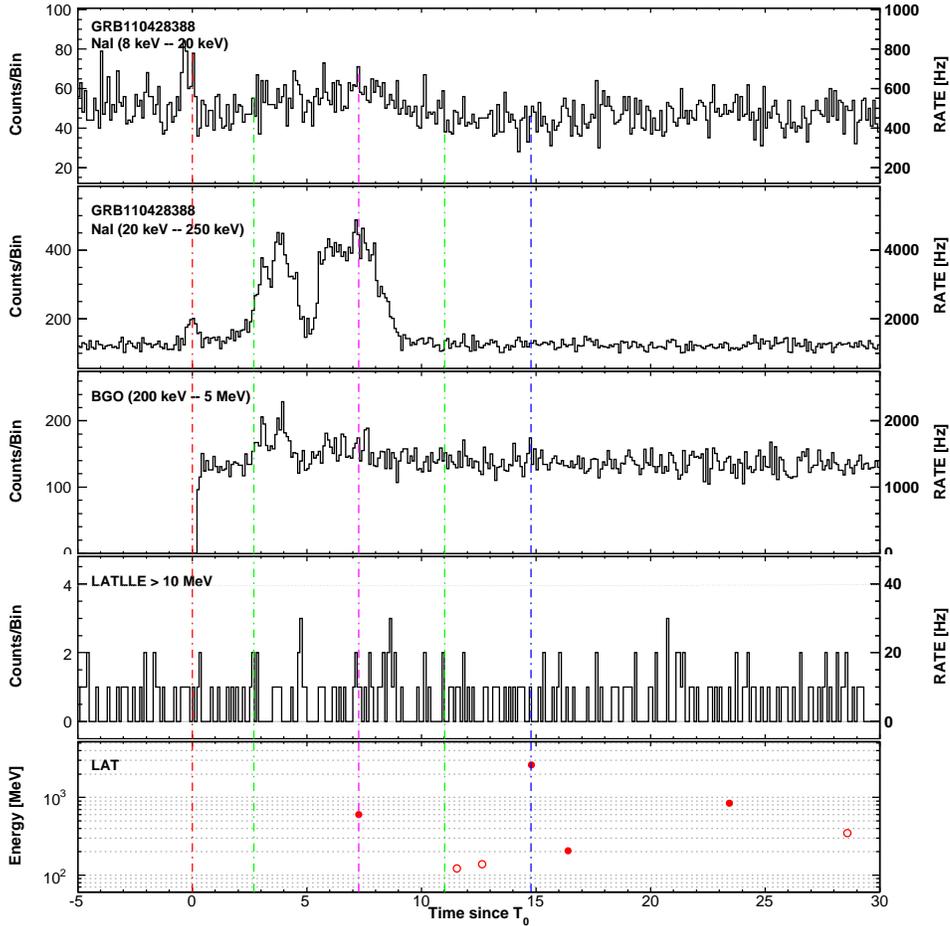}
\caption{Composite light curve for GRB\,110428A: summed GBM/NaI detectors (first two panels), GBM/BGO (third panel), LLE (fourth panel) and LAT Transient-class events above 100~MeV within a 12\de ROI  (last panel). See $\S$~\ref{sec_fermi_lat_grb_intro} for more information on lines and symbols in the LAT panels.}
\label{compo_110428A}
\end{center}
\end{figure}

\begin{figure}[ht!]
\begin{center}
\includegraphics[width=2.2in]{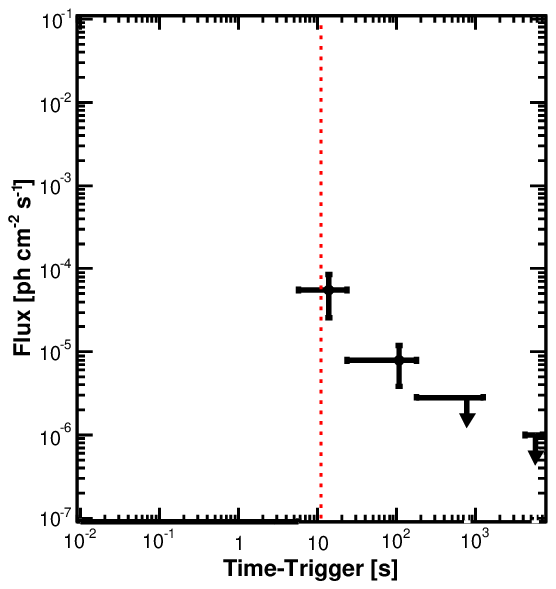}
\includegraphics[width=2.2in]{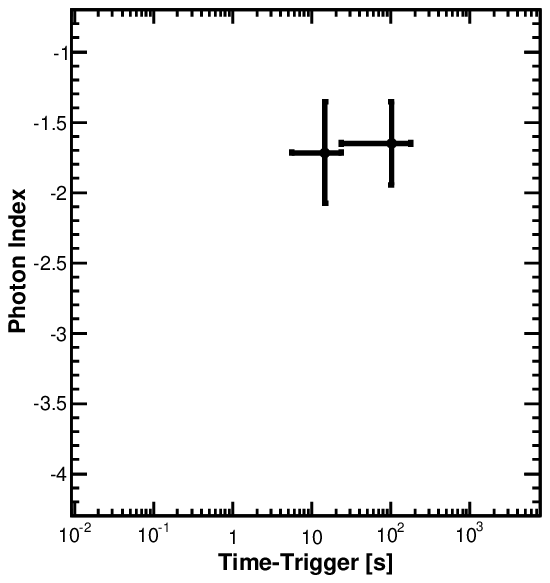}
\caption{Likelihood light curve for GRB\,110428A (flux above 100~MeV on the left, photon index on the right). See $\S$~\ref{sec_fermi_lat_grb_intro} for more information on lines and symbols.}
\label{like_110428A}
\end{center}
\end{figure}

The GBM light curve of GRB\,110428A consists of several overlapping pulses (Fig.~\ref{compo_110428A}).
No significant emission was detected in the LLE light curve. The highest-energy event (2.62~GeV) is detected at
T$_{0}$+14.79~s and does not coincide with any noticeable feature in the GBM light curve.
Although the LAT T$_{95}$=408$^{+93}_{-336}$~s suffers from a large uncertainty due to the relatively small statistics ($\sim$16 events),
the burst was detected up to this time with high significance by the LAT likelihood analysis of the Transient-class data above 100~MeV.
The LAT time-resolved likelihood analysis returned a significant flux in two time bins only, up to T$_{0}$+178~s (Fig.~\ref{like_110428A}).
More interestingly, the time-integrated spectrum of GRB\,110428A during the GBM T$_{90}$ is best represented by a Band function, with a steep value for
the high-energy spectral slope $\beta=-2.90\pm0.10$.
This value is very different from the hard photon index of -1.73$\pm$0.20 which is found by the likelihood
analysis at late times (Fig.~\ref{like_110428A}). In the catalog, GRB\,110428A is thus among the bursts which show the strongest spectral evolution
between the prompt and late emission phases.


\FloatBarrier \subsection{GRB\,110529A}
\begin{figure}[ht!]
\begin{center}
\includegraphics[width=5.0in]{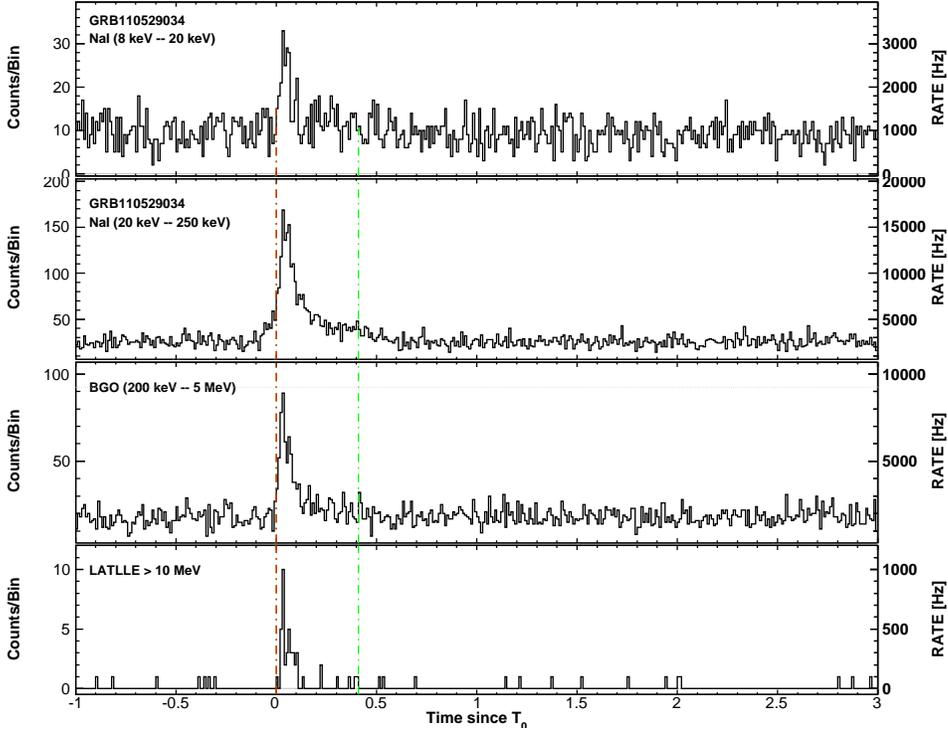}
\caption{Composite light curve for GRB\,110529A: summed GBM/NaI detectors (first two panels), GBM/BGO (third panel), LLE (fourth panel) and LAT Transient-class events above 100~MeV within a 12\de ROI  (last panel). See $\S$~\ref{sec_fermi_lat_grb_intro} for more information on lines and symbols in the LAT panels.}
\label{compo_110529A}
\end{center}
\end{figure}

The short GRB\,110529A triggered the GBM at T$_{0}$=00:48:42.87 UT on 29 May 2011 \citep[trigger 328322924,][]{2011GCN..12047...1B}.
Only a few LAT Transient-class events above 100~MeV are compatible with the burst position, therefore no significant emission was found in the
likelihood analysis. 
The burst was detected in the LLE data only \citep{2011GCN..12044...1M}, and the light curve consists of a short spike in coincident with the GBM
emission (Fig.~\ref{compo_110529A}).


\FloatBarrier \subsection{GRB\,110625A}
\begin{figure}[ht!]
\begin{center}
\includegraphics[width=5.0in]{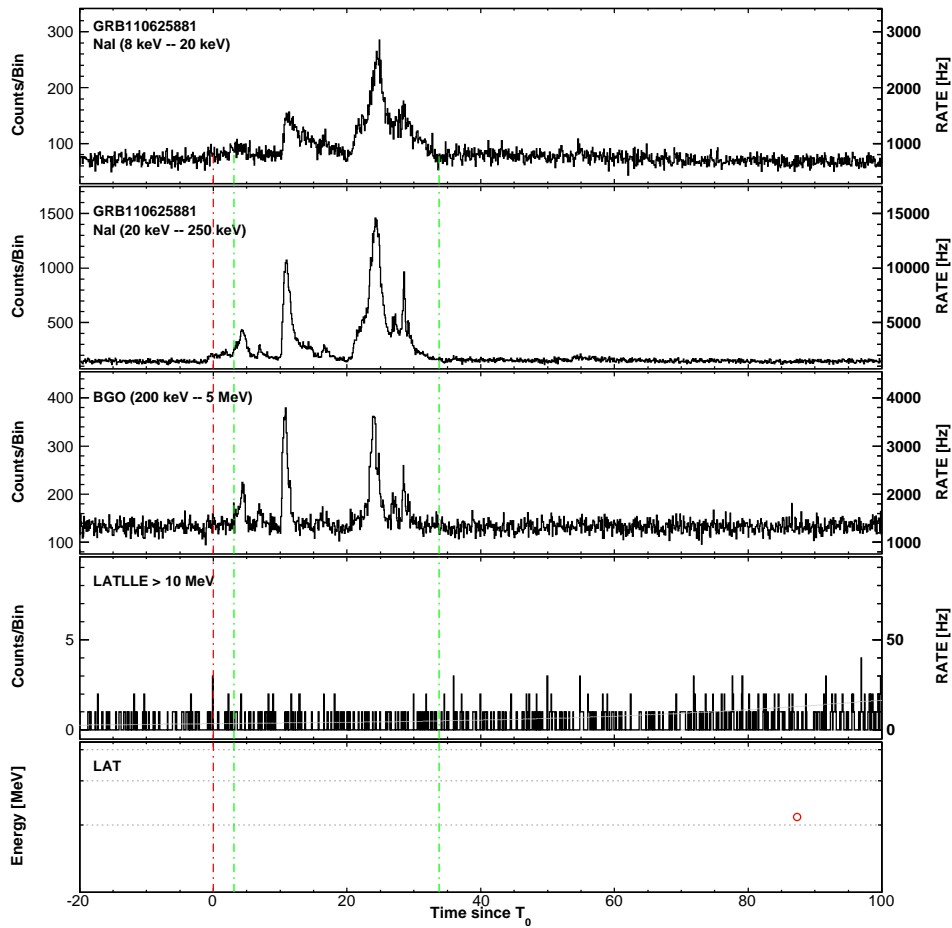}
\caption{Composite light curve for GRB\,110625A: summed GBM/NaI detectors (first two panels), GBM/BGO (third panel), LLE (fourth panel) and LAT Transient-class events above 100~MeV within a 12\de ROI  (last panel). See $\S$~\ref{sec_fermi_lat_grb_intro} for more information on lines and symbols in the LAT panels.}
\label{compo_110625A}
\end{center}
\end{figure}

\begin{figure}[ht!]
\begin{center}
\includegraphics[width=2.2in]{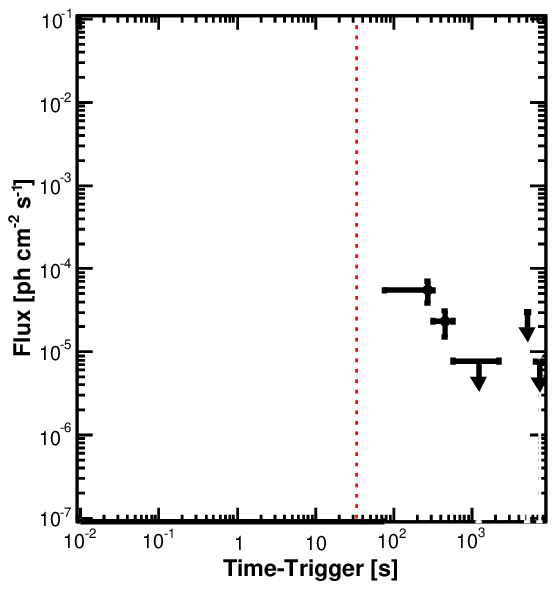}
\includegraphics[width=2.2in]{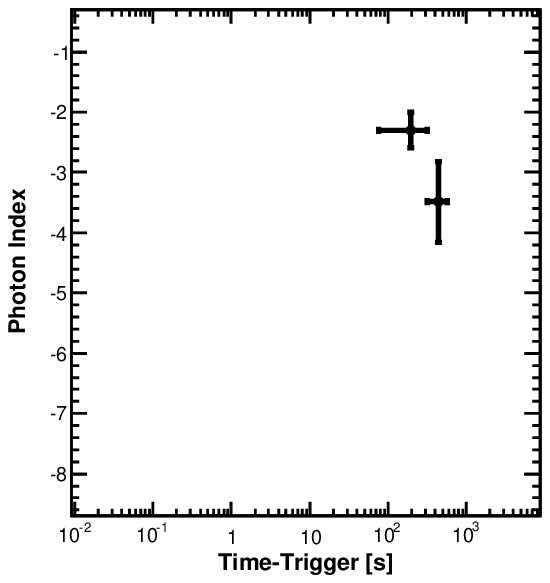}
\caption{Likelihood light curve for GRB\,110625A (flux above 100~MeV on the left, photon index on the right). See $\S$~\ref{sec_fermi_lat_grb_intro} for more information on lines and symbols.}
\label{like_110625A}
\end{center}
\end{figure}

The long GRB\,110625A triggered the GBM at T$_{0}$=21:08:18.24 UT on 25 June 2011 \citep[trigger 330728900,][]{2011GCN..12100...1G} and the
  \Swift-BAT at 21:08:28 UT \citep{2011GCN..12088...1P}.
\Swift-XRT observations started 140.3~s after the BAT trigger and a bright, fading and uncataloged X-ray source was immediately located
\citep{2011GCNR..336....1P}.
Further analysis refined the position of the X-ray source \citep{2011GCN..12091...1P,2011GCN..12092...1P}, enabling optical follow-up observations
\citep{2011GCN..12094...1K,2011GCN..12095...1I,2011GCN..12096...1F,2011GCN..12098...1G,2011GCN..12099...1H,2011GCN..12113...1G}, but no redshift could
be measured.
GRB\,110625A was bright enough to trigger an ARR of the \Fermi spacecraft. However, its initial off-axis angle of 87\ded9 in the LAT resulted in a very
poor photon statistics above 100~MeV (Fig.~\ref{compo_110625A}) and no LAT T$_{90}$ could be derived.
In addition, the \Fermi spacecraft continued its maneuver toward the GBM flight software reconstructed position, which was off by 68\de from the
enhanced \Swift-XRT position \citep{2011GCN..12092...1P}, providing non-optimal exposure for LAT follow-up observations.
Accumulating signal in the LAT time-resolved likelihood analysis allowed us to detect a significant flux in two time bins, up to T$_{0}$+562~s
(\citet{2011GCN..12100...1G} and Fig.~\ref{like_110625A}), confirming the earlier detection by \citet{2011GCN..12097...1T}. The highest-energy event
(2.42~GeV) is detected at T$_{0}$+272.44~s.

\FloatBarrier \subsection{GRB\,110709A}
\begin{figure}[ht!]
\begin{center}
\includegraphics[width=5.0in]{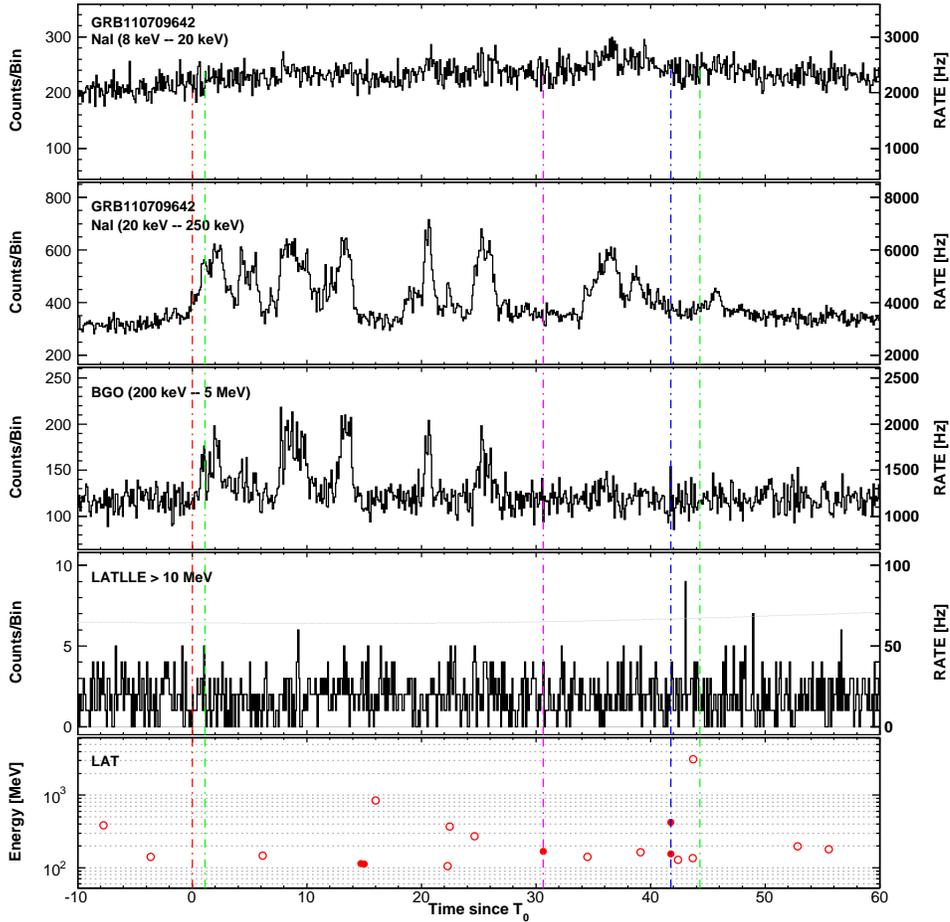}
\caption{Composite light curve for GRB\,110709A: summed GBM/NaI detectors (first two panels), GBM/BGO (third panel), LLE (fourth panel) and LAT Transient-class events above 100~MeV within a 12\de ROI  (last panel). See $\S$~\ref{sec_fermi_lat_grb_intro} for more information on lines and symbols in the LAT panels.}
\label{compo_110709A}
\end{center}
\end{figure}

\begin{figure}[ht!]
\begin{center}
\includegraphics[width=2.2in]{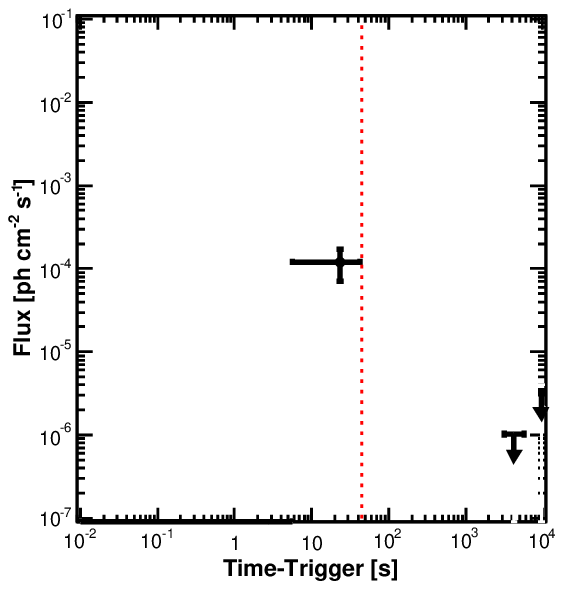}
\includegraphics[width=2.2in]{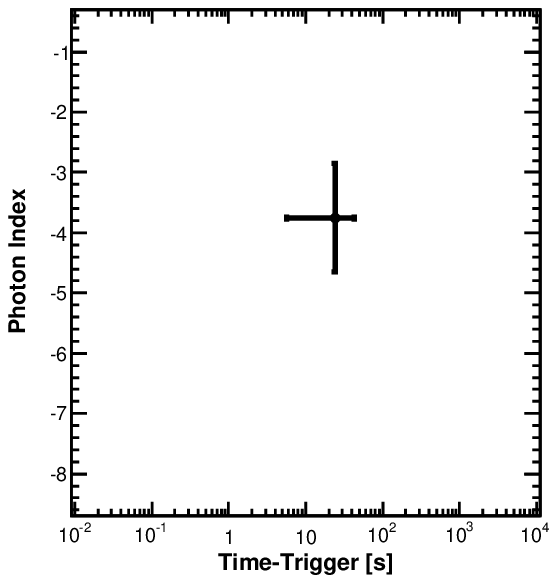}
\caption{Likelihood light curve for GRB\,110709A (flux above 100~MeV on the left, photon index on the right). See $\S$~\ref{sec_fermi_lat_grb_intro} for more information on lines and symbols.}
\label{like_110709A}
\end{center}
\end{figure}

The long GRB\,110709A triggered the GBM at T$_{0}$=15:24:27.37 UT on 09 July 2011 \citep[trigger 331917869,][]{2011GCN..12133...1C}  and the
  \Swift-BAT at 15:24:29 UT \citep{2011GCN..12118...1H}.
\Swift-XRT observations started 65.6~s after the BAT trigger and a bright, uncataloged X-ray source was immediately located \citep{2011GCNR..339....1H}.
Further analysis refined the position of the X-ray source \citep{2011GCN..12119...1E,2011GCN..12123...1O}.
In spite of numerous follow-up observations
\citep{2011GCN..12120...1I,2011GCN..12125...1X,2011GCN..12134...1T,2011GCN..12139...1K,2011GCN..12146...1K,2011GCN..12148...1H}, no optical afterglow
was detected.
GRB\,110709A was bright enough to trigger an ARR of the \Fermi spacecraft. However, its initial off-axis angle of 53\ded4 in the LAT resulted in a very
poor photon statistics above 100~MeV (Fig.~\ref{compo_110709A}).
The LAT time-resolved likelihood analysis returned a significant flux in one time bin only, ending at T$_{0}$+42~s (Fig.~\ref{like_110709A}).
In addition, no LAT T$_{90}$ could be derived due to the large Zenith angle of the burst.



\FloatBarrier \subsection{GRB\,110721A}
The long GRB\,110721A triggered the GBM at T$_{0}$=04:47:43.75 UT on 21 July 2011 \citep[trigger 332916465,][]{2011GCN..12187...1T}.
It had an initial off-axis angle of 40\ded3 in the LAT and the ARR triggered by the GBM brought it down to $\sim$10\de after $\sim$240~s.
The LAT preliminary localization was delivered via GCN \citep{2011GCN..12188...1V}, with a statistical error of 0\ded51.
A low-significance faint candidate afterglow was found by \citet{2011GCN..12192...1G} analyzing the \Swift-XRT data and GROND data.
Using the GMOS spectrograph mounted on the Gemini South Telescope, \citet{2011GCN..12193...1B} found two clear absorption features at 5487 and 5436
\AA, matching CaII H\&K at a redshift of z=0.382, with a significant decline in flux at shorter wavelengths, but to a non-zero level.
However, the triangulation of the burst by the IPN provided a position with a 3$\sigma$ error box area of 2250 square arc-minutes, 
excluding the position of the candidate afterglow \citep{2011GCN..12195...1H}.
Moreover, further observations with \Swift-XRT did not confirm the afterglow detection \citep{2011GCN..12212...1G} and radio observations with the
Expanded Very Large Array (EVLA) suggested that the X-ray candidate was instead associated with the radio-loud AGN PKS 2211-388 \citep{2011GCN..12245...1C}.
As a result, we used the IPN position in our analysis and we did not assume any redshift for this burst.

\begin{figure}[ht!]
\begin{center}
\includegraphics[width=5.0in]{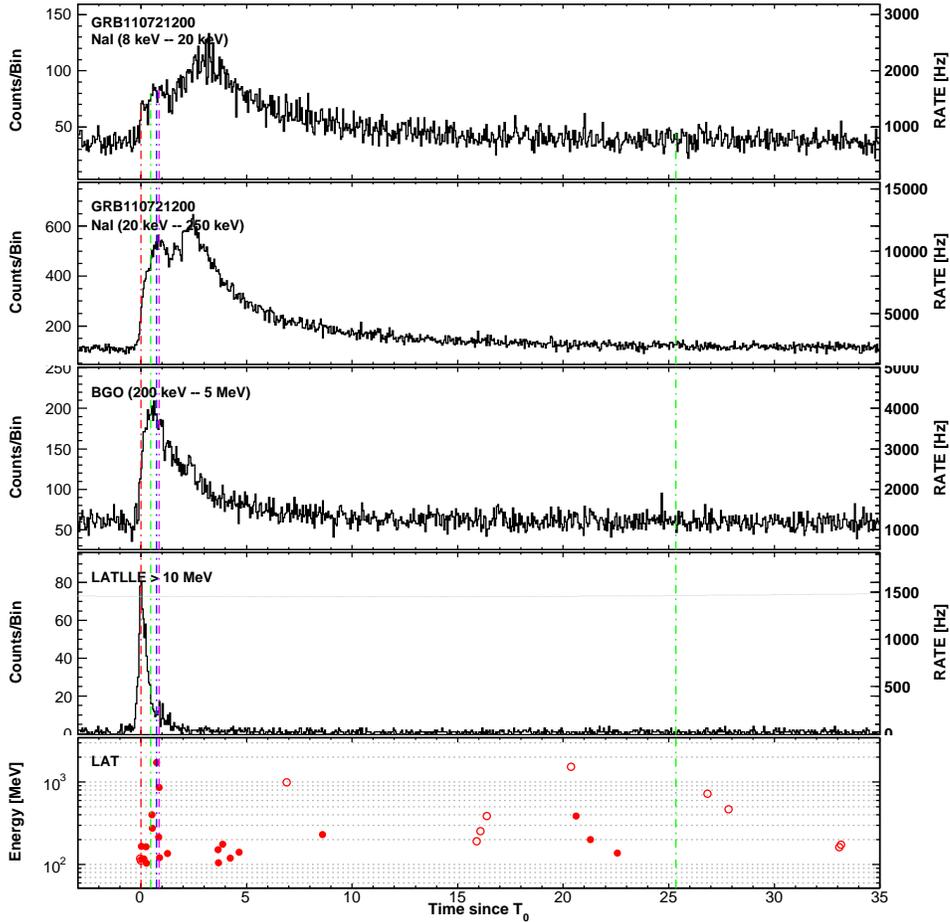}
\caption{Composite light curve for GRB\,110721A: summed GBM/NaI detectors (first two panels), GBM/BGO (third panel), LLE (fourth panel) and LAT Transient-class events above 100~MeV within a 12\de ROI  (last panel). See $\S$~\ref{sec_fermi_lat_grb_intro} for more information on lines and symbols in the LAT panels.}
\label{compo_110721A}
\end{center}
\end{figure}

\begin{figure}[ht!]
\begin{center}
\includegraphics[width=2.2in]{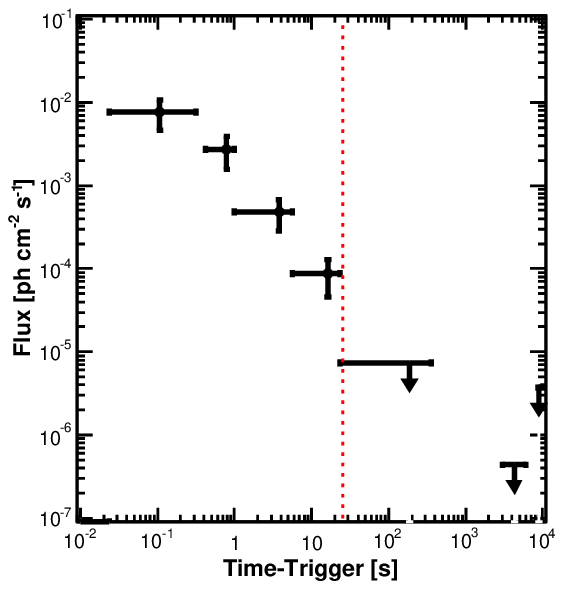}
\includegraphics[width=2.2in]{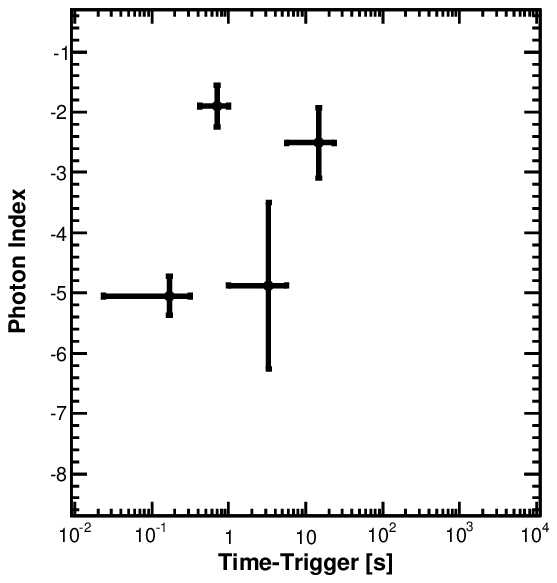}
\caption{Likelihood light curve for GRB\,110721A (flux above 100~MeV on the left, photon index on the right). See $\S$~\ref{sec_fermi_lat_grb_intro} for more information on lines and symbols.}
\label{like_110721A}
\end{center}
\end{figure}

A dedicated analysis of the prompt emission spectrum of GRB\,110721A is presented in \cite{Axelsson+12}.
The NaI light curve of GRB\,110721A consists of two overlapping pulses. Whereas only the first pulse is visible in the BGO and
  LLE light curves, the second pulse is much softer and is detected down to 8--20~keV (Fig.~\ref{compo_110721A}).
The LLE pulse starts and peaks earlier than the GBM emission. It appears narrower and the highest-energy event (1.73~GeV) is detected at T$_{0}$+0.74~s.
However, the LAT emission above 100~MeV could last longer, potentially up to T$_{0}$+239~s or later. Due to the large Zenith angle of the burst after this
time and to the paucity of events after the end of the GBM emission, we could not perform a good measurement of the LAT T$_{95}$ though.
The LAT time-resolved likelihood analysis actually returned a significant signal up to T$_{0}$+24~s only (Fig.~\ref{like_110721A}).



\FloatBarrier \subsection{GRB\,110731A}
The long GRB\,110731A triggered the GBM at T$_{0}$=11:09:29.94 UT on 31 July 2011 \citep[trigger 333803371,][]{2011GCN..12221...1G} and the
  \Swift-BAT at 11:09:30 UT \citep{2011GCN..12215...1O}.
The LAT preliminary localization was delivered via GCN \citep{2011GCN..12218...1B}, with a statistical error of 0\ded2.
GRB\,110731A was bright enough to trigger an ARR of the \Fermi spacecraft. Its initial off-axis angle was 3\ded4 in the LAT, thus the repointing had
little impact on the prompt emission phase observations and permitted excellent observations of the extended emission for 2.5~hours after the trigger time.
High quality continuous observations of the burst are available until the first \Fermi passage into the SAA at
$\sim$T$_{0}$+1400~s. The ARR continued for another 90 minutes after \Fermi had exited the SAA.
\Swift-XRT observations started 56~s after the BAT trigger \citep{2011GCNR..343....1O}. A bright, uncataloged X-ray source was found and \Swift-UVOT
detected a bright afterglow candidate consistent with the XRT localization \citep{2011GCN..12215...1O}.
Further analyses refined the position of the burst \citep{2011GCN..12217...1K,2011GCN..12219...1B} and further observations confirmed the
existence of a fading X-ray \citep{2011GCN..12224...1L} and optical afterglow \citep{2011GCN..12222...1O,2011GCN..12242...1T}.
\citet{2011GCN..12225...1T} reported a spectroscopic redshift of z=2.83 based on observations of the optical afterglow using the GMOS
  spectrograph mounted on the Gemini North Telescope.
After a weather-induced delay, GROND detected GRB\,110731A at a mean time of 2.74 days after the trigger time.
A dedicated analysis of the near-infrared to GeV observations of GRB\,110731A in its prompt and afterglow phases using data from \Fermi, \Swift, MOA and
GROND is presented in \citet{GRB110731A:Fermi}.

\begin{figure}[ht!]
\begin{center}
\includegraphics[width=5.0in]{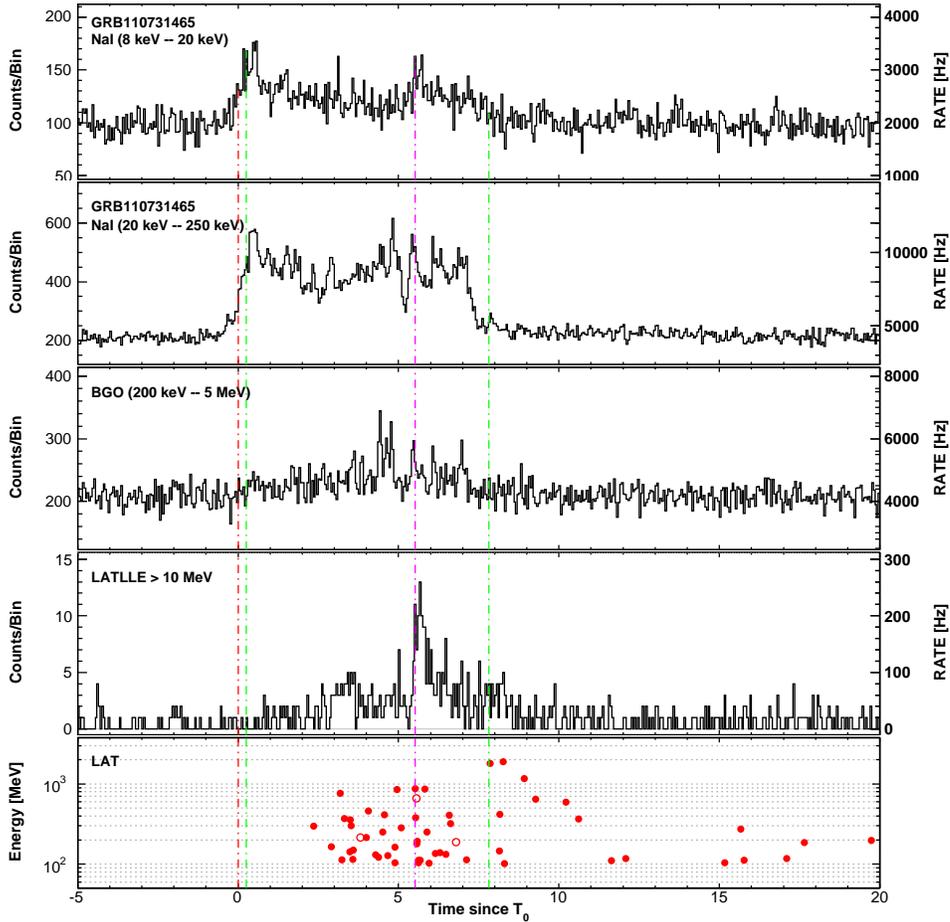}
\caption{Composite light curve for GRB\,110731A: summed GBM/NaI detectors (first two panels), GBM/BGO (third panel), LLE (fourth panel) and LAT Transient-class events above 100~MeV within a 12\de ROI  (last panel). See $\S$~\ref{sec_fermi_lat_grb_intro} for more information on lines and symbols in the LAT panels.}
\label{compo_110731A}
\end{center}
\end{figure}

\begin{figure}[ht!]
\begin{center}
\includegraphics[width=2.2in]{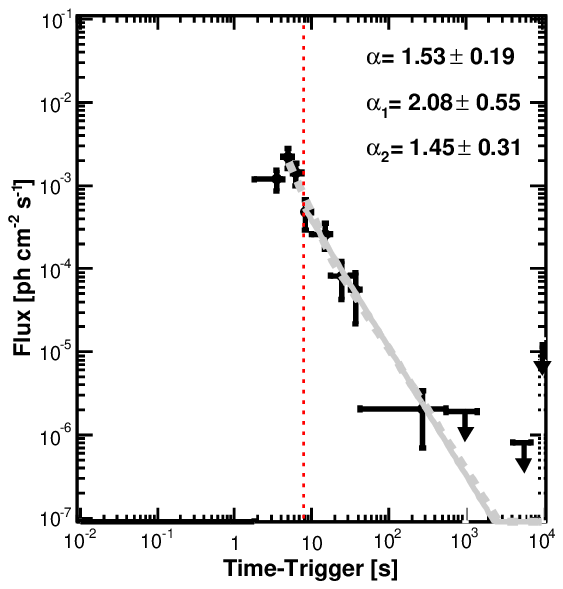}
\includegraphics[width=2.2in]{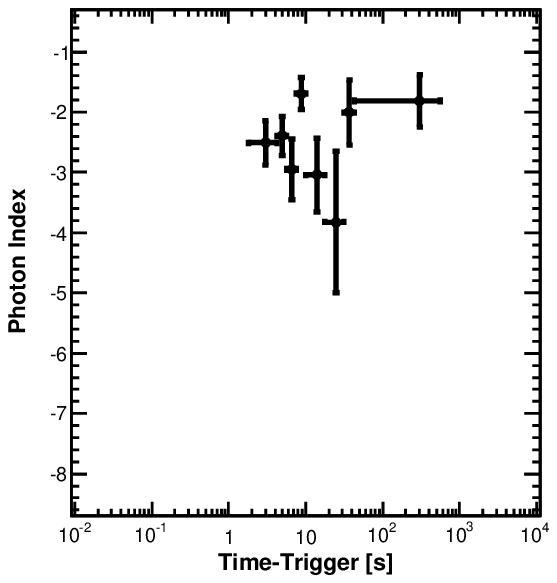}
\caption{Likelihood light curve for GRB\,110731A (flux above 100~MeV on the left, photon index on the right). See $\S$~\ref{sec_fermi_lat_grb_intro} for more information on lines and symbols.}
\label{like_110731A}
\end{center}
\end{figure}

The high-energy emission of GRB\,110731A lasts much longer than the GBM estimated duration.
A 1.90~GeV event is detected at T$_{0}$+8.27~s, right after the end of the GBM emission (Fig.~\ref{compo_110731A}).
The LAT time-resolved likelihood analysis resulted in a well sampled light curve of the high-energy flux up to $\sim$562~s (Fig.~\ref{like_110731A}).
The decay of the flux as a function of time follows a simple power law starting from the GBM T$_{95}$, with a decay index
  $\alpha$=1.53$\pm$0.19, in agreement with the result reported by \citet{GRB110731A:Fermi}.
This relatively steep decay is similar to the first part of the decay observed in GRBs~090510, 090902B and 090926A (Table~\ref{tab_extended}) for which a
  significant break was found in the flux light curve. This suggests that GRB\,110731A was observed during the transition from the prompt phase to the
  afterglow phase as discussed in $\S$~\ref{sub_interpretation_extended}.
Moreover, the time-integrated spectrum of GRB\,110731A is best represented by a Band function with an additional power-law component. As discussed in
$\S$~\ref{subsub_extracomponents}, the detection of this additional component is marginal in the ``GBM'' time interval but significant in the other
time interval (Tables~\ref{tab_JointFitGBMT90} and~\ref{tab_JointFitLATPROMPT}), in agreement with \citet{GRB110731A:Fermi}.



\end{document}